\pdfoutput=1
\documentclass[11pt,twoside,a4paper,cmspaper,final,collab]{cms-tdr}
\def\svnVersion{d8c7fe1}\def\svnDate{2024/11/15}\def\cmsCernNoTag{CERN-EP-2024-062}\def\cmsCernDate{\today}\def\cmsMessage{Published in Physics Reports as \href{http://dx.doi.org/10.1016/j.physrep.2024.09.004}{\doi{10.1016/j.physrep.2024.09.004}.}}
\begin{document}\cmsNoteHeader{B2G-23-002}

\newlength\cmsFigWidth
\ifthenelse{\boolean{cms@external}}{\setlength\cmsFigWidth{0.45\textwidth}}{\setlength\cmsFigWidth{0.49\textwidth}}
\providecommand{\cmsLeft}{left\xspace}
\providecommand{\cmsRight}{right\xspace}
\newlength\cmsTabSkip\setlength{\cmsTabSkip}{2ex}
\providecommand{\cmsTable}[1]{\resizebox{\textwidth}{!}{#1}}

\newcommand{\pp}{\ensuremath{\Pp\Pp}\xspace}
\newcommand{\PY}{\ensuremath{{\HepParticle{Y}{}{}}}\xspace}
\newcommand{\PA}{\ensuremath{{\HepParticle{A}{}{}}}\xspace}
\newcommand{\PR}{{\HepParticle{R}{}{}}\Xspace}
\newcommand{\PS}{{\HepParticle{S}{}{}}\Xspace}
\newcommand{\tautau}{\ensuremath{\PGt\PGt}\xspace}
\newcommand{\etauh}{\ensuremath{\Pe\tauh}\xspace}
\newcommand{\mutauh}{\ensuremath{\PGm\tauh}\xspace}
\newcommand{\tauhtauh}{\ensuremath{\tauh\tauh}\xspace}
\newcommand{\bb}{\ensuremath{\PQb\PQb}\xspace}
\newcommand{\cc}{\ensuremath{\PQc\PQc}\xspace}
\newcommand{\qq}{\ensuremath{\PQq\PQq}\xspace}
\newcommand{\bbtautau}{\ensuremath{\PQb\PQb\PGt\PGt}\xspace}
\newcommand{\Irel}{\ensuremath{I_{\text{rel}}}\xspace}
\newcommand{\ptem}{\ensuremath{\pt^{\Pell}}\xspace}
\newcommand{\sumCharged}{\ensuremath{\sum\pt^{\text{charged}}}\xspace}
\newcommand{\sumNeutral}{\ensuremath{\sum\et^{\text{neutral}}}\xspace}
\newcommand{\sumGamma}{\ensuremath{\sum\et^{\gamma}}\xspace}
\newcommand{\widthGamma}{\ensuremath{\Gamma_{\PX}}\xspace}
\newcommand{\widthRatio}{\ensuremath{\Gamma_{\PX}/m_{\PX}}\xspace}
\newcommand{\couplingLambda}{\ensuremath{\lambda_{\PH\PH\PX}}\xspace}
\newcommand{\kappaLambda}{\ensuremath{k_{\lambda}}\xspace}
\newcommand{\Rint}{\ensuremath{R_{\text{int}}}\xspace}
\newcommand{\ptPU}{\ensuremath{\pt^{\text{PU}}}\xspace}
\newcommand{\yDT}{\ensuremath{y^{\text{DT}}}\xspace}
\newcommand{\Dj}{\ensuremath{D_{\text{jet}}}\xspace}
\newcommand{\De}{\ensuremath{D_{\Pe}}\xspace}
\newcommand{\Dm}{\ensuremath{D_{\PGm}}\xspace}
\newcommand{\mj}{\ensuremath{m_{\mathrm{SD}}}\xspace}
\newcommand{\tauDDT}{\ensuremath{\tau_{21}^{\mathrm{DDT}}}\xspace}
\newcommand{\mjj}{\ensuremath{m_{\mathrm{jj}}}\xspace}
\newcommand{\nn}{\ensuremath{\PGn\PGn}\xspace}
\newcommand{\lep}{\ensuremath{\Pell\Pell}\xspace}
\newcommand{\mll}{\ensuremath{m_{\lep}}\xspace}
\newcommand{\Wjets}{\PW{}+jets\xspace}
\newcommand{\Zjets}{\PZ{}+jets\xspace}
\newcommand{\Gjets}{\Pgg{}+jets\xspace}
\newcommand{\GGjets}{\GamGam{}+jets\xspace}
\newcommand{\Ht}{\ensuremath{H_{\mathrm{T}}}\xspace}
\newcommand{\Htmiss}{\ensuremath{\Ht^{\text{miss}}}\xspace}
\newcommand{\Wln}{\ensuremath{\PW(\Pell\PGn)}\xspace}
\newcommand{\Zll}{\ensuremath{\PZ(\lep)}\xspace}
\newcommand{\Znn}{\ensuremath{\PZ(\nn)}\xspace}
\newcommand{\Vqq}{\ensuremath{\PV(\PQq\PQq)}\xspace}
\newcommand{\Wqq}{\ensuremath{\PW(\PQq\PQq)}\xspace}
\newcommand{\mjjAKEight}{\ensuremath{\mjj^{\mathrm{AK8}}}\xspace}
\newcommand{\mgg}{\ensuremath{m_{\GamGam}}\xspace}
\newcommand{\mggjj}{\ensuremath{m_{\GamGam\mathrm{jj}}}\xspace}
\newcommand{\mXtilde}{\ensuremath{\tilde{m}_{\PX}}\xspace}
\newcommand{\Aboson}{{\HepParticle{A}{}{}}\Xspace}
\newcommand{\cosba}{\ensuremath{\cos(\beta-\alpha)}\xspace}
\newcommand{\mZH}{\ensuremath{m_{\PZ\PH}}\xspace}
\newcommand{\mtZH}{\ensuremath{m_{\PZ\PH}^{\text{T}}}\xspace}
\newcommand{\mlltt}{\ensuremath{m_{\Pell\Pell\PGt\PGt}}\xspace}
\newcommand{\mtt}{\ensuremath{m_{\PGt\PGt}}\xspace}
\newcommand{\mAvis}{\ensuremath{m^\text{vis}_{\Pell\Pell\PGt\PGt}}\xspace}
\newcommand{\ellss}{\ensuremath{\Pell\text{ss}}\xspace}
\newcommand{\HH}{\ensuremath{{\PH\PH}}\xspace}
\newcommand{\YH}{\ensuremath{{\PY\PH}}\xspace}
\newcommand{\bbbb}{\ensuremath{\bb\bb}\xspace}
\newcommand{\bbtt}{\ensuremath{\bb\PGt\PGt}\xspace}
\newcommand{\bbgg}{\ensuremath{\bb\PGg\PGg}\xspace}
\newcommand{\WWWW}{\ensuremath{\PW\PW\PW\PW}\xspace}
\newcommand{\WWtt}{\ensuremath{\PW\PW\PGt\PGt}\xspace}
\newcommand{\tttt}{\ensuremath{\PGt\PGt\PGt\PGt}\xspace}
\newcommand{\GamGam}{\ensuremath{\PGg\PGg}\xspace}
\newcommand{\mHH}{\ensuremath{m_{\HH}}\xspace}
\newcommand{\Hbb}{\ensuremath{\PH\to\bb}\xspace}
\newcommand{\Htt}{\ensuremath{\PH\to\PGt\PGt}\xspace}
\newcommand{\ggF}{\ensuremath{\Pg\Pg\mathrm{F}}\xspace}
\newcommand{\VH}{\ensuremath{\PV\PH}\xspace}
\newcommand{\ttH}{\ensuremath{\ttbar\PH}\xspace}
\newcommand{\bbH}{\ensuremath{\bbbar\PH}\xspace}
\newcommand{\mt}{\ensuremath{m_\PQt}\xspace}
\newcommand{\mH}{\ensuremath{m_{\PH}}\xspace}
\newcommand{\MhScen}{\ensuremath{M^{125}_{\text{h}}}\xspace}
\newcommand{\MhEFTScen}{\ensuremath{M^{125}_{\text{h,EFT}}}\xspace}
\newcommand{\mX}{\ensuremath{m_{\PX}}\xspace}
\newcommand{\mY}{\ensuremath{m_{\PY}}\xspace}
\newcommand{\MX}{\mX}
\newcommand{\MY}{\mY}
\newcommand{\mA}{\ensuremath{m_{\PA}}\xspace}
\newcommand{\mHpm}{\ensuremath{m_{\Hpm}}\xspace}
\newcommand{\ma}{\ensuremath{m_{\Pa}}\xspace}
\newcommand{\PVpr}{\ensuremath{\PV^{\prime}}\xspace}
\newcommand{\ZH}{\ensuremath{\PZ\PH}\xspace}
\newcommand{\WH}{\ensuremath{\PW\PH}\xspace}
\newcommand{\WZ}{\ensuremath{\PW\PZ}\xspace}
\newcommand{\VV}{\ensuremath{\PV\PV}\xspace}
\newcommand{\cF}{\ensuremath{c_\mathrm{F}}\xspace}
\newcommand{\cH}{\ensuremath{c_{\PH}}\xspace}
\newcommand{\gV}{\ensuremath{g_{\PV}}\xspace}
\newcommand{\gH}{\ensuremath{g_{\PH}}\xspace}
\newcommand{\gF}{\ensuremath{g_\mathrm{F}}\xspace}
\newcommand{\PG}{{\HepParticle{G}{}{}}\Xspace}
\newcommand{\MVpr}{\ensuremath{m_{V^{\prime}}}\xspace}
\newcommand{\MWpr}{\ensuremath{m_{\PWpr}}\xspace}
\newcommand{\MZpr}{\ensuremath{m_{\PZpr}}\xspace}
\newcommand{\AMpl}{\ensuremath{\overline{\mathrm{M}}_\mathrm{Pl}}\xspace}
\newcommand{\LambdaR}{\ensuremath{\Lambda_\mathrm{R}}\xspace}
\newcommand{\BR}{\ensuremath{\mathcal{B}}\xspace}
\newcommand{\SD}{s.d.\xspace}
\newcommand{\CP}{\ensuremath{CP}\xspace}

\cmsNoteHeader{B2G-23-002}

\title{Searches for Higgs boson production through decays of heavy resonances}

\date{\today}

\abstract{
The discovery of the Higgs boson has led to new possible signatures for heavy resonance searches at the LHC. Since then, search channels including at least one Higgs boson plus another particle have formed an important part of the program of new physics searches. In this report, the status of these searches by the CMS Collaboration is reviewed. Searches are discussed for resonances decaying to two Higgs bosons, a Higgs and a vector boson, or a Higgs boson and another new resonance. All analyses use proton-proton collision data collected at $\sqrt{s}=13\TeV$ in the years 2016--2018. A combination of the results of these searches is presented together with constraints on different beyond-the-standard model scenarios, including scenarios with extended Higgs sectors, heavy vector bosons and extra dimensions. Studies are shown for the first time by CMS on the validity of the narrow-width approximation in searches for the resonant production of a pair of Higgs bosons. The potential for a discovery at the High Luminosity LHC is also discussed.
}

\hypersetup{
pdfauthor={CMS Collaboration},
pdftitle={Searches for Higgs boson production through decays of heavy resonances},
pdfsubject={CMS},
pdfkeywords={Higgs boson, search for new physics}}

\maketitle

\tableofcontents

\section{Introduction}\label{Sec:Introduction}

The discovery of the Higgs (\PH) boson in
2012~\cite{Higgs-Discovery_ATLAS,Higgs-Discovery_CMS,CMS:2013btf}, at
a mass of about 125\GeV, represented a breakthrough for elementary
particle physics. Its observation brought a direct confirmation of the
principle of spontaneous symmetry breaking, which lies at the origin
of elementary particle masses, and is a cornerstone of today's
standard model~(SM) of particle physics.

Since the \PH boson discovery, subsequent research has shown that the 
measured production and decay modes of the \PH 
boson are in agreement with the SM expectations~\cite{CMS:2022dwd}. 
Thanks to the high luminosity of the CERN Large Hadron Collider (LHC), 
the very sizable data set of proton-proton (\pp) collisions at a 
center-of-mass energy of 13\TeV accumulated during Run~2 (2015--2018), 
and the considerable advances in analysis methodology, 
the very elusive SM Higgs boson pair (\HH)
production process gradually comes into reach. \HH 
production allows the probing of the trilinear Higgs coupling and thus
the investigation of the shape of the Higgs potential.

The \PH boson also provides an excellent instrument to probe
hitherto unknown physics beyond the SM (BSM). A
very striking feature of such extensions would be the existence of
heavy resonances coupling directly to the \PH boson, which might,
even dominantly,
decay into final states involving \PH bosons. This would lead to 
an additional source of \PH boson production in resonant topologies,
which is absent in the SM. Given the small cross section of SM \HH production 
due to the destructive interference between several contributing processes, 
an excess of \HH production with respect to the SM prediction 
could reveal the existence of heavy BSM resonances. Similarly, the associated production of \PH and vector
bosons could be significantly modified by resonant contributions.

Extended Higgs sectors, which comprise more than the
single complex Higgs doublet of the SM, would provide natural
candidates for heavy scalar bosons that decay into final states with
one or more \PH bosons. Examples of such models are two-Higgs-doublet
models (2HDM)~\cite{PhysRevD.8.1226,Branco:2011iw,Haber:2015pua,Kling:2016opi}, 
2HDMs extended with a scalar
singlet (2HDM+S)~\cite{Chalons:2012qe,Chen:2013jvg,Muhlleitner:2016mzt}, 
and the two-real-singlet model (TRSM)~\cite{Robens:2019kga}. Supersymmetry naturally
incorporates extended Higgs sectors. The minimal supersymmetric model
(MSSM)~\cite{Gunion:1984yn,Gunion:1986nh,Degrassi:2002fi,Djouadi:2005gj} 
features a 2HDM-type Higgs sector, while the next-to-minimal
supersymmetric model (NMSSM)~\cite{Maniatis:2009re,Ellwanger:2009dp,King:2012is} includes a 2HDM+S-type Higgs
sector. 

Models of warped extra dimensions (WED)~\cite{Randall:1999ee,Goldberger:1999uk,DeWolfe:1999cp,Csaki:1999mp,Davoudiasl:1999jd,Csaki:2000zn, Agashe:2007zd,Fitzpatrick:2007qr,Oliveira:2014kla,Giudice:2000av} 
predict the existence of an
additional spatial dimension in which the field quanta of gravity, 
the gravitons, propagate. The Randall--Sundrum bulk model gives
rise to heavy resonances such as the spin-0 radion, and a tower of
Kaluza--Klein excitations of the spin-2 graviton, which might have
sizable branching fractions into \HH. In certain models,
which might potentially solve the hierarchy problem,
heavy vector resonances, like \PWpr and \PZpr bosons, form a heavy vector triplet (HVT)~\cite{Pappadopulo:2014qza}. 
They could manifest themselves through decays into \ZH and \WH, 
where \PW and \PZ denote the electroweak (EW) gauge bosons.

This report is organized as follows: Section 1 provides a brief
introduction to the \PH boson and the theoretical concepts
underlying the resonant production of \PH bosons. Section 2
summarizes the techniques of the respective CMS
analyses on 13\TeV data. Section 3 presents a coherent picture of the results of these
analyses, as well as combinations of these results. Section 4 is dedicated to
interpretations of the results in the various models. Section 5
discusses projections towards higher integrated luminosities, 
including the potential for a discovery at the High-Luminosity LHC (HL-LHC). 
A summary is given in Section 6.

Tabulated results unique to this report are provided in a HEPData record~\cite{hepdata}.

\subsection{The Higgs boson at the LHC}\label{Sec:The_Higgs_boson_at_the_LHC}

The Higgs boson was first proposed in the 1960s~\cite{Higgs:1964ia, PhysRev.145.1156, PhysRevLett.13.321}, 
and was finally discovered at the LHC in 2012, by the ATLAS~\cite{Higgs-Discovery_ATLAS} and CMS~\cite{Higgs-Discovery_CMS, CMS:2013btf} collaborations. 

In the SM, the Higgs field has a nonzero vacuum expectation value, 
which breaks the EW symmetry and generates the masses of the \PW and \PZ bosons, while leaving the photon massless. 
This process is called the EW symmetry breaking of the Brout--Englert--Higgs 
mechanism.

\begin{figure}[tb]
  \centering
    \includegraphics[width=\cmsFigWidth]{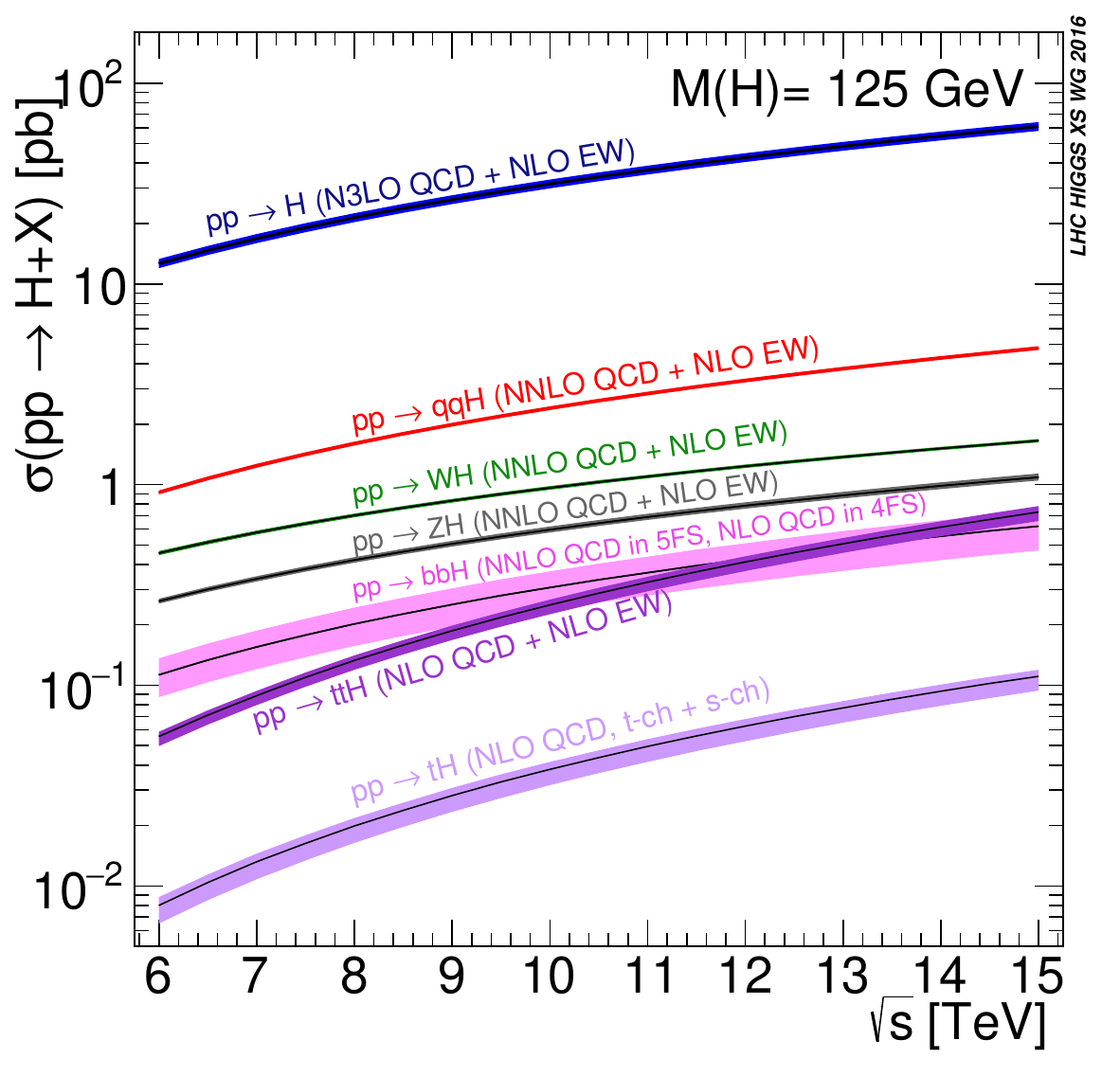}
    \includegraphics[width=\cmsFigWidth]{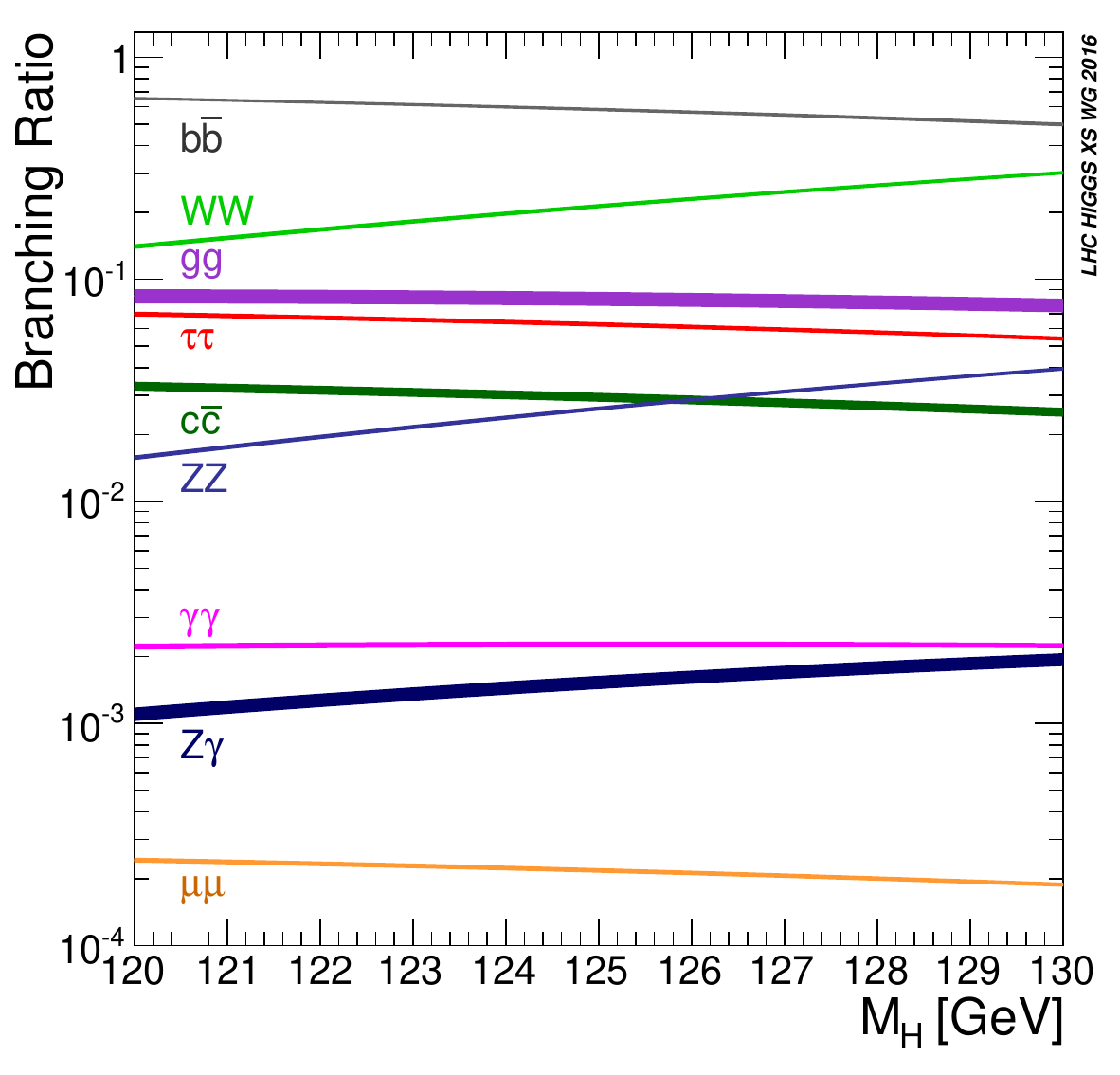}
    \caption{
      Higgs boson production cross sections in the SM as a function of the 
      collider center-of-mass energy (\cmsLeft), and Higgs boson branching 
      fractions in the SM as a function of the Higgs boson mass (\cmsRight). 
      Both figures are taken from Ref.~\cite{deFlorian:2016spz}. 
    }
    \label{H_crosssections_BR}
\end{figure}
There are several ways that \PH bosons can be produced at the LHC in
\pp collisions, each with their own unique experimental signature. The
production cross sections are shown in Fig.~\ref{H_crosssections_BR}
(\cmsLeft).
 
The dominant \PH boson production mechanism is through the ``gluon fusion'' process (\ggF), 
which involves the fusion of two gluons from the colliding protons. 
This process accounts for around 88\% of all \PH bosons produced at the LHC. 
``Vector boson fusion'' (VBF) describes the scattering of two vector bosons \PV (\PW or \PZ bosons) 
exchanged between the colliding protons, and accounts for around 8\% of all Higgs bosons produced at the LHC. 
Other processes include associated \PH boson production with vector bosons (\VH), with top quarks (\ttH), 
or with bottom quarks (\bbH). 

Once produced, Higgs bosons can decay in various ways, each producing a different final state of particles.
The most common and experimentally accessible \PH boson decay modes are to a pair of bottom quarks, 
\PW bosons, \PGt leptons, and \PZ bosons as shown in Fig.~\ref{H_crosssections_BR} (\cmsRight). 
The \PH boson does not directly couple to gluons or photons, but can decay into them via fermion 
or \PW boson loops.

The properties and couplings of the \PH boson have been extensively studied at the LHC. 
The SM does not predict the mass of the \PH boson, but once the mass is given, all its other properties are defined. 
The most recent and precise measurement of the \PH boson mass is $\mH=125.22 \pm 0.14\GeV$~\cite{ATLAS:2023owm}. 
The CMS Collaboration has measured the \PH boson width to be $\Gamma_{\PH}=3.2^{+2.4}_{-1.7}\MeV$ 
via off-shell production in the $4\Pell$ and $2\Pell+2\PGn$ final states~\cite{HiggsWidth}. 
This value is in agreement with the SM prediction of 4.1\MeV~\cite{LHCHiggsCrossSectionWorkingGroup:2013rie}.

\begin{figure}[tb]
  \centering
  \includegraphics[width=\cmsFigWidth]{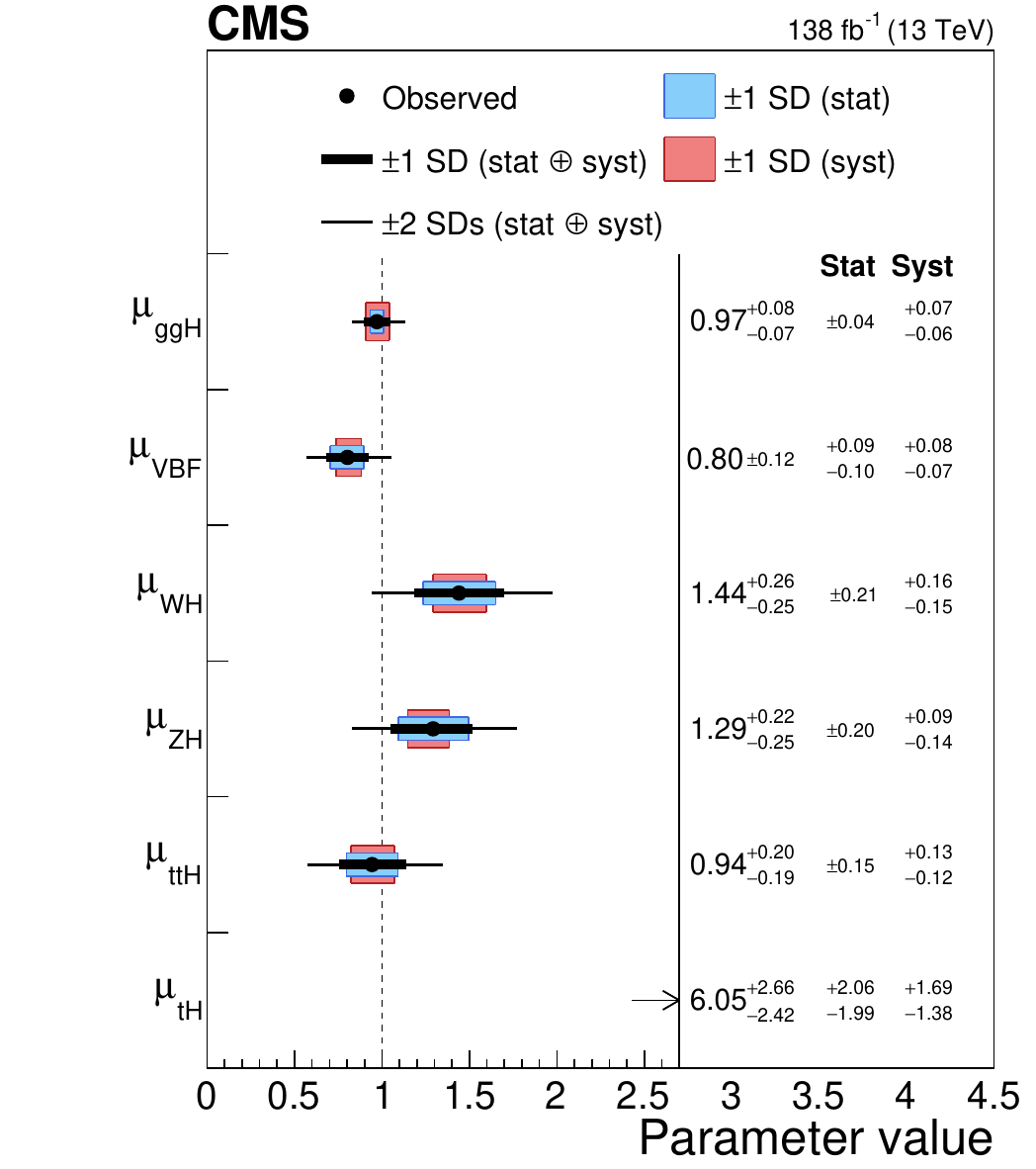}
  \includegraphics[width=\cmsFigWidth]{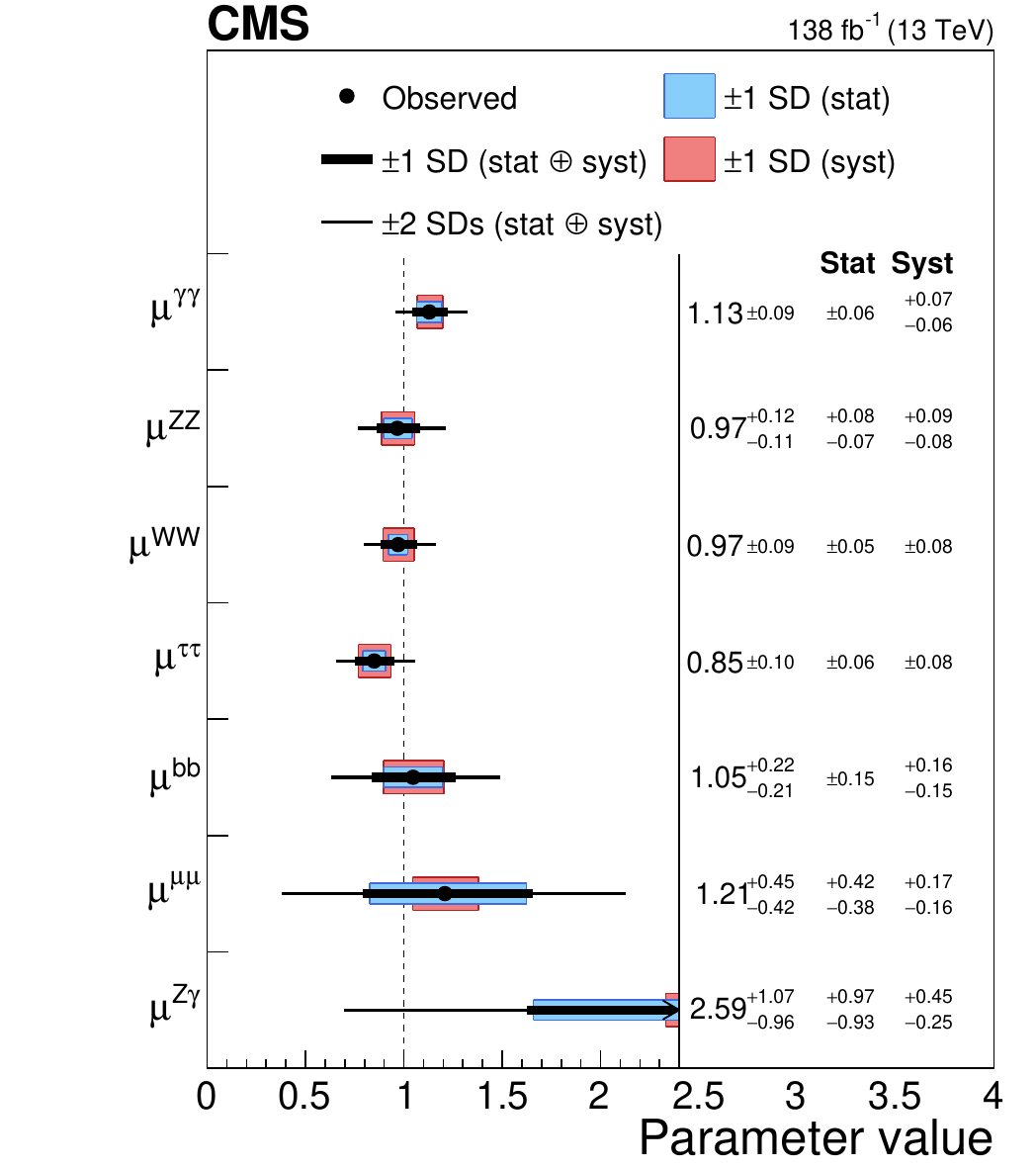}
  \caption{
    Signal strength parameters extracted for various production modes $\mu_i$, 
    assuming the branching fractions $\BR^f = \BR^\text{f}_\text{SM}$ (\cmsLeft), and 
    decay channels $\mu^\text{f}$, assuming the production cross sections as 
    predicted by the SM (\cmsRight). The thick and thin black lines indicate the 
    one and two \SD confidence intervals (labelled by SD in the figures), with 
    the systematic and statistical components of the former indicated by the red 
    and blue bands, respectively. The vertical dashed line at unity represents 
    the values of $\mu_i$ (resp. $\mu^\text{f}$) in the SM. Taken from 
    Ref~\cite{CMS:2022dwd}.
  }
  \label{nature_mu}
\end{figure}

To quantify the agreement between the data and the SM predictions, the concept of a signal strength can be used, 
which is defined as $\mu = \sigma/\sigma_\text{SM}$, where $\sigma$ is the measured production cross section and 
$\sigma_\text{SM}$ is the SM prediction. 
After performing a combined fit of all the data collected at $\sqrt{s}=13\TeV$, in all production modes and decay channels, 
the observed \PH signal strength is $\mu = 1.002 \pm 0.057$~\cite{CMS:2022dwd}, in agreement with the SM prediction. 
Figure~\ref{nature_mu} shows the CMS measurements of the signal strength parameter for each production 
mode, $\mu_i = \sigma_i /\sigma_{i,\text{SM}}$, and several decay channels, 
$\mu^\text{f}=\BR^\text{f} / \BR^\text{f}_\text{SM}$, where \BR is the branching fraction. 
The production modes \ggF, VBF, \WH, \ZH, and \ttH are all observed with a 
significance of five standard deviations (\SD) or above~\cite{CMS:2022dwd}.

\begin{figure}[tbh!]
  \centering
  \includegraphics[width=0.75\textwidth]{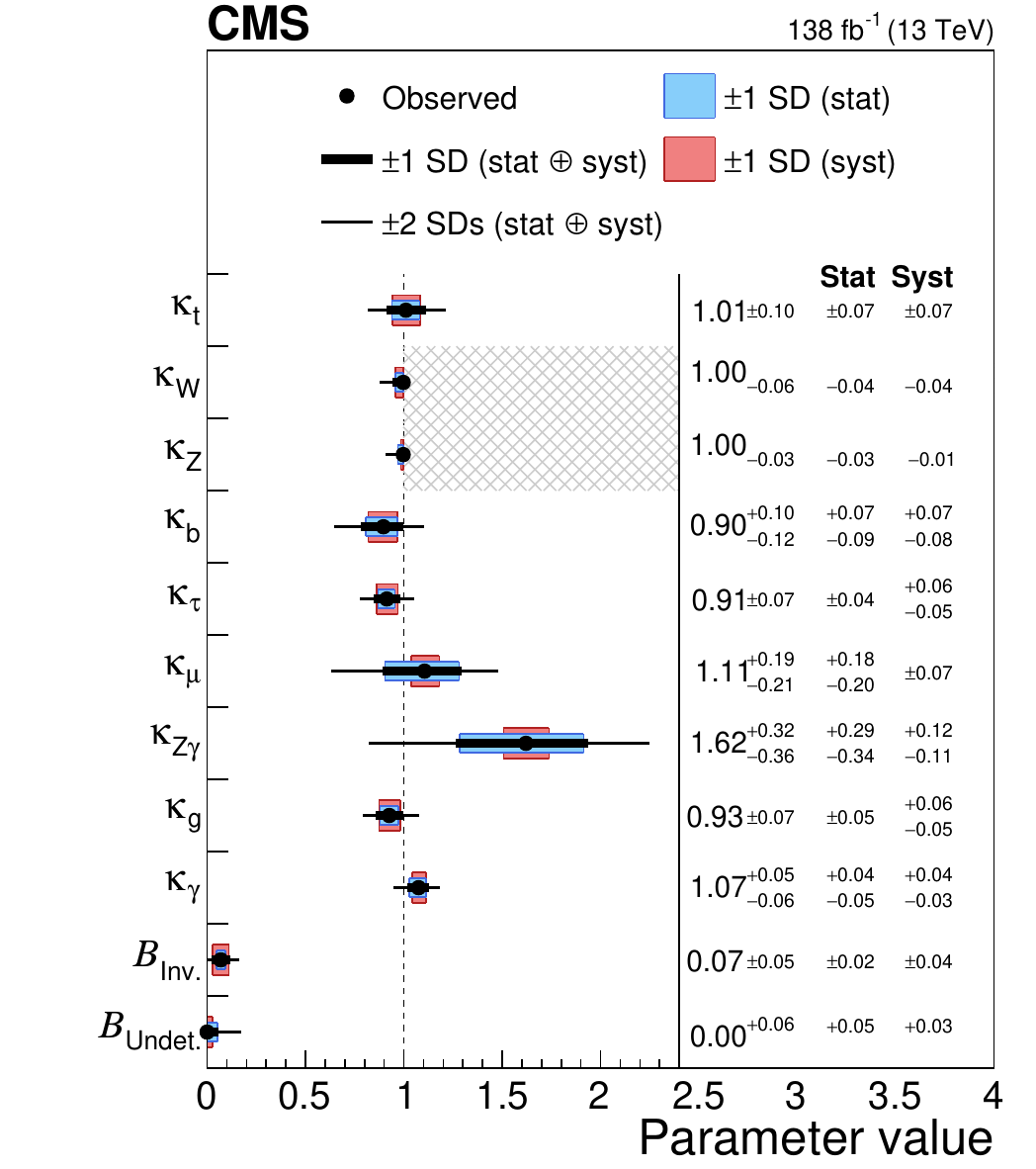}
  \caption{
    Measurements of the coupling modifiers $\kappa_{i}$, allowing both invisible 
    and undetected decay modes, with the SM value used as an upper bound on both 
    $\kappa_\PW$ and $\kappa_\PZ$. The thick and thin black lines indicate the 
    $\pm1$ and $\pm2$ \SD confidence intervals, respectively, with the systematic and 
    statistical components of the $\pm1$~\SD interval indicated by the red and blue 
    bands. The resulting branching fractions for invisible 
    and undetected decay modes are also displayed. Taken from Ref.~\cite{CMS:2022dwd}.
  }
  \label{nature_kappa}
\end{figure}

To quantify the impact of new physics on the interaction between the \PH boson and other particles, 
we scale the coupling strengths as predicted in the SM by a factor $\kappa_i$ 
(the coupling modifiers)~\cite{LHCHiggsCrossSectionWorkingGroup:2013rie}, 
where $i$ is the coupling which the modifier corresponds to. 
Two coupling modifiers are introduced, $\kappa_\PV$ and $\kappa_\text{f}$, to scale the couplings to 
the EW gauge bosons and fermions, respectively. 
When differentiating between the two heavy gauge bosons \PW and \PZ, we can define $\kappa_\PW$ and $\kappa_\PZ$. 
Equivalently, for the fermions we define $\kappa_{\PQt}$, $\kappa_{\PQb}$, $\kappa_{\PGt}$, and $\kappa_{\PGm}$. 
In extensions of the SM with new particles, loop-induced processes may receive additional contributions, therefore 
introducing additional modifiers for the effective couplings of the \PH boson to gluons ($\kappa_{\Pg}$), 
photons ($\kappa_{\PGg}$), and $\PZ\PGg$ ($\kappa_{\PZ\PGg}$). 
According to the latest combined results on these effective couplings shown in Fig.~\ref{nature_kappa}, 
the measured coupling modifiers are compatible with the SM expectations within 1.5~\SD, 
with uncertainties around 10\% for most couplings~\cite{CMS:2022dwd}. 
The invisible and undetected decays of the Higgs boson are also considered, where the latter expression 
refers to Higgs boson decays into final states that can not be distinguished from background processes, 
at the LHC.

\begin{figure}[tb]
    \centering
    \includegraphics[width=0.3\textwidth]{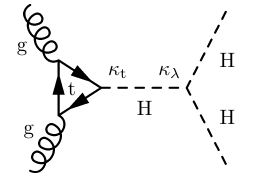}
    \includegraphics[width=0.3\textwidth]{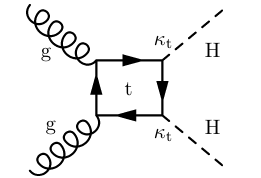}
    \includegraphics[width=0.3\textwidth]{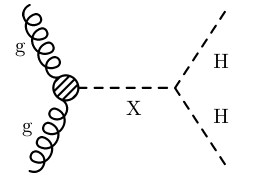}
    \caption{
      Leading order Feynman diagrams of Higgs boson pair production via gluon 
      fusion. The left and middle parts of the figure show the ``triangle'' 
      and ``box'' diagrams, respectively for nonresonant \PH production, as 
      expected from the SM. The right part of the figure shows a diagram for 
      \PH boson production through a new resonance labeled as \PX.
    }
    \label{dihiggs-production-diagrams-ggf_SM}
\end{figure}

The \PH boson trilinear coupling is a measure of the Higgs field's self-interaction strength and
determines the shape of its potential. 
The Higgs boson self-coupling can be accessed directly at the LHC via Higgs boson pair production. 
This rare process has a very low SM cross section because of the destructive interference of the two 
contributing processes at leading order (LO), the box and triangle diagrams shown in 
Fig.~\ref{dihiggs-production-diagrams-ggf_SM} (left and middle). 
Only the triangle diagram is sensitive to the \PH trilinear coupling. 
As a result, detecting the production of Higgs boson pairs and determining the trilinear self-coupling is a major experimental challenge.

The CMS Collaboration has constrained the coupling modifier for the trilinear \PH boson self-coupling
to be within $-1.24$ and $6.49$ at 95\% confidence level (\CL) using \pp data corresponding to an 
integrated luminosity of 138\fbinv, and assuming SM values for the Higgs boson couplings to top quarks and vector bosons. 
The production cross section for inclusive \HH production has been constrained to be smaller than 
3.4 times the value predicted in the SM at 95\%~\CL~\cite{CMS:2022dwd}. 

To understand the Higgs sector is essential for particle physics and cosmology. 
Determining the exact behavior of the Higgs field will help us understand the formation of structures 
in the early universe immediately after the Big Bang, and the stability of the vacuum. 
Some theories of inflation involve a scalar field as the inflaton, the field responsible for driving the 
expansion of the universe~\cite{Bezrukov:2015dty,Bezrukov:2007ep}. 
The fact that the Higgs field is the only scalar field currently known motivates a theory based on the idea 
that the Higgs field caused inflation. 
In addition, the behavior of the Higgs field during the phase transition in the early universe can explain 
leptogenesis~\cite{Fukugita:1986hr} and baryogenesis~\cite{Cohen:1993nk,Morrissey:2012db} 
and could potentially shed light on the matter-antimatter asymmetry we observe today~\cite{Hamada:2016gux}. 
However, the Higgs sector does not have to be minimal. The presence of additional fields would have an 
impact on the mechanisms of inflation, leptogenesis, and baryogenesis. 
Studying the Higgs interactions in detail and exploring the Higgs sector more broadly provide a 
promising avenue for discovering BSM physics and understanding the evolution of the early universe.

\subsection{Resonant Higgs boson production in models beyond the SM}\label{Sec:X_to_HY_in_models_beyond_the_SM} 

The production of \PH bosons through the decay of heavy resonances is not possible within the SM. 
A generic example of a Feynman diagram for such a process is shown in
Fig.~\ref{dihiggs-production-diagrams-ggf_SM} (\cmsRight), where \PX denotes
a sufficiently heavy resonance. 
In the following, we briefly review BSM models in which such production mechanisms occur.

\subsubsection{Extended Higgs sectors}\label{Sec:Extended_Higgs_Sectors}

The SM Higgs sector consists of one complex Higgs doublet and leads to the prediction of one physical \PH boson.
However, there is no guarantee that the Higgs sector is minimal.
The SM Higgs sector can be extended with additional singlets, doublets, or triplets, 
or combinations thereof~\cite{Branco:2011iw,Ivanov:2017dad,Dawson:2018dcd,Steggemann:2020egv,PDG2022}.
Extended Higgs sectors imply the presence of additional Higgs bosons: Neutral Higgs bosons appear in singlet, 
doublet, and triplet extensions; charged Higgs bosons $\PH^\pm$ appear in doublet and triplet extensions;
and doubly charged Higgs bosons, like \eg, $\PH^{++}$, appear in triplet extensions.
In this report, we focus on extensions with singlets and doublets since these lead to final states with SM-like \PH bosons.
Furthermore, it is assumed that the singlets and doublets acquire a vacuum expectation value and couple to the SM particles.

The phenomenology of the additional neutral bosons also depends on their \CP structure. 
Here, we only consider \CP eigenstates: pure scalars are denoted by either \PX or \PY, 
with $\mX > \mY$, and pseudoscalars are denoted by \PA or \Pa, with $\mA > \ma$.
Depending on the structure of the extended Higgs sector and on the masses of the additional scalars and pseudoscalars, 
the following decays involve SM-like \PH bosons in the decay chain:
\begin{enumerate}
    \item decays of a heavy neutral scalar to two SM-like Higgs bosons, $\PX\to\HH$,
    \item decays of a heavy neutral scalar to an SM-like Higgs boson and another scalar, $\PX\to\PY\PH$,
    \item decays of a heavy pseudoscalar to an SM-like Higgs boson and another pseudoscalar, $\PA\to\PH\Pa$, and
    \item decays of a heavy pseudoscalar to an SM-like Higgs boson and a \PZ boson, $\PA\to\PZ\PH$.
    \item For large \mX and if $\mY \gtrsim 250\GeV$, there can also be chained decays leading to 
    the production of multiple \PH bosons, $\PX\to\PY\PY\to\HH\HH$ and $\PX\to\PY\PH\to\HH\PH$.
\end{enumerate}

In general, the additional singlets and doublets mix with the SM doublet and therefore modify the 
couplings of the SM-like scalar \PH~\cite{Englert:2014uua}.
The couplings of the observed \PH boson agree well with the SM predictions, 
leading to significant constraints on the parameter space of extended Higgs 
models~\cite{CMS:2018uag,ATLAS:2019nkf,CMS:2022dwd,ATLAS:2022vkf}.

\paragraph*{Additional singlets}\label{Sec:AdditionalSinglets}

The most straightforward extension of the SM Higgs sector is to introduce an additional real-singlet 
field \PS~\cite{Datta:1997fx,Barger:2008jx,Costa:2015llh,Robens:2015gla}.
The ratio of the vacuum expectation values $v$ of the SM complex doublet and of the singlet, $\langle \PS \rangle$, 
defines the parameter $\tan\beta = v/\langle \PS\rangle$.
The real-singlet model leads to one additional scalar \PX, which can be heavier or lighter than \PH. 
When applying a $\mathbb{Z}_2$ symmetry that requires invariance under transformation of the field $\PS \to -\PS$, 
the scalar \PX obtains its couplings to SM particles from mixing with the SM-like \PH boson, with a mixing angle $\alpha$.
The couplings to SM particles hence correspond to those of the \PH boson, 
albeit suppressed by a factor $\sin\alpha$ if $\mX > \mH$.

For this reason, also the branching fractions of the scalar \PX equal those of an SM-like Higgs boson, 
unless $\mX > 2 \mH$, in which case the decay $\PX\to\HH$ becomes kinematically possible.
In this case, the decays to other SM particles are suppressed depending on the partial width of the \PX decay to \HH, 
which depends on $\tan\beta$.
In summary, the extension with a real additional scalar with $\mathbb{Z}_2$ symmetry involves three new free 
parameters --- the mass of the additional scalar \mX, the mixing angle $\alpha$, 
and the ratio of the two vacuum expectation values $\tan\beta$ --- and leads to $\PX\to\HH$ decays if kinematically possible, 
with a relative rate compared to the other decay modes depending on $\tan\beta$.

Adding a second real singlet field, again imposing $\mathbb{Z}_2$ symmetry, 
defines the two real singlet model (TRSM)~\cite{Robens:2019kga}. 
Compared to the real-singlet extension, the TRSM introduces four new parameters: the mass of a second new scalar \PY, 
the mixing angles between the second real singlet field and the complex SM doublet as well as the other real singlet field, 
and the vacuum expectation value of the second real singlet field.
The main difference in terms of phenomenology is that, depending on the masses of the additional Higgs bosons, 
decays of type $\PX\to\PY\PH$, $\PX\to\HH$ (or $\PY\to\HH$), and $\PX\to\PY\PY$ become possible, 
including the chained decays $\PX\to\PY\PY\to\HH\HH$ and $\PX\to\PY\PH\to\HH\PH$. 
All of the decays involving \PH bosons can have large branching fractions and are hence of experimental interest. 
The final states that provide the largest sensitivity after the decays of the \PH bosons depend on the branching 
fractions of \PH (discussed in Section~\ref{Sec:The_Higgs_boson_at_the_LHC}) and \PY. 
The \PY branching fractions correspond to those of an SM-like Higgs boson of mass \mY. 
Therefore, decays into \PW and \PZ bosons dominate above the $\PW\PW$ and $\PZ\PZ$ production thresholds, 
and decays to \ttbar become relevant above the \ttbar production threshold.
This is different from 2HDMs where decays into fermions are dominant for a large fraction of the parameter space.

\paragraph*{Additional doublets} 

In 2HDMs, the SM is extended with a second Higgs doublet leading to the emergence of three neutral 
and two charged Higgs bosons~\cite{PhysRevD.8.1226,Branco:2011iw,Haber:2015pua,Kling:2016opi}. 
Most commonly it is required that each Higgs doublet only couples to charged fermions of one type 
(up-type quarks, down-type quarks, or charged leptons) and \CP-violating terms at the tree level are forbidden 
to evade constraints from flavor-changing neutral currents and the negative results of searches for \CP violation in the Higgs sector.
These constraints give rise to four types of 2HDMs that are distinguished by which of the Higgs doublet fields 
couples to which type of fermions:
\begin{enumerate}
  \item Type~I, with all charged fermions coupled to the second Higgs doublet, 
  \item Type~II, with only up-type quarks coupled to the second Higgs doublet, 
  \item Lepton-specific (Type~X), with only up-type and down-type quarks coupled to the second Higgs doublet, and
  \item Flipped (Type~Y), with only up-type quarks and charged leptons coupled to the second Higgs doublet.
\end{enumerate}
The free parameters of the model can be expressed in terms of the masses of the Higgs bosons 
(\mH, with \PH being the SM-like Higgs boson, as well as \mX, \mA, and \mHpm), 
the vacuum expectation value of the SM-like doublet $v$, the ratio of the two vacuum expectation values $\tan\beta$, 
the mixing angle $\alpha$, and a parameter $m_{12}$ that softly breaks the $\mathbb{Z}_2$ symmetry. 
Since \mH and $v$ are known, there are six free parameters in these 2HDM scenarios. 
However, there are important constraints on the parameter space.
Constraints from EW precision data require the masses of at least 
two of the additional Higgs bosons to be close to each other~\cite{Haller:2018nnx}, 
which is why mass-degenerate scenarios are studied most frequently. 
In addition, flavor observables lead to strong constraints in the overall parameter space, 
and in particular to lower bounds on \mHpm of around 600\GeV in Type~II and Type~Y 
models~\cite{Hermann:2012fc,Misiak:2015xwa,Misiak:2017bgg,Misiak:2020vlo}. 
Furthermore, in the so-called alignment limit with $\cos(\beta - \alpha) \to 0$ the \PH boson becomes SM-like. 
In turn, the measurements of the \PH boson couplings, which are consistent with the SM predictions 
within uncertainties~\cite{CMS:2022dwd,ATLAS:2022vkf}, lead to constraints on $\abs{\cos(\beta - \alpha)}$ 
between 0.02 and 0.3, depending on the type of 2HDM 
and on $\tan\beta$~\cite{CMS:2018uag,ATLAS:2019nkf,Haller:2018nnx,ATLAS:2022vkf}.

\begin{figure}[tbh!]
    \centering
    \includegraphics[width=\cmsFigWidth]{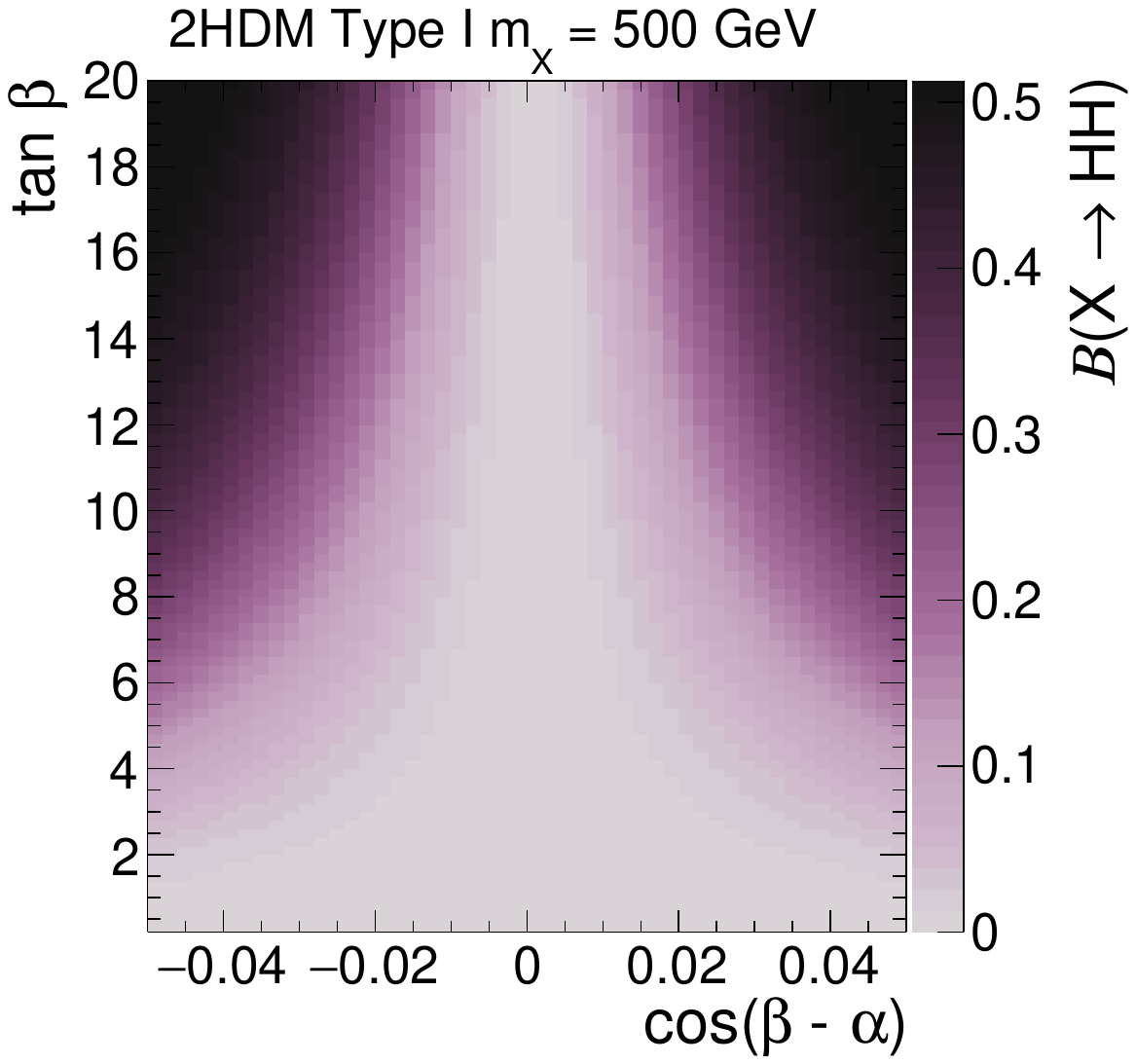}
    \includegraphics[width=\cmsFigWidth]{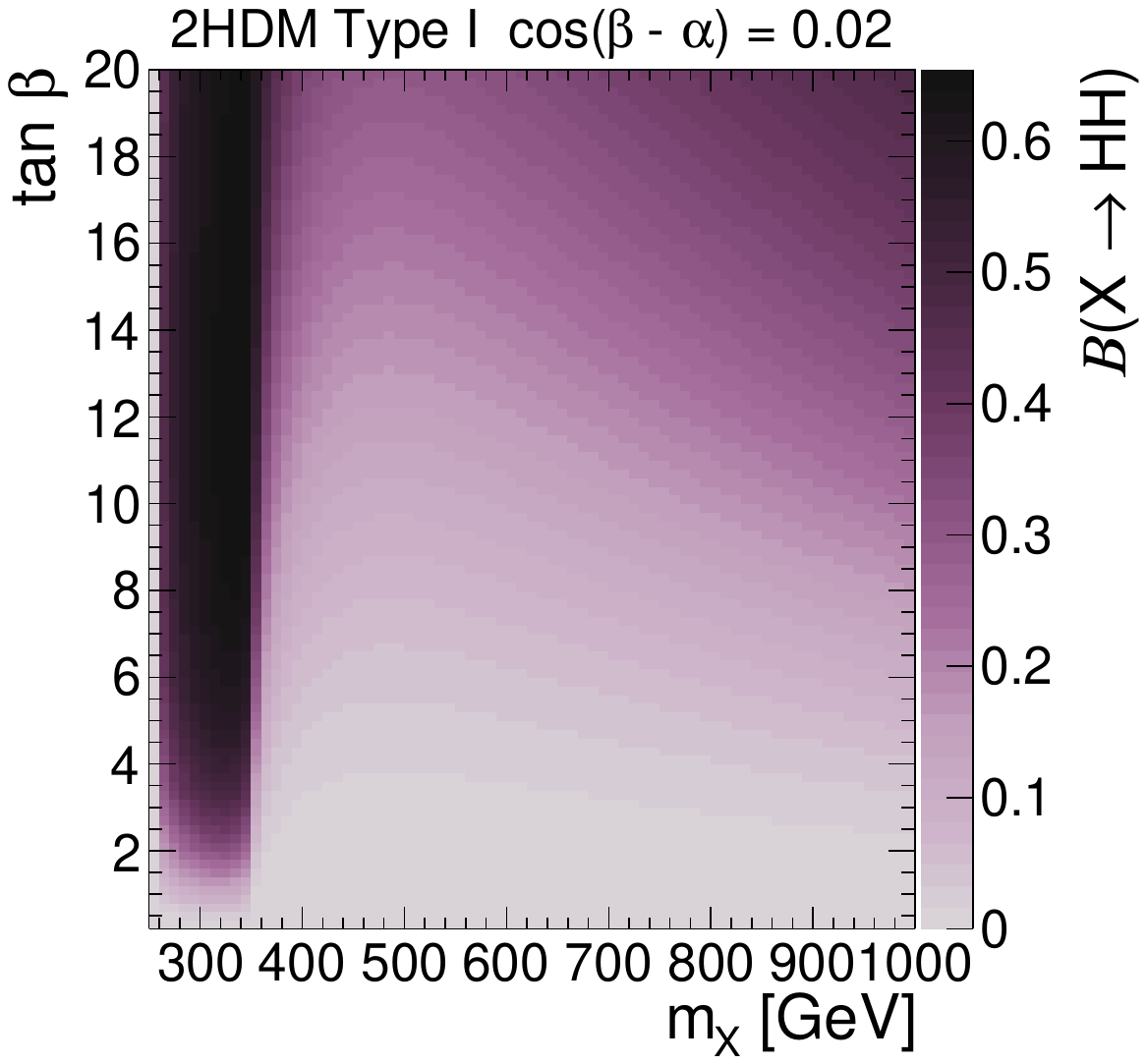} \\ 
    \includegraphics[width=\cmsFigWidth]{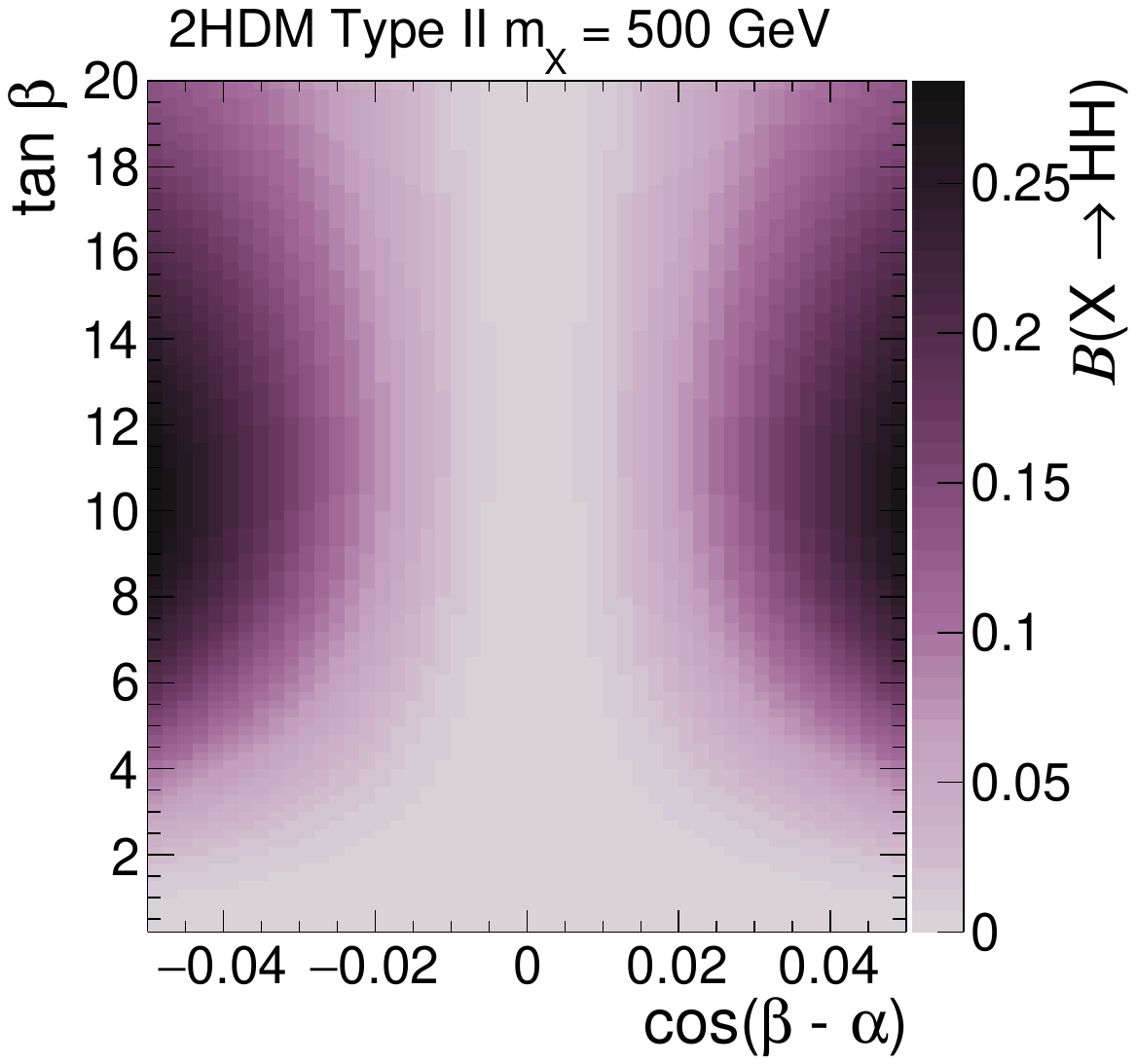}
    \includegraphics[width=\cmsFigWidth]{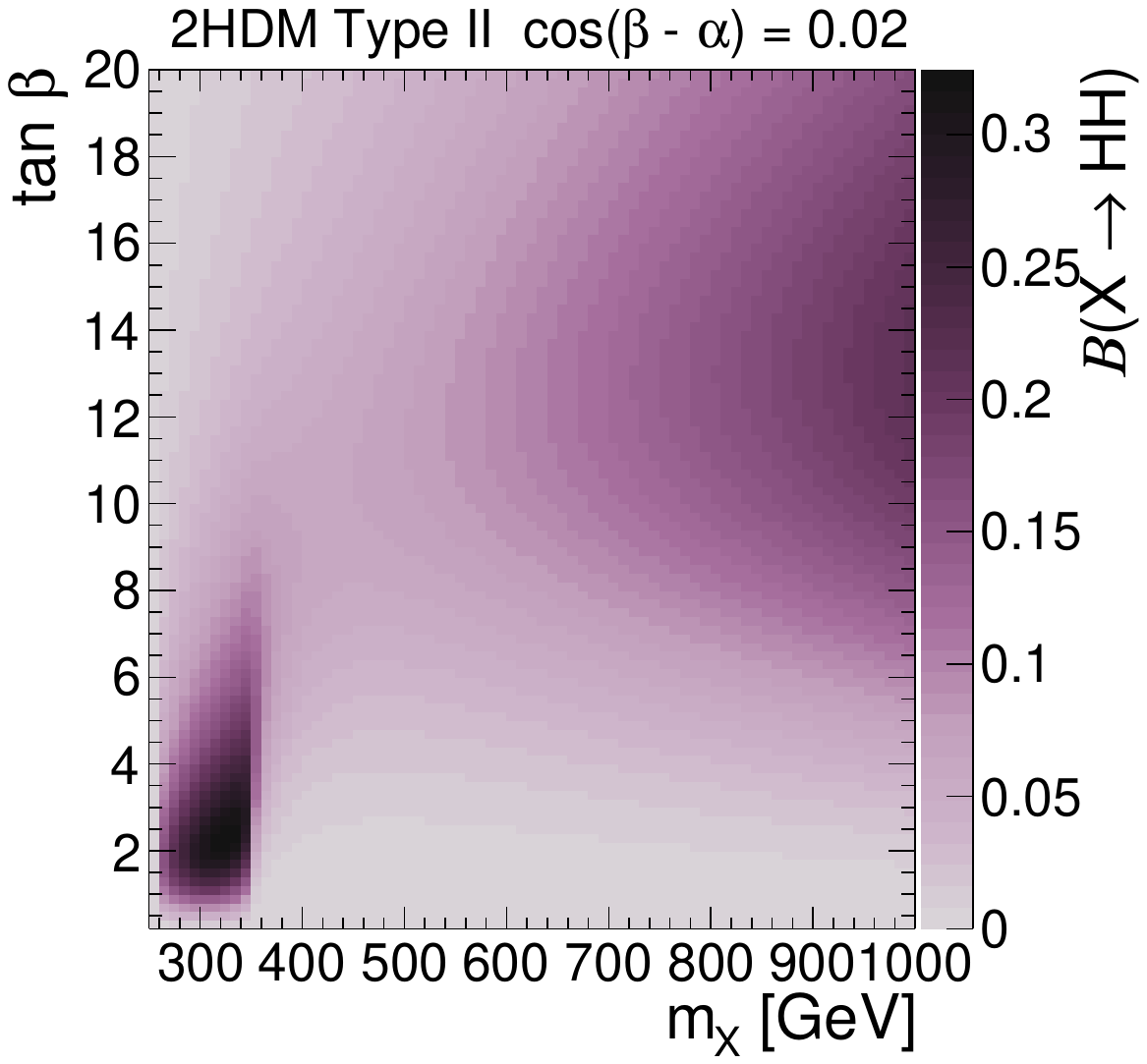}
    \caption{
      Branching fractions of $\PX\to\HH$ decays in 2HDMs of Type~I (upper) and 
      Type~II (lower) in the $\cos(\beta - \alpha)$--$\tan\beta$ plane for 
      $\mX = 500\GeV$ (\cmsLeft) and in the \mX--$\tan\beta$ plane for 
      $\cos(\beta - \alpha) = 0.02$ (\cmsRight). The masses of all non-SM-like 
      Higgs bosons are set to be the same, $\mX = \mA$, and $m_{12}^2 = \mA^2 
      \tan\beta/(1 + \tan^2\beta)$. The branching fractions have been calculated 
      with 2HDMC~v1.8.0~\cite{Eriksson:2009ws,Harlander:2013qxa}.
    }
    \label{fig:2hdm_branching_fractions}
\end{figure}
The couplings of additional heavy Higgs bosons \PX and \PA that involve an \PH boson crucially depend on $\cos(\beta - \alpha)$. 
For $\cos(\beta - \alpha)\to 0$, the branching fractions for the decays $\PX\to\PH\PH$ and $\PA\to\PZ\PH$ vanish. 
While the couplings remain the same for different types of 2HDMs, the branching fractions can be different 
because these depend on the partial widths of all decay modes.
Example branching fractions for 2HDMs of Type~I and Type~II are shown in Fig.~\ref{fig:2hdm_branching_fractions}.
For non-mass-degenerate 2HDMs, decays of type $\PA\to\PZ\PX$ and $\PX\to\PZ\PA$ are possible. 
If allowed, these decays usually have large branching fractions that are not suppressed in the alignment limit.

An important special case of a 2HDM of Type~II is the Higgs sector of the minimal 
supersymmetric standard model (MSSM)~\cite{Gunion:1984yn,Gunion:1986nh,Degrassi:2002fi,Djouadi:2005gj}. 
At the tree level, the MSSM Higgs sector can be described by two parameters, \mA and $\tan\beta$. 
The MSSM naturally predicts an \PH boson that has nearly SM-like couplings, in particular when \mA is large. 
Furthermore, the mass of the SM-like \PH boson follows from the MSSM parameter values. 
For small values of $\tan\beta$, \ie, $\tan\beta \gtrsim 1$, a large supersymmetry (SUSY) breaking scale is needed to 
achieve $\mH = 125\GeV$. This is, however, consistent with the nonobservation of SUSY partners at the LHC. 

\begin{figure}[tbh!]
  \centering
  \includegraphics[width=\cmsFigWidth]{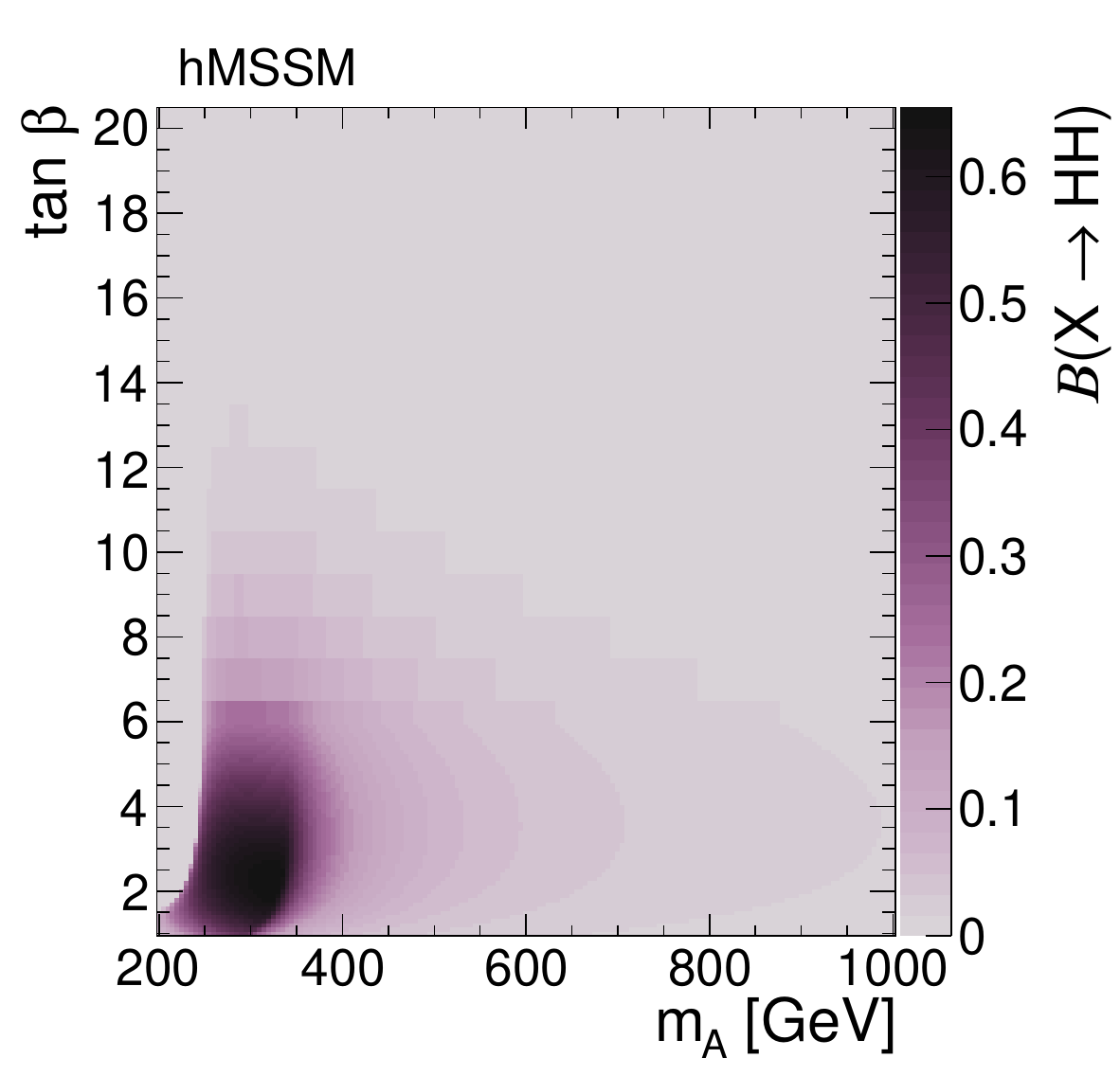}
  \includegraphics[width=\cmsFigWidth]{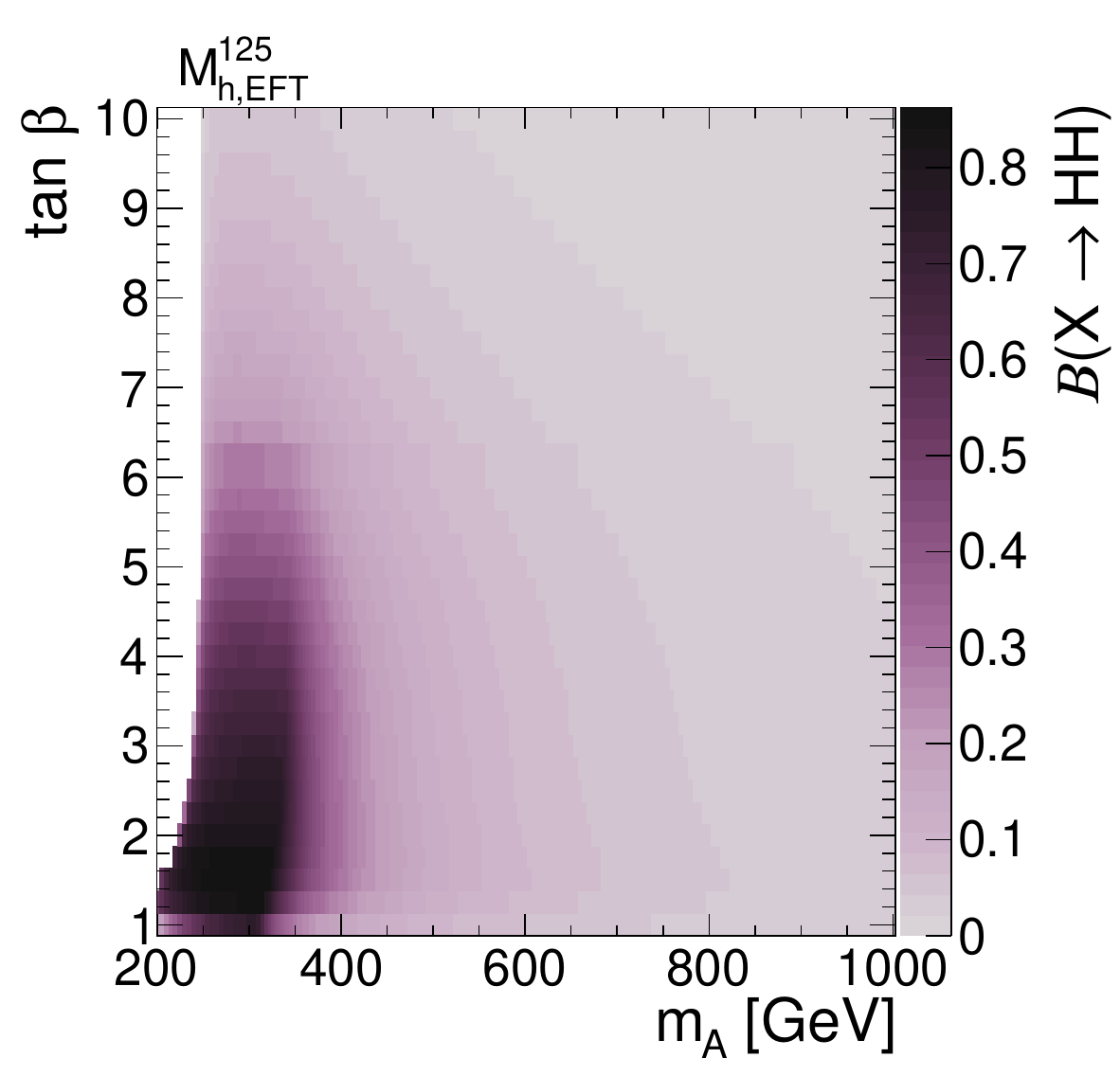}
  \caption{
    Branching fraction of $\PX\to\HH$ decays in the MSSM, for the 
    hMSSM~\cite{Djouadi:2013vqa,Djouadi:2013uqa,Djouadi:2015jea} (\cmsLeft) 
    and the \MhEFTScen~\cite{Bahl:2019ago} benchmarks, in the $\mA$--$\tan
    \beta$ plane. The branching fractions are taken from benchmark files 
    produced by the MSSM subgroup of the LHC Higgs Working 
    Group~\cite{Bagnaschi:2791954,lhc_higgs_working_group_mssm_subgroup_2022_6793918}.
  \label{fig:mssm_branching_fractions}}
\end{figure}
Various MSSM benchmark scenarios have been 
proposed~\cite{Djouadi:2013vqa,Djouadi:2013uqa,Djouadi:2015jea,deFlorian:2016spz,Bahl:2018zmf,Bahl:2019ago}.
Searches for $\PA/\PX\to\tautau$ exclude a large fraction of parameter space at medium to high $\tan\beta$~\cite{CMS:2022goy}, 
such that the phenomenologically interesting parameter space is at low to medium $\tan\beta$. 
Figure~\ref{fig:mssm_branching_fractions} shows the $\PX\to\PH\PH$ branching fractions in two scenarios 
that are particularly designed to give non-excluded predictions at low $\tan\beta$. 
The branching fractions for masses below the \ttbar threshold can reach values of more than 80\%, 
making this channel very important, though measurements of the \PH boson couplings indicate 
that $\mA > 400$--500\GeV~\cite{CMS:2018uag,ATLAS:2019nkf}.
For intermediate values of $\tan\beta$, the branching fraction $\BR(\PX\to\HH)$ can still be of order 10\% 
and the search remains important, whereas $\BR(\PA\to\PZ\PH)$ is found to become negligible.

\paragraph*{Models with additional singlets and doublets}

Models that combine singlets and doublets include the next-to-minimal 2HDM (N2HDM), 
which extends the 2HDM with a real singlet, and the 2HDM+S, which extends the 2HDM 
with a complex singlet~\cite{Chalons:2012qe,Chen:2013jvg,Muhlleitner:2016mzt}.
Considering only \CP-conserving versions of the model, the N2HDM predicts three \CP-even neutral Higgs 
bosons (again denoted as \PH, \PX, and \PY), one \CP-odd neutral Higgs boson \PA, and two charged Higgs bosons \Hpm.
Compared to 2HDMs, the N2HDM adds four additional parameters~\cite{Muhlleitner:2016mzt}: 
two mixing angles $\alpha_2$ and $\alpha_3$, responsible for the mixing of the additional 
singlet with the two doublets, the mass of the additional \CP-even Higgs boson \mY, 
and the vacuum expectation value of the real singlet $v_\PS$. 
The N2HDMs can be categorized in the same four scenarios as 2HDMs, depending on the Yukawa couplings. 

Phenomenologically, N2HDMs share many similarities and constraints with 2HDMs, 
in particular, those related to EW constraints and flavor observables.
However, the singlet admixture can affect the couplings of the SM-like \PH boson.
Most importantly, the additional scalar \PY can either be produced directly or in decays of 
heavier Higgs bosons, $\PX\to\PY\PH$ and $\PX\to\PY\PY$.
Unlike decays to two SM-like Higgs bosons like $\PX\to\PH\PH$, these decays are not suppressed in the alignment limit, which 
is at least approximately realized given the SM-like nature of the \PH boson. 
The decays $\PX\to\PY\PH$ and $\PX\to\PY\PY$ can hence be dominant if kinematically allowed.
The branching fractions of the \PY boson to SM particles depend on the Yukawa type and the other model parameters. 
This leads to a variety of experimentally relevant final states~\cite{Engeln:2018mbg,Abouabid:2021yvw}.

The 2HDM+S model adds a second \CP-odd Higgs boson, leading to the additional $\PA\to\PH\Pa$ decay, 
which is experimentally consistent with the $\PX\to\PH\PY$ signature~\cite{Baum:2018zhf}. 
The Higgs sector of the next-to-minimal MSSM (NMSSM) is a 2HDM+S of Type~II~\cite{Maniatis:2009re,Ellwanger:2009dp,King:2012is}. 
The NMSSM generally leads to the same signatures as the N2HDM and 2HDM+S models, but, like the MSSM, 
is more constrained due to the characteristics of SUSY, 
and these constraints differ from the MSSM because of the additional particle content~\cite{Costa:2015llh,Ellwanger:2017skc,Baum:2019uzg}.

\subsubsection{Warped extra dimensions (radion and Randall--Sundrum graviton)} \label{Sec:Warped_Extra_Dimensions}

Models with a warped extra dimension (WED), as proposed by Randall and Sundrum (RS)~\cite{Randall:1999ee}, 
postulate the existence of one extra spatial dimension compactified between two fixed ``branes''. 
The region between the branes is referred to as bulk and possesses an exponential metric. 
The gap between the two fundamental scales of nature, such as the Planck scale (\Mpl) and the EW scale, 
is controlled by a warp factor ($k$) in the metric, which corresponds to one of the fundamental parameters of the model. 
The brane where the extra-dimensional metric's density is localized is called the ``Planck brane'', 
while the other, where the Higgs field is localized, is called the ``TeV brane''. 
This class of models predicts the existence of new particles that can decay to \HH, 
such as the spin-0 radion \PR~\cite{Goldberger:1999uk,DeWolfe:1999cp,Csaki:1999mp}, 
and the first spin-2 Kaluza-Klein (KK) excitation of the graviton, \PG~\cite{Davoudiasl:1999jd,Csaki:2000zn, Agashe:2007zd}.
The radion is an additional element of WED models, needed to stabilize the size of the extra dimension $l$, 
where the distance between the branes is connected to its vacuum expectation value~\cite{Goldberger:1999uk}.

\begin{figure}[tb]
  \centering
  \includegraphics[width=\cmsFigWidth]{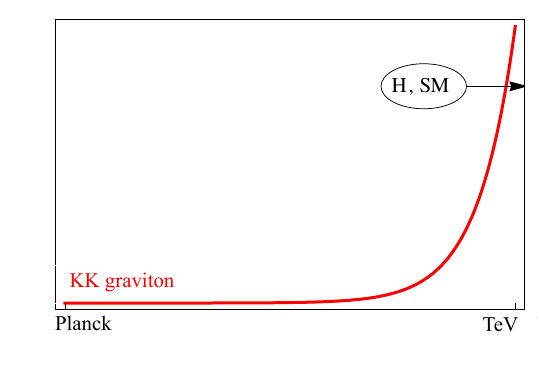}
  \includegraphics[width=\cmsFigWidth]{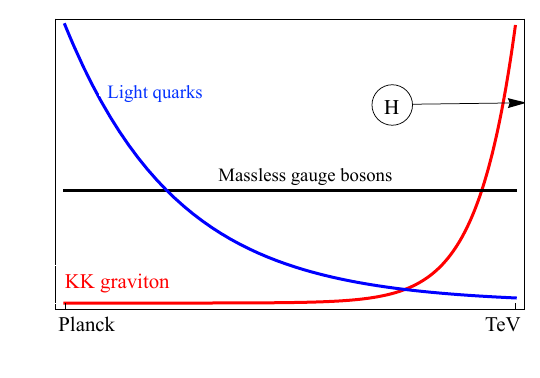}
  \caption{
    Localization of fields on the branes, in different types of the Randall-Sundrum 
    (RS) model: RS1 (\cmsLeft) and bulk-RS (\cmsRight). The $x$-axis represents 
    the 5th dimension with the Planck brane on the left and the \TeV brane on the 
    right. The $y$-axis is the probability density. Adapted from Ref.~\cite{Oliveira:2014kla}.
  }
  \label{fig:RS}
\end{figure}

There are two possible ways of describing a KK graviton in WED models, that depend on the localization 
choice for the SM matter fields, as shown in Fig.~\ref{fig:RS}.
In the RS1 model, only gravity is allowed to propagate in the extra-dimensional bulk. 
In this model, the KK graviton couplings to matter fields are controlled by 
the dimensionless quantity $\tilde{k} = k/\AMpl$~\cite{Randall:1999ee}, 
with the reduced Planck mass $\AMpl$ defined by $\Mpl /\sqrt{8\pi}$. 
The second possibility is to have SM particles propagate in the bulk, the bulk-RS model. 
In this scenario, the KK graviton couplings to the matter fields depend on the localization of the SM fields in the bulk. 
This report uses the phenomenology of Ref.~\cite{Fitzpatrick:2007qr}. 
The SM particles are allowed to propagate in the bulk and follow the SM gauge group's characteristics, 
with the right-handed top quark localized very near the \TeV brane (the so-called elementary top quark hypothesis). 

It is common practice to express the benchmark points of the model in terms of $\tilde{k}$, and the mass scale 
$\LambdaR = \sqrt{6}\,e^{-kl} \AMpl$, where the latter is interpreted as the ultraviolet cutoff of the model~\cite{Giudice:2000av}.
The addition of a scalar-curvature term can induce mixing between the 
scalar \PR and the \PH boson~\cite{Giudice:2000av,Dominici:2002jv,Antoniadis:2002ut}. 
Precision EW studies suggest that this mixing is small~\cite{Desai:2013pga}, 
so we neglect the possibility of \PR-\PH mixing in this report.

The choice of localization of the SM matter fields for the KK-graviton resonance 
impacts the kinematics of the signal and drastically modifies its production and 
decay properties~\cite{Oliveira:2010uv}, so that a distinction of the RS1 and 
bulk-RS models is necessary for the \PG phenomenology.
In contrast, the physics of the radion depends only very weakly on the choice of the model~\cite{Giudice:2000av}, 
which obviates the need to distinguish the RS1 and bulk-RS possibilities in this case. 
More details on WED models can be found in Ref.~\cite{Oliveira:2014kla}.

\begin{figure}[tbh!]
  \centering
  \includegraphics[width=\cmsFigWidth]{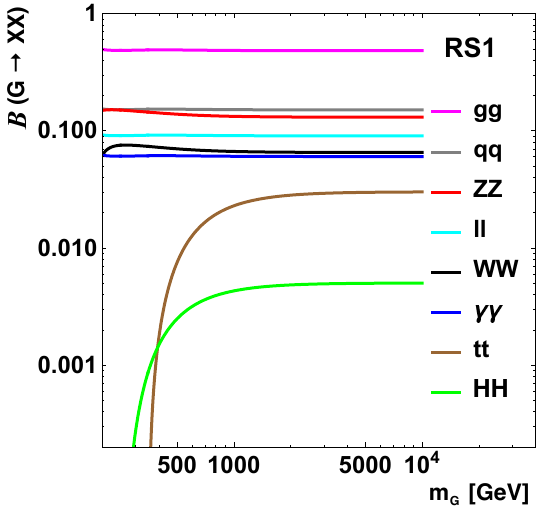}
  \includegraphics[width=\cmsFigWidth]{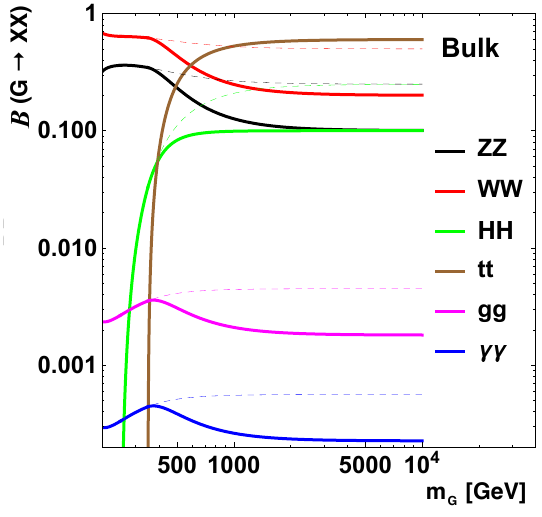}
  \includegraphics[width=\cmsFigWidth]{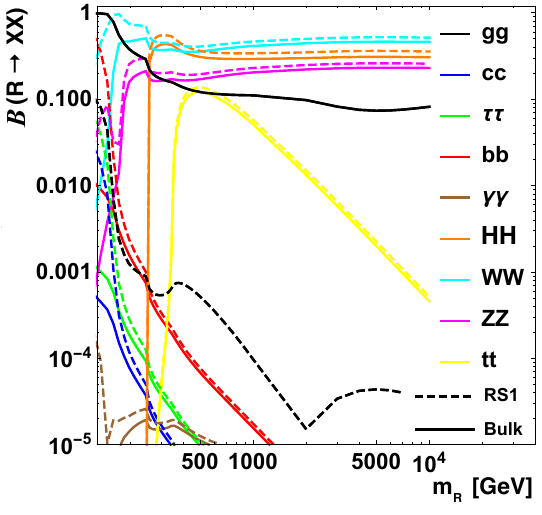}
  \caption{
    The decay branching fractions of an RS1 graviton (upper left), bulk graviton 
    (upper right), and radion (lower). Solid lines assume a fully elementary 
    top quark, while the dashed lines ignore the coupling of the graviton to top 
    quarks. Adapted from Ref.~\cite{Oliveira:2014kla}. 
  }
  \label{fig:WEDbranching}
\end{figure}

In RS1, with all the SM fields localized on the \TeV brane, a heavy graviton would decay to a wide range of final states with significant branching fractions
 as shown in Fig.~\ref{fig:WEDbranching} (upper left), 
and constraints on the RS1 model are mainly obtained from fermionic channels~\cite{Sirunyan:2018exx}. 
In the bulk-RS model, the maximum branching fraction to a pair of Higgs bosons is below 10\% 
under the hypothesis of an elementary \PH boson, as shown in Fig.~\ref{fig:WEDbranching} (upper right). 
Accordingly, the \HH final state is usually not the most important one for placing constraints on the bulk-RS model, 
where the largest sensitivity arises from searches in $\PW\PW$, $\PZ\PZ$, or \ttbar signatures. 
However, the branching fraction to \HH can reach 25\% if the top quark coupling becomes small, 
such that investigations of \HH signatures are necessary in the context of bulk-RS models, 
because the branching fractions are very model dependent.

The dominant \PR decay modes are into pairs of massive gauge bosons, \PH bosons, and top quarks, 
as shown in Fig.~\ref{fig:WEDbranching} (lower). 
Since the couplings are determined by the masses of the final-state particles, and these masses arise from the \PH 
boson localized on the \TeV brane, the RS1 and bulk-RS couplings are identical at LO. 
For large resonance mass \mX, the corresponding widths are 
\begin{equation}
\Gamma(\PR \to \HH ) = \Gamma(\PR \to \PZ\PZ) = \Gamma(\PR \to \PW\PW ) /2 = \frac{1}{32 \pi} \frac{\mX^3}{\LambdaR^2} 
\end{equation}
and
\begin{equation}
\Gamma(\PR \to \ttbar ) = \frac{3}{8 \pi} \left(\frac{m_\PQt}{\mX} \right)^2\frac{\mX^3}{\LambdaR^2}. 
\end{equation}
For large radion masses, the branching fraction to \HH is approximately
25\%, independent of \LambdaR, because the contribution from decays to \ttbar is suppressed by $(\mt / \mX )^2$. 
This makes $\PR\to\HH$ an important channel in the search for a radion resonance.

\subsubsection{Heavy vector triplet (\texorpdfstring{\PWpr}{W'} and \texorpdfstring{\PZpr}{Z'})} \label{Sec:Heavy_Vector_Triplet}

A class of models extending the gauge groups of the SM predicts new force-carrying vector bosons.
They may form a heavy vector triplet (HVT)~\cite{Pappadopulo:2014qza} consisting of \PWpr and \PZpr, 
in analogy with the carriers of the weak force.
Examples of such theories include weakly coupled \PWpr and \PZpr models~\cite{Grojean:2011vu,Salvioni:2009mt,Altarelli:1989ff}, 
little Higgs models~\cite{Schmaltz:2005ky,ArkaniHamed:2002qy}, and 
composite Higgs scenarios~\cite{Bellazzini:2014yua,Contino:2011np,Marzocca:2012zn,Greco:2014aza,Lane:2016kvg}.
The latter are of particular interest as they offer a potential solution to the hierarchy problem.
In these scenarios, the \PH boson is a strongly coupled bound state, 
a pseudo-Nambu--Goldstone boson, sharing constituents with new heavy vector bosons. 
One possible signature of such models at the LHC is \PH boson production through new heavy vector resonances.

The phenomenology of models including an HVT can be deduced from a simplified La\-gran\-gian~\cite{Pappadopulo:2014qza}.
The HVT model is characterized in terms of four parameters:
the masses of the \PWpr and \PZpr resonances, which are degenerate, 
a coefficient \cF, which scales the \PWpr and \PZpr couplings to fermions, 
another coefficient \cH, which scales the \PWpr and \PZpr couplings to the \PH boson and longitudinally polarized SM vector bosons, 
and \gV, which represents the typical strength of the new vector boson interaction.

\begin{figure}[tb!]
  \centering
  \includegraphics[width=0.8\cmsFigWidth]{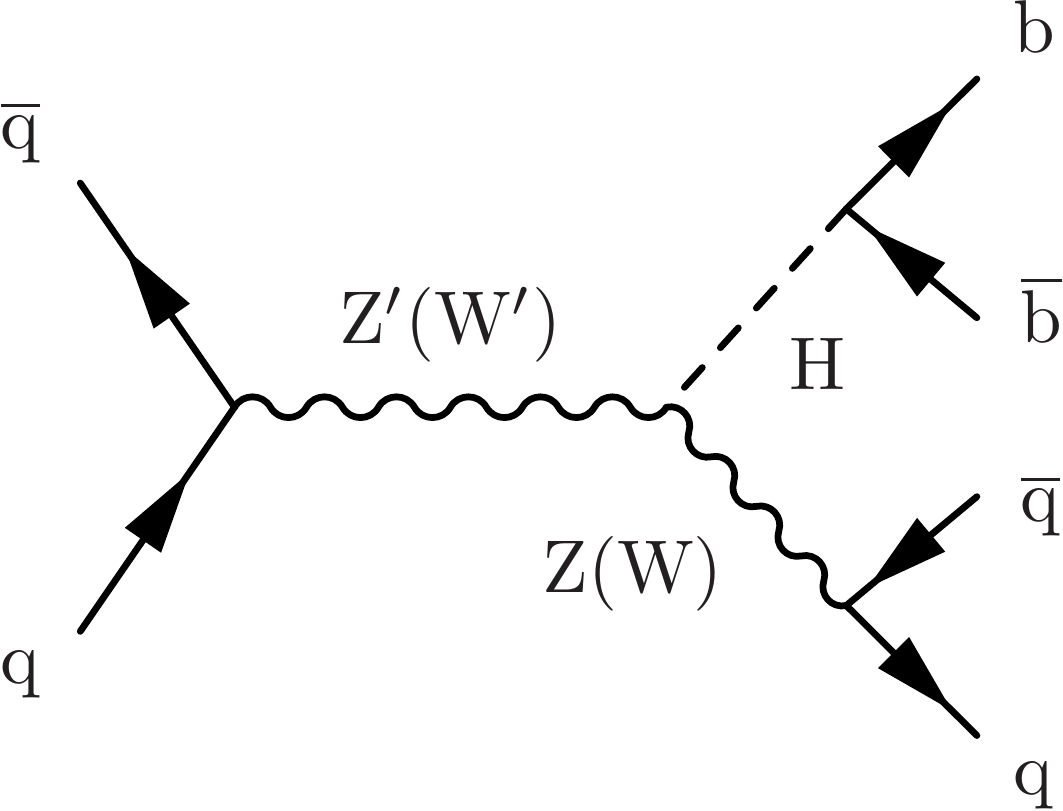}\qquad
  \includegraphics[width=0.7\cmsFigWidth]{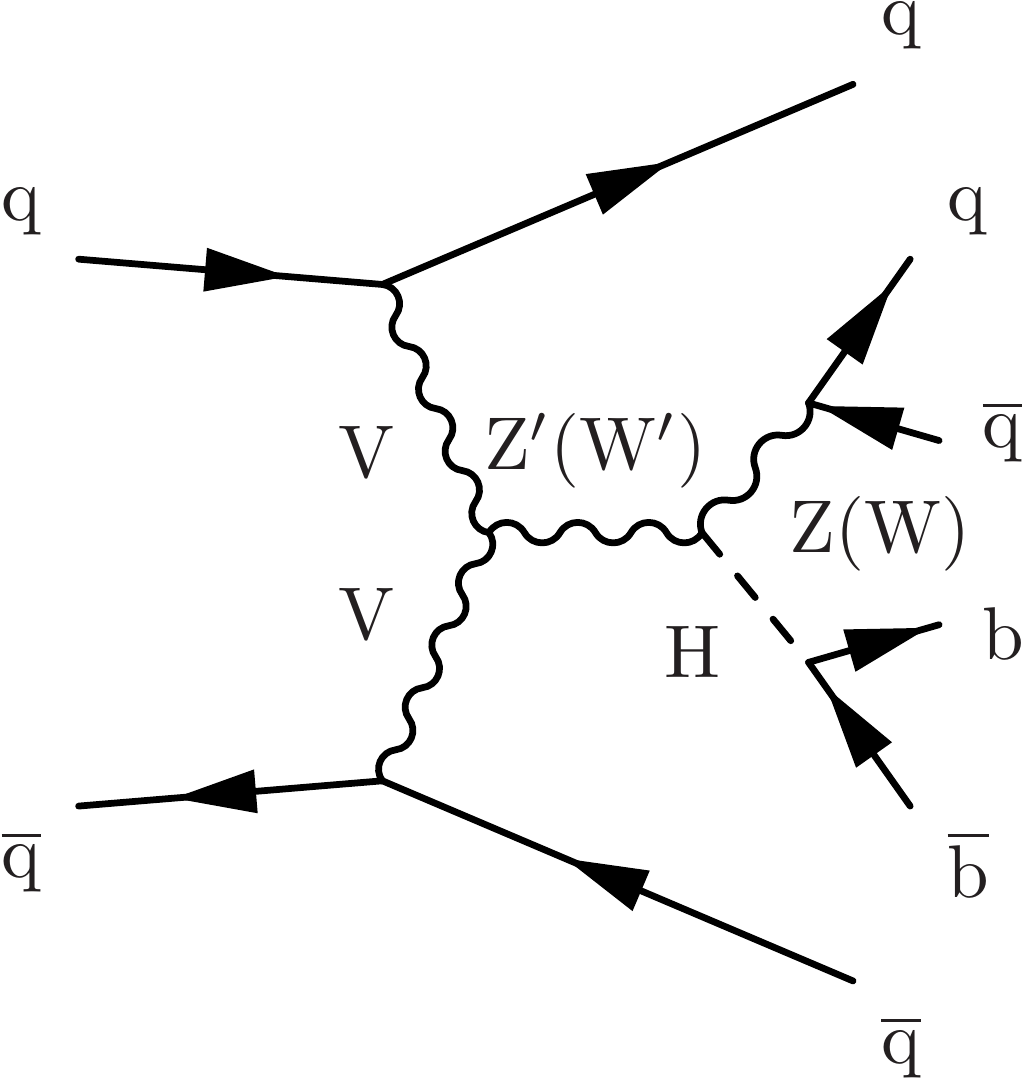}
  \caption{
    Feynman diagrams for the production of \PZpr and \PWpr bosons produced through 
    the (\cmsLeft) Drell--Yan and (\cmsRight) vector boson fusion process. The 
    \PZpr (resp. \PWpr) boson subsequently decays into \ZH and \WH, respectively.
  }
  \label{fig:feynman_HVT}
\end{figure}
The two main \PWpr and \PZpr production modes at the LHC and their decays to \VH are shown in Fig.~\ref{fig:feynman_HVT}.
The triplet field, which mixes with the SM gauge bosons, couples to the fermionic current 
through the combination of parameters $\gF = g^2 \cF / \gV$ and to the \PH and \PV bosons 
through $\gH = \gV \cH$, where $g$ is the $\text{SU}(2)_\text{L}$ gauge coupling, taken to be 
$2 m_{\PW}/ v = 0.6534$ with the \PW boson mass $m_{\PW}$ and the vacuum expectation value 
$v$ from Ref.~\cite{PDG2022}.
We will derive the constraints on the couplings \gH and \gF for several values of \gV below.

Three benchmark scenarios are typically considered in searches.
\begin{itemize}
  \item Model A, with $\gV=1$, $\cH=-0.556$, and $\cF=-1.316$, corresponding to $\gF=-0.562$ and $\gH=-0.556$. 
  This scenario reproduces a model with a weakly coupled extended gauge theory~\cite{Barger:1980ix}.
  \item Model B, with $\gV=3$, $\cH=-0.976$, and $\cF=1.024$, corresponding to $\gF=0.146$ and $\gH=-2.928$. 
  It mimics a minimal strongly coupled composite Higgs model~\cite{Contino:2011np}.
  \item Model C, with $\gV=1$, $\cH=1 - 3$, and $\cF=0$, is a model where couplings to fermions are suppressed, 
such that no production via a Drell--Yan (DY) process is possible at the LHC and the production of \PWpr and \PZpr bosons 
happens exclusively via VBF.
\end{itemize}
In all three scenarios, fermion universality is assumed. 
In model A, the vector resonances have larger couplings to fermions than to bosons, 
with the branching fractions to quarks enhanced by the color factor in QCD. 
Thus, searches for resonances in fermion pair production are most sensitive.
Models B and C have large branching fractions to boson pairs, while the fermionic couplings are suppressed.

\begin{figure}[bt]
  \centering
  \includegraphics[width=\cmsFigWidth]{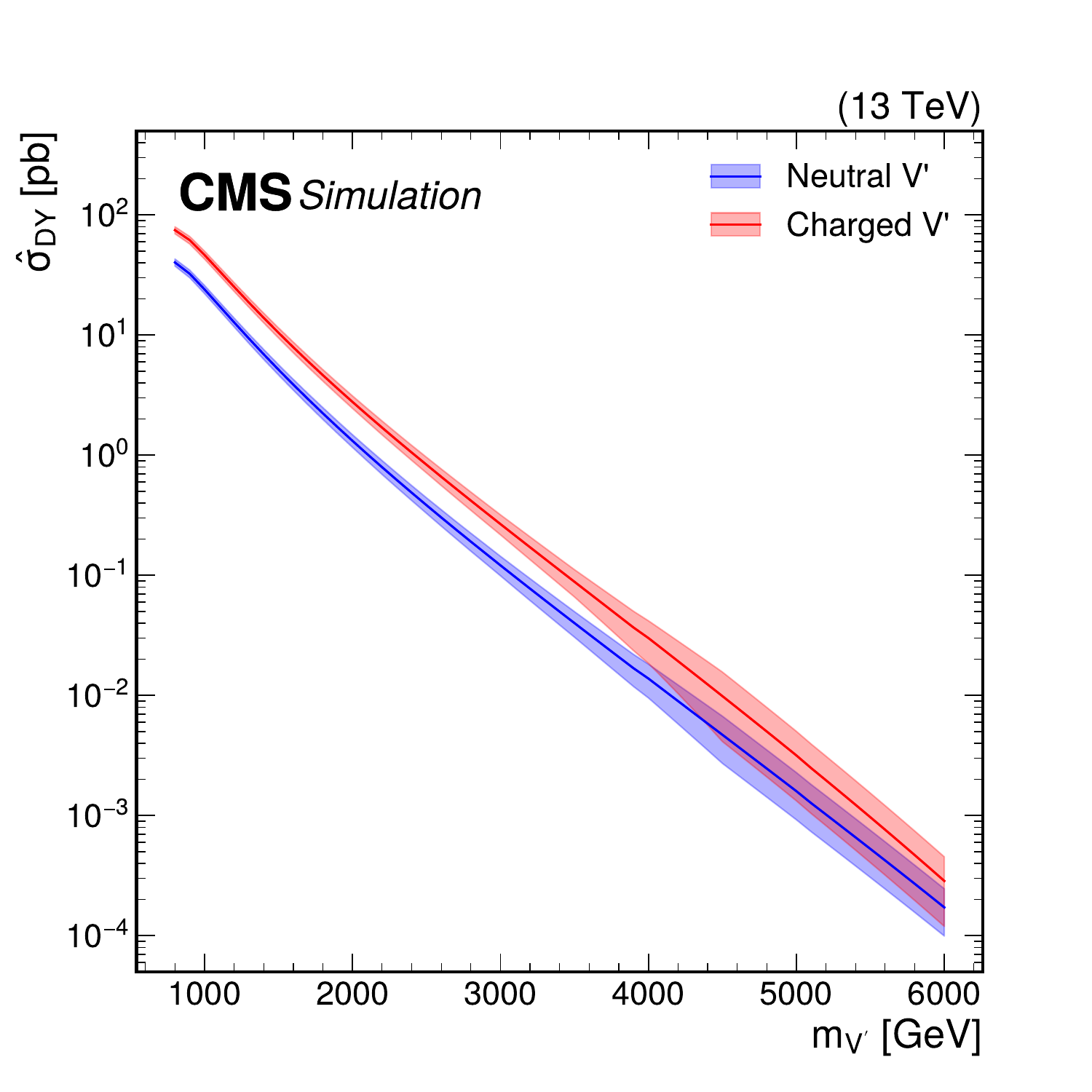}
  \includegraphics[width=\cmsFigWidth]{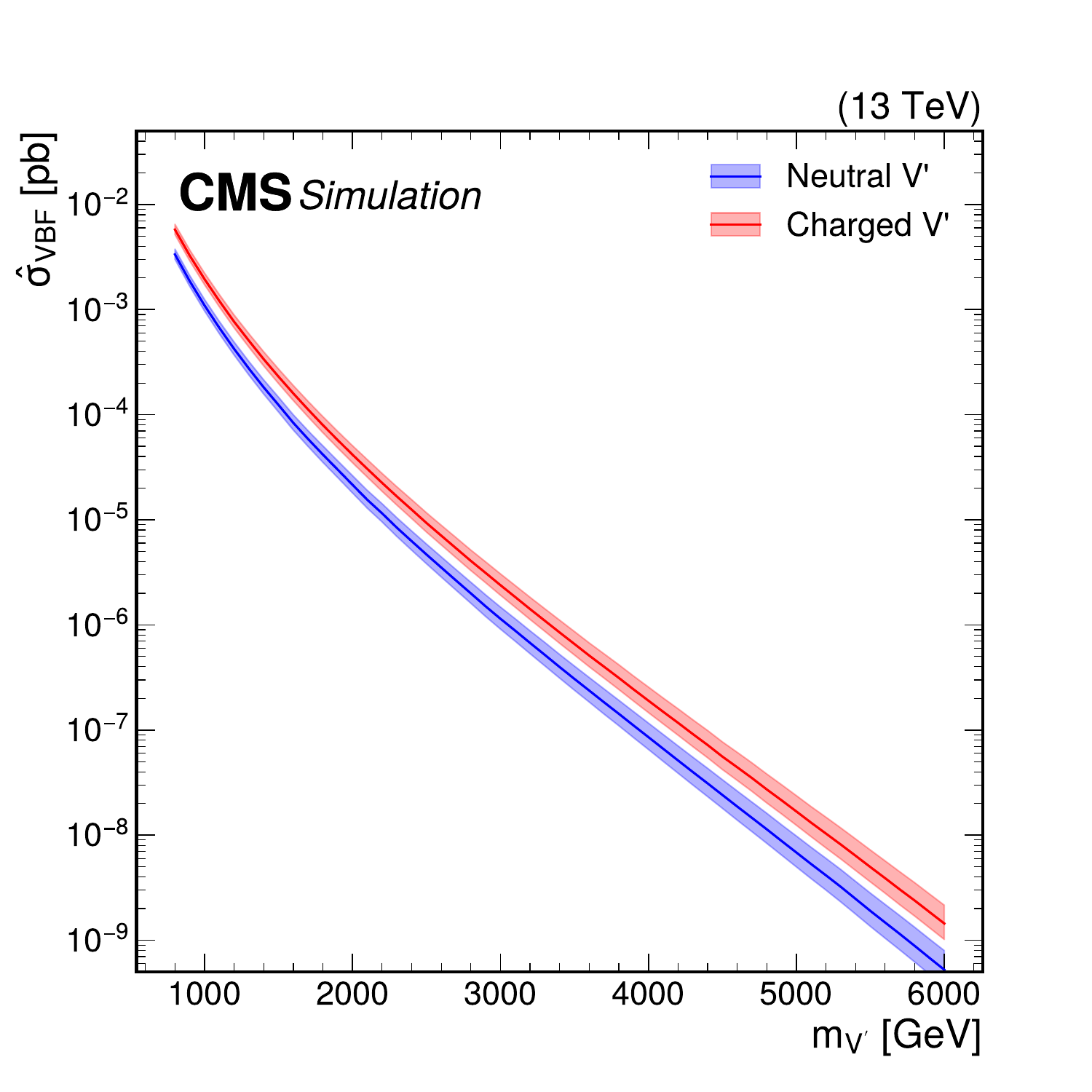}
  \caption{
    Cross sections for (\cmsLeft) Drell--Yan production ($\hat{\sigma}_{\mathrm{DY}}$) 
    and (\cmsRight) production through vector boson fusion ($\hat{\sigma}_{\mathrm{
    VBF}}$), as defined in Eqs.~\eqref{eq:DY_HVT} and~\eqref{eq:VBF_HVT}, for 
    \PZpr and \PWpr bosons in the heavy vector triplet (HVT) model B at $\sqrt{s} = 13\TeV$.
    Calculations are based on the work of Ref.~\cite{Pappadopulo:2014qza}. 
  }
  \label{fig:Crosssection_HVT}
\end{figure}

The analyses discussed in this paper derive results using the narrow-width approximation. 
In this approximation, the production of a certain final state through the \PWpr and \PZpr resonances 
can be factorized into the production of the \PWpr and \PZpr resonances, followed by the decay with 
the respective branching fractions to the final state under consideration. 
We parametrize the \PWpr and \PZpr production cross sections as 
\begin{equation}
\sigma_{\mathrm{DY}} (\gF, \gH, \gV) = \gF^2\, \hat{\sigma}_{\mathrm{DY}}  
\label{eq:DY_HVT}
\end{equation}
\begin{equation}
\sigma_{\mathrm{VBF}} (\gF, \gH, \gV) = \gH^2\, \hat{\sigma}_{\mathrm{VBF}}
\label{eq:VBF_HVT}
\end{equation}
where $\hat{\sigma}_{\mathrm{DY}}$ and $\hat{\sigma}_{\mathrm{VBF}}$ for model B are shown in Fig.~\ref{fig:Crosssection_HVT}.
Compared to the production cross sections of heavy scalar resonances discussed in Section~\ref{Sec:Extended_Higgs_Sectors}, 
where the smallness of the cross sections restricts the sensitivity at the LHC to masses of order 1\TeV, 
the \PWpr and \PZpr cross sections are large enough to probe masses of multiple \TeV.

\begin{figure}[tb!]
  \centering
  \includegraphics[width=\cmsFigWidth]{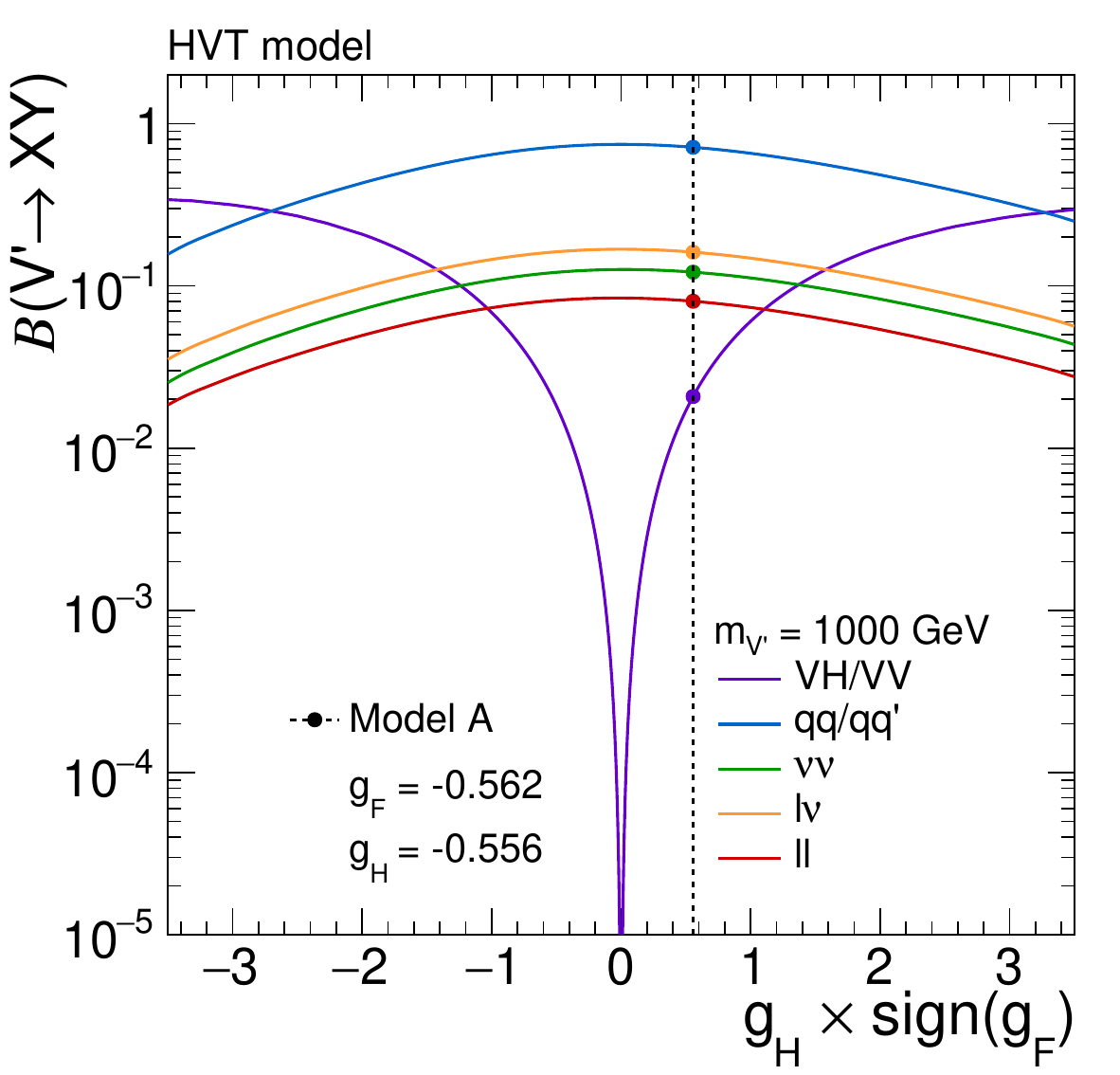}
  \includegraphics[width=\cmsFigWidth]{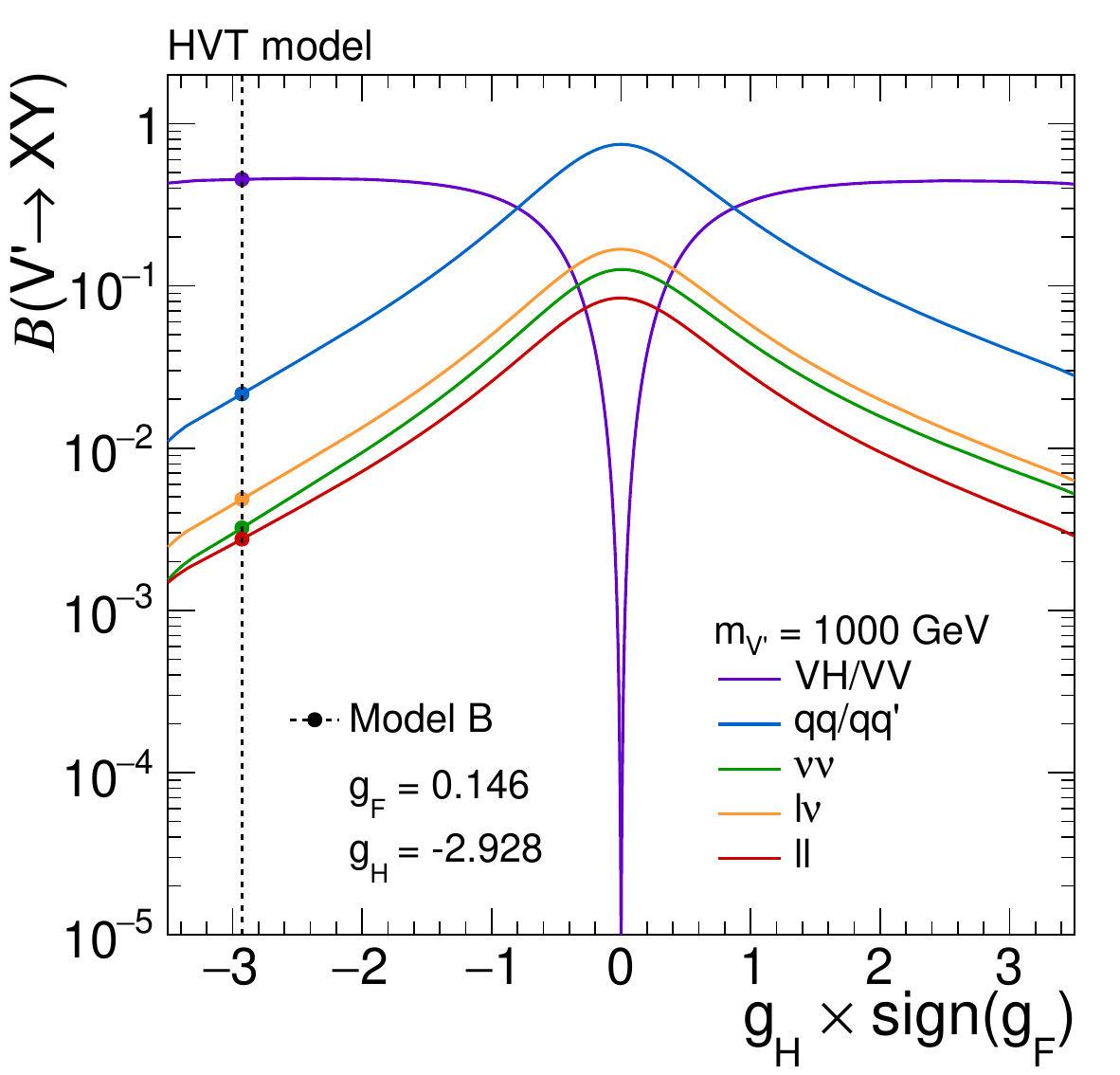}
  
  \includegraphics[width=\cmsFigWidth]{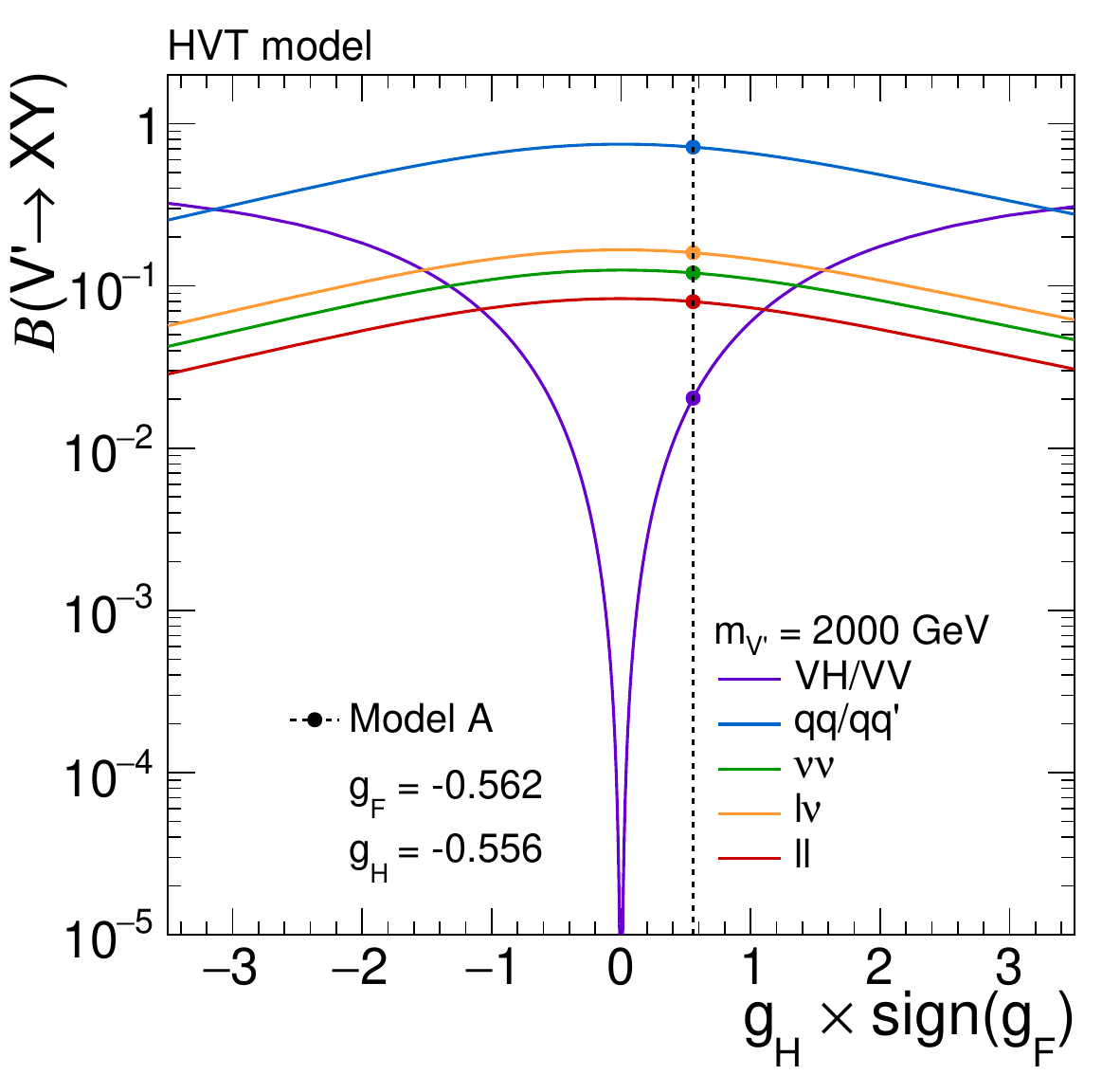}
  \includegraphics[width=\cmsFigWidth]{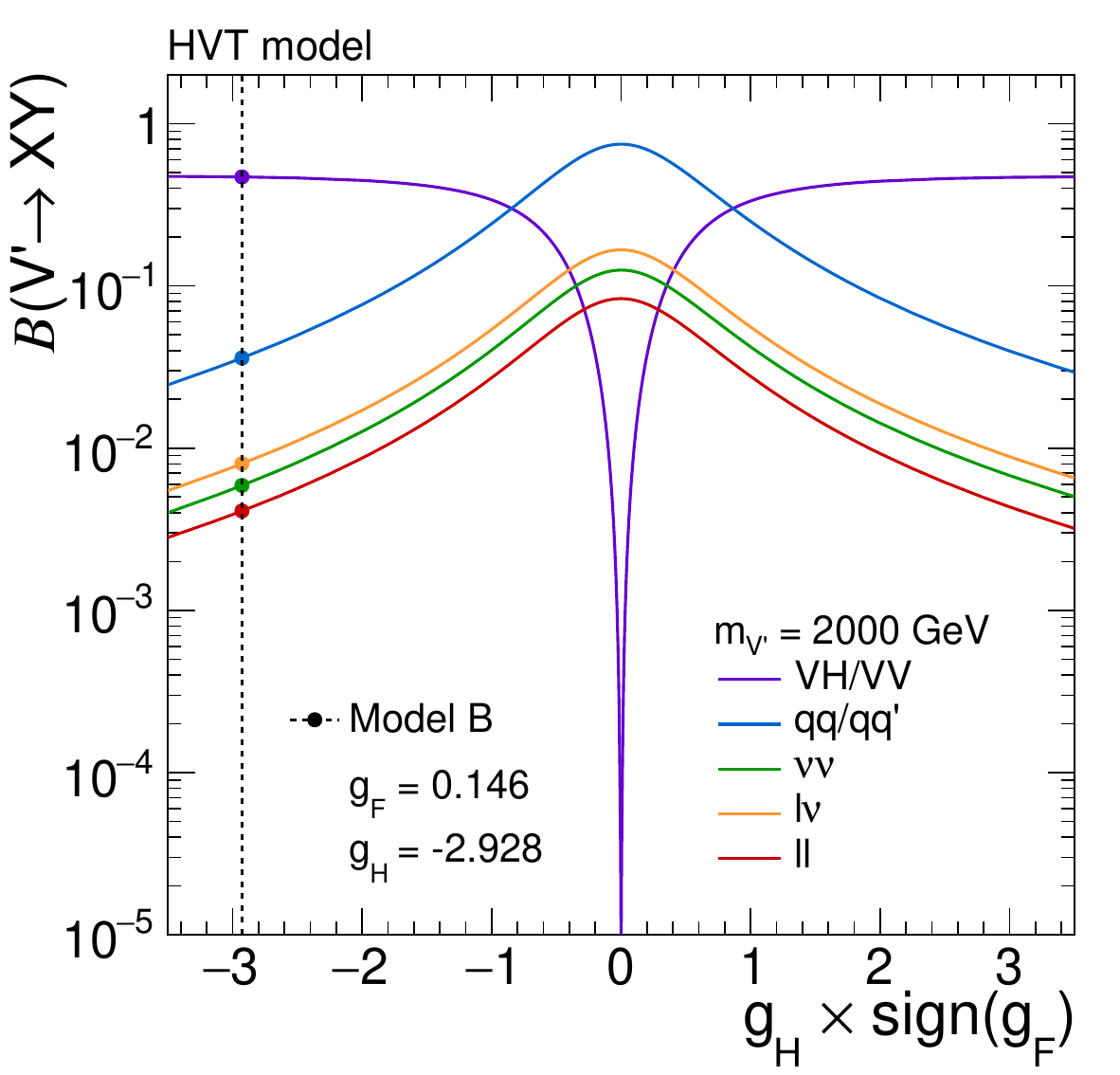}
  
  \caption{
    Branching fractions for heavy vector triplet (HVT) bosons with masses of 
    (upper) 1 and (lower) 2\TeV for values of the parameter \gF corresponding to 
    models (\cmsLeft) A and (\cmsRight) B. The exact branching fractions of each 
    model are indicated by the crossing points of the individual curves with 
    the dashed vertical lines. Calculations are based on the work of 
    Ref.~\cite{Pappadopulo:2014qza}. 
    \label{fig:BR_HVT_all}
  }
\end{figure}

The branching fractions of \PWpr and \PZpr bosons as functions of the coupling parameter \gH times 
the sign of \gF are shown in Fig.~\ref{fig:BR_HVT_all}.
These are computed for \gF values corresponding to the benchmarks models A and B, and for two distinct resonance masses 
of 1 and 2\TeV. For a resonance mass of 1\TeV, a subtle distinction between positive and negative values of \gF is observed, 
whereas branching fractions are symmetric with respect to the sign of \gF for higher masses.
When \gH is small, the branching fractions for decays into quark final states are large. 
The leptonic decay modes are suppressed due to the QCD color factors. 
Conversely, for large values of \gH, the bosonic decay modes dominate the branching fractions, 
indicating that the searches for \VH resonances have the best sensitivity together with searches for \VV resonances. 

\begin{figure}[tb!]
  \centering
  \includegraphics[width=\cmsFigWidth]{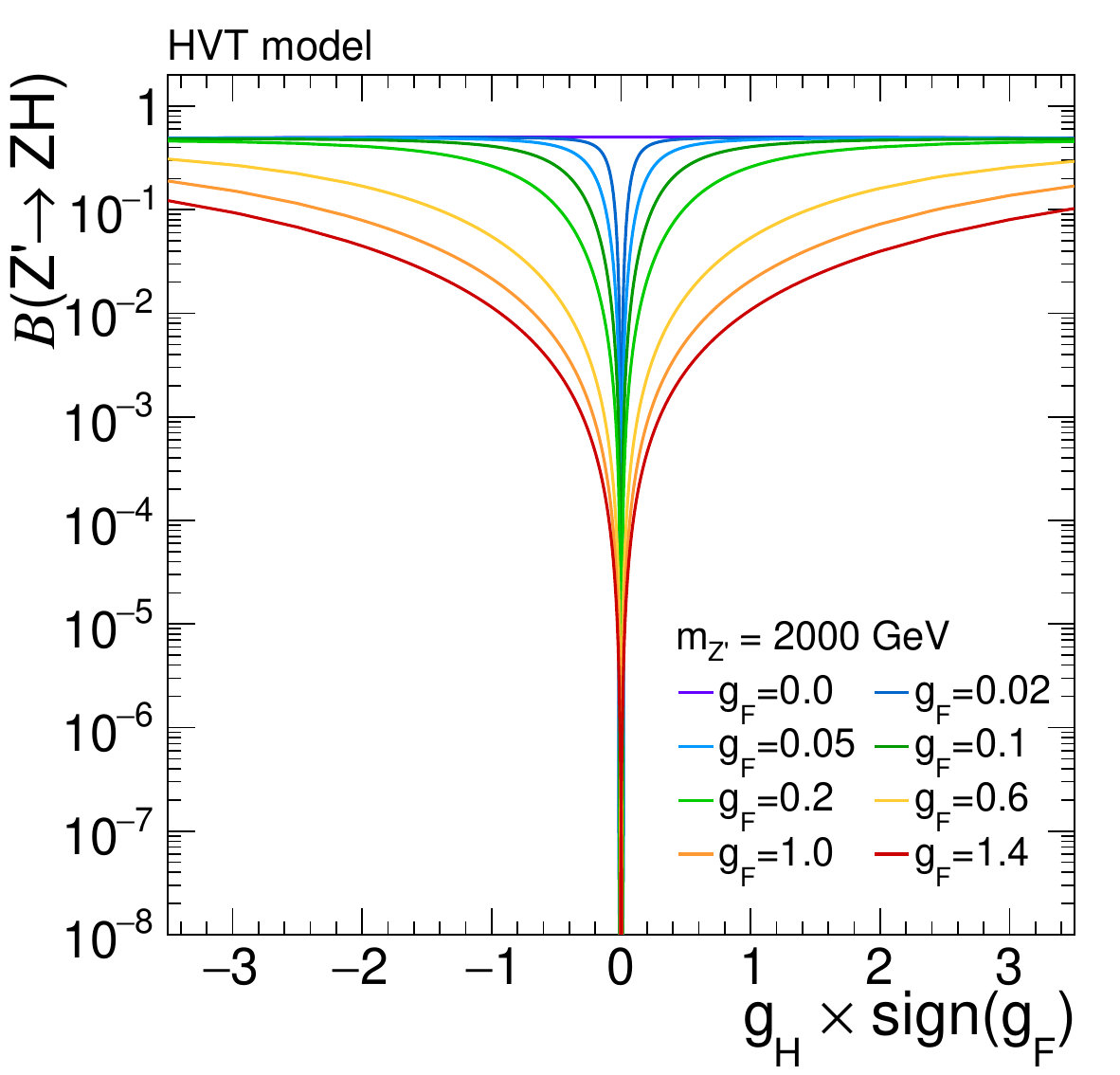}
  \includegraphics[width=\cmsFigWidth]{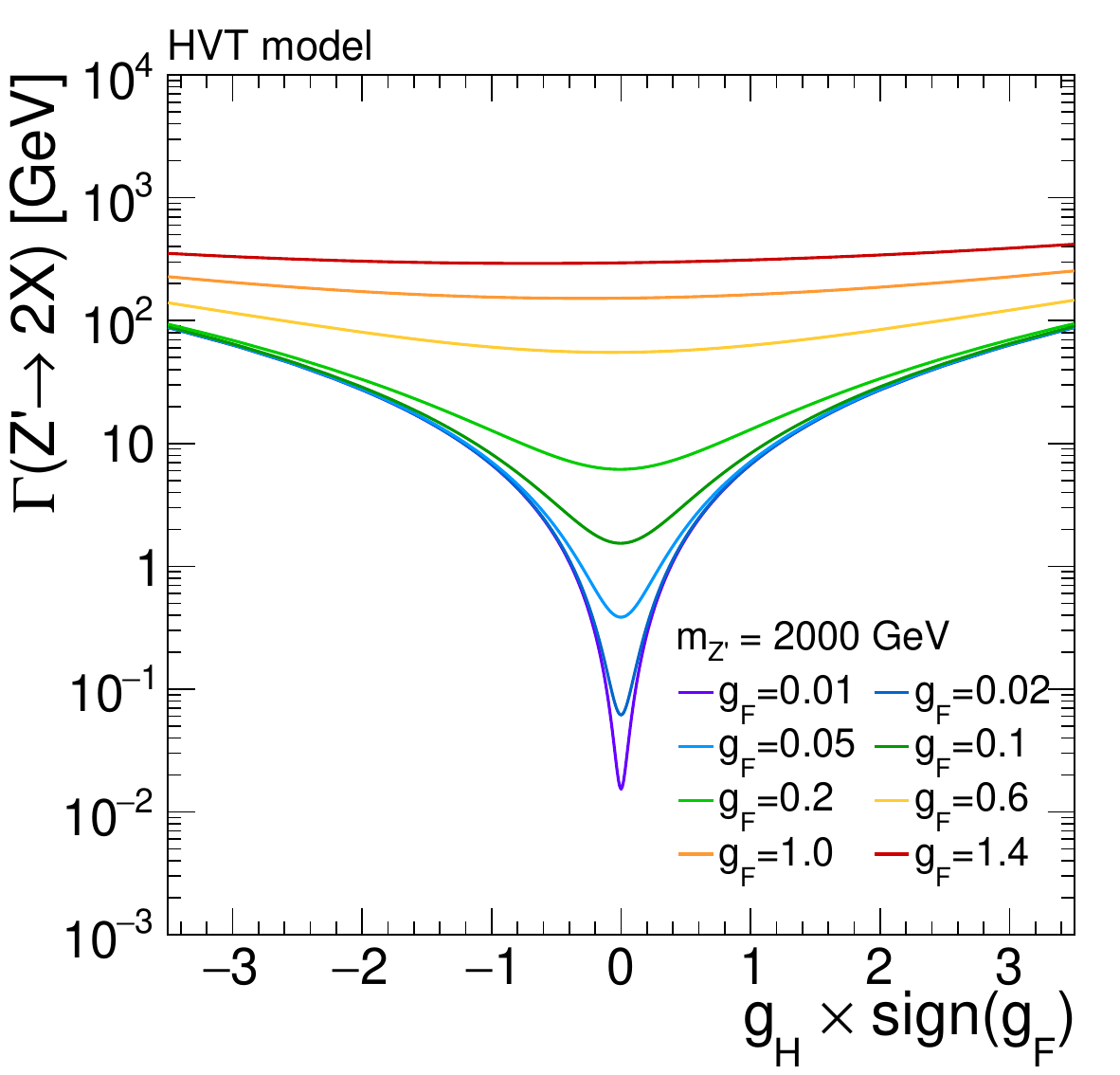}
  \caption{
    (\cmsLeft) Branching fraction for the decay $\PZpr\to\PZ\PH$, and (\cmsRight) 
    total width of the \PZpr boson, for a resonance with 2\TeV mass, for different 
    values of the parameter \gF. Calculations are based on the work of 
    Ref.~\cite{Pappadopulo:2014qza}. 
  }
  \label{fig:BR_HVT_VH}
\end{figure}

The dependence of the branching fraction of the decay $\PZpr\to\VH$ on the parameter \gF is shown in 
Fig.~\ref{fig:BR_HVT_VH} (\cmsLeft) for a resonance with a mass of 2\TeV. 
This branching fraction increases for decreasing \gF, asymptotically approaching the maximum value of about 50\% 
as \gH increases.  
The total width of the \PZpr boson is shown in Fig.~\ref{fig:BR_HVT_VH} (\cmsRight) for a resonance mass of 2\TeV. 
The width increases for increasing values of \gF and \gH. For small values of \gF, the width changes more rapidly as 
a function of \gH. The \PWpr boson branching fractions and decay widths exhibit very similar behavior as a function 
of \gF and \gH to those of \PZpr bosons, and are not shown here.
Previous searches by the CMS and ATLAS Collaborations in the \VH channel have not observed significant deviations
from the SM~\cite{Sirunyan:2018fuh,Sirunyan:2018qob,Sirunyan:2017wto,Khachatryan:2016cfx,Khachatryan:2016yji,Khachatryan:2015bma,Khachatryan:2015ywa, ATLAS:2022enb,ATLAS:2020qiz,ATLAS:2018sbw,ATLAS:2018sxj,ATLAS:2017xel,Aad:2015yza,ATLAS:2016hpd,Aad:2015owa}.

\section{Detector and analysis techniques}\label{Sec:AnalysisTechniques}

The analyses described in the following sections are searches for a new heavy 
resonance \PX in the $\PX\to\HH$, $\PX\to\VH$, and $\PX\to\YH$ decays, where \PV 
denotes a \PW or \PZ boson, \PH the observed Higgs boson, and \PY another new 
particle like an additional Higgs boson, as predicted in several extensions of 
the SM Higgs sector. This report focuses on analyses by the CMS Collaboration; 
related results from the ATLAS Collaboration can be found 
in Refs.~\cite{ATLAS:2020gxx,ATLAS:2020qiz,ATLAS:2021ifb,ATLAS:2022xzm,ATLAS:2022hwc,
ATLAS:2023tkl,ATLAS:2023azi}.

In most analyses, the intermediate particles \PV or \PY are targeted in decay modes involving
\PQb quarks, leptons or photons. This choice is made to
profit from the accuracy with which these particles can be 
reconstructed and identified to distinguish the resulting final states from the
overwhelming amount of purely hadronic processes at the LHC. 
In all searches the natural decay widths of the \PX and \PY particles
are assumed to be small compared to the experimental resolution.
Implications of this assumption are discussed in
Section~\ref{Sec:Effects_finite_width_and_interference}.

The search signatures are rich in characteristic features, most notably there are three 
resonance masses in the full decay chain, of which the masses of the \PV bosons and \mH 
are known to a precision of better than 1\%. To include this information in the process 
description, we introduce a notation where the final-state particles are 
associated with the \PV, \PH, or \PY particles. For example, with this notation the decay $\PX\to
\PY(\bb)\PH(\tautau)$ indicates the subsequent decays of $\PY\to\bb$ and $\PH\to\tautau$. 
An overview is given in Table~\ref{tab:summary} of all the analyses discussed 
in the following sections and the kinematic ranges of their sensitivity. 
\begin{table}[tb]
  \centering
  \topcaption{
    Summary of all analyses discussed in Section~\ref{Sec:AnalysisTechniques}. 
    Note that the list of sub-channels is not exhaustive in all cases. 
    All analyses listed under YH also contribute to the HH measurements.
  }
  \cmsTable{
    \begin{tabular}{cccr@{\,--\,}ll@{\,--\,}ll}
      \multicolumn{2}{c}{Target final state} & Ref. & \multicolumn{4}{c}{Mass coverage (\GeVns)} & Comment \\ [\cmsTabSkip] 
      \PV & \PH & & \multicolumn{2}{c}{\mX} & \multicolumn{2}{c}{} \\
      \hline
      \Zll                   & \tautau & \cite{CMS:2019kca} &   220 & 400               & \multicolumn{2}{c}{} & \\
      $\PZ(\Pell\Pell+\PGn\PGn)$ & \bb        & \cite{CMS:2019qcx} &   225 & 1000              & \multicolumn{2}{c}{} & resolved jets \\
      $\PW(\Pell\PGn)$         & \bb        & \cite{CMS:2021klu} &  1000 & 4500              & \multicolumn{2}{c}{} & $\PW\to\Pell\PGn$ and merged \bb jet \\ 
      \Zll                   & \bb        & \cite{CMS:2021fyk} &   800 & 4600              & \multicolumn{2}{c}{} & $\PZ\to\Pell\Pell/\PGn\PGn$ and merged \bb jet \\ 
      $\PZ(\PQq\PQq)$        & \bb        & \cite{CMS:2022pjv} &  1300 & 6000              & \multicolumn{2}{c}{} & two merged jets \\ [\cmsTabSkip] 
      \PH & \PH & & \multicolumn{2}{c}{\mX} & \multicolumn{2}{c}{} & \\
      \hline
      \bb   & $\PW(\Pell\PGn)\PW(\Pell\PGn+\qq)$ & \cite{CMS:2024rgy}        & 250 &  900            & \multicolumn{2}{c}{} & resolved\\ 
      \bb   & $\PW(\Pell\PGn)\PW(\Pell\PGn+\qq)$ & \cite{CMS:2021roc}        & 800 & 4500            & \multicolumn{2}{c}{} & merged\\ 
      $\PW\PW+\tautau$ & $\PW\PW+\tautau$  & \cite{CMS:2022kdx}        & 250 & 1000            & \multicolumn{2}{c}{} & multilepton final state\\ [\cmsTabSkip] 
      \PY  & \PH & & \multicolumn{2}{c}{\mX} & \multicolumn{2}{c}{\mY} & \\
      \hline
      \bb  & \tautau   & \cite{CMS:2021yci} & 240 & 3000  & $\hphantom{0}60$ & 2800              & resolved jets and \PGt leptons \\ 
      \bb  & $\PGg\PGg$   & \cite{CMS:2023boe} & 300 & 1000  & $\hphantom{0}90$ & 800               & resolved jets and photons \\ 
      \bb  & \bb          & \cite{CMS:2022suh} & 900 & 4000  & $\hphantom{0}60$ & 600               & two merged \bb jets\\ 
    \end{tabular}
  }
  \label{tab:summary}
\end{table}

The analyses span resonance mass ranges of $90<\mX<6000\GeV$ and $60<\mY<2800\GeV$.  
Such large ranges in mass require different reconstruction techniques even for the same final state~\cite{Gouzevitch:2013qca}. 
In the case of small \mX, the decay products of \PY and \PH are produced with large angular separation and can 
be reconstructed as separate objects. 
This characterizes a regime with fully resolved final states. 
In contrast, for large \mX and sufficiently small \mY, 
the decay products of \PY and \PH are strongly collimated because of the large Lorentz boost; such kinematic regimes 
are referred to as boosted throughout this report. 
Final state objects being part of such collimated decays are referred to as merged. 
In the boosted regime, the hadronic decay products of the intermediate resonances, \eg, $\PH(\bb)$, are reconstructed as a 
single large-$R$ jet with substructure identification and grooming of soft and large-angle 
radiation~\cite{Larkoski:2017jix, Asquith:2018igt, Kogler:2021kkw},
where the parameter $R$ denotes the distance parameter of the jet finding algorithm, as described in 
Section~\ref{Sec:JetMET}. 
Typically for the values of $R$ used in this paper, the resolved regime corresponds to $\mX \lesssim 1\TeV$, and the boosted 
regime to $\mX \gtrsim 1\TeV$ 
with $\mY \ll \mX$. For $\mX \approx 1\TeV$ both merged and resolved final states might be encountered depending on the 
helicity angles of the boson decays. 

Unless stated otherwise, all analyses are based on \pp collision 
data collected between 2016 and 2018, at a center-of-mass energy of 13\TeV, 
corresponding to an integrated luminosity of 137--138\fbinv.

\subsection{The CMS Detector}
\label{Sec:Detector}
The central feature of the CMS apparatus is a superconducting solenoid
of 6\unit{m} internal diameter, providing a magnetic field of
3.8\unit{T}. Within the solenoid volume are a silicon pixel and strip
tracker, a lead tungstate crystal electromagnetic calorimeter (ECAL),
and a brass and scintillator hadron calorimeter (HCAL), each composed
of a barrel and two endcap sections. Forward calorimeters extend the
pseudorapidity coverage provided by the barrel and endcap
detectors. Muons are measured in gas-ionization detectors embedded in
the steel flux-return yoke outside the solenoid. A more detailed
description of the CMS detector, together with a definition of the
coordinate system used and the relevant kinematic variables, can be
found in Ref.~\cite{CMS:2008xjf}.

Events of interest are selected using a two-tiered trigger system. The
first level, composed of custom hardware processors, uses
information from the calorimeters and muon detectors to select events
at a rate of around 100\unit{kHz} within a fixed latency of
4\mus~\cite{CMS:2020cmk}. The second level, known as the high-level
trigger, consists of a farm of processors running a version of
the full event reconstruction software optimized for fast processing,
and reduces the event rate to around 1\unit{kHz} before data
storage~\cite{CMS:2016ngn}.

\subsection{Physics objects} 
\label{Sec:Physics_objects}

The reconstruction of the \pp collision products is based on the particle-flow 
(PF) algorithm~\cite{CMS:2017yfk}, combining the available information from all 
CMS subdetectors to reconstruct individual particle candidates, categorized into 
charged and neutral hadrons, electrons, photons, and muons. 
The average number of interactions per bunch crossing was 23 in the year 2016 
and 32 in the years 2017--2018. The fully recorded detector data of a 
bunch crossing define an event for further processing. 
The primary interaction vertex (PV) is taken to be the vertex 
corresponding to the hardest scattering in an event, evaluated using tracking 
information alone, as described in Ref.~\cite{CMS-TDR-15-02}. Secondary vertices, 
which are detached from the PV, might be associated with decays of long-lived 
particles emerging from the PV. Any other collision vertices in an event are 
associated with additional mostly soft inelastic \pp collisions referred to as 
pileup (PU).

\subsubsection{Jets and missing transverse momentum}
\label{Sec:JetMET}

\paragraph*{AK4 and AK8 jets} 
All PF candidates are clustered into jets using the anti-\kt clustering 
algorithm~\cite{Cacciari:2008gp} as implemented in the \FASTJET 
software package~\cite{Cacciari:2011ma}. By default, a distance parameter of $R=0.4$ 
is used. The resulting jet collection will be referred to as AK4 jets. 
Ideally, the kinematic properties of AK4 jets resemble those of 
the single quarks or gluons initiating them. 
In the boosted regime the fragmentation products of the individual quarks, resulting 
from hadronic \PV, \PH, or \PY decays, start to overlap and cannot be properly reconstructed as 
AK4 jets. For this purpose, a second collection of large-$R$ jets 
is obtained with a distance parameter of 0.8, referred to as AK8 jets. 
The larger jet radius allows the inclusion of all hadronic decay products 
in a single jet, and subsequently jet substructure techniques can be applied
to identify the boosted decay within this jet, as explained further down.

In each case, the jet momenta are determined from the vectorial sum of the momenta 
of all PF candidates contained in a jet. The value of this sum is measured to be 
within 5 to 10\% of the same quantity calculated from the momenta of stable 
particles inside equally clustered generated jets in simulation, which holds over the 
entire transverse momentum (\pt) spectrum and geometrical detector acceptance. 
Jet energy corrections to the stable-particle level are obtained from simulation, 
and are confirmed with \textit{in situ} measurements of the energy balance in dijet, 
multijet, \Gjets, and \Zjets events, where the \PZ boson decays into light 
leptons~\cite{CMS:2016lmd}. 
Residual corrections to the simulated energies to match the observed spectra usually 
amount to no more than 2--3\%.
When combining information from the entire detector, the jet energy resolution typically amounts to 15--20\% at 30\GeV, 10\% at 100\GeV, and 5\% at 1\TeV~\cite{CMS:2016lmd}.

The AK4 jets are restricted to $\pt>30\GeV$. Depending on the analysis, 
these are either used in a range of $\abs{\eta}<2.4$, well contained in the coverage 
of the tracker, or in a range of $\abs{\eta}<4.7$, where the extension 
is based on the calorimeters but not covered by the tracker.
The AK8 jets are restricted to $\pt>200\GeV$ and $\abs{\eta}<2.4$ or 2.5, depending on the analysis, where the higher 
value in $\abs{\eta}$ has been used from 2017 on after the upgrade of the silicon pixel detector. 
Ideally, their properties match those of the decaying resonance. 
In a first step, AK8 jets are used for the selection of events of interest. 
In a second step, their specific properties are used to distinguish signal from 
background processes, based on dedicated algorithms, as discussed below. To reduce 
the dependence of the related observables on PU, the pileup-per-particle identification (PUPPI) 
algorithm~\cite{Bertolini:2014bba, Sirunyan:2020foa} is applied to the AK8 jets, 
weighting all PF candidates by their probability to originate from the 
PV~\cite{Sirunyan:2019kia}. For AK4 jets the charged-hadron subtraction (CHS) 
technique, as described in Refs.~\cite{CMS:2017yfk,Sirunyan:2020foa} is used.
In addition, both AK4 and AK8 jets are required to pass tight identification 
requirements~\cite{CMS-PAS-JME-16-003} to remove jets originating from 
calorimetric noise and track misreconstruction. Several properties of the selected 
jets are of importance for the analyses described in this report and will be discussed in more 
detail in the following. 

\paragraph*{Identification of \texorpdfstring{\PQb}{b} jets}
To identify jets resulting from the hadronization of \PQb quarks (\PQb jets) 
several strategies and tagging algorithms are used, depending on the analysis. 
These comprise the \textsc{DeepCSV}~\cite{CMS:2017wtu} and 
\textsc{DeepJet}~\cite{CMS:2017wtu,Bols:2020bkb} algorithms, which are 
applied either to AK4 jets or to the subjets of the selected AK8 jet. 
The subjets are obtained with the soft drop algorithm, where the 
Cambridge--Aachen algorithm~\cite{Dokshitzer:1997in,Wobisch:1998wt} 
is reversed until the soft drop condition~\cite{Larkoski:2014wba} is fulfilled. The two 
resulting clusters are identified as subjets of the AK8 jet.
Alternatively, the ``double-\PQb tagger''~\cite{CMS:2017wtu}, the \textsc{DeepAK8} 
jet tagging~\cite{Sirunyan:2020lcu}, or the \textsc{ParticleNet}~\cite{Qu:2019gqs} algorithms 
are used to identify AK8 jets that are consistent with being initiated by two 
\PQb quarks, a process referred to as ``double-\PQb tagging''. These algorithms usually result in a multiclass output from which 
a suitable discriminant is built. This discriminant is used to distinguish AK8 jets 
produced from light-flavor quarks or gluons versus AK8 jets from two near-collinear \PQb quarks. 
All algorithms use secondary vertex and impact parameter information, 
as well as the multiplicities and kinematic properties of the 
clustered PF candidates. The resonance tagging algorithms are usually trained to 
be insensitive to the mass of the corresponding resonance. For all algorithms, 
predefined working points corresponding to an expected \PQb jet identification 
efficiency for a given misidentification rate for jets initiated by light-flavor 
quarks or gluons, as defined in Ref.~\cite{CMS-DP-2018-058}, are used, 
exhibiting efficiencies of 70--90\% for misidentification rates of 1--10\%. 

\paragraph*{Mass of AK8 jets}
The mass of the hadronically decaying boson resonance associated with
an AK8 jet is estimated using the ``soft-drop'' mass \mj~\cite{Larkoski:2014wba}, 
obtained with the soft-drop algorithm with an angular exponent $\beta = 0$, 
soft-cutoff threshold $z_{\mathrm{cut}} = 0.1$, and
characteristic radius $R_{0} = 0.8$. The soft-drop algorithm is a
generalization  
of the modified mass-drop algorithm~\cite{Dasgupta:2013ihk, Butterworth:2008iy}, 
which is identical to soft drop for $\beta = 0$. The reconstruction of \mj is 
tested and calibrated in \ttbar-enriched event selections, where the mass of hadronically 
decaying \PW bosons can be reconstructed with a resolution of 10\%.

\paragraph*{Jet substructure}
To exploit the substructure of a selected AK8 jet, the ratio $\tau_{21} = \tau_{2}
/\tau_{1}$ is used, where $\tau_{1}$ and $\tau_{2}$ are the 1- and 2-subjettiness 
observables~\cite{Thaler:2010tr}, respectively. 
The quantity $\tau_{21}$ takes lower values for jets originating from two-prong 
\PV, \PH, or \PY decays and larger values for one-prong jets initiated, \eg, by 
single quarks or gluons. However, a selection on $\tau_{21}$ alters the distribution in \mj, 
such that the monotonically falling distribution might feature a resonant structure 
after selecting jets with a minimum value of $\tau_{21}$. This feature prevents 
typical background estimation methods from working for important background processes, like \Wjets 
production~\cite{CMS:2021klu}. To overcome this drawback, 
we use the ``designing decorrelated tagger'' (DDT) procedure~\cite{ddt} leading to 
\begin{equation}
    \tauDDT = \tau_{21}+0.08\log\left(\frac{\mj^{2}}{\pt\,\mu}\right),
\end{equation}
where \pt refers to the transverse momentum of the AK8 jet and $\mu=1\GeV$. A selection 
based on \tauDDT does not alter the distribution in \mj, such that the shape of 
this distribution can be derived from control regions (CRs) to estimate the background 
from SM processes in signal regions (SRs). 

\paragraph*{Event observables}
The two-dimensional vector in the $x$ and $y$ coordinates, \ptvecmiss, 
describes the missing transverse momentum and 
is computed as the negative \ptvec sum of all PF candidates in the event~\cite{Sirunyan:2019kia}. 
The magnitude of \ptvecmiss is referred to as \ptmiss. It is used for the event 
selection and in the calculation of the transverse mass 
\begin{equation}
    \mT = \sqrt{ 2 \pt \ptmiss \left( 1 - \cos\Delta\phi \right) },
    \label{eq:mT}
\end{equation} 
where $\Delta\phi$ refers to the azimuthal angular difference between the transverse 
momentum vector \ptvec of the visible decay product of the particle whose 
transverse mass is to be estimated, and \ptvecmiss.
Depending on the analysis, sometimes the PUPPI algorithm is used to mitigate 
PU effects on \ptvecmiss. Alternatively, the quantity \Htmiss is used, defined as the 
magnitude of the \ptvec sum of all AK4 jets with $\pt>30\GeV$ and $\abs{\eta}<3.0$.
In the same context also the observable \Ht is used, which corresponds to the 
scalar \pt-sum of all selected AK4 jets. It indicates the overall magnitude of 
hadronic activity in an event.

\subsubsection{Leptons and photons}

\paragraph*{Electrons, muons, and photons}
Electron candidates are reconstructed from matching clusters of energy deposits in the 
ECAL with tracks, which are fitted to form hits in 
the tracker~\cite{Khachatryan:2015hwa,CMS:2020uim}. To increase their purity, 
reconstructed electrons are required to pass a multivariate electron 
identification discriminant, which combines information on the track quality, 
shower shape, and kinematic quantities~\cite{CMS:2020uim}. 
Predefined working points corresponding to electron identification 
efficiencies of 70--90\% and misidentification rates of 2--15\% are used.
Energy deposits in the ECAL that are not linked to any charged-particle trajectory 
associated with a \pp collision are identified as photons.

Muons in an event are reconstructed by performing a simultaneous track fit to hits 
in the tracker and in the muon chambers~\cite{CMS:2012nsv,CMS:2018rym}. The 
presence of hits in the muon chambers already leads to a strong suppression of 
particles misidentified as muons. Additional identification requirements on the 
track-fit quality and the compatibility of individual track segments with the 
fitted track reduce the misidentification rate further. 

The contributions from backgrounds to the electron and muon selections are 
usually further reduced by requiring the corresponding lepton to be isolated from 
any hadronic activity in the detector. This property is quantified by an isolation 
variable
\begin{equation}
    \Irel=\frac{1}{\ptem}\left(
    \sumCharged + \max\left(0, \sumNeutral+\sumGamma-\ptPU\right)\right),
\end{equation}
where \ptem corresponds to the measured electron or muon \pt. 
The variables \sumCharged, \sumNeutral, and \sumGamma 
are calculated from the sum over \pt or transverse energy \et 
of all charged particles, neutral hadrons, and photons, respectively. 
These sums include all particles in a predefined cone of radius $\Delta R = \sqrt{
\smash[b]{\left(\Delta\eta\right)^{2}+\left(\Delta\phi\right)^{2}}}$ around the 
lepton direction at the PV, where $\Delta\eta$ and $\Delta\phi$ 
are the angular distances between the corresponding particle and the lepton in 
the $\eta$ and azimuthal $\phi$ directions. 
The lepton itself is excluded from the calculation of \Irel. 
Typical values for cone sizes are $\Delta R=0.3$ and 0.4 for electrons and muons, respectively. 
To mitigate effects from PU on \Irel, only charged particles with tracks 
associated with the PV are taken into account. Since for neutral hadrons and 
photons an unambiguous association with the PV or PU is not possible, an estimate 
of the contribution from PU (\ptPU) is subtracted from the sum of \sumNeutral 
and \sumGamma. This estimate is obtained from tracks not associated with the PV in 
the case of muons, and from the mean energy flow per unit area in the case of 
electrons. If the sum is negative, the result is set to zero. 
The \Irel selection threshold is optimized for each analysis separately. 
If a lepton isolation requirement is imposed in an analysis, leptons failing 
this requirement are not considered further.

\paragraph*{Hadronic \texorpdfstring{\PGt}{tau} lepton decays}
The reconstruction of hadronic \PGt lepton decays (\tauh) starts from AK4 jets, further 
exploiting their substructure with the hadrons-plus-strips 
algorithm~\cite{Sirunyan:2018pgf}. This algorithm acts like a tagger, with 
different working points as defined in~\cite{Sirunyan:2018pgf}. The decays 
into one or three charged hadrons with up to two neutral pions with $\pt>2.5\GeV$ 
are used. The neutral pions are reconstructed as strips in the ECAL with a 
dynamic size in $\eta$-$\phi$ from reconstructed electrons and photons 
contained in a jet, where the strip size varies as a function of the \pt of the 
electron or photon candidate. Electrons, which may emerge from photon conversions, 
are considered in the reconstruction of the strips. The \tauh decay mode is then 
obtained by combining the charged hadrons with the strips. 

To distinguish \tauh candidates from jets originating from the hadronization of 
quarks or gluons, and from electrons or muons, the \textsc{DeepTau} (DT) 
algorithm~\cite{CMS:2022prd} is used. This algorithm exploits basic tracking 
and clustering information in the tracker, ECAL, and HCAL, 
the kinematic and object-identification properties of both the PF candidates forming the \tauh 
candidate and all remaining PF candidates in the vicinity of the \tauh 
candidate, and several characterizing quantities of the whole event. It results 
in a multiclassification output $\yDT_{\alpha}$, where 
$\alpha=\PGt,\text{jet},\Pe,\Pgm$. The output can be identified with a Bayesian probability of the \tauh candidate to 
originate from a genuine \PGt lepton decay, the hadronization of a quark or gluon, an 
isolated electron, or an isolated muon. From this output three discriminants 
are built according to 
\begin{equation}
    D_{i} = \frac{\yDT_{\PGt}}{\yDT_{\PGt}+\yDT_{i}} \quad \text{with} \quad i=\text{jet},\Pe,\text{or} \, \Pgm.
\end{equation}
For the identification of \tauh candidates based on $D_{i}$ predefined 
working points with varying efficiencies for given misidentification rates are 
used as described in Ref.~\cite{CMS:2022prd}. For \Dj the efficiencies vary within 
50--70\% for misidentification rates of ${<}0.5\%$. For \De and \Dm, the 
efficiencies vary within 80--99\% for misidentification rates between 0.05 and 
2.50\%.

\subsection{Search for resonances in \texorpdfstring{\VH}{VH} channels}\label{Sec:Analysis_X_to_VH}

Searches for \VH resonances are optimized either for the mass range up to about 1\TeV, 
motivated by neutral members of an extended
Higgs sector, or masses greater than 1\TeV, where heavy vector bosons might be found. 
The latter case corresponds to the boosted regime, where the \PV and \PH bosons are 
strongly Lorentz-boosted and dedicated reconstruction and identification techniques 
are applied.

All searches presented in the following target the \Hbb or \Htt
decays. The \PV bosons are reconstructed either through the leptonic decays 
\Wln~\cite{CMS:2021klu}, \Zll or \Znn~\cite{CMS:2019qcx,CMS:2019kca,CMS:2021fyk}, or
by the hadronic decays \Vqq~\cite{CMS:2022pjv}. Where allowed by the
signal signature, the trigger selection is based on the presence of a
single high-\pt electron or muon within the geometrical acceptance of
$\abs{\eta}<2.4\,(2.5)$ for electrons (muons), large \ptmiss, or
large \Htmiss. Also, dilepton triggers are used. Otherwise, triggers
based on the presence of high-\pt AK4 jets or large values of \Ht are
required. 
Offline, search regions are defined by the presence of exactly one 
($1\Pell$ category), two ($2\Pell$ category), or no ($0\Pell$ category)
charged leptons. This categorization ensures that the analyses are mutually 
exclusive, allowing for a combined statistical interpretation subsequently  
to the completion of all analyses.

\subsubsection{The sub-TeV mass region}\label{Sec:VH_sub_TeV}

The analyses searching for \CP-odd Higgs bosons through their
$\PA\to\PZ\PH$ decay, in the mass region below 1\TeV, 
are based on the 2016 data set. 
The \PH boson is reconstructed in the \Hbb~\cite{CMS:2019qcx} and
\Htt~\cite{CMS:2019kca} decay channels.

In the \Hbb case, the two \PQb jet candidates with the highest
\PQb tagging scores are selected to form the \PH boson
candidate. Both gluon-gluon fusion and \PQb quark-associated production
are considered. In the
$0\Pell$ category, which targets decays of the \PZ boson into
neutrinos, the mass of the \PA resonance cannot be reconstructed
directly. In this case, its mass is estimated by computing the
transverse mass \mtZH using Eq.~\eqref{eq:mT}. It is calculated using \ptvecmiss 
and the four-momentum of the \PH boson candidate, and must be larger than 500\GeV, 
where triggers are fully efficient. 
The resulting efficiency for signal events with $\mA \lesssim 500\GeV$ is small 
because the \pt of the \PZ boson is not sufficient to produce a
\ptmiss large enough to pass this selection. Therefore, the sensitivity of
the $0\Pell$ category is significant only for large values of \mA.

In the $2\Pe$ and $2\mu$ categories, events are required to have at
least two isolated electrons or muons within the detector's geometrical
acceptance. The \PZ boson
candidate is formed from the two highest \pt, opposite-sign,
same-flavor leptons, and must have an invariant mass \mll between 70
and 110\GeV. The \mll selection lowers the contamination from \ttbar
dileptonic decays and significantly reduces the contribution from
$\PZ\to\tautau$ decays. The \Aboson boson candidate is reconstructed from the invariant
mass \mZH of the \PZ and \PH boson candidates.

\begin{figure}[tb]\centering
  \includegraphics[width=\cmsFigWidth]{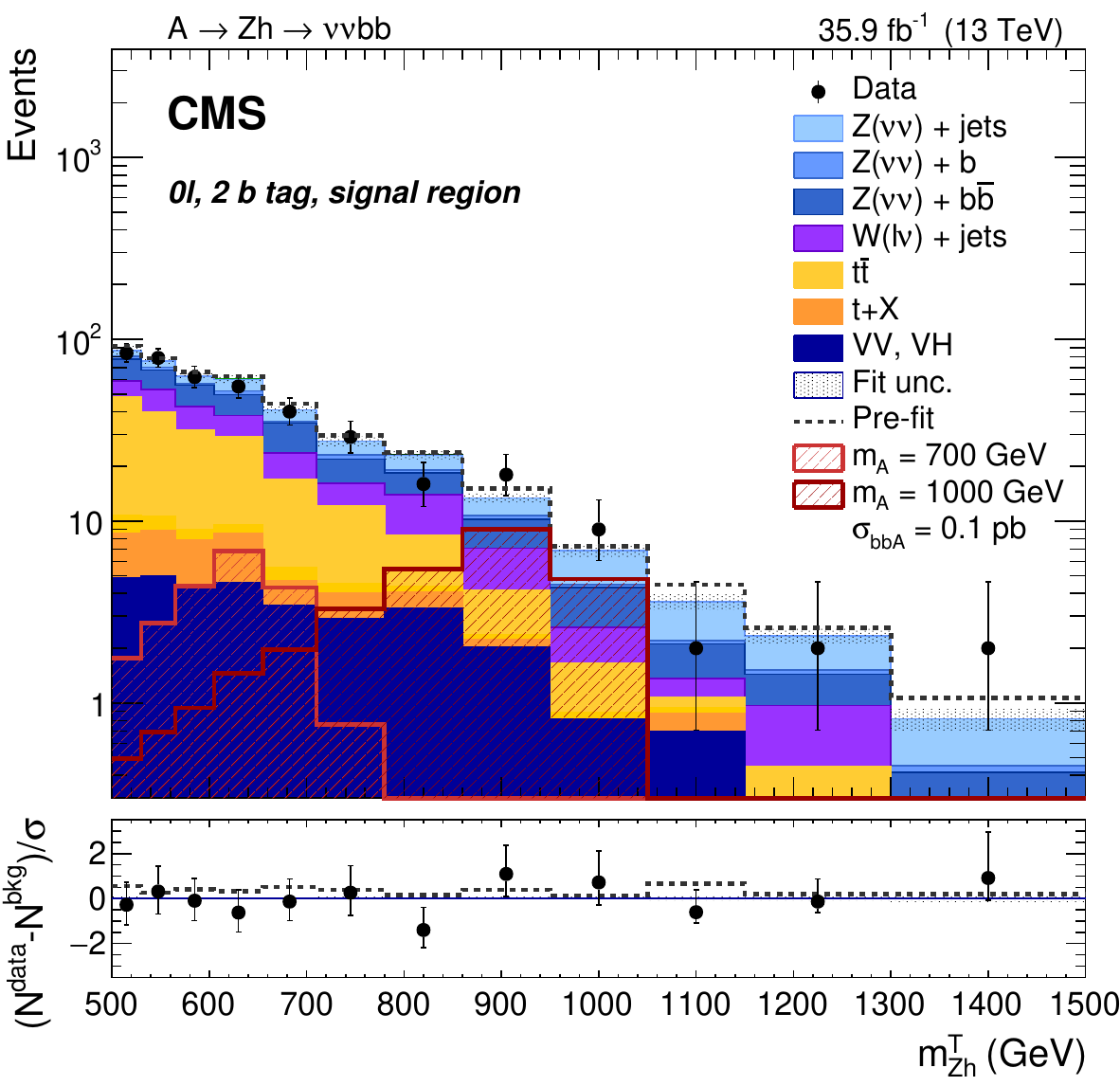}
  \includegraphics[width=\cmsFigWidth]{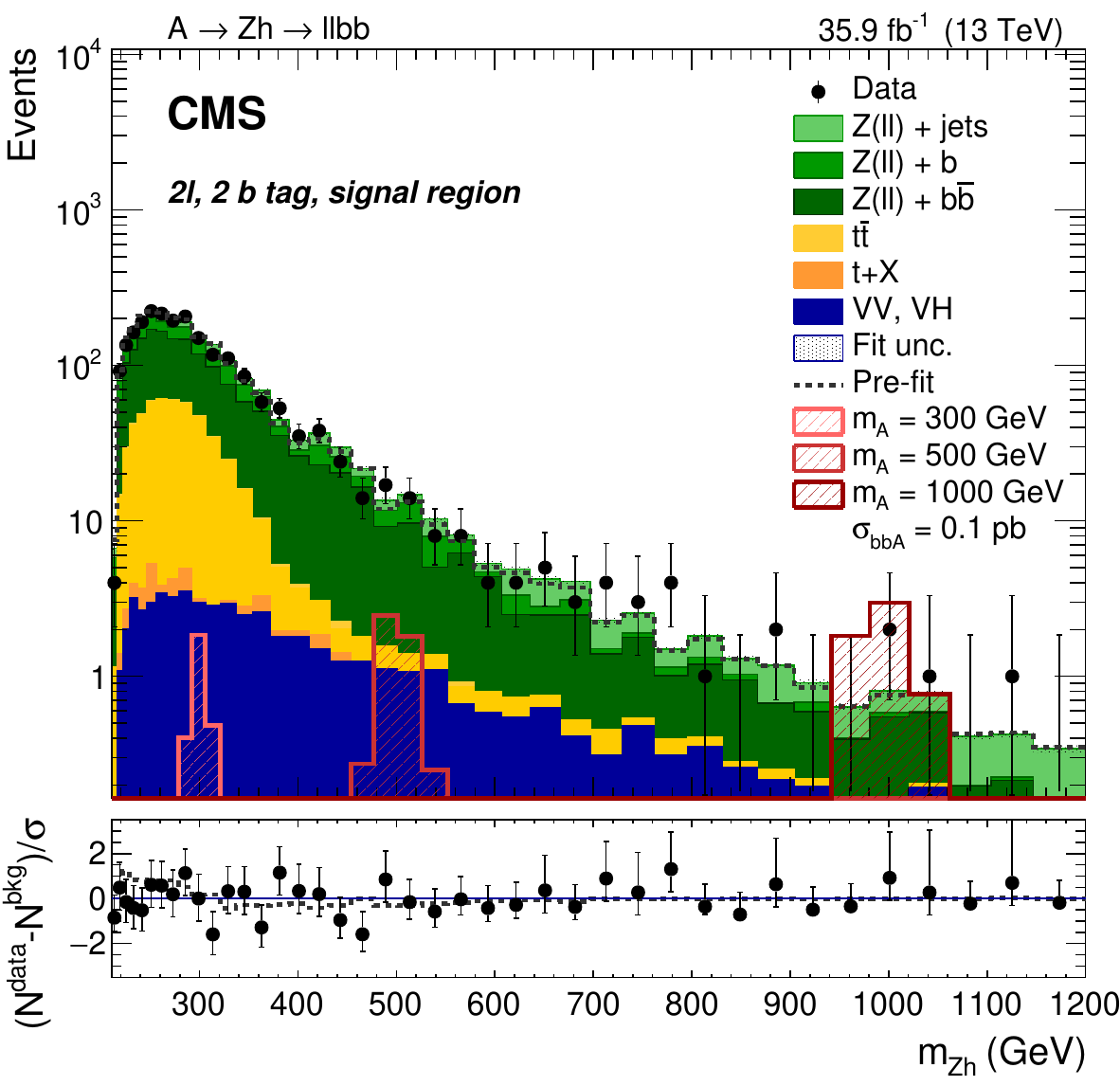}
  \caption{
    Search for $\PA\to\PZ\PH(\bb)$: Distributions of the \mtZH and \mZH variables, as introduced in the text, in 
    the (\cmsLeft) $0\Pell$ and (\cmsRight) $2\Pell$ categories, in the 2 \PQb tag 
    signal region of the $\Aboson\to\PZ\PH(\Pb\Pb)$ analysis~\cite{CMS:2019qcx}. In 
    the $2\Pell$ categories, the contributions of the $2\Pe$ and $2\PGm$ channels 
    have been summed. The gray dotted line represents the sum of all background 
    processes before the fit to data; the shaded area represents the post-fit 
    uncertainty. The hatched red histograms represent signal hypotheses for \PQb 
    quark associated \PA production corresponding to $\sigma_{\Aboson}\BR(\Aboson\to\PZ\PH)
    \BR(\PH\to\Pb\Pb)=0.1\unit{pb}$. The lower panels depict $(N^\text{
    data}-N^\text{bkg})/\sigma$ in each bin, where $\sigma$ refers to the statistical 
    uncertainty in the given bin. Figure from Ref.~\cite{CMS:2019qcx}.
  }
  \label{fig:ZHbb_fit}
\end{figure}

If the two jets originate from an \PH boson, their invariant mass is
expected to peak close to 125\GeV. Events with a dijet invariant mass
\mjj between 100 and 140\GeV enter the SRs;
otherwise, if $\mjj<400\GeV$, they fall in the dijet mass sidebands, which
are used as CRs to estimate the contributions of the
main backgrounds. The SRs are further divided by the number of
jets passing the \PQb tagging requirement (1, 2, or at least 3
\PQb tags). The 3 \PQb tag category has been defined to select
the additional \PQb quarks from \PQb quark associated
production. In this region, at least one additional jet, other than
the two used to reconstruct the \PH boson, has to pass the kinematic
selections and \PQb tagging requirements. Comparisons between data and 
background predictions, together with examples for signal distributions 
are shown in Fig.~\ref{fig:ZHbb_fit} in the 2 \PQb tag
SR for the $0\Pell$ and $2\Pell$ categories. The data are
well described by the backgrounds expected from SM processes. The
$0\Pell$ and $2\Pell$ categories can be combined using the known
branching fractions of the \PZ boson.

In the \Htt case, only the dielectron and dimuon
decays of the \PZ boson are used. For the \PH boson reconstruction, the
$\Pe\tauh$, $\PGm\tauh$, $\Pe\PGm$, and $\tauh\tauh$ topologies are
considered.  The leptons associated with the \PH boson decay are
required to have opposite signs. In case of the $\Pe\tauh$,
$\PGm\tauh$, and $\Pe\PGm$ decay channels, tighter selection criteria
are applied to the light leptons to decrease the background
contributions from \Zjets and other reducible
backgrounds. These four \Htt decay patterns are combined
with the \PZ boson decays into two light leptons, \ie, 
$\PZ\to\Pell\Pell$ with $\Pell=\Pe,\PGm$, resulting in eight distinct
final states of the \PA boson decay.

The mass resolution of the reconstructed \PA boson candidate can be
significantly improved by accounting for the neutrinos associated with
the leptonic and hadronic \PGt decays.  We use the \textsc{svfit}
algorithm~\cite{Bianchini:2016yrt} to estimate the mass of the \PH
boson, denoted as $\mtt^{\mathrm{fit}}$. The \textsc{svfit} algorithm
combines the \ptvecmiss with the four-vectors of both \PGt
candidates (\Pe, \PGm, or \tauh), resulting in an improved
estimate of the four-vector of the \PH boson, which is further improved by
giving the measured mass of the Higgs boson (125\GeV) as an input to
the \textsc{svfit} algorithm. This yields a constrained estimate of
the four-vector of the \PH boson, which results in an even more
precise estimate of the \PA boson candidate mass, denoted as
$\mlltt^{\mathrm{c}}$. The resolution of $\mlltt^{\mathrm{c}}$ is as good as
3\% at 300\GeV, which improves the expected 95\% \CL 
model-independent limits by approximately 40\% compared to using
the visible mass of the \PA boson \mAvis as the discriminating
variable. 

\begin{figure}[tb]
  \centering
  \includegraphics[width=\cmsFigWidth]{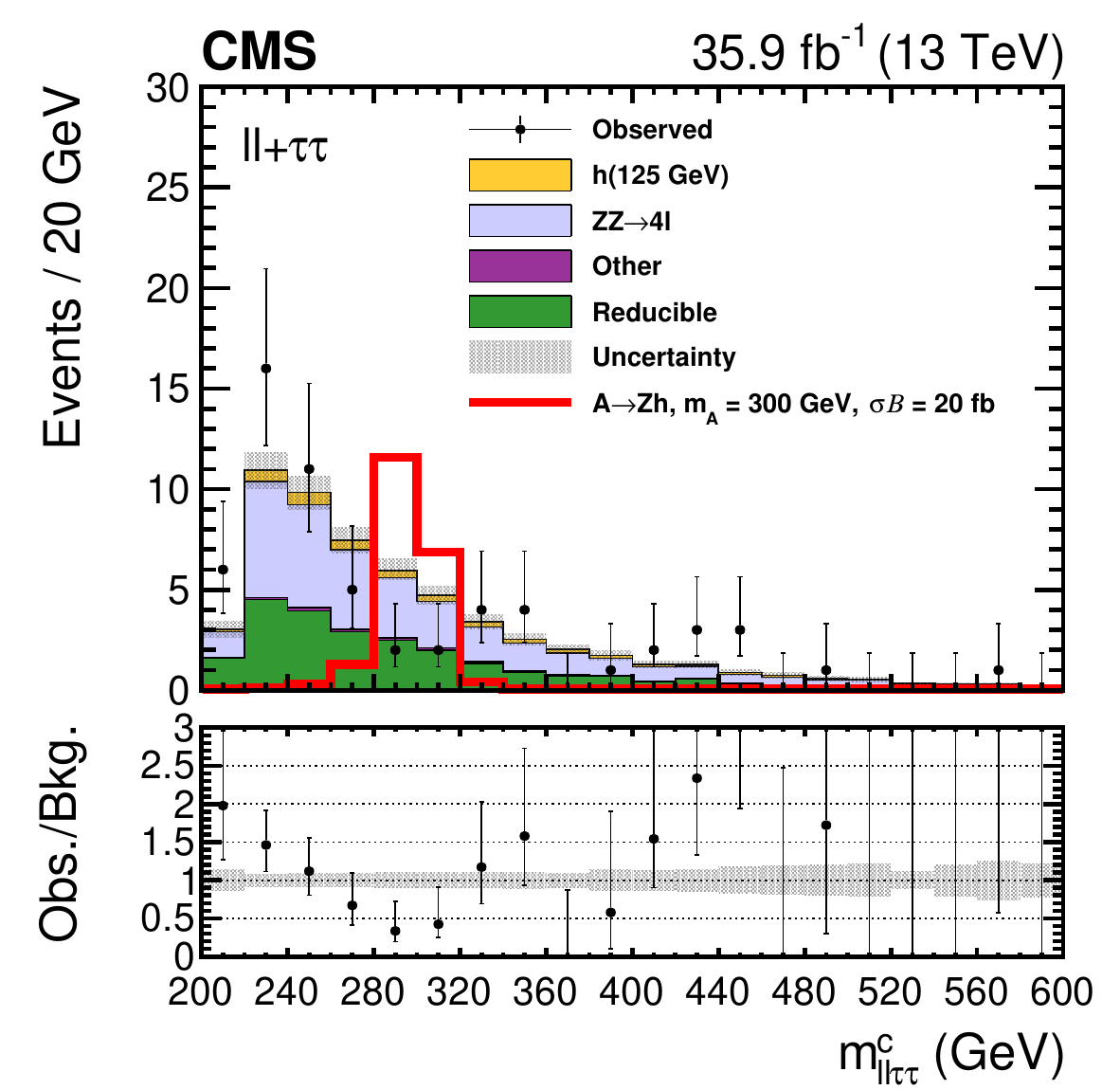}
  \caption{
    Search for $\PA\to\PZ\PH(\tautau)$: Distribution of the $\mlltt^{\mathrm{c}}$ variable, as introduced in the text, 
    of the $\Aboson\to\PZ\PH(\tautau)$ analysis~\cite{CMS:2019kca}, after a fit of 
    the background-only hypothesis in all eight final states. While the fit is 
    based on corresponding distributions, for each final state individually, 
    these have been combined into a single distribution, for visualization 
    purposes for this figure. Uncertainties include both statistical and systematic 
    components. The expected contribution from the $\PA\to\PZ\PH$ signal process 
    is shown for a pseudoscalar Higgs boson with $\mA = 300\GeV$ with the product 
    of the cross section and branching fraction of 20\unit{fb}. 
    Figure from Ref.~\cite{CMS:2019kca}.
  }
  \label{fig:azhtautau_results_All}
\end{figure}

The resulting distribution in $\mlltt^{\mathrm{c}}$, summed over all 
eight categories, is shown in Fig.~\ref{fig:azhtautau_results_All},
together with the expected signal shape for $\mA = 300\GeV$. No excess
above the SM background expectation is observed in the data.

\subsubsection{The high-mass region: leptonic \texorpdfstring{\PV}{V} boson decays} 
\label{Sec:VlepHbb}

In the high-mass analyses targeting leptonic \PV boson decays, 
the presence of an isolated electron with $\pt>115\GeV$ and $\abs{\eta}<2.4$
or muon with $\pt>55\GeV$ and $\abs{\eta}<2.5$ is required. The high \pt thresholds 
are imposed to guarantee a sufficiently high selection efficiency at the trigger 
level. 

For \Wln decays, this selection is complemented by the requirement of 
$\ptmiss>80\GeV$ in the electron channel and $\ptmiss>40\GeV$ in the muon channel. 
Events with additional leptons 
fulfilling looser selection criteria are discarded from the selection. To 
reconstruct a \Wln boson candidate, the \ptvecmiss is used as an estimate of the 
\ptvec of the neutrino. The longitudinal component $p_{z}$ of the 
neutrino momentum is estimated by imposing the constraint that the mass of the 
system formed from the selected lepton and the neutrino should equal $m_{\PW}$. 
This leads to a quadratic equation in $p_{z}$, of which the solution with the 
smallest magnitude is chosen. When no real solution is found, only the real part 
of the two complex solutions is considered.

For \Zll decays, a second lepton with the same flavor and $\eta$ range as the 
first lepton, and $\pt>20\GeV$ is required. The system formed by the two 
leptons is used to reconstruct the \PZ boson candidate. It must 
have $\pt^{\lep}>200\GeV$ and $70<\mll<110\GeV$. The window in \mll has not 
been chosen tighter, since a narrowing would reduce the signal
efficiency while not improving the relation of the signal to the main background from \Zjets 
events. In all cases, all leptons are required to have a large 
enough distance in \DR{} to any AK4 jet.
For the \Znn channel, the absence of leptons and a value of $\ptmiss>250\GeV$ are required. 

In addition to the identification of the reconstructed objects forming the \PV 
boson candidate, all events are required to have an AK8 jet with $\pt>200\GeV$ 
and $\abs{\eta}<2.4$~\cite{CMS:2021klu} or 2.5~\cite{CMS:2021fyk}. In the 
case that more than one AK8 jet is found in an event, the leading one in \pt is 
interpreted to originate from the \Hbb decay, hence referred to as the \PH 
candidate.

In the \Wln channel~\cite{CMS:2021klu}, the \PH candidate is required
to have a distance of $\DR>\pi/2$ from the selected lepton and a
distance larger than 2 in $\Delta\phi$ from both \ptvecmiss and the \Wln
candidate.  Furthermore, it is required to have a value of $\tauDDT <
0.8$. All accepted events are assigned to 24 high-purity (HP) and low-purity (LP) 
event categories according to the flavor of the selected
lepton (\Pe or \Pgm), the value of the double-\PQb-tagger output of the \PH
candidate, the distance between the \Wln and the \PH candidate in
rapidity $\abs{\Delta y} < 1$ (LDy) or $\abs{\Delta y}>1$ (HDy), a
tightened requirement of $\tauDDT < 0.5$ (defining the HP and LP
categories), and a VBF tag requiring the pseudorapidity difference of
the two VBF jets $\abs{\Delta \eta}>4$ and the dijet mass larger than
500\GeV.  The VBF tag is imposed to increase the sensitivity of the
search to the production of the \VH system through VBF. The
categorization in $\Delta y$ is imposed to tag different spin
hypotheses for \PX. A two-dimensional (2D) discriminant composed of
\mj of the \PH candidate and an estimate of \mX obtained from the
four-vectors of the \PH and the \Wln candidates is used for the final 
signal extraction in these categories. We note that this analysis is
not restricted to the \VH process, which is the subject of this report,
but extends towards \VV final states as well. The data are interpreted
under both signal hypotheses each time considering all event
categories, including those enriched and depleted in \VH and \VV by the
value of the double-\PQb tagger of the \PH candidate.

\begin{figure}[htbp]
  \centering
  \includegraphics[width=\cmsFigWidth]{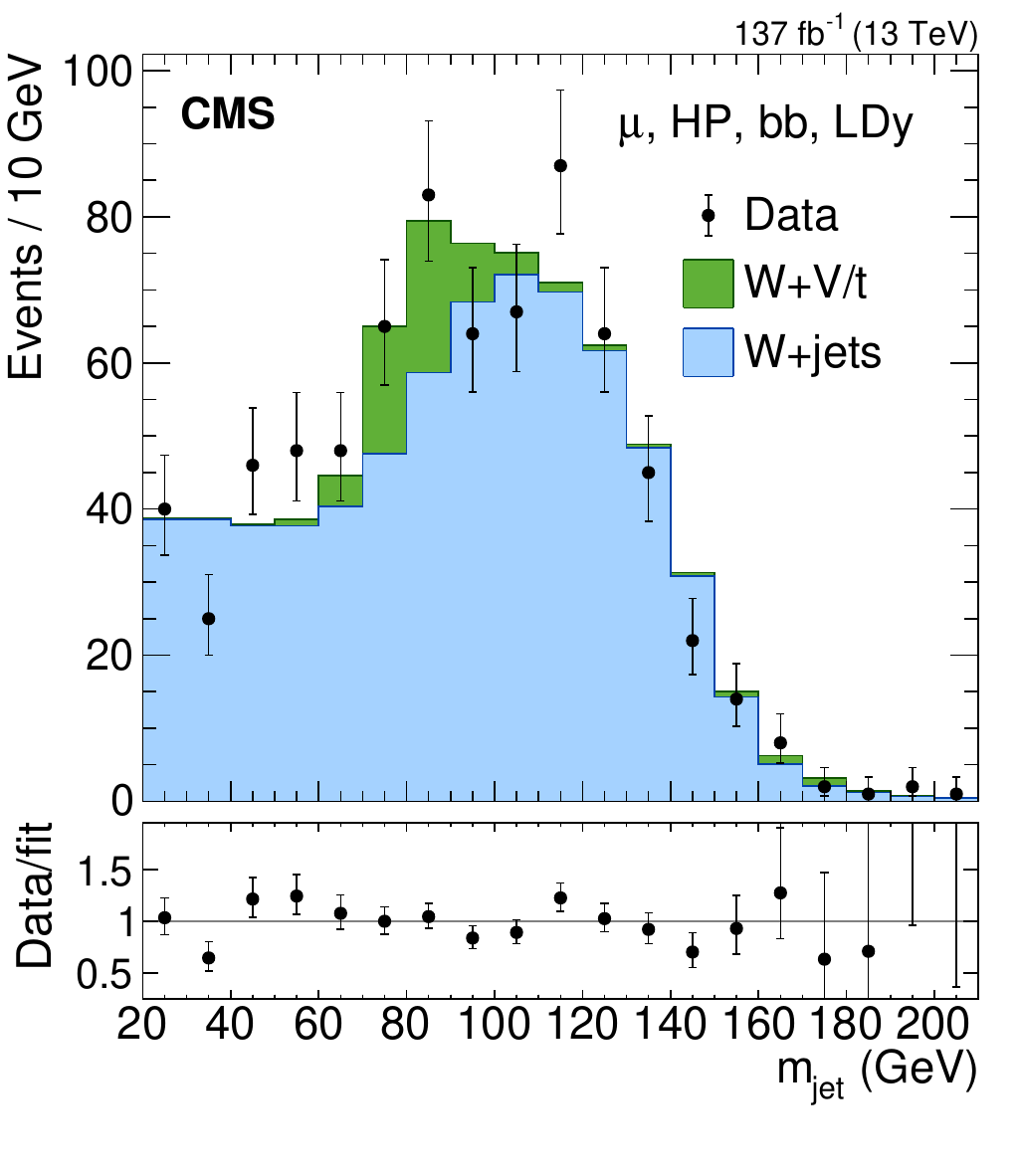}
  \includegraphics[width=\cmsFigWidth]{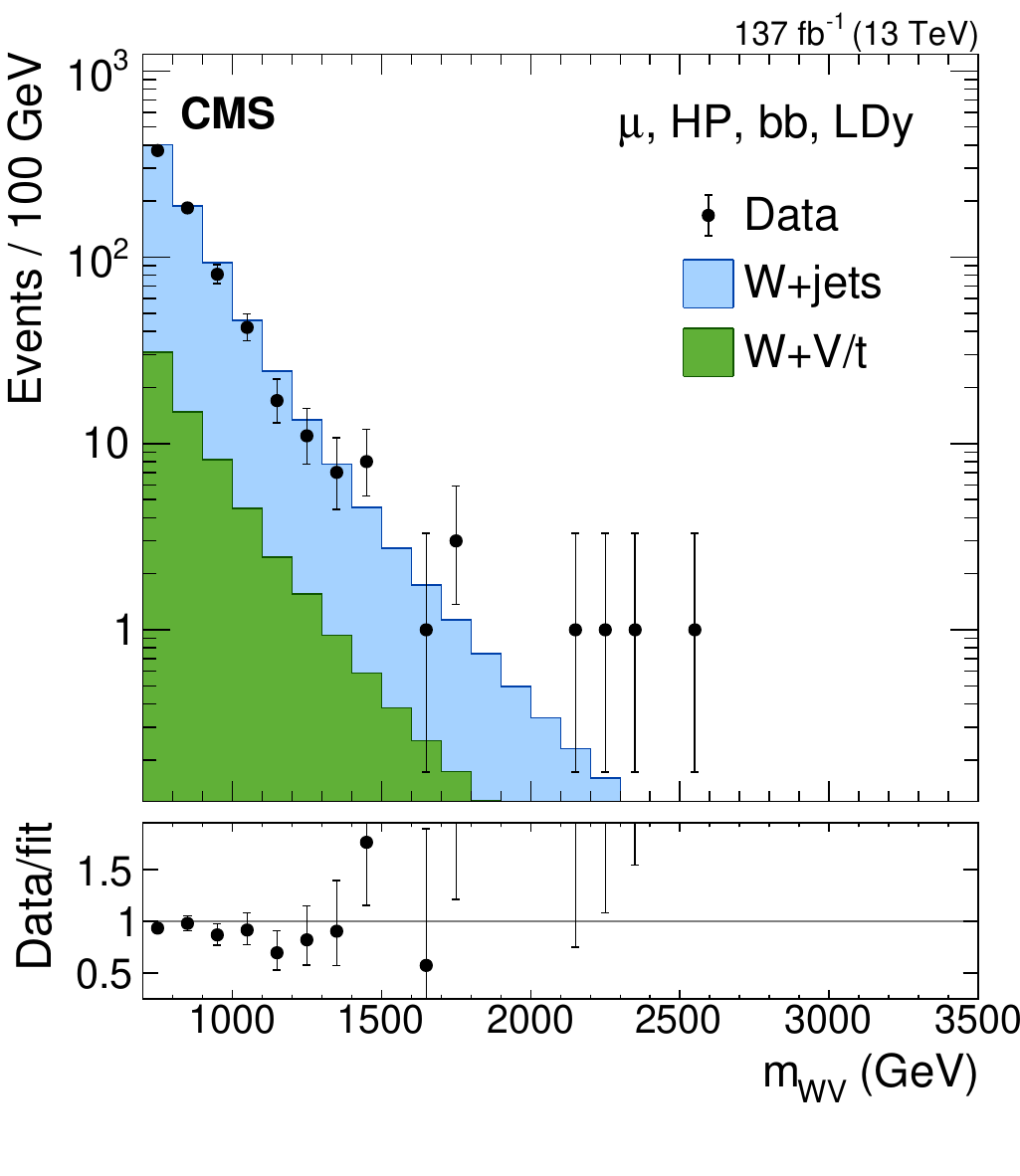}
  \caption{ Search for $\PX\to\PV\PH(\bb)$: Distributions of (\cmsLeft) the jet
    soft drop mass of a boosted Higgs boson candidate, labeled
    $m_{\mathrm{jet}}$, and  (\cmsRight) the mass of the \PX resonance
    candidate, labeled $m_{\mathrm{WV}}$ in the
    $\Wln\PH(\bb)$ channel.  The notation
    $m_{\mathrm{WV}}$ is used as a shorthand since the analysis also
    searches for resonances in the $\PW\PW$ and $\PW\PZ$ final states.  Figures
    from Ref.~\cite{CMS:2021klu}.
  }
  \label{fig:VlepHbb-W-discriminators}
\end{figure}

Distributions of \mj and the reconstructed \mX in the event category with a selected 
muon, $\tauDDT < 0.5$, double-\PQb tag, and $\abs{\Delta y} > 1$ are shown in 
Fig.~\ref{fig:VlepHbb-W-discriminators}. The background processes 
are split into two categories: (i) processes exhibiting resonant structures close 
to the \PQt quark and/or \PV boson mass in \mj ($\PW+\PV/\PQt$), comprising 
\ttbar, single-\PQt quark, and diboson production, and (ii) processes without resonant 
structure in the \mj distribution, dominated by \Wjets events. All backgrounds 
are characterized by a falling spectrum in \mX. These are estimated with the help of simulation, 
where the smoothness of the background shape is ensured by using conditional 
probability density functions in \mj and the reconstructed \mX, as detailed in 
Ref.~\cite{CMS:2021klu}.

\begin{figure}[htbp]
  \centering
  \includegraphics[width=\cmsFigWidth]{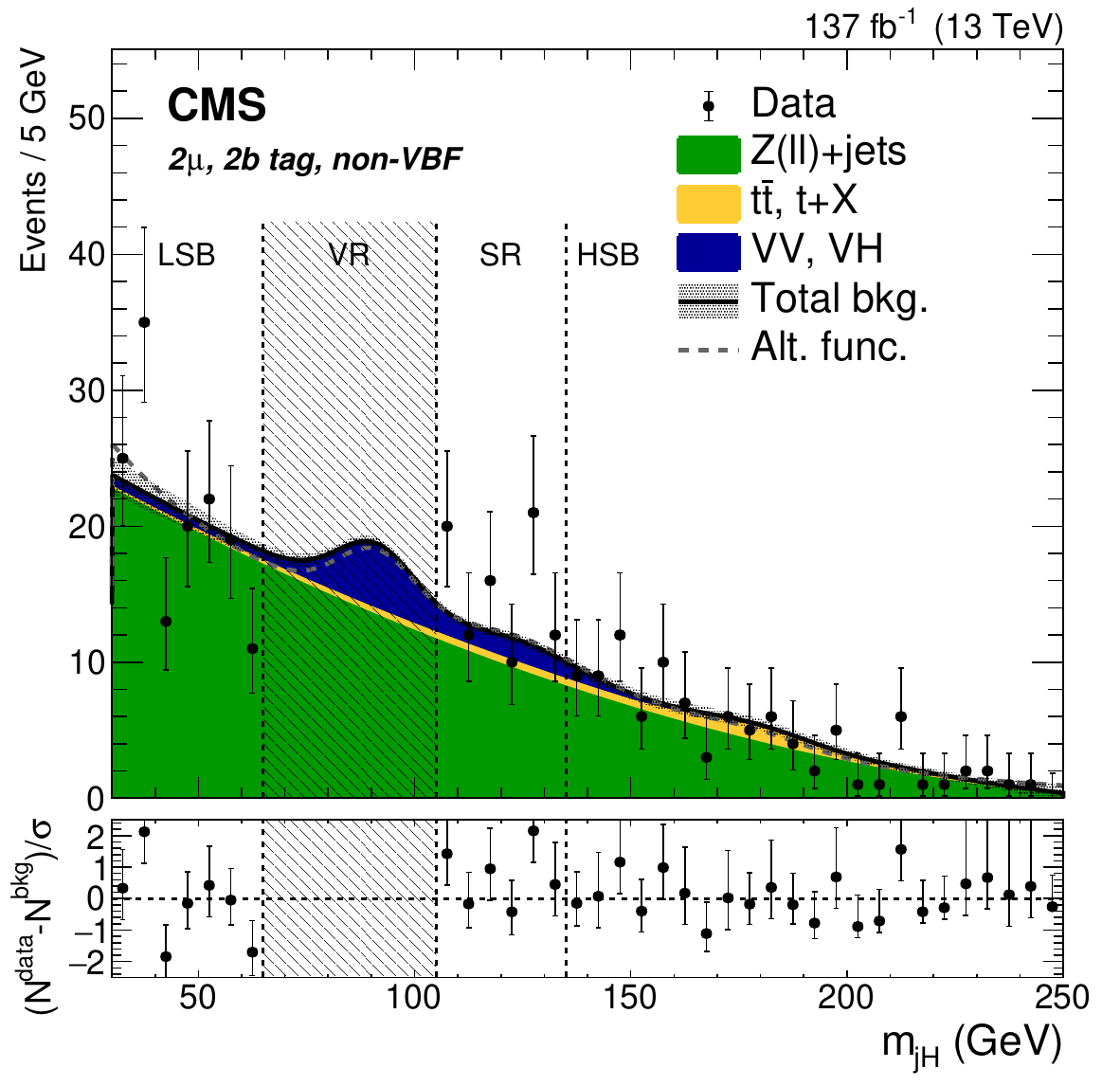}
  \includegraphics[width=\cmsFigWidth]{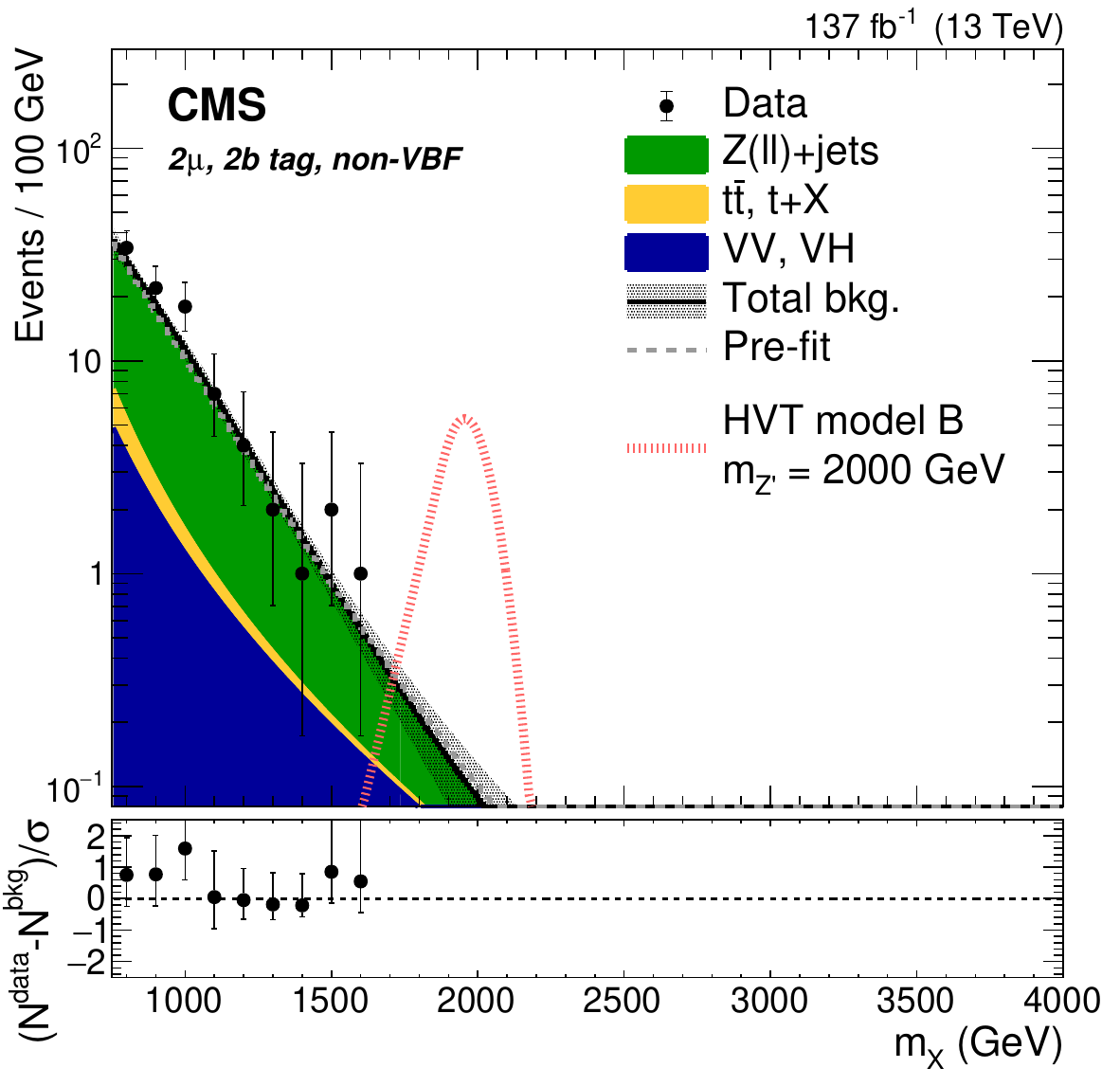}
  
  \includegraphics[width=\cmsFigWidth]{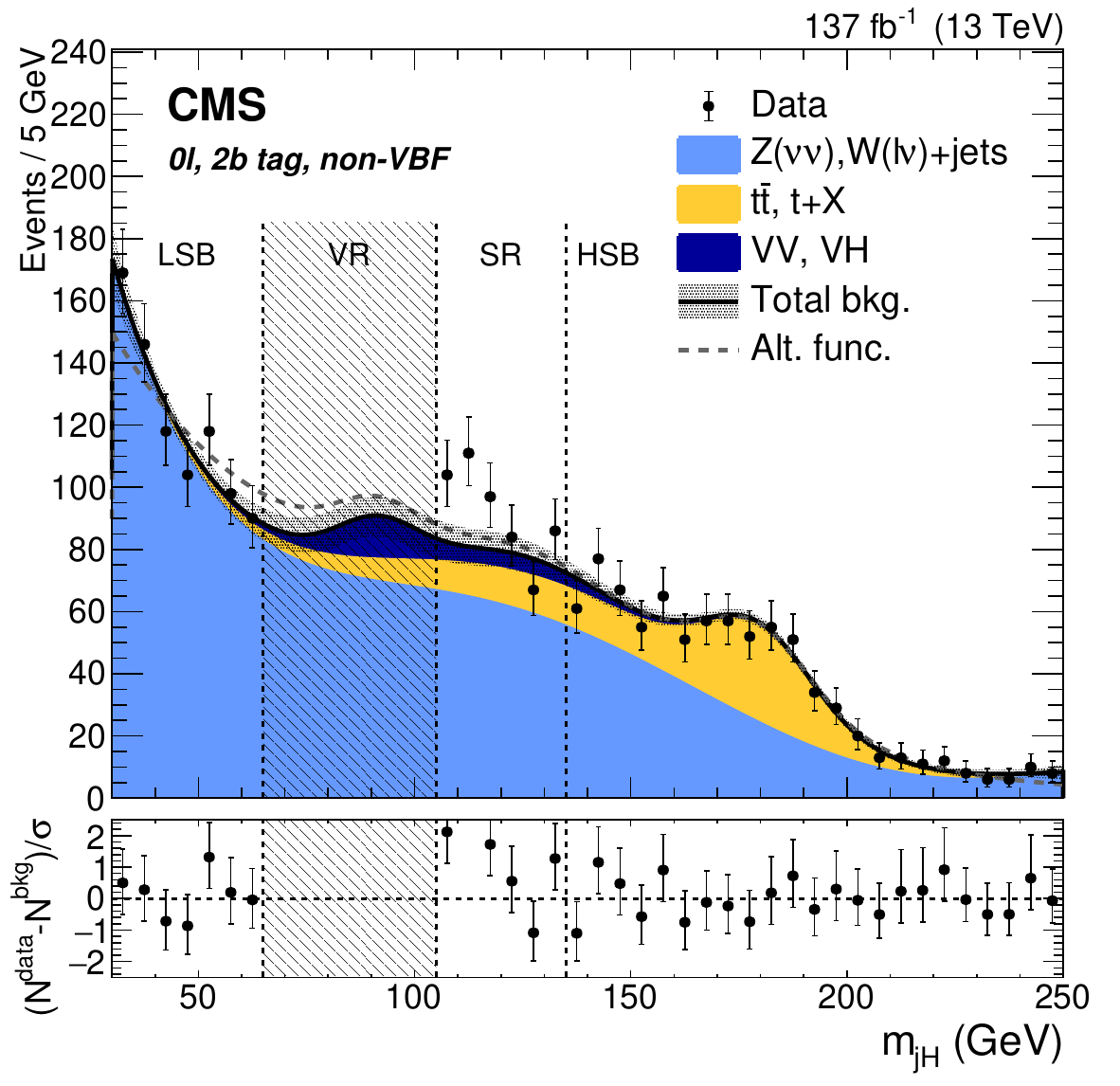}
  \includegraphics[width=\cmsFigWidth]{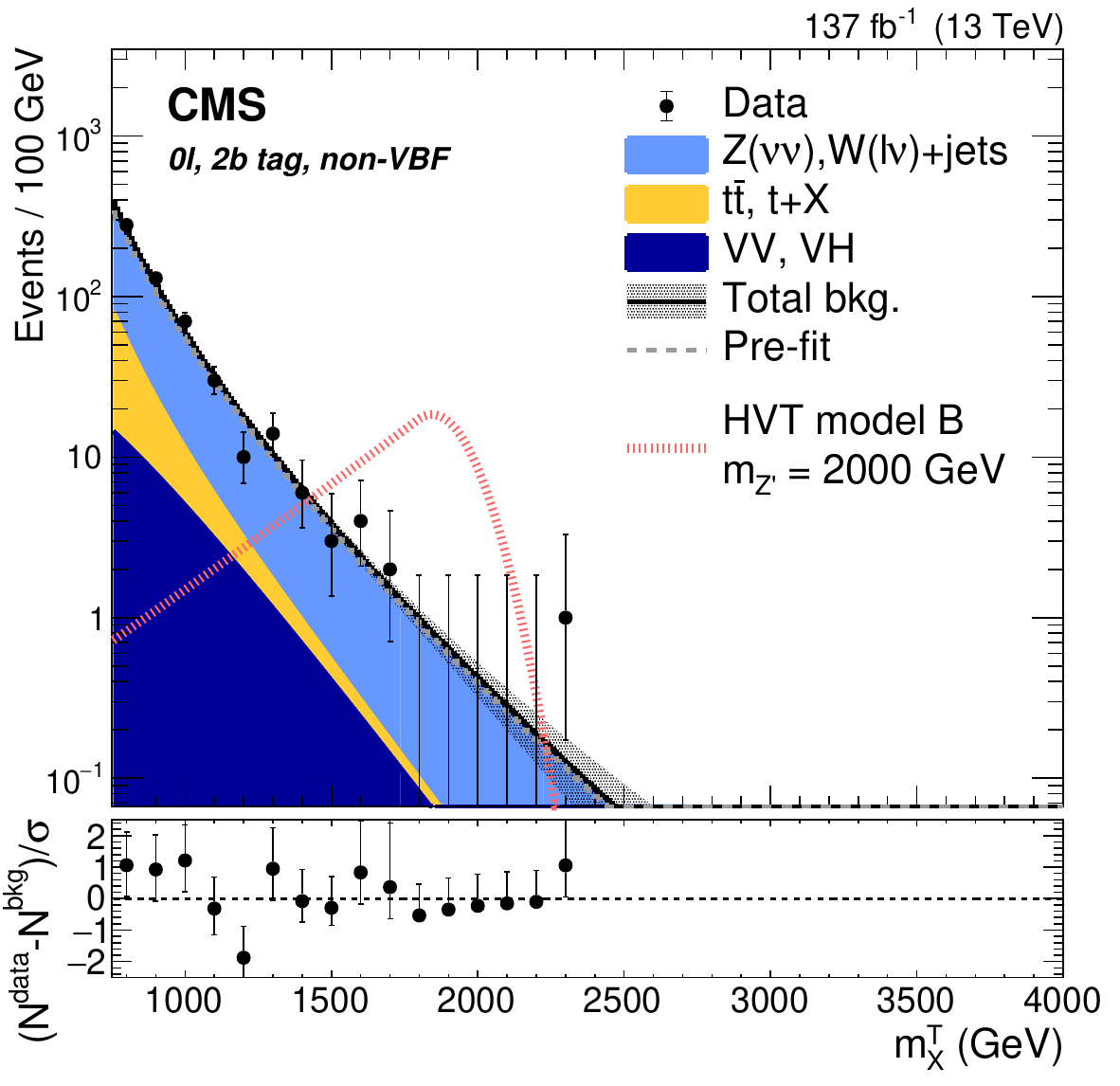}

  \caption{
    Search for $\PX\to\PV\PH(\bb)$: Distributions of (\cmsLeft) the
    jet soft drop mass of a boosted Higgs boson candidate, labeled
    $m_{\text{j}\PH}$, and (\cmsRight) the mass or transverse mass of the X
    resonance candidate, labeled \MX and $m_{\PX}^{\mathrm{T}}$,
    respectively, in the $\Zll\PH(\bb)$ (upper) and $\Znn\PH(\bb)$ channels (lower). 
    The shaded area depicts a veto region excluded from the analysis 
    to minimize the event overlap with dedicated searches in the 
    \VV decay channel. Figures from Ref.~\cite{CMS:2021fyk}.
  }
  \label{fig:VlepHbb-Z-discriminators}
\end{figure}

In the \Zll channel, the \PH candidate is required to have a distance of $\DR>0.8$ 
from each selected lepton. For the \Znn channel $\DR(\ptvecmiss, \PH)>2$ 
and a ratio of $\ptmiss/\pt^{\PH}>0.6$ are required. Finally, the value of \mj 
of the \PH candidate is required to lie within $105<\mj<135\GeV$, compatible with 
\mH. The \textsc{DeepCSV} algorithm is applied to the two subjets of the AK8 jet 
with the highest \pt. 
All remaining events are assigned to 12 event categories based on the flavor 
of the selected leptons ($\PGm\PGm$, $\Pe\Pe$, or $\PGn\PGn$), the number of 
\PQb-tagged subjets of the \PH candidate (${\leq}1$, or 2), and a VBF tag 
identical to the one defined above. The signal is extracted 
in the \Zll channel from the distribution in \mX, obtained from the four-vectors of 
the \PH and the \PZ candidates.
In the \Znn channel the variable \mT, as obtained from $\ptvec^{\PH}$ and \ptvecmiss,
is chosen. Distributions of \mj, \mX, and \mT in the $\PGm\PGm$ and 
$\PGn\PGn$ event categories with two \PQb-tagged subjets, and no VBF tag are shown 
in Fig.~\ref{fig:VlepHbb-Z-discriminators}. The 
dominant background in this search is from \Zjets events, which is 
estimated from data using low-mass (LSB) and high-mass (HSB) CRs in \mj. 
A veto region with $65<\mj<105\GeV$ is excluded from the CRs to minimize 
the event overlap with dedicated searches in the \VV decay 
channel~\cite{CMS:2019kaf,CMS:2018ygj,CMS:2018sdh}. Minor backgrounds originate 
from \ttbar, single \PQt quark, and SM \VH production. In all cases, analytical 
functions are fitted to the observed distributions in \mX in the LSB and HSB CRs, 
where the functions are predefined with the help of simulation. For the minor backgrounds 
from diboson (including \VH) production these functions are purely determined 
from simulation. For \ttbar production and for the dominating background from 
\Zjets events, these are fitted to the data in dedicated sideband regions, after 
subtracting the estimates of the minor backgrounds in these regions. The functions 
are then extrapolated to the SR through transfer functions, which have 
been obtained from simulation. This method has been validated with slightly 
modified sideband definitions providing independent control regions close to the 
SRs. Several functional forms have been tested and a bias test has been 
conducted to ensure that the method is capable of describing the data in the 
validation regions and that no spurious signal may emerge this way.  

\subsubsection{The high-mass region: hadronic \texorpdfstring{\PV}{V} boson decays} 

In the high-mass analysis targeting hadronic \PV boson decays~\cite{CMS:2022pjv}, 
the presence of two AK8 jets with $\pt>200\GeV$ and $\abs{\eta}<2.5$ is required. 
In cases where more than two AK8 jets are found in an event, the leading ones in 
\pt are interpreted to originate from the $\PV\to\PQq\PQq$ 
and \Hbb decays. To guarantee an 
efficiency of ${>}99\%$ of the trigger selection, the invariant mass of the 
selected AK8 jets is required to be larger than $\mjjAKEight>1250\GeV$. 
Furthermore, the jets are required to have a distance $\DR>0.8$ 
from any reconstructed electron or muon with $\pt>20$ or
30\GeV satisfying identification criteria optimized for high-momentum
leptons, respectively.

To reduce the background from events with purely QCD-induced light-quark and 
gluon jets, referred to as QCD multijet production in the following, the 
selected AK8 jets are required to be separated by no more than $\abs{\Delta
\eta_{\text{jj}}^{\text{AK8}}}<1.3$. In addition, $55<\mj<215\GeV$ and a loose 
requirement of
\begin{equation}
  \rho = \log(\mj^{2}/\pt^{2})<1.8
\end{equation}
are imposed. The requirement on $\rho$ is imposed to prevent 
high values for \mj while the \pt of the AK8 jet is low. In those cases, the cone 
size of $\DR = 0.8$ is too small to contain the full jet, affecting both the \mj 
resolution and the efficiency to assign each AK8 jet to a \PW, \PZ, or \PH boson. 

The \textsc{DeepAK8} jet tagging algorithm~\cite{Sirunyan:2020lcu} is used to 
identify AK8 jets originating from a \PW, \PZ, or \PH boson. 
The algorithm consists of a staggered 
two-step deep neural network (NN) architecture, based on the properties of the 
clustered PF candidates, such as the \pt, charge, and angular distance to the jet 
axis, as well as track and secondary vertex information. The latter is used to 
infer whether a jet contains heavy-quark decays or not. The input features of 
both, the PF candidates and the secondary vertices and tracks are processed in 
two independent convolutional NNs. The outputs of these NNs are passed to a third, 
fully connected deep NN (DNN) to assign a jet to one of the following classes: 
single quark or gluon, \PW boson decaying to $\PQc\PQq$ or $\PQq\PQq$, and 
\PH or \PZ boson, each decaying into \bb, \cc, or light flavor quarks. 
For the reported analysis two discriminants are of relevance, both discriminating 
between signal jets and jets from single quarks or gluons. A \qq discriminant is built 
to identify the decay of a \PW or \PZ boson into light quarks, and a \bb discriminant 
is built to identify the \PH and \PZ boson decays into \bb. 

The strategy of the search is to distinguish the signal, which is peaking in 
three distributions, namely \mj of each selected AK8 jet and the dijet mass of the two AK8 jets \mjjAKEight, 
from the considered backgrounds that exhibit a smoothly falling spectrum in at least 
one of these observables. To guarantee that the \mj distributions of the considered 
background processes are not kinematically biased by the selection based on the 
\textsc{DeepAK8} tagger, an adversarial training of the NNs is performed that 
leads to a reduced correlation between \mj and the output of the 
tagger~\cite{Sirunyan:2020lcu}. The remaining correlations are further suppressed by 
a DDT method~\cite{ddt} applied on the tagger output. Eventually, the selection of the output of 
the \textsc{DeepAK8} tagger is chosen such that it yields a constant tagging rate 
as a function of \pt and \mj for the quark or gluon jets with the highest \pt in 
QCD multijet production, based on simulation. 

The \qq tagging efficiency, and the probability of hadronic \PQt quark decays to 
be misidentified by the \qq tagger are calibrated in a \ttbar event selection 
enriched in hadronic \PW boson decays. The \bb discriminant is calibrated in an 
event sample enriched with jets from $\Pg\to\bb$ splitting using a double-muon 
tag, as described in Ref.~\cite{CMS:2017wtu}.

\begin{figure}[htbp]
  \centering
  \includegraphics[width=\cmsFigWidth]{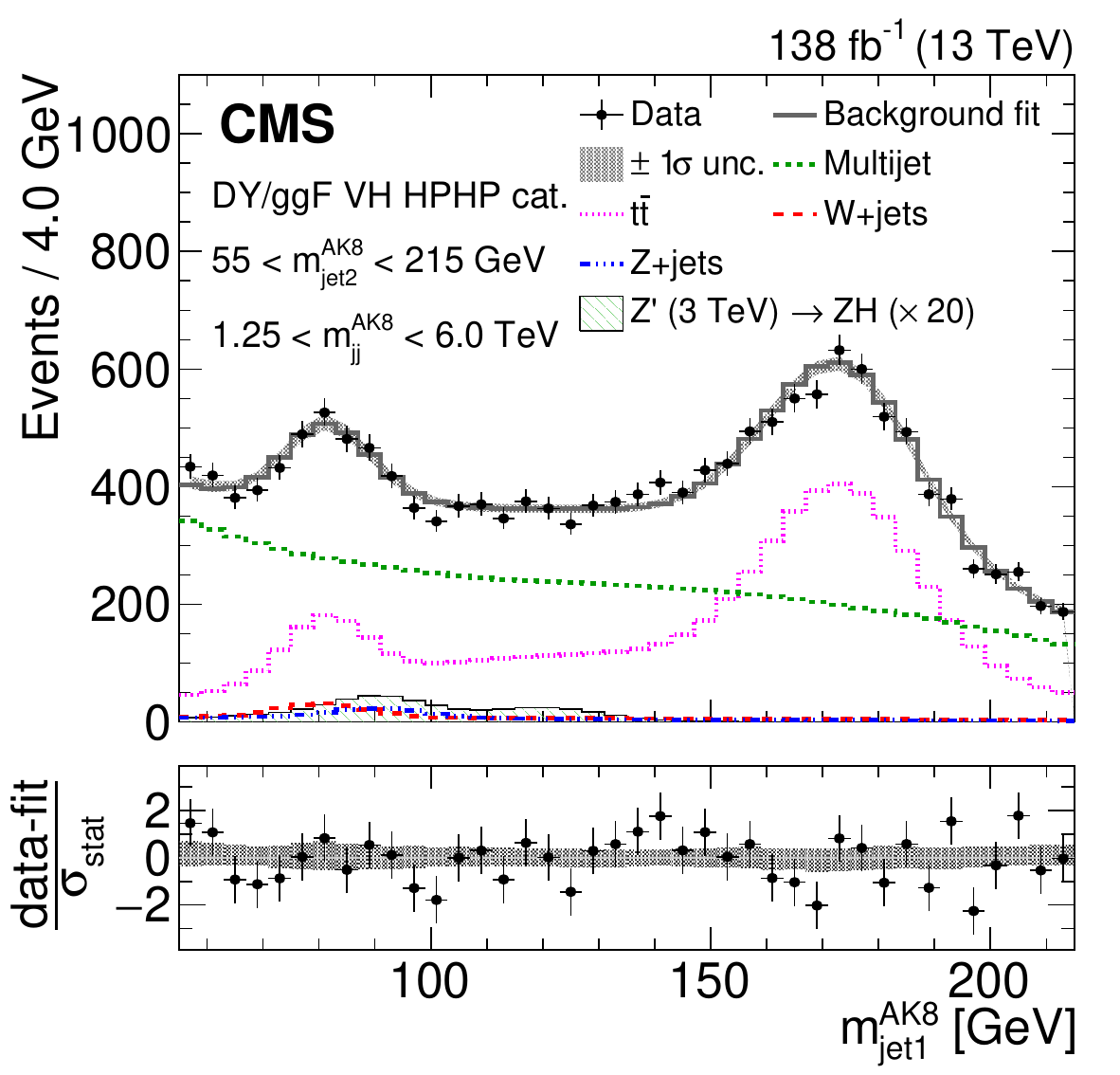}
  \includegraphics[width=\cmsFigWidth]{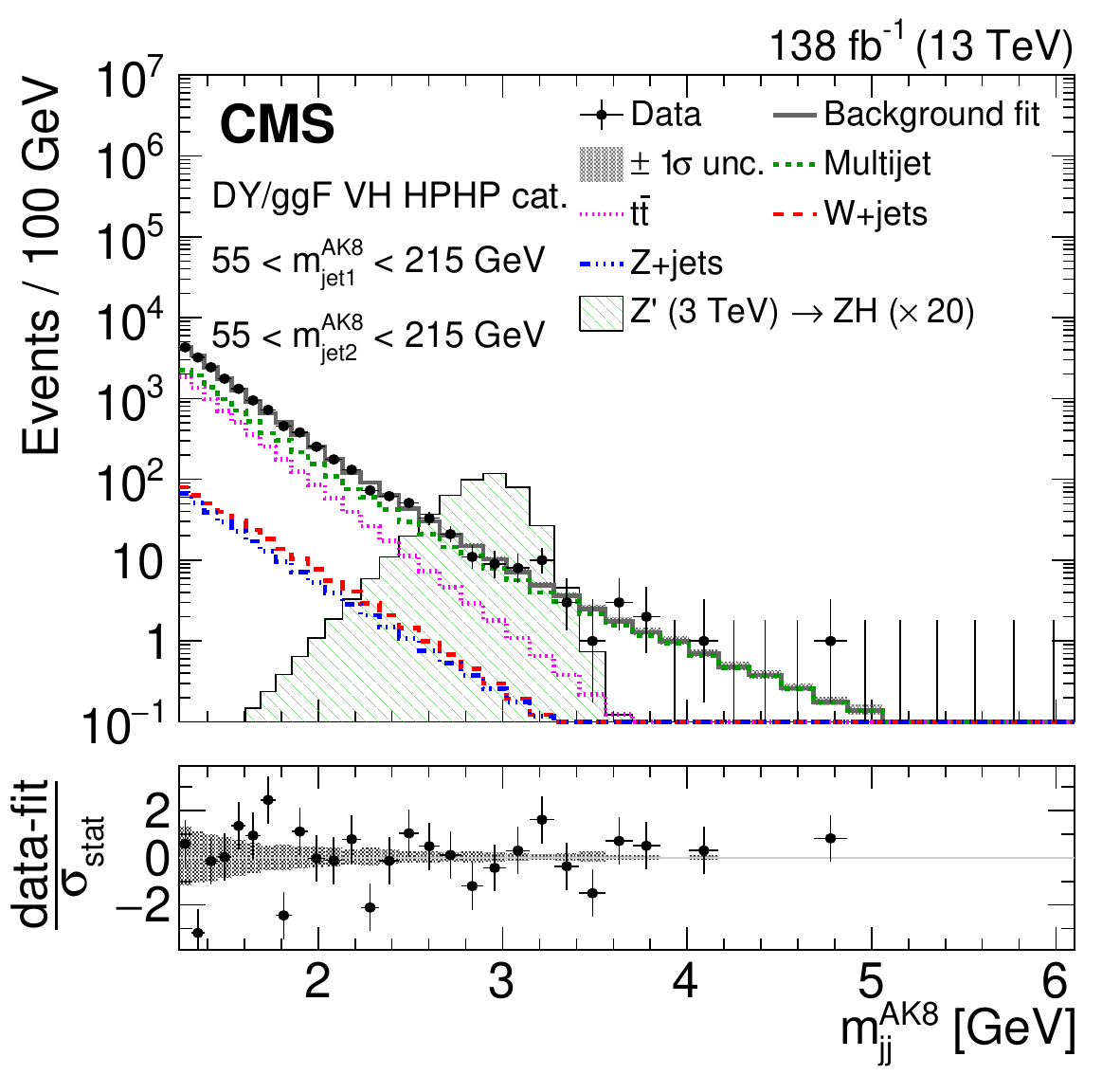}
  \caption{
    Search for $\PX\to\PV(\text{qq})\PH(\bb)$: Distributions of (\cmsLeft) the soft drop mass \mj variable, 
    labelled as $m_{\text{jet1}}^{\text{AK8}}$, and (\cmsRight) the dijet mass \mjjAKEight in the $\Vqq\PH(\bb)$
    channel~\cite{CMS:2022pjv}. The individual contributions of the background 
    model are shown by open histograms with different colors and line styles. The 
    signal of a \PZpr boson with a mass of 3\TeV decaying via $\PZ\to\text{qq}$ and 
    $\PH\to\bb$ is also shown, by a green filled histogram.
    Figure from Ref.~\cite{CMS:2022pjv}.
  }
  \label{fig:VhadHbb-discriminators}
\end{figure}
To increase the sensitivity of the search after selection, all remaining events 
are assigned to one of 10 mutually exclusive event categories based on a VBF tag 
and the outputs of the \qq and \bb discriminants from each of the selected AK8 
jets. The VBF tag is defined by requiring at least two AK4 jets with $\pt>30\GeV$ 
and $\abs{\eta}<5.0$ that do not overlap with the selected AK8 jets within $\DR<
1.2$. For the two AK4 jets, which are leading in \pt, $\mjj>800\GeV$ and a 
separation of $\abs{\Delta \eta}>4.5$ are required. The event categories target \VV and 
\VH decays. 

Distributions of \mj and \mjjAKEight in the non-VBF \VH event category, where one 
of the selected AK8 jets (the \PV candidate) exhibits a high-purity classification 
score in the \qq discriminant, while the other one (the \PH candidate) exhibits a 
high purity classification score in the \bb discriminant, are shown in 
Fig.~\ref{fig:VhadHbb-discriminators}. To simplify the modelling of the 
three-dimensional (3D) (\mjjAKEight, $\mj^{\text{Jet 1}}$, $\mj^{\text{Jet 2}}$) 
shapes, the assignment of AK8 jets to ``Jet 1'' and ``Jet 2'' is performed randomly, 
so that both \mj distributions exhibit the same shape. The largest background 
originates from QCD multijet production, which is estimated from a parametric 
functional form obtained through a forward-folding technique applied to a set of 
simulated samples at the level of stable particles~\cite{CMS:2019qem}. This functional form can be constrained 
through nuisance parameters attached to physics-motivated alternative shapes. 
All other processes are obtained from simulation. 
The production of \ttbar events is expected to contribute 40\%, the expected 
background from \Wjets and \Zjets production amounts to up to 4\% of the selected 
events. Backgrounds from single \PQt quark and diboson production are expected to 
contribute less than 1.5\%. 
While all backgrounds are nonresonant in \mjjAKEight, some exhibit one or two 
resonant structures at different values in the \mj distribution of only one or 
both AK8 jets. The signal is obtained from analytic functional forms as in the 
case of the analyses described in Section~\ref{Sec:VlepHbb}.

\subsection{Search for resonances in the \texorpdfstring{\HH}{HH} channel}\label{Sec:Analysis_X_to_HH}

This section describes searches for resonances decaying into two \PH bosons 
with a mass of 125\GeV. Various further decay mode combinations are
covered by the more general searches in the $\PX\to\PY\PH$ analyses, 
and are discussed in Section~\ref{Sec:Analysis_X_to_YH}.

\subsubsection{The \texorpdfstring{$\PX\to\PH(\bb)\PH(\PW\PW)$}{X->H(bb) H(WW)} decay in resolved jet topology}\label{Sec:Analysis_X_to_HH_to_bbWW_resolved} 

For a pair of Higgs bosons, the $\bb\PW\PW$ decay channel has
the second-largest branching fraction of all \HH decay modes, of about
24\%. The analysis described in this section 
focuses on the single lepton (SL) $\bb\Pell\PGn \qq$ and
dilepton (DL) $\bb\Pell\PGn \Pell\PGn$ final states~\cite{CMS:2024rgy}.

The data have been collected with a combination of single- and double-lepton
triggers. The event selection requires one or two isolated leptons, in the
second case with opposite signs. The analysis considers $\PH
\to \bb$ in both resolved and merged jet topologies, and
requires suitable numbers of AK4 and AK8 jets. 
Jets associated with the $\PH \to \bb$ candidate are
\PQb tagged by passing the medium working point of the
\textsc{DeepJet} algorithm~\cite{Bols:2020bkb} in the AK4 case, or the medium
working point of the \textsc{DeepCSV} algorithm~\cite{CMS:2017wtu} in case of
AK8 subjets.

The selection vetoes pairs of leptons, which according to their
invariant mass are likely to originate from quarkonia or \PZ boson
decays. Overlap with events selected by the analysis in the
$\bb\tautau$ channel (discussed in Section~\ref{Sec:HIG-20-014}) is removed by
vetoing events containing at least one \tauh candidate
passing the selection described in Ref.~\cite{CMS:2021yci}.

The events are classified into processes based on the output
of multiclass deep neural networks (DNNs), separately trained for the SL and DL cases.
The DNNs feature output nodes for a number of backgrounds and 
one signal node. The DNNs are trained on all signal samples; they are
parameterized in the nominal signal mass and contain five 
background nodes for the SL and seven for the DL category.
Depending on the highest scoring node, events are subdivided into signal and background
categories. The network inputs include a number of high-level
features, such as invariant masses and the hadronic activity of the
event, but in addition also the output of a Lorentz-boost
network~\cite{Erdmann:2018shi} performing automated feature engineering
based on the four-momenta of selected leptons and jets. Due to
similarity of the DNN score distributions for various background
classes, they are merged into process groups prior to the signal
extraction.

The signal categories are further divided into subcategories according
to the \PQb jet topology and multiplicity, referred to as
resolved 1\PQb, resolved 2\PQb, and merged. The merged
jet subcategories are excluded in the \HH combination
(Section~\ref{Sec:Combination}) as they would overlap with the analysis
described in Section~\ref{Sec:Analysis_X_to_HH_to_bbWW_boosted}.
\begin{figure}[tbp]
  \centering
  \includegraphics[width=0.75\textwidth]{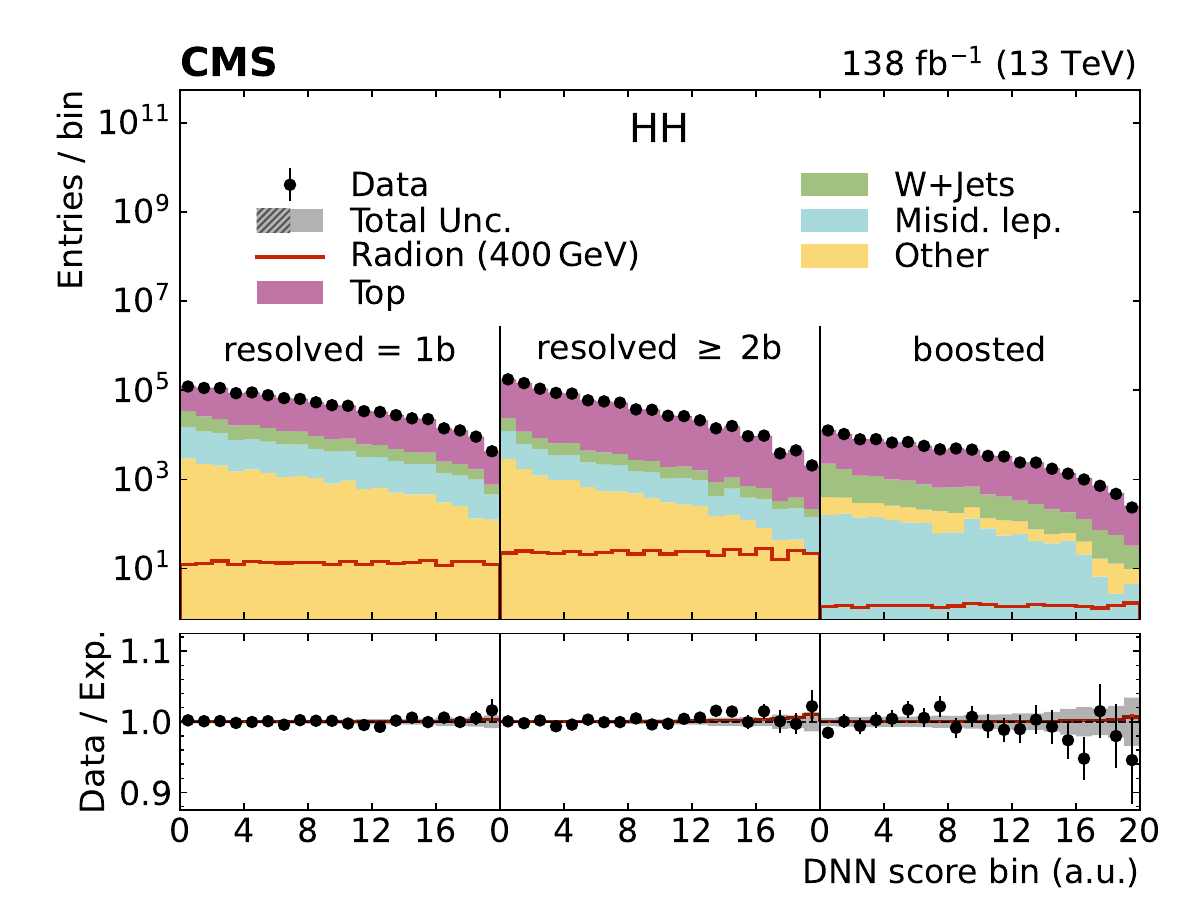}
  \includegraphics[width=0.75\textwidth]{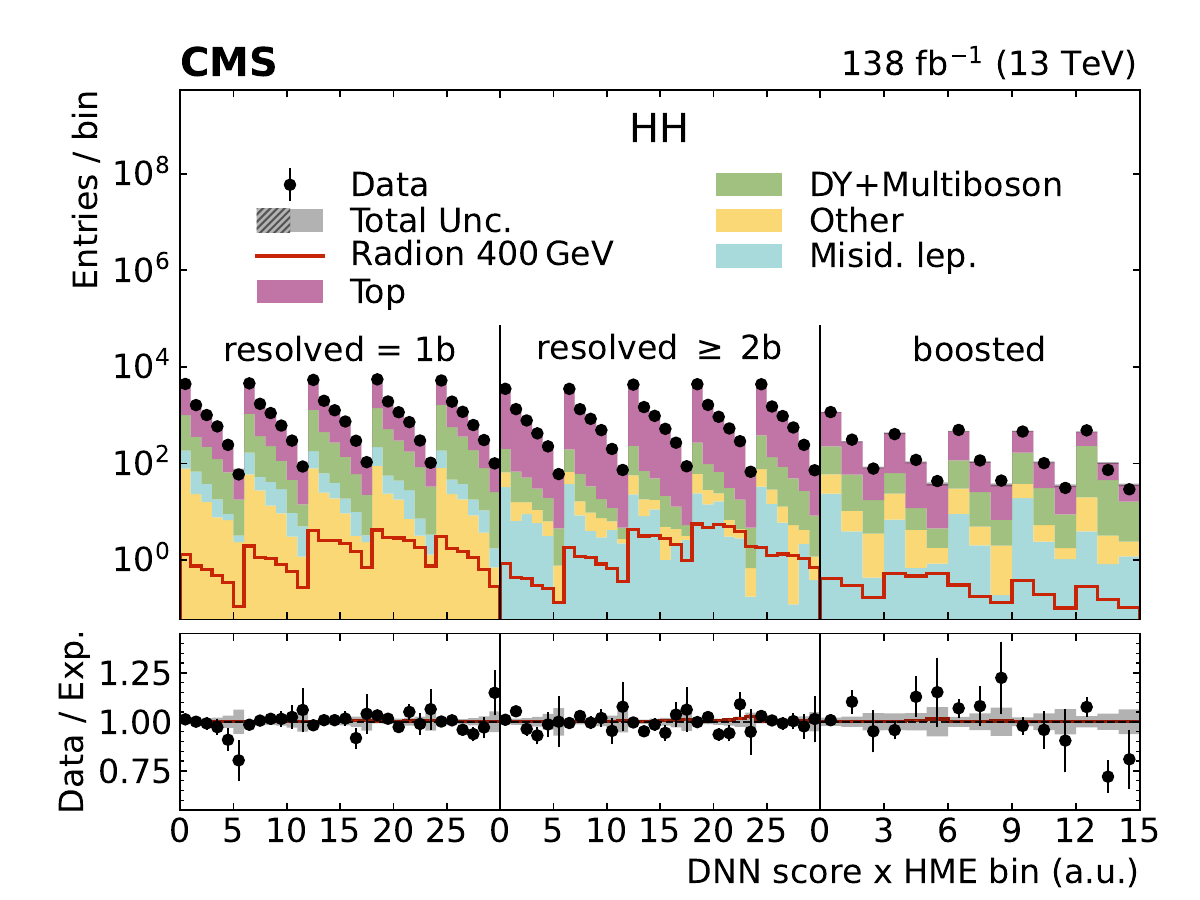}
  \caption{
    Search for $\PX\to\PH(\bb)\PH(\PW\PW)$: 
    Distributions of the DNN output for events in the signal nodes of the 
    (upper) SL and (lower) DL categories of the $\PH(\bb)\PH(\PW
    \PW)$ analysis based on merged and resolved jets~\cite{CMS:2024rgy}. 
    The distributions for a signal of a resonant radion with a mass of 400\GeV 
    are also shown, by open red histograms. Figure from Ref.~\cite{CMS:2024rgy}.
  }
  \label{fig:HIG_21_005_DNN_resonant}
\end{figure}
Background contributions dominantly originating from $\ttbar$
production are estimated from simulation, with two exceptions:
contributions arising from jets misidentified as leptons are estimated
with the misidentification-factor method~\cite{CMS:2018fdh}, while the
DY background is addressed with a different method using the
0\PQb CR in data and transfer factors determined in the \PZ boson peak region.

In the SL case, the signal extraction is performed by a simultaneous
maximum likelihood fit to the distributions of the DNN outputs of the
signal and the background process groups. In the DL case, the DNN
output of the signal category is combined into an unrolled 2D variable
with the output of a heavy-mass estimator (HME)~\cite{Huang:2017jws},
a variable that estimates the most likely invariant mass of the \HH
system considering the presence of two neutrinos and the \ptvecmiss measurement.

Examples of the signal extraction are shown in
Fig.~\ref{fig:HIG_21_005_DNN_resonant}. The distribution in the DNN score
(SL case) and the unrolled combination of the DNN score and the HME bin (DL case) are
shown for the SR for the three event categories, for
an assumed resonance mass of $\mX=400\GeV$. The signal expected
for a radion of this mass with a cross section of 1\unit{pb} is also
displayed. The observed distributions agree well with the expectation from
SM backgrounds and no significant signal is observed.

\subsubsection{The \texorpdfstring{$\PX\to\PH(\bb)\PH(\PW\PW)$}{X->H(bb)H(WW)} decay in merged-jet topology}\label{Sec:Analysis_X_to_HH_to_bbWW_boosted}

This study extends the search in the $\bb\PW\PW$ channel towards
higher \PX masses~\cite{CMS:2021roc}. As in Section~\ref{Sec:Analysis_X_to_HH_to_bbWW_resolved},
the SL selection targets the 
$\HH\to\bb\PW\PW\to\bb\Pell\PGn\qq$ decay mode.
The DL selection covers both the
$\HH \to \bb\PW\PW \to \bb \Pell \PGn \Pell\PGn$ 
and the $\HH \to \bbtt \to \bb \Pell \PGn \PGn \Pell \PGn \PGn$ decay modes, 
and the latter comprises 30--35\% of the total expected DL signal yield. 
As this analysis targets resonance masses of $\mX > 0.8\TeV$, the \PH bosons emerge
with a large Lorentz boost with respect to the laboratory rest frame, and their decay
products are contained in collimated cones. For this reason, the
hadronically decaying \PH and \PW bosons are each reconstructed with a
single AK8 jet.

The signal is searched for in the two-dimensional distribution in the
($m_{\bb}$, \mHH) mass plane, hence the kinematics of both
Higgs boson candidates need to be reconstructed.

In the SL final state, the highest {\pt} lepton in the
event is selected as the lepton candidate from the leptonic \PW decay.
The \Wqq decay is reconstructed with a high-\pt AK8 jet. The AK8 jet nearest to the lepton
satisfying $\DR < 1.2$ is taken to be the \PW candidate. The $\PH \to
\PW\PW$ decay chain is reconstructed using a likelihood-based
technique, which provides an estimate of the neutrino momentum vector
and also a correction to the \pt of the \Wqq candidate jet.  

In the DL final state, two opposite-sign light leptons
with the highest \pt are taken to be the leptons arising from the $\PH
\to \PW\PW$ decay. 
Due to the collimation of the Higgs boson decay products, the polar
angles of the dineutrino and dilepton systems can be assumed to be the
same. Using this assumption, and by approximating the invariant mass
of the two neutrinos with its expected mean value of 55\GeV, the sum
of the four-momenta of the two neutrinos is estimated using
\ptmiss. These assumptions are guided by simulation studies. 
The
leptons are required to be close by satisfying $\DR < 1$, and their
invariant mass is required to be in the range of 6--75\GeV to reduce
contamination from DY production. Events are required to have large
\ptmiss, pointing in approximately the same direction in the
transverse plane as the dilepton system, satisfying $\Delta\phi <
\pi/2$.

In all considered final states, the $\PH \to \bb$ decay is
reconstructed as a single AK8 jet with two-prong substructure and high \pt.
The double-\PQb tagging is performed with the
\textsc{DeepAK8} algorithm.
The SR is defined by requiring \mj of the $\PH \to \bb$
candidate to be within 30--210\GeV. This allows us to also capture the neighboring
background, which is important for the signal extraction described
below.  Events containing any \PQb-tagged AK4 jets outside the $\PH
\to \bb$ candidate AK8 jet are vetoed, as they are likely to arise
from \ttbar production.

Events with the production of one or more top quarks constitute the
majority of the background, where particles from a hadronic top quark decay
are captured in the $\PH \to \bb$ candidate jet, particularly in
the SL final state. The invariant mass distributions of
such backgrounds may resonate in $m_{\PQt}$, $m_{\PW}$, or neither, 
depending on which daughter partons are captured, and the
backgrounds are classified accordingly. Events originating from other
processes, primarily from \Wjets and QCD multijet production in
the SL channel and DY production in the DL channel, are taken
as a separate component.

\begin{figure}[htb]\centering
  \includegraphics[width=\cmsFigWidth]{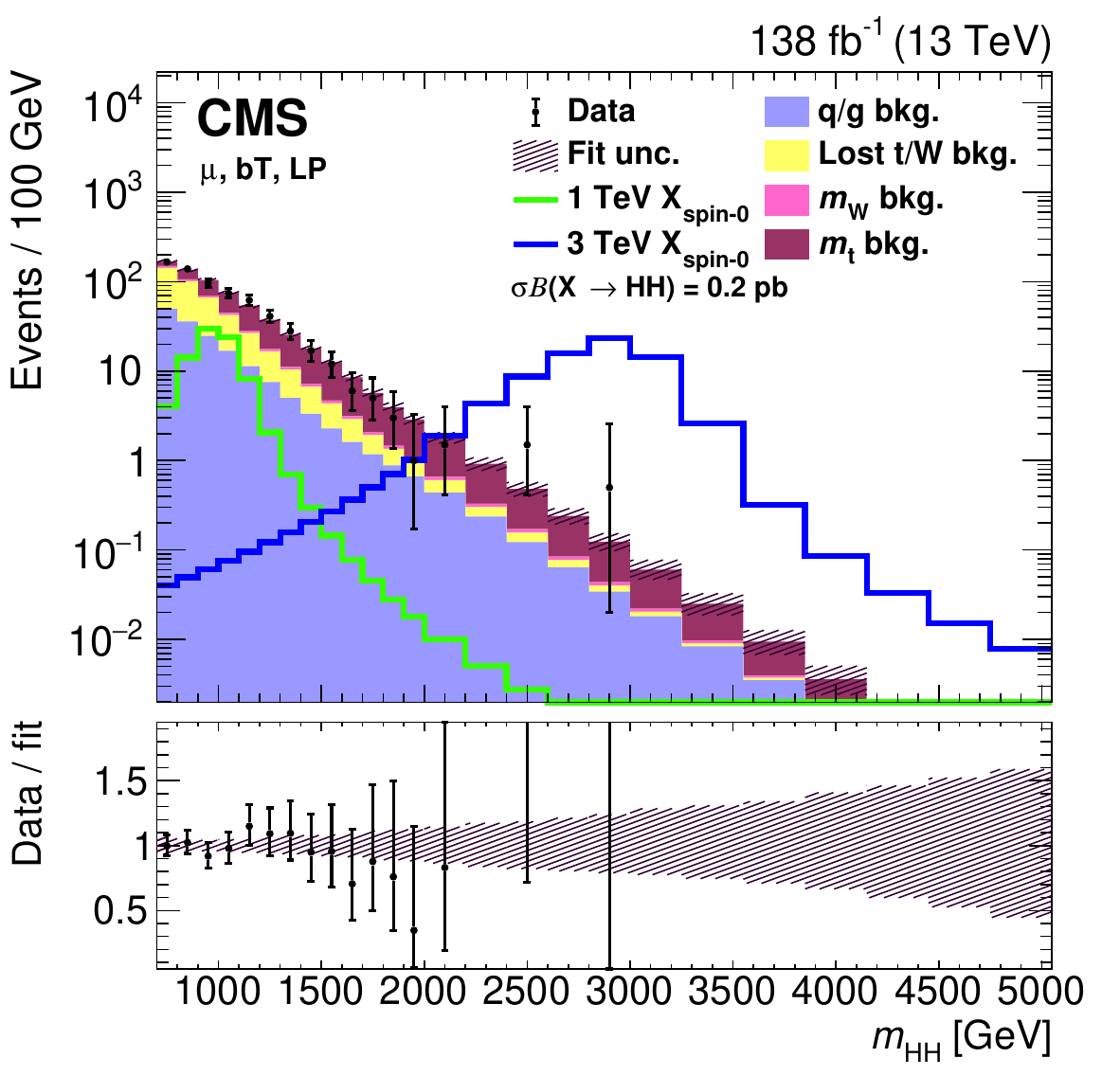}
  \includegraphics[width=\cmsFigWidth]{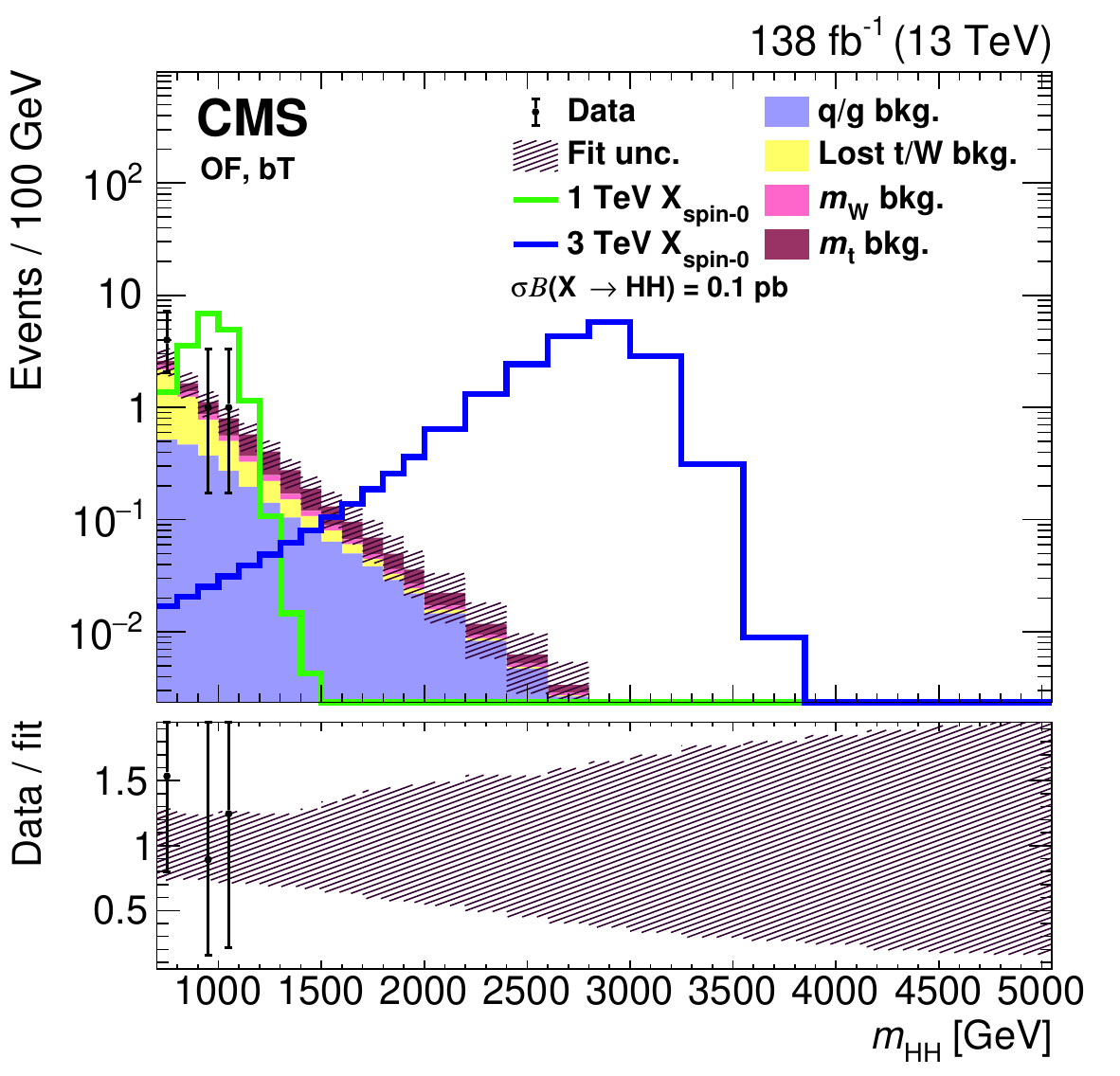}
  \caption{
    Search for $\PX\to\PH(\bb)\PH(\PW\PW)$:
    Distributions of the \mHH variable, in the (\cmsLeft) SL and (\cmsRight) 
    DL categories of the $\PH(\bb)\PH(\PW\PW)$ analysis with merged 
    jets~\cite{CMS:2021roc}. Expected signal distributions from a spin-0 resonance 
    with a mass of 1 or 3\TeV are also shown, by the open green and blue histograms. Figure from Ref.~\cite{CMS:2021roc}.
  }
  \label{fig:B2G-20-007_postfit}
\end{figure}
Events are divided into twelve categories according to the lepton
flavor, the purity of the $\PH \to \bb$ flavor tagging, and, for
the SL final states, the purity of the $\PH \to \PW \PW$ decay
reconstruction.  Results are extracted by performing a maximum
likelihood fit in the 2D ($m_{\bb}$, \mHH)
mass plane. The background-only model is found to describe the observed
distributions well. The distributions of events projected into \mHH for two
selected categories in SL and DL final states are shown in
Fig.~\ref{fig:B2G-20-007_postfit}. In both cases high purity of the
$\PH \to \bb$ flavor tagging (\PQb\PQT) is required. The chosen SL
category is further characterized by a muon and a low purity (LP) 
requirement of the $\PH \to \PW \PW$ decay reconstruction. The DL
category is characterized by different flavors.

\subsubsection{The \texorpdfstring{$\PX\to\HH$}{X->HH} decays into multilepton final states}
\label{Sec:Analysis_X_to_HH_multilepton}
\label{Sec:HIG-21-002}

The analysis of multilepton final states~\cite{CMS:2022kdx} does not assume 
at least one $\PH\to\bb$ decay, unlike all other \HH analyses discussed in this report, 
and thus gains access to various hitherto uncovered \HH signatures. 
This search is focused on resonant \HH production in the
\WWWW, \WWtt, and \tttt decay modes. The chosen final states provide
a good compromise between a relatively large \HH branching fraction
and a clean leptonic event signature. The latter provides a series of
almost background-free categories with low event counts, which are
especially sensitive at low resonance masses. The data are collected
using triggers combining single and multiple lepton and \tauh signatures.

The selected events are split into seven event categories with
different multiplicities of reconstructed leptons and hadronically
decaying taus, denoted as $2\ellss$ (same sign),
$3\Pell$, $4\Pell$, $3\Pell+1\tauh$, $2\Pell+2\tauh$,
$1\Pell+3\tauh$, and $4\tauh$. The lower multiplicity $2\ellss$ and $3\Pell$ categories also require additional jets, as
expected from hadronic \PW boson decays.  To reduce the impact of
backgrounds involving the decay of top quarks, such as \ttbar production, and to
avoid overlap with other \HH analyses, events with \PQb jets identified with
the \textsc{deepJet} algorithm~\cite{Bols:2020bkb} are explicitly vetoed.
Similarly, events with two opposite-sign same-flavor lepton pairs and a
combined invariant mass below 140\GeV are vetoed
to avoid overlap with the $\HH\to\bb\PZ\PZ$ signature, the analysis of which 
is not included in this Report. To exclude phase
space regions enriched in low mass resonances, which are not well
modelled in simulation, and to reduce the impact of backgrounds
involving \PZ boson decays, events containing DL pairs with a
mass of $m_{\Pell\Pell}<12\GeV$ or in the vicinity of the \PZ
boson mass are vetoed as well.

The main backgrounds arise from genuine multiboson processes,
such as $\PZ\PZ$ production in all categories, as well as $\PW\PZ$ production 
in the $2\ellss$ and $3\Pell$ categories. These
backgrounds, as well as smaller backgrounds, such as contributions from
DY, \ttbar, and single \PH production together with backgrounds from photon 
conversions in the detector material are estimated from the simulation. 
Background events from charge misidentification in the $2\ellss$ category, however, are also estimated from data extrapolating
events with opposite-sign electrons from $\PZ\to\Pe\Pe$ decays into the
SR. In most categories, backgrounds arising from jets
misidentified as either leptons or \tauh also play an important 
role; these are estimated with the misidentification-rate factor method~\cite{CMS:2018fdh}.

The signal extraction is based on maximum likelihood fits to the
distributions of boosted decision tree (BDT) discriminants. The BDTs are trained to
discriminate signal from background in each of the seven categories
and for each resonance spin assumption (spin 0 or 2). The BDTs
are parameterized in the nominal resonance mass. 

The selection of BDT input variables is optimized for each signal
category, and includes kinematic variables of \Pell and \tauh candidates
as well as angular separations and invariant masses of their
combinations. Also, the reconstructed \HH mass is used.
The fits also include two kinematic distributions in control regions
enriched in the dominant prompt backgrounds from $\PZ\PZ$ and $\PW\PZ$
production.

\begin{figure}[tb]
    \centering\includegraphics[width=\cmsFigWidth]{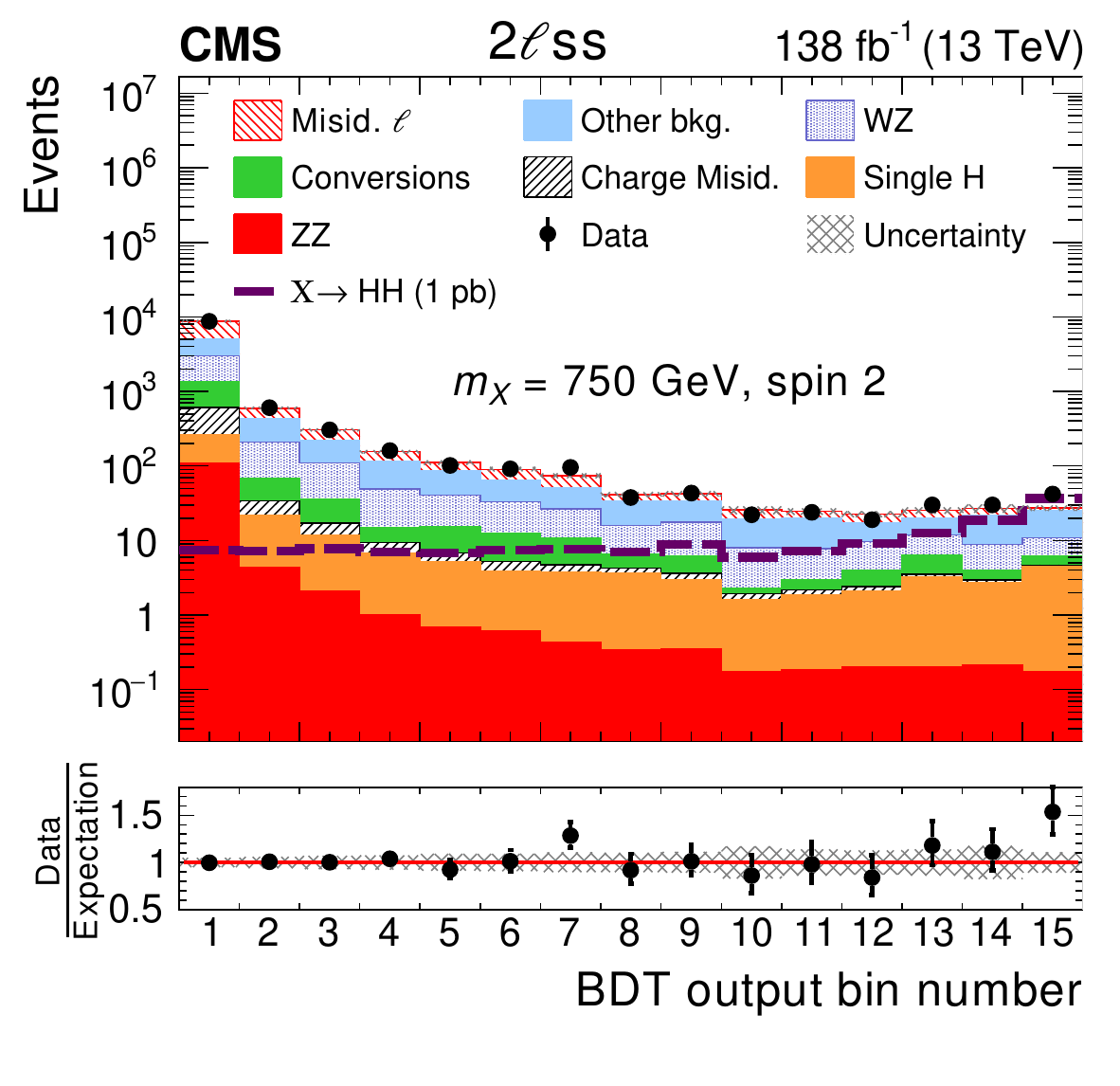}
    \centering\includegraphics[width=\cmsFigWidth]{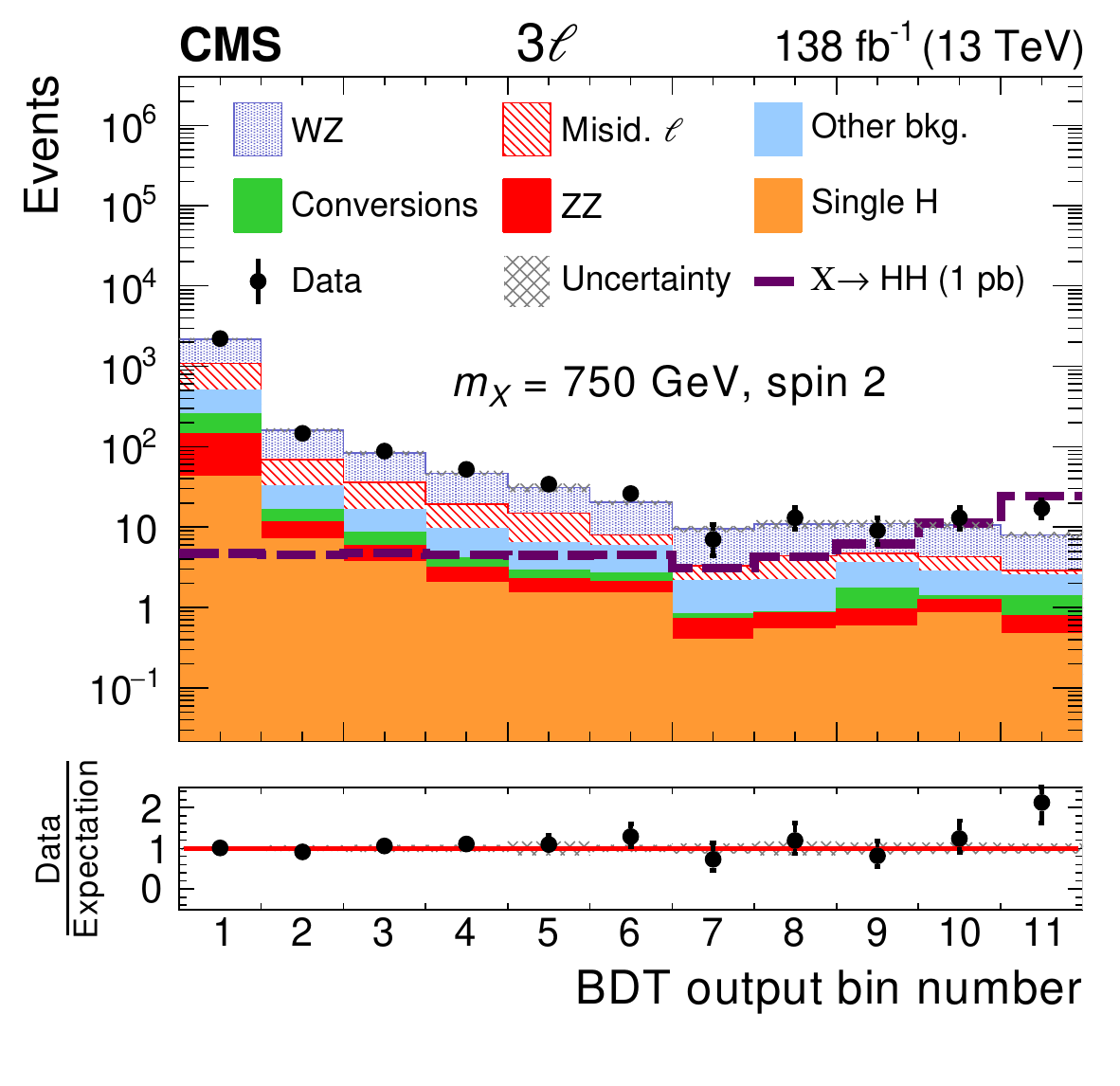}
    \caption{ 
      Search for $\PX\to\PH\PH$ in multi-lepton final states:
      Distributions of the BDT classifier output for events in the (\cmsLeft) 
      $2\ellss$ and (\cmsRight) $3\Pell$ categories of the
      $\PH(\PW\PW+\tautau)\PH(\PW\PW+\tautau)$ analysis in multilepton final 
      states~\cite{CMS:2022kdx}. The expected signal for a spin-2 resonance 
      with a mass of 750\GeV is shown by the open dashed 
      histogram. The signal is normalized to a a cross section of 1\unit{pb}.  
      The distributions of the estimated background processes and corresponding 
      uncertainties are shown after a fit of the signal plus background hypothesis 
      to the data. Figure from Ref.~\cite{CMS:2022kdx}.
    }
    \label{fig:HIG_21_002_BDT_resonant}
\end{figure}

As an example of the SR composition,
Fig.~\ref{fig:HIG_21_002_BDT_resonant} shows the BDT output 
in the two highest signal yield categories $2\ellss$
and $3\Pell$ for a spin-2 particle with $\mX=750\GeV$. 
A small excess is observed in the rightmost bin
of both distributions, amounting to a local significance of about 2.1
standard deviations in both categories. This leads to a mild local excess of
the observed limits corresponding to about 1.5--2 standard deviations for
masses above 600\GeV, which is visible in Fig.~\ref{fig:Limits_on_HH}.

\subsection{Search for resonances in the \texorpdfstring{\YH}{YH} channel} 
\label{Sec:Analysis_X_to_YH}

Recently, direct searches for new physics in the Higgs sector have been 
extended towards \YH decays, where \PY denotes another unknown bosonic resonance. 
Such decays are expected, \eg, in 2HDM+S models like the NMSSM, where \PX and \PY 
can be identified with additional heavy or light Higgs bosons. In cases where 
\PY carries a large singlet component its couplings to SM particles are suppressed 
and its dominant production at the LHC proceeds via the decay $\PX\to\PY\PH$. The first 
analysis presenting such a search was performed in $\PY(\bb)\PH(\tautau)$ 
final states~\cite{CMS:2021yci}, covering mass ranges of $260<\mX<3000\GeV$ and 
$60<\mY<2800\GeV$. Later on, this analysis was complemented by similar 
searches in the $\PY(\bb)\PH(\bb)$~\cite{CMS:2022suh} and 
$\PY(\bb)\PH(\Pgg\Pgg)$~\cite{CMS:2023boe} final states. The search in the $\PY(\bb)\PH(\bb)$ 
final state covers mass ranges of $900<\mX<4000\GeV$ and $60<\mY<600\GeV$, 
targeting kinematic regimes where both \bb decays are reconstructed as large-$R$ 
AK8 jets. The $\PY(\bb)\PH(\Pgg\Pgg)$ analysis covers mass ranges of $300
<\mX<1000\GeV$ and $90<\mY<800\GeV$, where the upper bound on \mX is implied by 
the requirement that it should be possible to resolve the \PQb quarks originating 
from the \PY decay as two distinct AK4 jets. All final states have the $\PY(\bb)$ 
decay in common. The $\PH(\bb)$ decay utilizes the large branching fraction 
of the \PH boson to \PQb quarks; the $\PH(\tautau)$ decay comprises the second 
largest branching fraction with advantageous reconstruction and identification 
properties; the $\PH(\GamGam)$ decay contributes through the excellent mass 
resolution of the ECAL. 

\subsubsection{The \texorpdfstring{$\PX\to\PY(\bb)\PH(\tautau)$}{X->Y(bb)H(tau tau)} decays}\label{Sec:HIG-20-014}

The analysis in $\PY(\bb)\PH(\tautau)$ final states~\cite{CMS:2021yci} evolved from an analysis of 
$\PH\to\tautau$~\cite{CMS:2022kdi}, adding loose requirements on the $\PY\to\bb$ decay 
where the \PQb quarks are reconstructed as AK4 jets. 
For the $\PH(\tautau)$ decay the \etauh, \mutauh, and \tauhtauh 
final states are considered. These final states have been shown to have the 
largest sensitivity to this signature, while the contribution from 
\tautau decays into an \Pe and \Pgm, mostly due to the large background from 
\ttbar production in the kinematic phase space with additionally selected \PQb 
jets is marginal. 

The trigger selection of the events proceeds via the presence of high-\pt electrons, 
muons, or \tauh decays, or combinations of those at the trigger level. 
Due to the trigger requirements evolving with time, the offline \pt thresholds range from
25--33\GeV for electrons, from 20--25\GeV for muons, and from
30--40\GeV for \tauh candidates, the latter depending also on the \tautau final state.
The \tauh candidates are 
identified using the \textsc{DeepTau} algorithm~\cite{CMS:2022prd}, as discussed 
in Section~\ref{Sec:Physics_objects}. Jets are \PQb tagged using the 
\textsc{DeepJet} algorithm, with a working point of 80\% efficiency with a 
misidentification rate for light-flavor or gluon jets of 
${\approx}1\%$~\cite{CMS:2017wtu,Bols:2020bkb}. For the event selection, a \tautau pair in 
one of the targeted \tautau final states and at least one \PQb jet are required. 
Events that contain only one \PQb jet and no other jet are discarded from the 
analysis. If more than two \PQb jets are found, the $\PY(\bb)$ decay is built 
from those jets that are leading in \pt. If only one \PQb jet exists, the \PY 
candidate is built from this \PQb jet and the jet with the highest \PQb 
jet score of the \textsc{DeepJet} algorithm, even if this lies below the threshold 
of the chosen working point. The energies of the jets used to form the \PY 
candidate are corrected using the multivariate energy-momentum regression~\cite{Sirunyan:2019wwa}. 

Depending on the \tautau final state, all selected events are then passed through 
one of three NNs exploiting multiclass classification to distinguish signal, for 
given values of \mX and \mY, from four background classes: (i) events with genuine 
\PGt pairs in the final state; (ii) events with quark- or gluon-induced jets 
misidentified as \tauh candidates; (iii) \ttbar events where the intermediate \PW bosons in 
the decay chain decays into any combination of electrons and muons or into a 
single \PGt lepton and an electron or muon (not included in (i) or (ii)); (iv) events 
from remaining background processes that are of minor importance for the analysis 
and not yet included in any of the previous classes. This last class comprises single 
\PH, single \PQt quark, and diboson production, as well as \PZ boson decays into 
electrons or muons. For single \PH production, rates and branching fractions as 
predicted by the SM are assumed.

\begin{figure}[tbp]
  \centering
  \includegraphics[width=\cmsFigWidth]{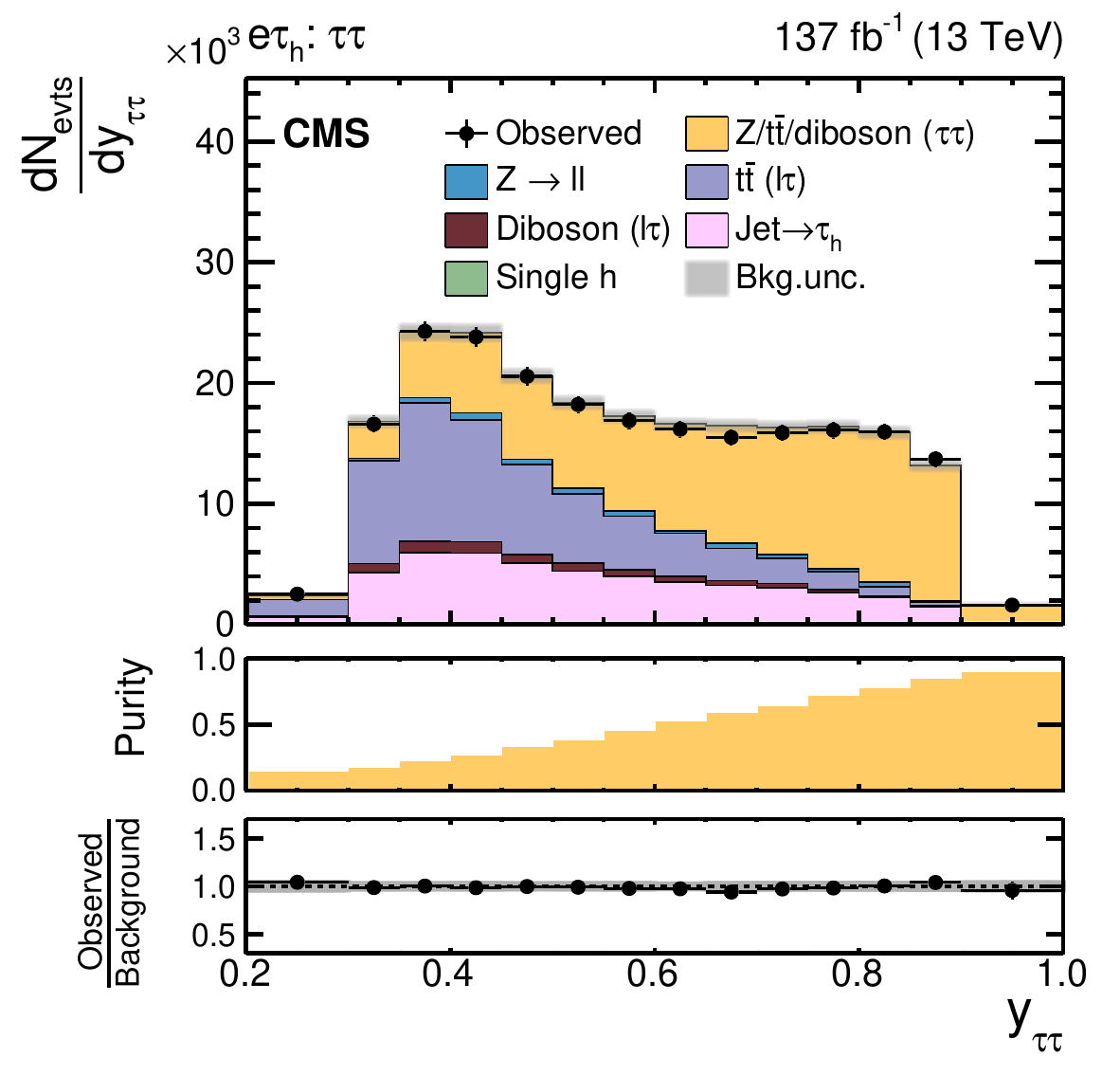}
  \includegraphics[width=\cmsFigWidth]{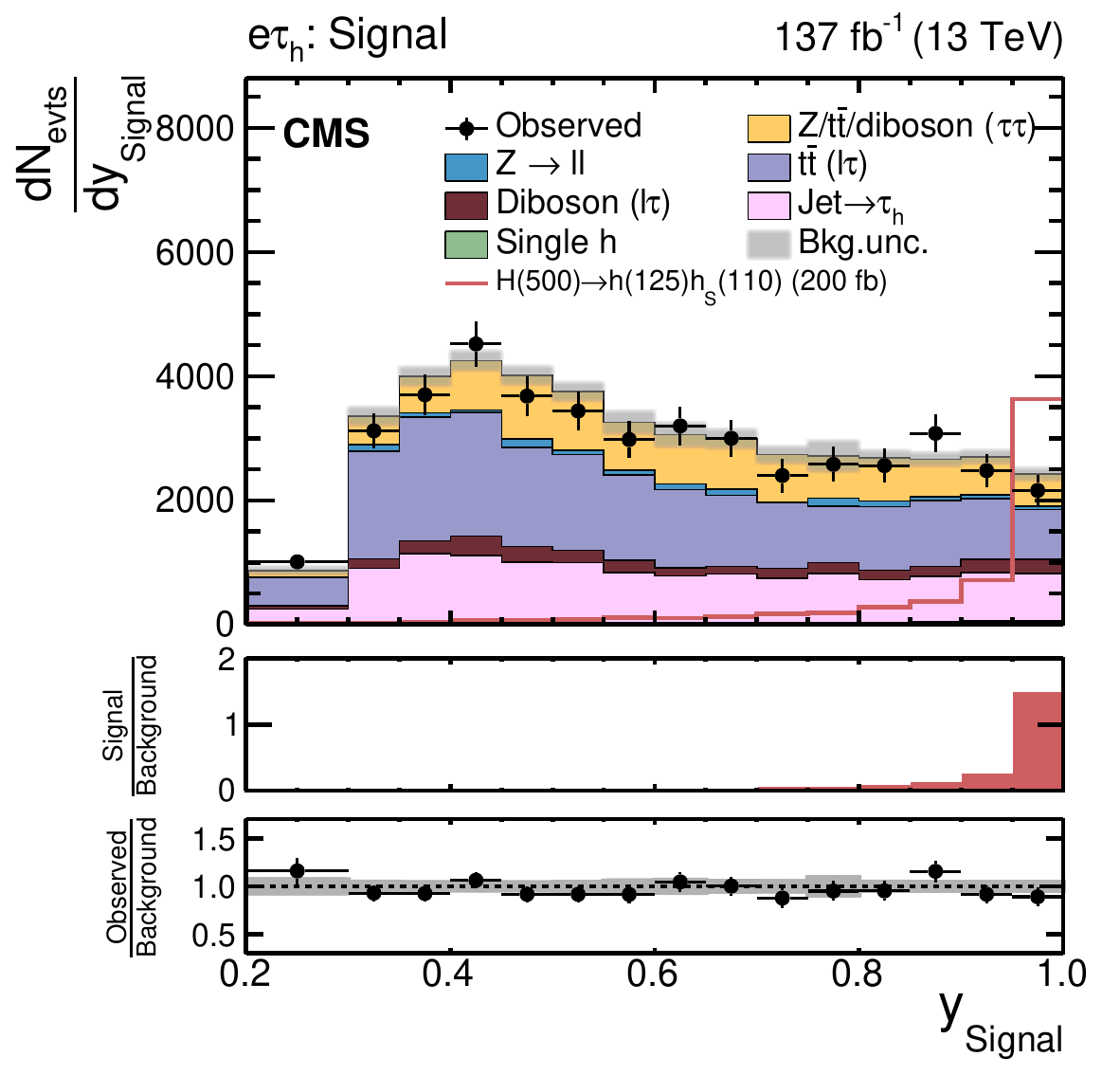}
  \caption{
    Search for $\PX\to\PY(\bb)\PH(\tautau)$:
    Distributions of the NN output scores $y_{i}$, in different event categories 
    after NN classification, based on a training for a resonance \PX with $\mX=
    500\GeV$ and a resonance \PY with $100\leq\mY<150\GeV$ in the $\etauh$ final 
    state of the $\PH(\tautau)\PY(\bb)$ analysis~\cite{CMS:2021yci}. Shown are 
    the (\cmsLeft) \tautau and (\cmsRight) signal categories. For these figures, 
    the data of all years have been combined. The uncertainty bands correspond to 
    the combination of statistical and systematic uncertainties after the fit of 
    the signal plus background hypothesis for $\mX=500\GeV$ and $\mY=110\GeV$ to 
    the data. In the lower panels of the figures the (\cmsLeft) purity and (\cmsRight) 
    fraction of the expected signal over background yields for a signal with a 
    cross section of $200\unit{fb}$, as well as the ratio of the obtained yields 
    in data over the expectation based on only the background model, are shown. 
    Figure from Ref.~\cite{CMS:2021yci}.
  }
  \label{fig:bbtt_postfit}
\end{figure}

Inputs to the NNs are 20--25 features, of which the following 
have been identified to be most discriminating, according to an unambiguous 
metric as given in Ref.~\cite{Wunsch:2018oxb}: the invariant masses of the \bb, \tautau, 
and \bbtautau systems, and the $\chi^{2}$-value of a kinematic fit to the data 
of the signal hypothesis for given values of \mX and \mY. Since the discrimination depends 
on the signal hypothesis, individual NNs have been trained for 68 groups of 
kinematically adjacent and similar signal hypotheses. Each event has been assigned 
to the class with the highest NN output score. Eventually, the NN output scores 
have been chosen as discriminating variables for a maximum likelihood fit in 45 
individual event categories, split by \tautau final state and data-taking year. 
Typical output distributions are shown in Fig.~\ref{fig:bbtt_postfit}. 

\subsubsection{The \texorpdfstring{$\PX\to\PY(\bb)\PH(\GamGam)$}{X->Y(bb)H(gamma gamma)} decays}\label{Sec:HIG-21-011}

As in the previous case, the analysis in the $\PY(\bb)\PH(\GamGam)$
final state~\cite{CMS:2023boe} starts from the well-identified
$\PH\to\GamGam$ decay complemented by the selection of two additional
AK4 jets to form the $\PY(\bb)$ candidate.

The trigger selection proceeds through the requirement of two photons with 
thresholds of $\pt>30\GeV$ for the leading ($\PGg_{1}$) and $\pt>18\GeV$  
for the subleading ($\PGg_{2}$) photon in \pt, for data taken in 2016. 
For data taken in the years 2017--2018, the requirement on $\PGg_{2}$ is raised to 
$\pt>22\GeV$ because of the larger amount of PU in data. 
The photons are required to pass identification and isolation 
criteria, already through the trigger selection, and the mass of the two photons 
is required to be $\mgg>90\GeV$.

In the offline selection, the photons from which the \PH candidate is formed 
are required to be well contained in the ECAL and tracker fiducial volumes of 
$\abs{\eta}<2.5$, excluding the transition region of $1.44<\abs{\eta}<1.57$ between 
the ECAL barrel and endcaps, and to fulfill kinematic requirements of $100<\mgg<
180\GeV$, $\pt^{\PGg_{1}}/\mgg>1/3$, and $\pt^{\PGg_{2}}/\mgg>1/4$. In addition 
to the photons at least two AK4 jets originating from the same PV as the photons 
are required, where the assignment of the photons to the PV is achieved with the 
help of an MVA technique, as described in Ref.~\cite{Khachatryan:2014ira}. The jets must 
fulfill identification requirements as described in Section~\ref{Sec:Physics_objects}, 
have $\pt>25\GeV$ and $\abs{\eta}<2.4$ or 2.5, for the data taken in 2016 and 
2017--2018, respectively, and be separated from each of the selected photons by $\DR>0.4$. 
Of all jets in an event that match these criteria, those with the highest sum of 
their \textsc{DeepJet} discriminant scores are chosen to form the \PY candidate 
with a requirement on the dijet mass of $70<\mjj<1200\GeV$. The lower bound on 
\mjj is implied by the kinematic turn-on in the \mjj distribution, the upper bound 
is defined by the transition towards Lorentz-boosted regimes, where the \bb system 
may not be resolved by two spatially separated AK4 jets. 

The \PX candidate is reconstructed from the jets and photons forming the \PH and 
\PY candidates. To improve the resolution, its mass \mX is estimated from 
\begin{equation}
    \mXtilde=\mggjj - \left(\mgg-\mH\right) - \left(\mjj-\mY\right),
\end{equation}
where \mggjj is the mass calculated from the two jets and the two photons. It 
is corrected by subtracting \mgg and \mjj, with their values replaced by the 
nominal values of the masses \mH and \mY. This estimate has been shown to lead to a 30 to 90\% improvement 
in signal resolution compared to \mggjj alone, in the high- and low-\mX regimes, respectively. 
For the signal extraction, events are required to be located inside a window in 
\mXtilde, depending on the \mX hypothesis under test. The width of this window 
has been defined such that it contains at least 60\% of the signal. 

Resonant backgrounds to the analysis originate from single \PH production, which 
is strongly suppressed already by the selection in \mXtilde. For hypotheses of 
$\mX<550\GeV$ a sizeable contribution from \ttH production is 
further reduced by an NN-based discriminant developed for the search for nonresonant 
$\PH(\GamGam)\PH(\bb)$ production~\cite{CMS:2020tkr}, exploiting the decays of 
\PW bosons arising from the \ttbar decay chains.     

\begin{figure}[tbp]
  \centering
  \includegraphics[width=\cmsFigWidth]{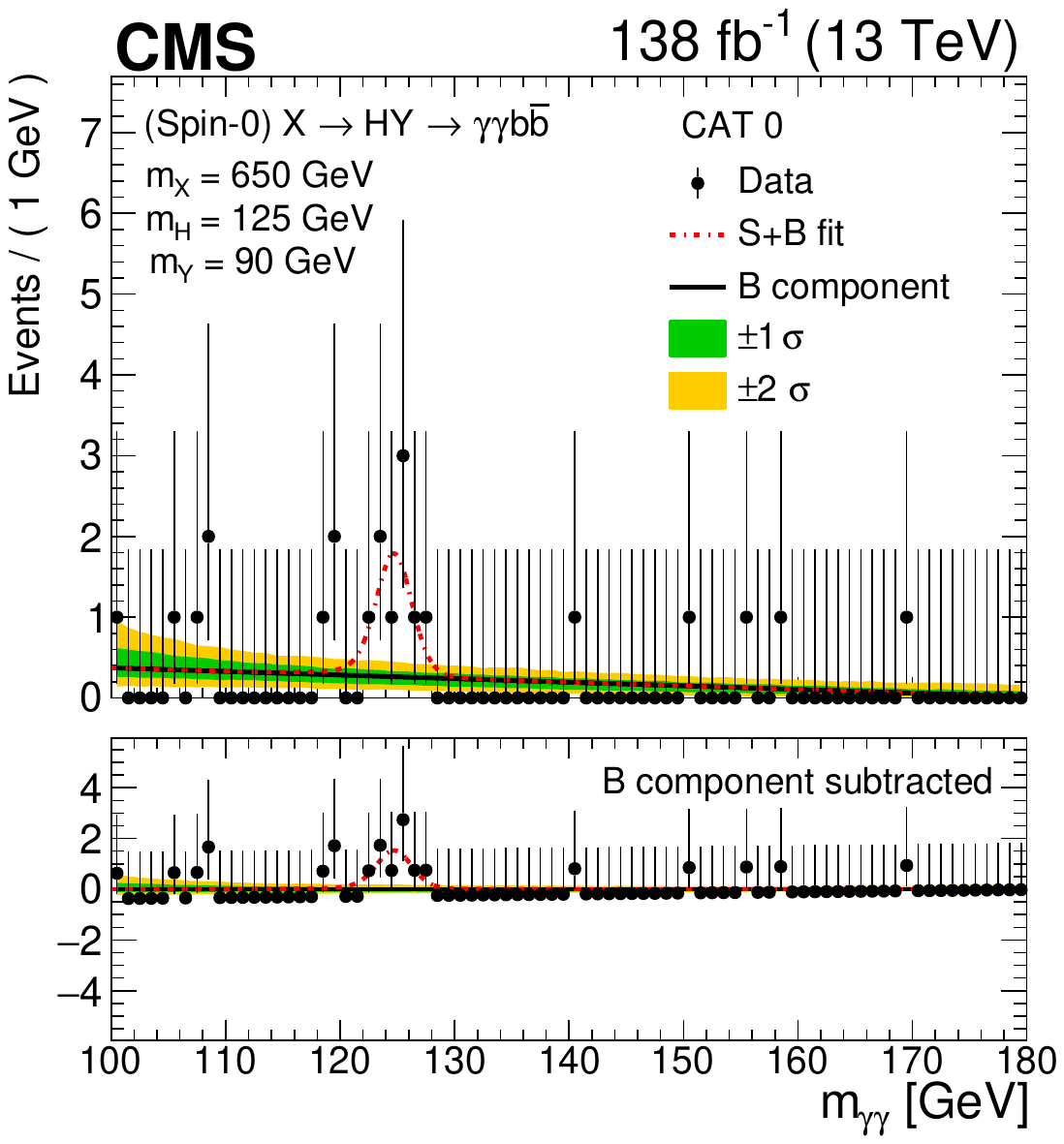}
  \includegraphics[width=\cmsFigWidth]{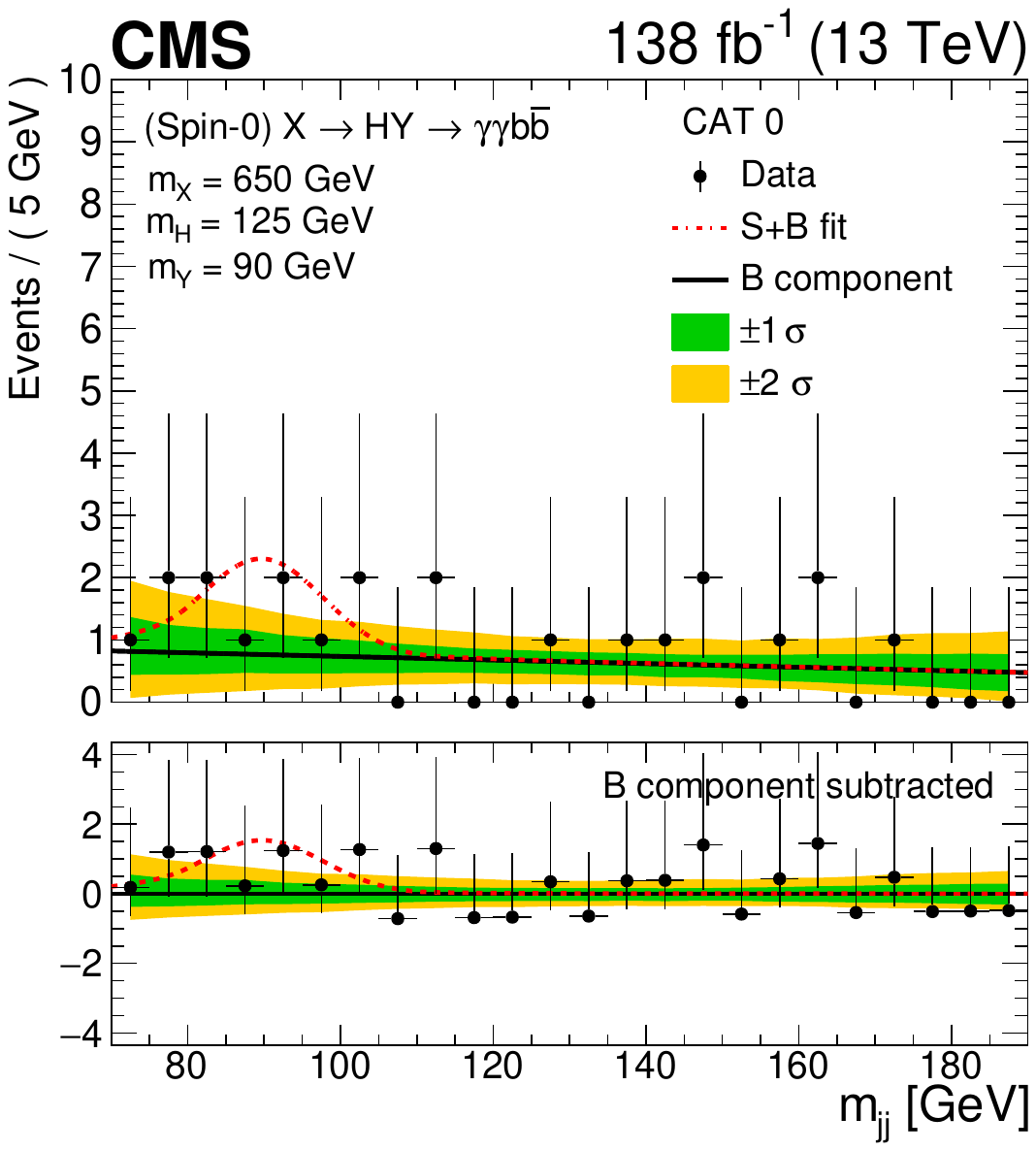}
  \caption{
    Search for $\PX\to\PY(\text{bb})\PH(\gamma\gamma)$:
    Marginal distributions of the (\cmsLeft) \mgg and (\cmsRight) \mjj variables, 
    in the high-purity SR (labeled ``CAT 0'') of the $\PY(\bb)\PH(\GamGam)$ 
    analysis~\cite{CMS:2023boe}. The figure is shown, for a hypothesis of $\mX=
    650\GeV$ and $\mY=90\GeV$, for which the largest excess of events over the 
    background model is observed. In the lower panels, the numbers of 
    background-subtracted events are shown after the fit of the background model 
    to the data. Figure from Ref.~\cite{CMS:2023boe}.
  }
  \label{fig:ggbb_postfit}
\end{figure}

Nonresonant backgrounds in this search mostly originate from the production of one (\Gjets) and 
two (\GGjets) photons in association with jets. To separate these backgrounds 
from the signal a BDT with three output classes, one for 
each background and one for signal, and 22 input features is used. The input 
features comprise kinematic and identification observables of the selected jets 
and photons, estimates of the mass, energy, and \pt resolutions, and an estimate 
of the \pt density from PU. For training, six exclusive kinematic regions are 
defined, based on the hypothesized values of \mX and \mY, where in each region 
all contained signal samples and the two background processes in question, as 
obtained from simulation, are used with equal weight. These training regions are 
defined to resemble similar kinematic properties for signals inside the given 
\mX--\mY window. For each kinematic region, three event categories are defined, 
based on the output of the corresponding BDT. These categories are introduced 
to indicate regimes of varying signal purity. For each \mX hypothesis the signal 
is inferred from an unbinned likelihood fit of a parametric model to the data in 
the 2D discriminating distributions given by the values of \mgg and \mjj, in each 
of the BDT categories. 
The data are found to be compatible with the SM predictions. In Fig.~\ref{fig:ggbb_postfit} the marginal distributions 
of \mgg and \mjj in the BDT category with the highest expected signal purity
for a selection corresponding to $\mX=650\GeV$ and $\mY=90\GeV$, are shown
together with the results of the fit to the data. For these mass
values, the largest deviation from the background-only hypothesis is
observed, with a local (global) significance of 3.8 (below 2.8) standard
deviations. A hypothesized signal for 
$\mX=650\GeV$ and $\mY=90\GeV$ with an arbitrary normalization is also shown.

\subsubsection{The \texorpdfstring{$\PX\to\PY(\bb)\PH(\bb)$}{X->Y(bb)H(bb)} decays in merged jet topology}
\label{Sec:B2G-21-003}

This analysis in the $\PY(\bb)\PH(\bb)$ final state~\cite{CMS:2022suh}
 explicitly targets ranges in \mX and \mY where both, the $\PY(\bb)$ and 
$\PH(\bb)$ decays can be reconstructed with AK8 jets, as 
described in Section~\ref{Sec:Physics_objects}.  

The trigger selection proceeds through a logical OR of a mixture of trigger 
paths requiring the presence of high-\pt AK8 jets, large values of \Ht, or 
combinations of those. In addition, \PQb tagging requirements and a requirement 
on the mass of the two leading jets, in cases where more than one AK8 jet is 
present in an event, are imposed. This setup aims at a trigger efficiency close 
to 100\% for the offline selected events. Residual corrections to the trigger 
efficiency have been derived from CRs. These corrections usually range 
below 5\%.   

In the offline selection, the events are required to contain at least two AK8 
jets with $\pt>350\GeV$ and $\abs{\eta}<2.4$ for the data taken in 
2016 and $\pt>400\GeV$ and $\abs{\eta}<2.5$ for data taken in 2017--2018. 
The dominant backgrounds for this analysis arise from 
\ttbar production in the all-hadronic decay of the intermediate \PW bosons and 
QCD multijet production. To further suppress the latter an additional pairwise 
requirement of $\abs{\Delta\eta}<1.3$ for the selected AK8 jets is imposed. 
Eventually the two leading jets in \pt are identified as the \PH and \PY 
candidates, where requirements of $110<\mj<140\GeV$ for the \PH candidate and 
$\mj>60\GeV$ for the \PY candidate are imposed. 
When both AK8 jets satisfy the first mass requirement, the \PY jet is chosen 
at random.
The \PH and \PY candidates are 
then passed to the \textsc{ParticleNet} algorithm~\cite{Qu:2019gqs} 
to discriminate the decays of a boosted resonance, $\PH(\bb)$ or $\PY(\bb)$, 
from light-flavor quark or gluon jets. 

\begin{figure}[tb]
  \centering
  \includegraphics[width=\cmsFigWidth]{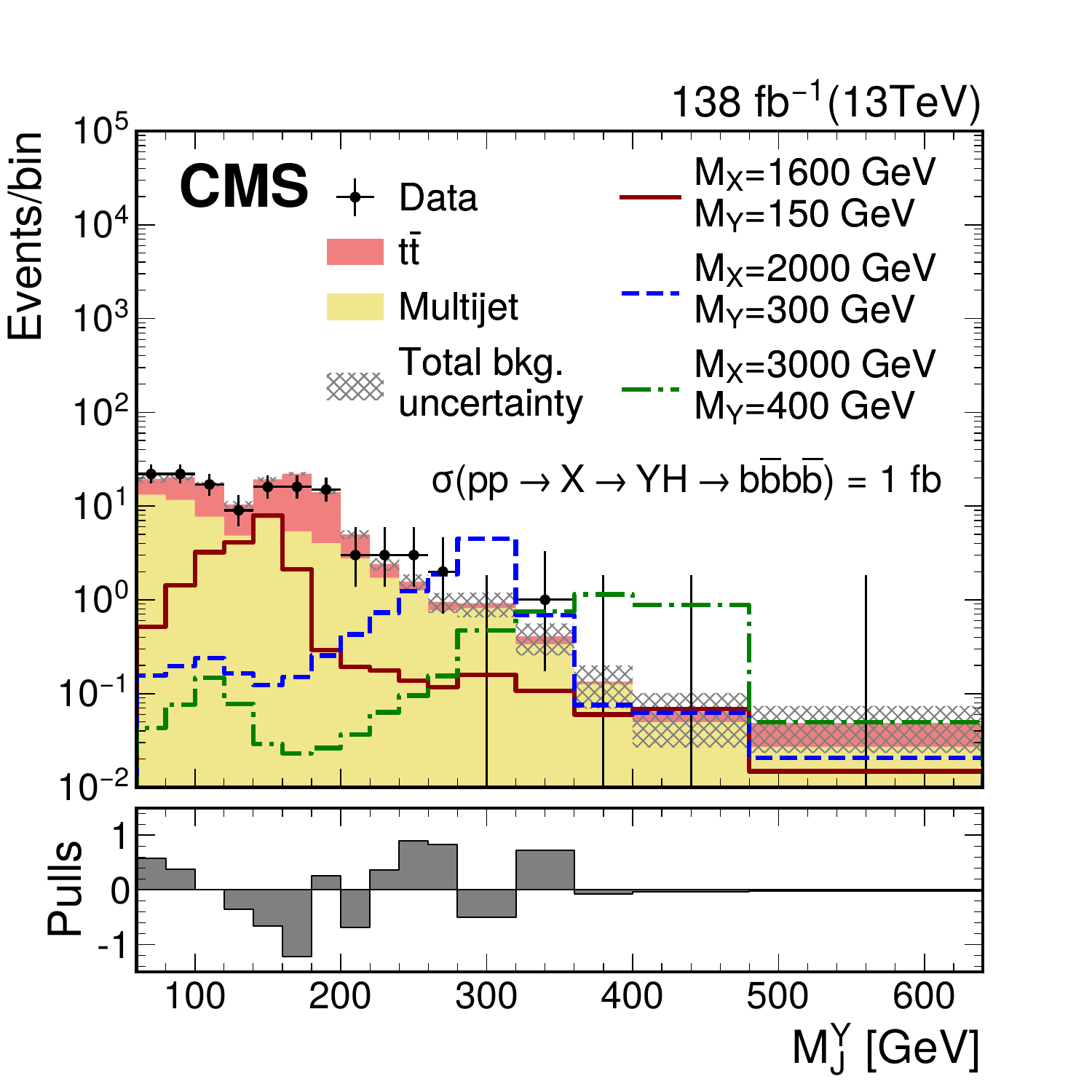}
  \includegraphics[width=\cmsFigWidth]{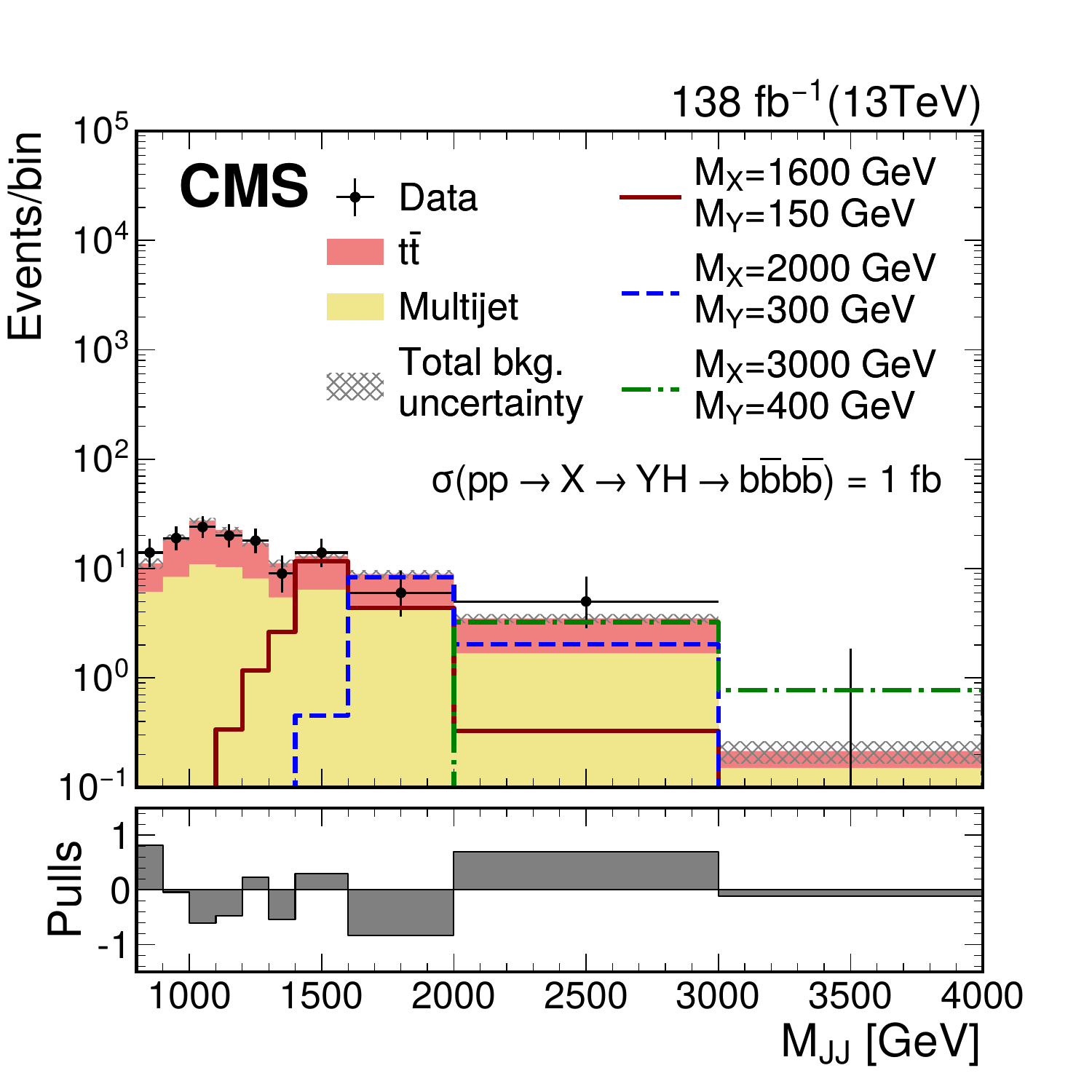}
  \caption{
    Search for $\PX\to\PY(\bb)\PH(\bb)$:
    Distributions of the (\cmsLeft) soft-drop mass of the boosted \PY
    candidate, labeled $M^\PY_\text{J}$, and (\cmsRight) the dijet
    mass of the \PY and \PH candidates, $M_\text{JJ}$, 
    in the high-purity SR of the $\PY(\bb)\PH(\bb)$ analysis 
    with two merged \bb jets~\cite{CMS:2022suh}. The distributions as expected 
    for signals with three different values of \mX and \mY (labeled
    $M^\PX$ and $M^\PY$) are also shown. In the 
    lower panels the statistical pull in each bin is displayed. 
    Figure from Ref.~\cite{CMS:2022suh}.
  }
  \label{fig:bbbb_postfit}
\end{figure}
Based on the output of the \textsc{ParticleNet} algorithm for each of the AK8 jets, a 
loose and a tight SR, with varying expected signal purity, are defined. 
In addition, corresponding sideband regions for the estimation of the background from QCD 
multijet production and a series of regions to validate this background estimate 
are constructed. The background from \ttbar production is estimated from simulation 
and monitored in a dedicated CR in data, obtained from a selection of 
either an isolated electron or muon with $\pt>40\GeV$ and $\abs{\eta}<2.4$, and 
an AK4 jet tagged as a \PQb jet with the \textsc{DeepJet} algorithm, with a 
distance of $\DR<1.5$ from the selected lepton. For this purpose, a working point 
of the \textsc{DeepJet} algorithm with an efficiency of 90\% with a 
misidentification rate of ${\approx}10\%$ has been chosen. In addition, 
requirements of $\ptmiss>60\GeV$ and $\Ht>500\GeV$ are imposed. The lepton, 
\ptmiss, and the \PQb-tagged jet provide the signature of a leptonically decaying 
\PQt quark. A hadronically decaying \PQt quark candidate is reconstructed from an 
AK8 jet fulfilling the same kinematic requirement as in the SR and 
$\mj>60\GeV$. In addition the selected AK8 jet must have a distance of $\DR>2$ 
from the selected lepton. This selection reaches a purity in \ttbar events of 
greater than 90\%. 
The signal is obtained from a fit of the signal and background models
to the observed 2D distribution in the ($m^\PY_\text{J}, m_\text{JJ}$) plane, where
$m^\PY_\text{J}$ is the soft-drop mass of the \PY boson candidate,
and $m_\text{JJ}$ is the invariant mass of the \PH and \PY boson
candidates.
The loose and tight SRs are fit jointly with the
corresponding CRs that constrain the QCD multijet
and \ttbar backgrounds.
Distributions in $m^\PY_\text{J}$ and $m_\text{JJ}$ in the high-purity signal region are shown in 
Fig.~\ref{fig:bbbb_postfit}. 

\subsection{Statistical combination} \label{Sec:Combination}

Combinations are performed based on the \HH and \YH decay channels presented in this report, 
and the results will be presented in Sec~\ref{Sec:ResultsLimits}. 
A combination of \VH decay channels is foreseen at a later date as various analyses are still ongoing. 
The combination is performed by integrating the signal extraction procedures of the respective 
decay channels into a combined likelihood analysis determining a single combined signal strength. 
In the \HH case, this combined signal strength measures the product 
$\sigma(\pp \to \PX) \BR (\PX \to \PH\PH)$. It is linked to the signal strength 
in each individual \HH decay mode combination by the product of the corresponding \PH 
branching fractions, where the SM values for a Higgs boson with $\mH=125\GeV$ are used. 
In the \YH case, the combined signal strength measures the product 
$\sigma(\pp \to \PX) \BR(\PX \to \YH) \BR (\PY \to \bb)$. 
It is connected to the signal strength in each \YH decay mode by the corresponding 
SM \PH branching fraction. The branching fraction $\BR(\PY \to \bb)$ is unknown, 
model dependent, and no attempt is made to correct for it. 
This way of combination is possible because all considered \YH channels 
share the $\PY \to \bb$ decay mode.

For this combined analysis, those systematic variations that should act in the same way for 
each individual 
search in consideration, are treated as correlated. A typical example of this kind 
is the uncertainty in the integrated luminosities of the used data sets.  
According to the values of \MX and \MY, different channels may contribute at 
different levels of relative sensitivity. This is due to differences in the 
selection efficiency, the acceptance of the CMS detector, the trigger efficiency 
and the branching fractions. Therefore a combination might be either dominated 
by one channel or benefit from the joint effect of many channels with similar 
sensitivity depending on the phase-space region.

For the $\PX \to \HH$ decay, the combination is performed separately for the spin-0 and 
spin-2 hypotheses on the \PX boson. For the $\PX \to \YH$ case, spin-0 is assumed for both 
the \PX and \PY bosons. For all measurements described in the following, the \PH boson mass is 
fixed to $\mH=125\GeV$. In case of the $\PY(\bb)\PH(\tautau)$ 
analysis (Section~\ref{Sec:HIG-20-014}), for which no limits at $\MY = 125\GeV$ are available, 
we use $\MY = 130\GeV$ instead to estimate the result for the $\HH$ case.  
This is justified by the limited mass resolution. We make this particular choice because a comparison 
of the limits with $\MY = 120\GeV$ and $\MY = 130\GeV$ shows that the latter choice 
yields more conservative limits. Theoretical uncertainties in the branching fractions and in the HH cross sections are taken into account~\cite{deFlorian:2016spz}.

The used grids in points of \MX and (\MX, \MY) can differ across the various analyses. 
In general, the combination is performed only for the points common to all analyses considered in the combination. 

As theoretical systematic uncertainties, we consider normalization uncertainties related to PDF, 
QCD scale, and $\alpha_\text{S}$ in the total cross section for the main backgrounds and for the 
single \PH production process, which follows the recommendations by the LHC cross section working group~\cite{deFlorian:2016spz}.
These uncertainties are considered to be uncorrelated across processes, 
and fully correlated across channels that share the same process.

\section{Upper limits on the cross sections}\label{Sec:ResultsLimits}

We now turn to the results of the searches for heavy resonances \PX decaying 
into $\PV\PH$, $\PH\PH$, and $\PY\PH$ channels.
Each search features at least one instance of the \PH boson at a mass of 
125\GeV originating from the decay of a heavy resonance \PX. 
The analyses are performed in a variety of final states with complementary 
sensitivity in the masses \mX and \mY. 

The observed data for all searches are found to be in agreement with the SM expectations 
in the corresponding SRs. We set upper limits on 
the product of the production cross section of the resonance and the branching fraction, $\sigma\BR$.
The upper limits are set at 95\%~\CL, using the \CLs criterion~\cite{Junk:1999kv,CLs1,Cowan:2010js}. Statistical combinations of the different 
\HH and \YH analyses are performed as described in Section~\ref{Sec:Combination} 
to extract the maximum information from the data.

\subsection{The \texorpdfstring{$\PX\to\PV\PH$}{X->VH} decays}\label{Sec:Results_X_to_VH}

Five searches for $\PV\PH$ resonances are presented in Section~\ref{Sec:Analysis_X_to_VH}. 
These target final states with 0, 1, and 2 leptons, originating from the decay of 
the vector boson (\PW or \PZ) produced together with the \PH boson.  
The \PH boson is assumed to either decay to \bb or \tautau. 
The results are shown as upper limits on $\sigma\BR$ as functions of 
the resonance mass \PX. This can either be a scalar particle, which can occur, \eg, in 2HDM 
models, or a vector boson resonance, like \PWpr and \PZpr, as predicted in the HVT models.

\begin{figure}[tb]\centering
    \includegraphics[width=1.5\cmsFigWidth]{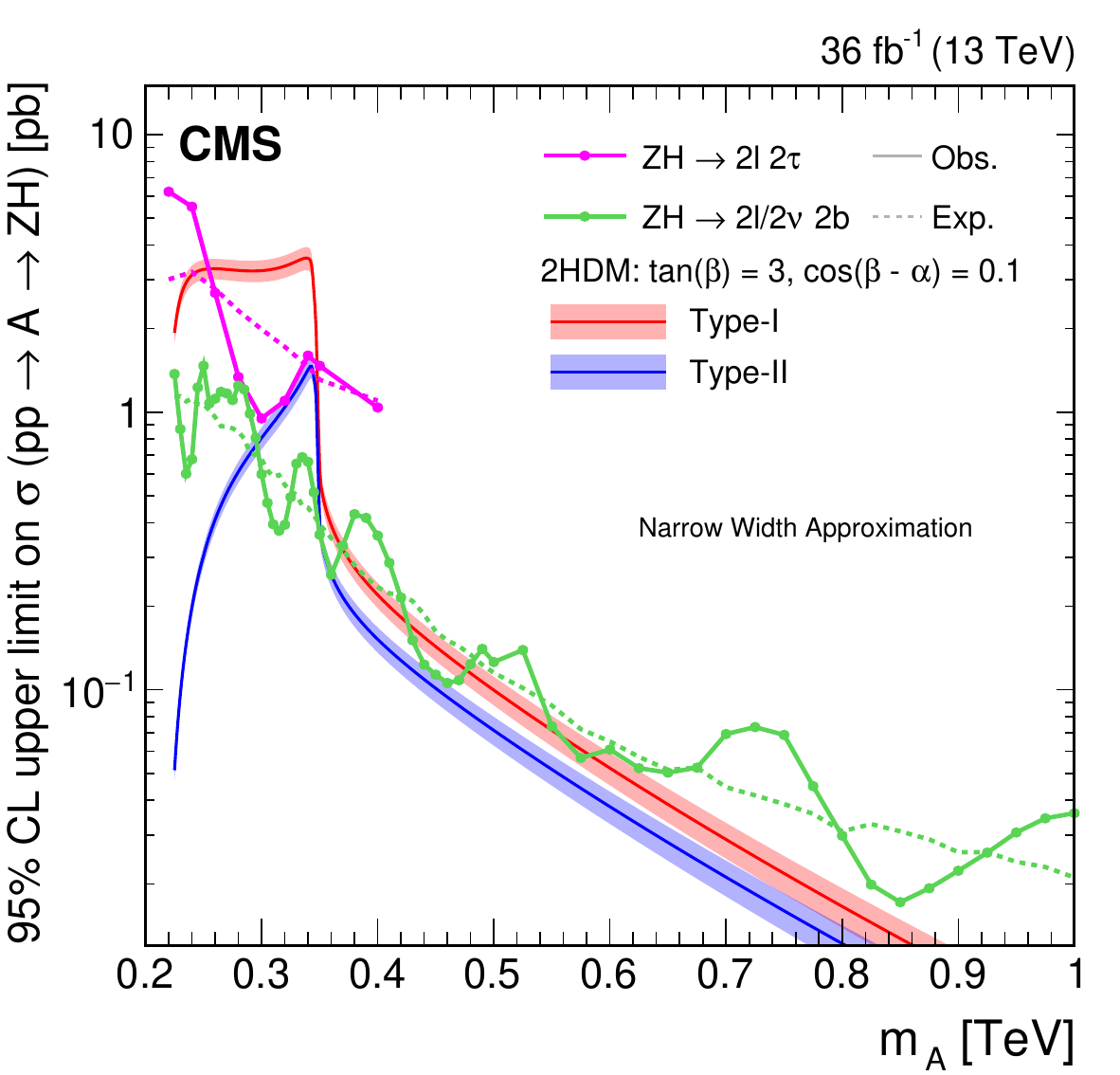}
    \caption{
      Search for $\PX\to\PZ\PH$: Observed and expected 95\% \CL upper limits on the product of the cross 
      section $\sigma$ for the production of an \PA boson, via gluon-gluon fusion 
      and the branching fraction \BR for the $\PA\to\PZ\PH$ decay. The limits 
      are given in \unit{pb} as functions of \mA. The markers connected with 
      solid lines (dashed lines) 
      indicate the observed (expected) limits. The green (magenta) lines refer 
      to the $\PZ(\lep+\nn)\PH(\bb)$~\cite{CMS:2019qcx} ($\PZ(\lep)\PH(\tautau
      )$~\cite{CMS:2019kca}) analysis. The red and blue solid lines indicate the 
      product $\sigma\BR$ as expected by the 2HDM Type~I and Type~II models, respectively, for 
      the parameters $\tan\beta=3$ and $\cosba=0.1$. The shaded areas associated 
      with these predictions indicate the corresponding model uncertainties. The 
      results and model predictions have been adapted from Refs.~\cite{CMS:2019qcx,
      CMS:2019kca}.
  }
  \label{fig:AHZ_limits}
\end{figure}
Figure~\ref{fig:AHZ_limits} shows the upper limits on $\sigma(\pp\to\PA)\BR(\PA\to\ZH)$
as functions of the mass of the \CP-odd Higgs boson \PA, using \PH decays to \bb~\cite{CMS:2019qcx} 
and \tautau~\cite{CMS:2019kca}, obtained with the data set recorded in 2016. 
This plot also shows the expected cross sections for \PA bosons in two typical 2HDM scenarios.  
These feature a drop beyond the \ttbar threshold because of the $\PA\to\ttbar$ channel opening up. 

\begin{figure}[tbp]\centering
  \includegraphics[width=\cmsFigWidth]{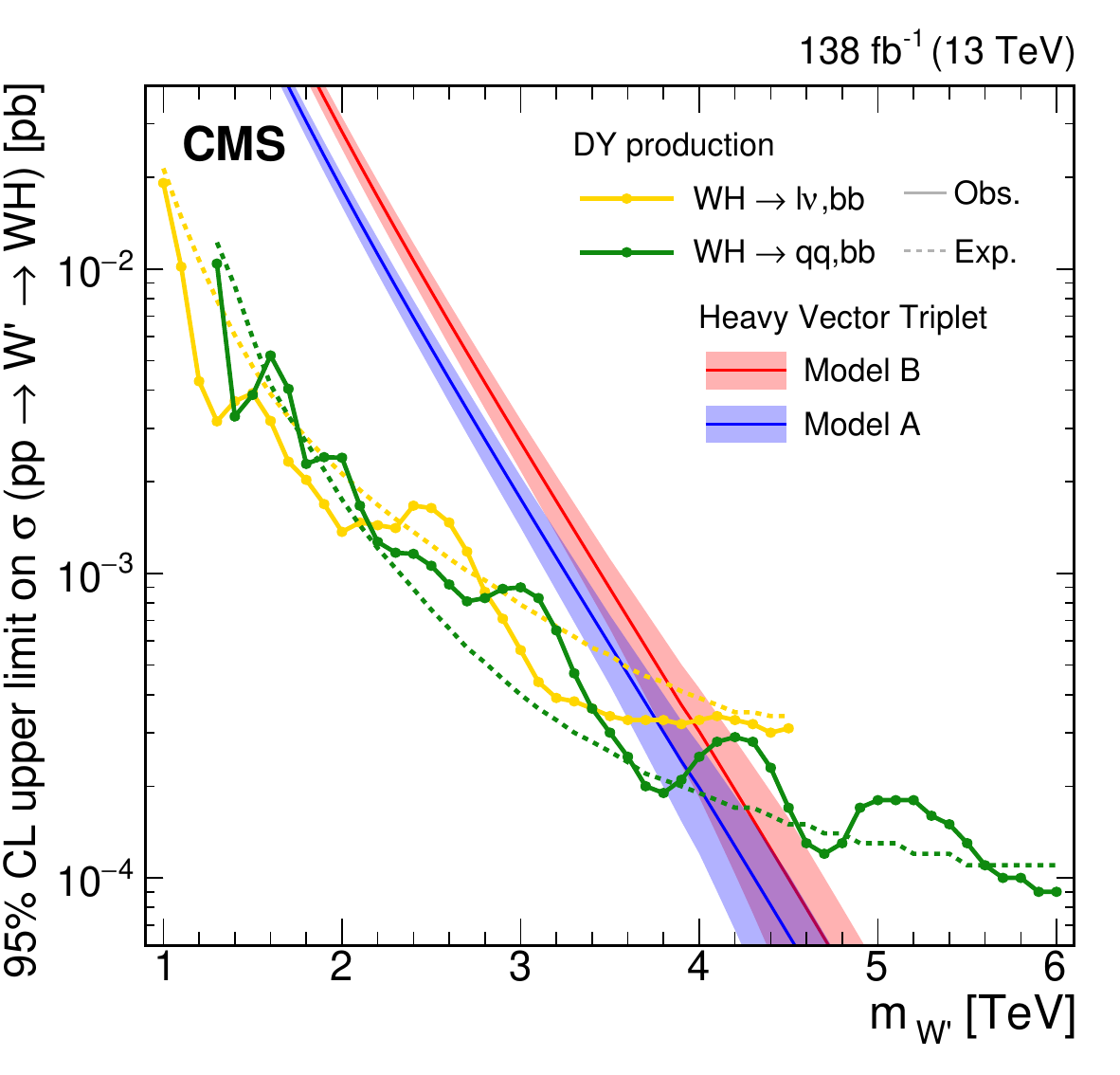}
  \includegraphics[width=\cmsFigWidth]{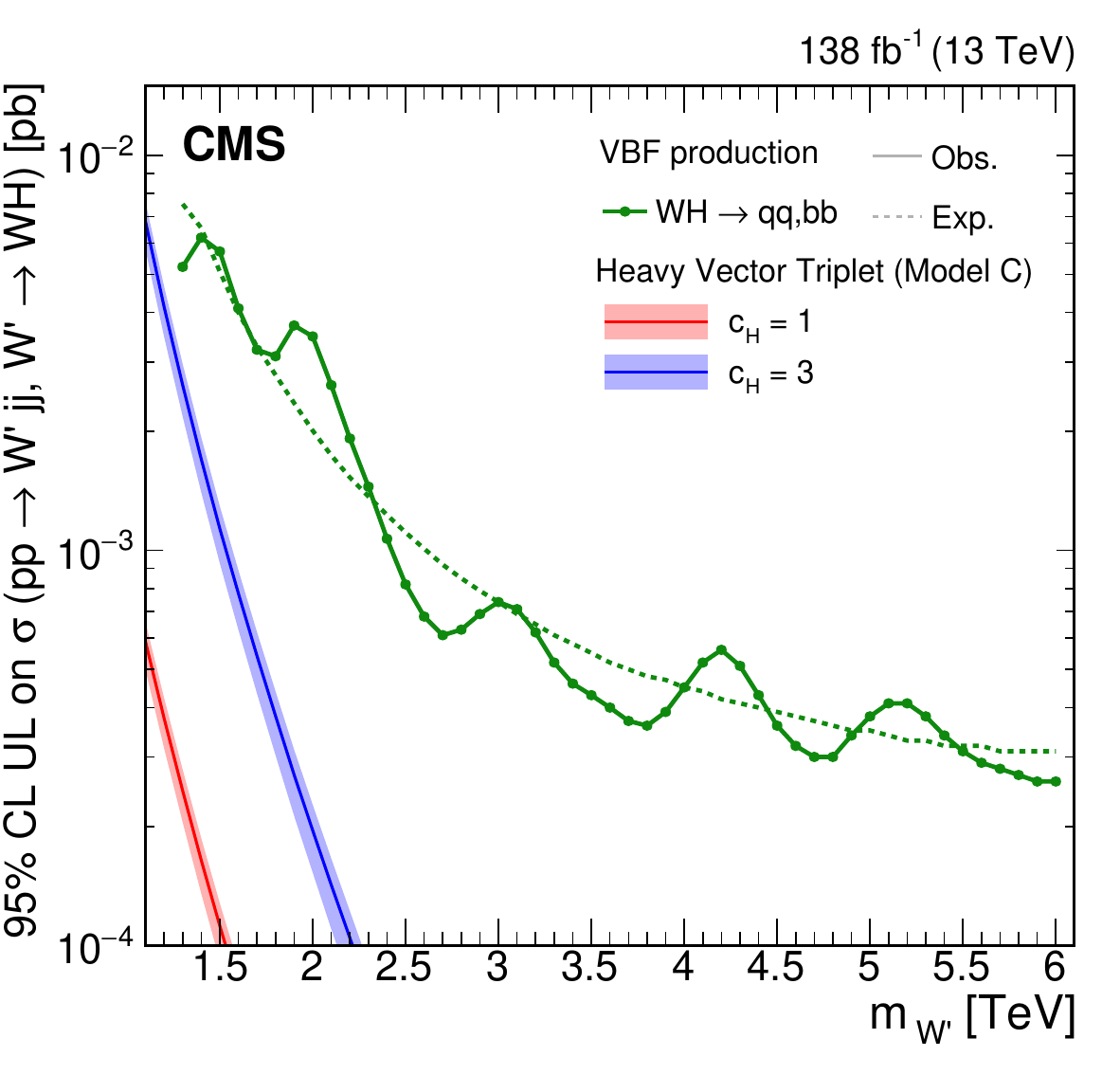}
  \caption{
    Search for $\PX\to\PW\PH$: Observed and expected 95\% \CL upper limits on the 
    product of the cross section $\sigma$ for the production of a \PWpr spin-1 
    resonance, via (\cmsLeft) DY production or (\cmsRight) vector boson 
    fusion and the branching fraction \BR for the $\PWpr\to\PW\PH$ decay. 
    The solid lines represent the observed and the dotted lines the expected limits.
    The theory predictions from the heavy vector triplet models A, B, and C are also 
    shown.
  }
  \label{fig:Limits_on_WH}
\end{figure}
\begin{figure}[tbp]\centering
  \includegraphics[width=\cmsFigWidth]{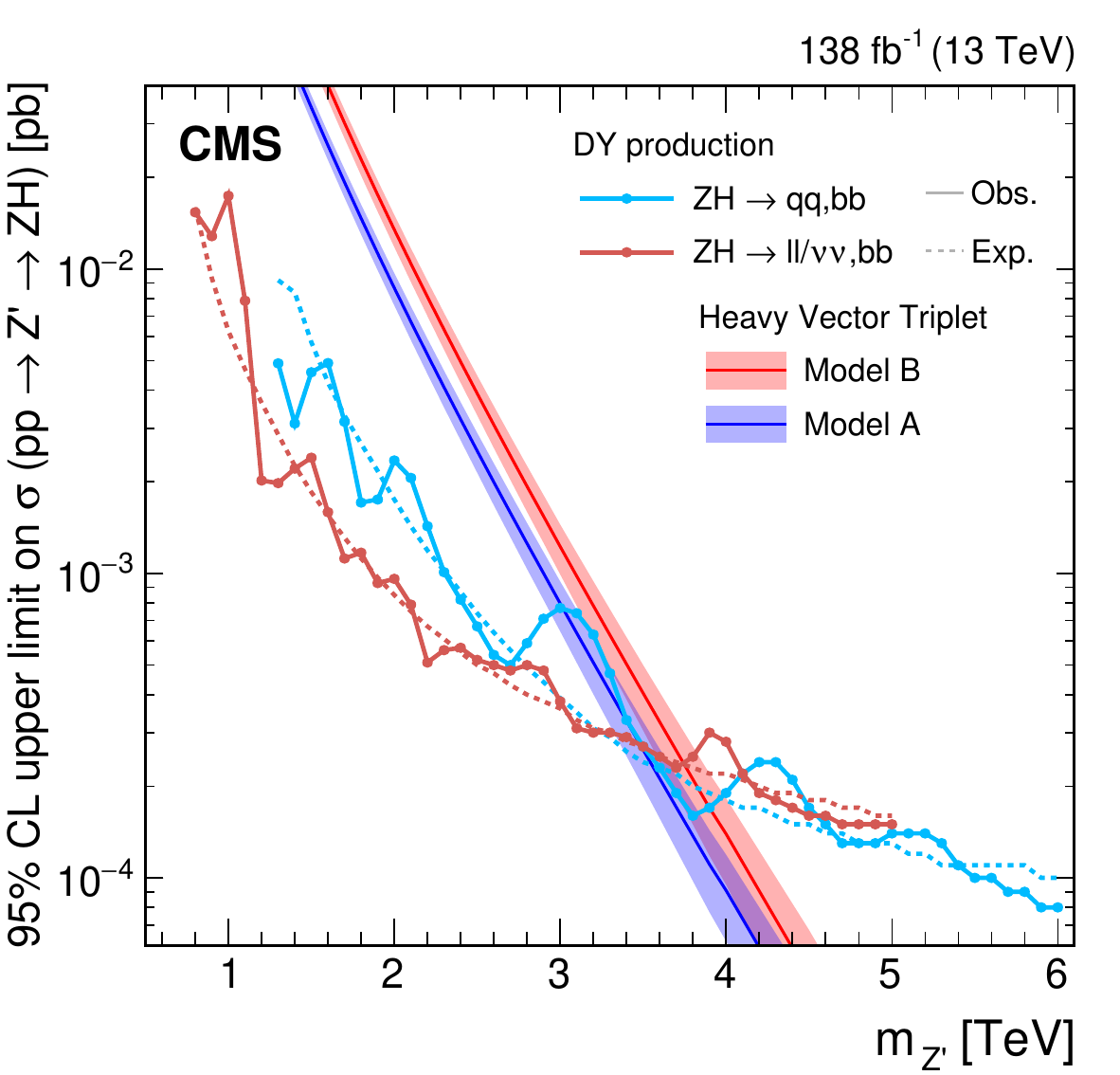}
  \includegraphics[width=\cmsFigWidth]{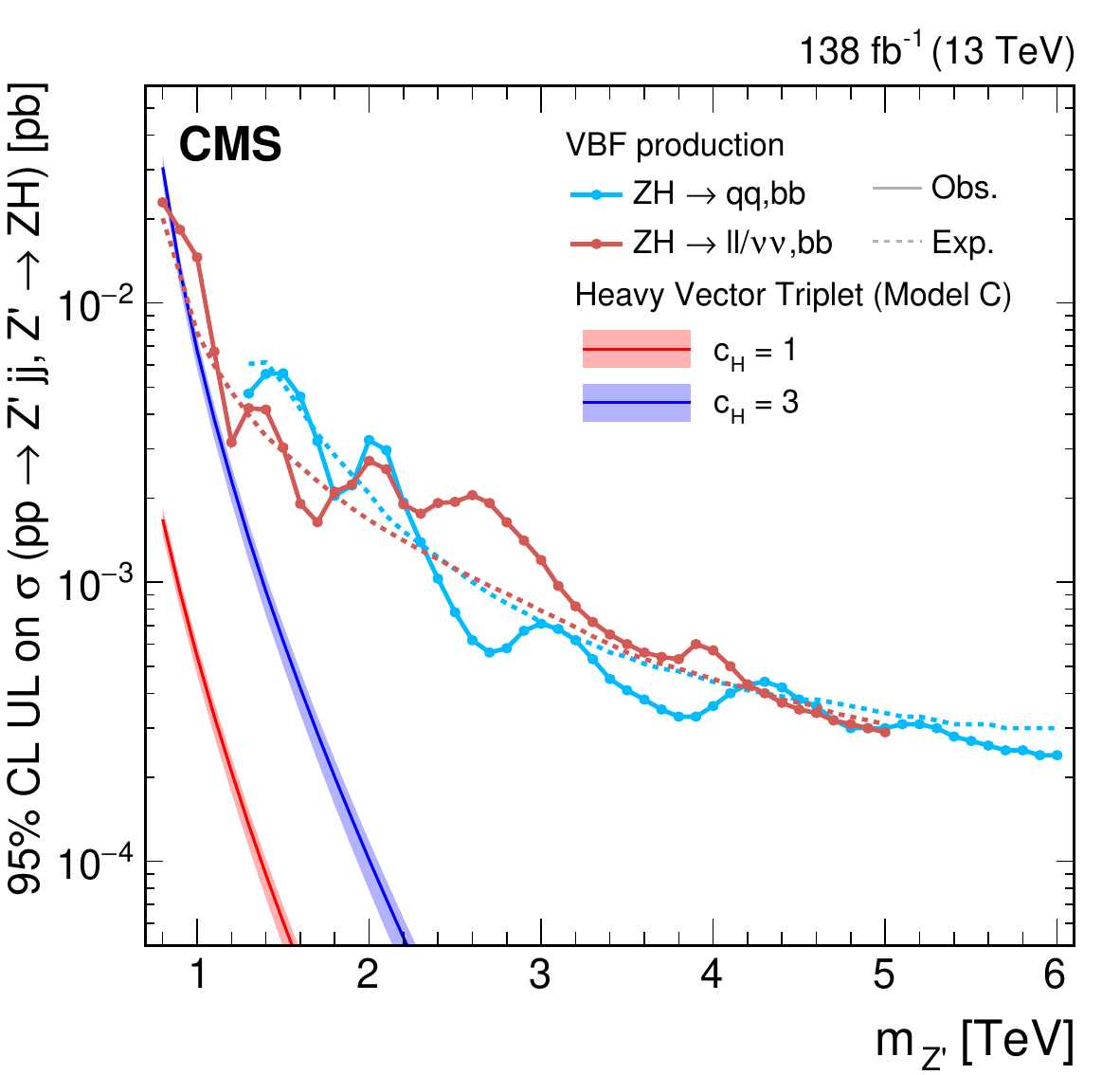}
  \caption{
    Search for $\PX\to\PZ\PH$: Observed and expected 95\% \CL upper limits on the 
    product of the cross section $\sigma$ for the production of a \PZpr spin-1 
    resonance, via (\cmsLeft) DY production or (\cmsRight) vector boson 
    fusion and the branching fraction \BR for the $\PZpr\to\PZ\PH$ decay. 
    The solid lines represent the observed and the dotted lines the expected limits. The 
    theory predictions from the heavy vector triplet models A, B and C are also 
    shown.
  }
  \label{fig:Limits_on_ZH}
\end{figure}
Figures~\ref{fig:Limits_on_WH} and \ref{fig:Limits_on_ZH} show the upper limits 
on $\sigma\BR$ for spin-1 \PWpr and \PZpr resonances, 
as a function of the masses \MWpr and \MZpr, respectively. 
The limits are derived for DY (\cmsLeft) and VBF (\cmsRight) production separately. 
The exclusion limits reach values of $\sigma\BR$ below 0.1 and 0.3\unit{fb} 
for the DY and VBF topologies, respectively.
In DY production the results from searches with leptons in the final state yield a stronger exclusion 
for \MWpr masses below 1.7\TeV and \MZpr below 3.2\TeV. 
For higher masses, the fully hadronic final state shows higher sensitivity. 
The interpretations of these upper limits on $\sigma\BR$ in HVT models will be 
discussed in detail in Section~\ref{Sec:Interp_in_Heavy_Vector_Triplet}.

\subsection{The \texorpdfstring{$\PX\to\HH$}{X->HH} decays}\label{Sec:Results_X_to_HH}

The six searches for $\PX\to\HH$ discussed in Section~\ref{Sec:Analysis_X_to_HH} 
target a variety of final states with \PQb jets, photons, light leptons, and \tauh leptons. 
The searches study spin-0 and spin-2 resonances in the mass range 0.28--4.5\TeV. We denote the spin-0
resonance as \PX since interpretations in warped extra dimension and extended Higgs sector
models are both possible. We denote the spin-2 resonance as \PG having a graviton in mind.

\begin{figure}[tbp]\centering
  \includegraphics[width=\cmsFigWidth]{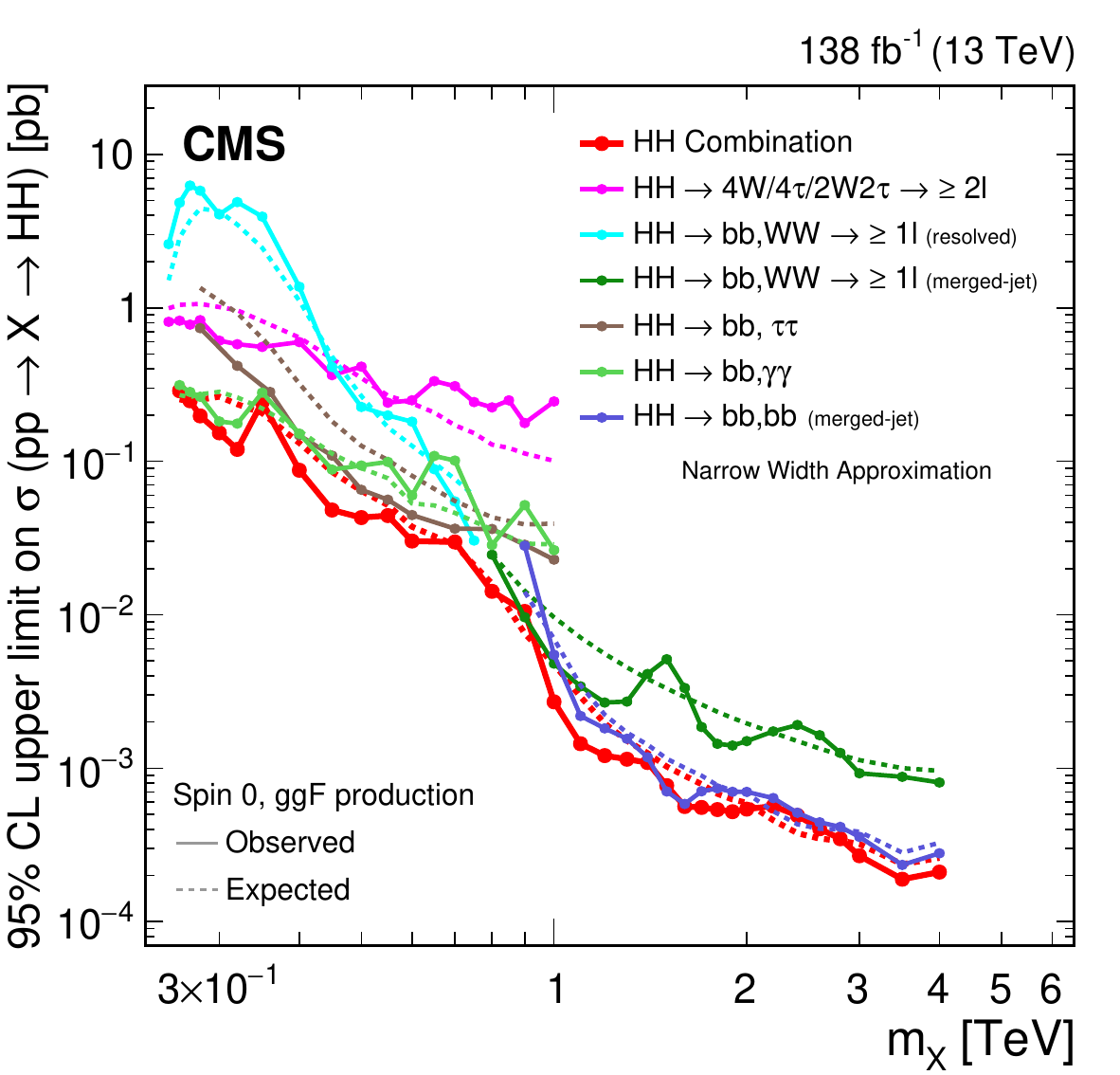}
  \includegraphics[width=\cmsFigWidth]{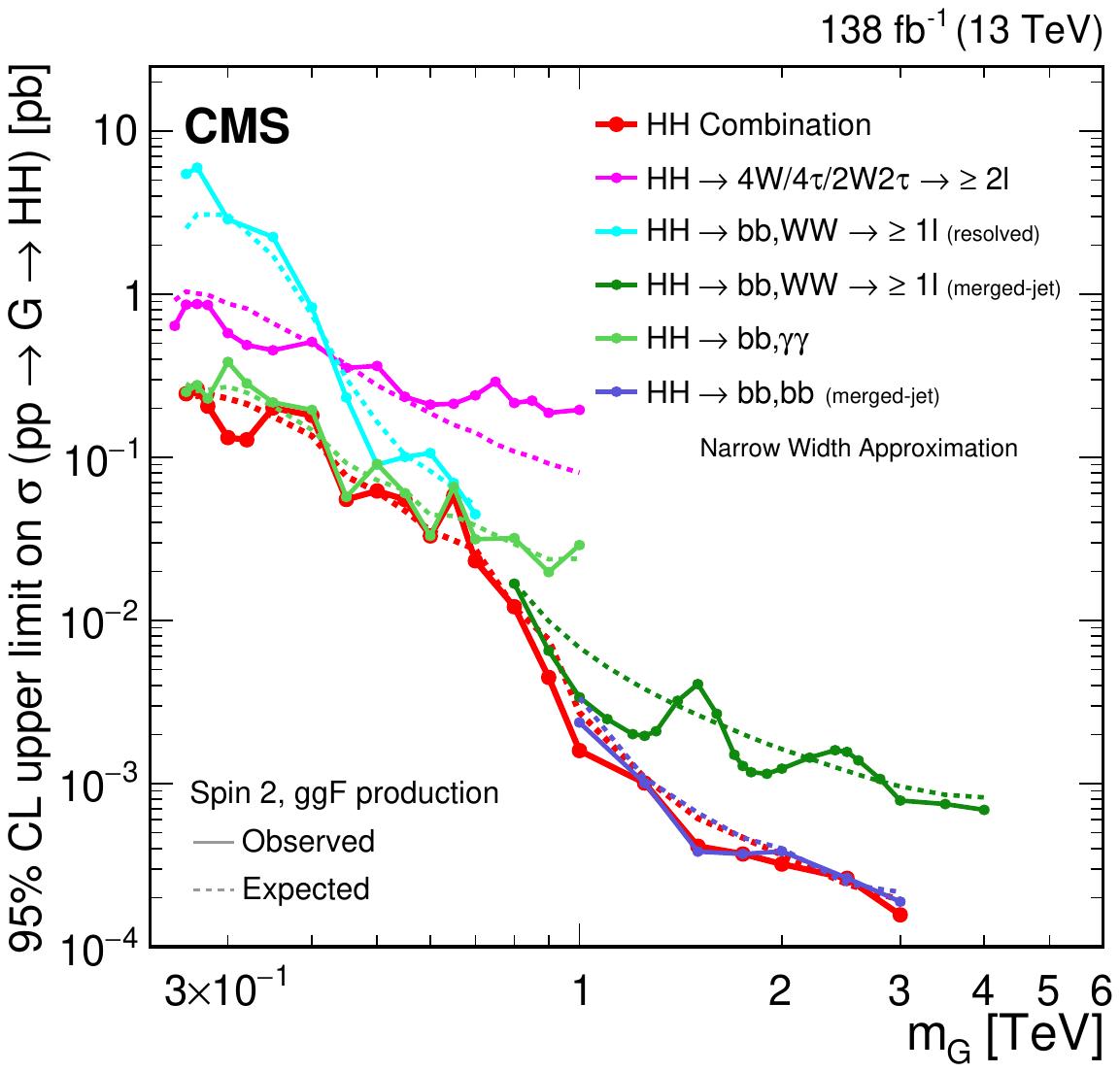}
  \caption{
    Search for $\PX\to\PH\PH$/$\PG\to\PH\PH$: Observed and expected 95\%~\CL upper 
    limits on the product of the cross section $\sigma$ for the production of a 
    (\cmsLeft) spin-0 resonance \PX and (\cmsRight) a spin-2 resonance \PG, via 
    gluon-gluon fusion and the branching fraction \BR for the corresponding 
    \HH decay. The results of the individual analyses presented in this report,
    corrected for the branching fractions of the respective H decay modes,  
    and the result of their combined likelihood analysis are shown. The observed 
    limits are indicated by markers connected with solid lines and 
    the expected limits by dashed lines. 
  }
  \label{fig:Limits_on_HH}
\end{figure}
\begin{figure}[tbp]\centering
  \includegraphics[width=\cmsFigWidth]{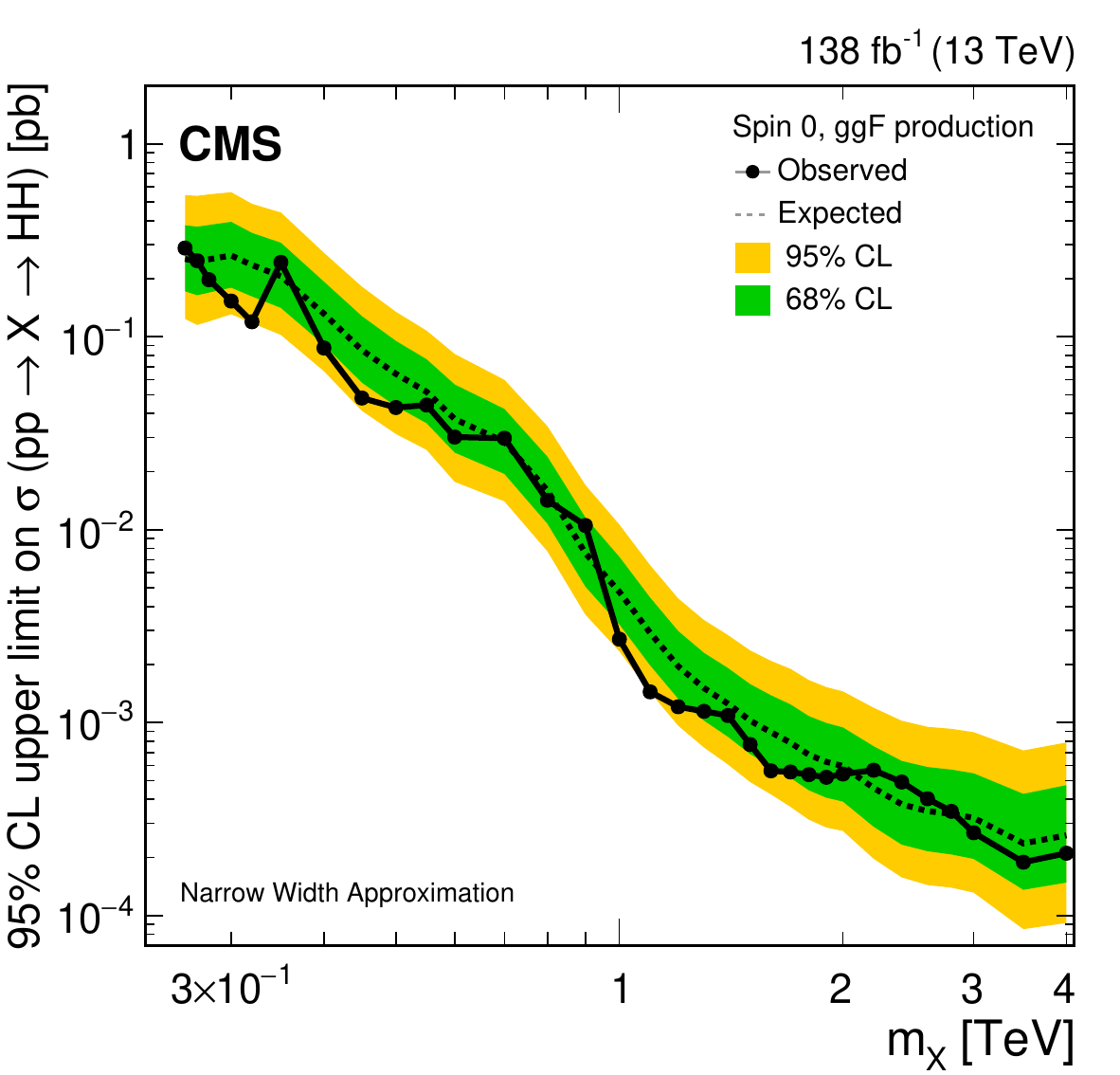}
  \includegraphics[width=\cmsFigWidth]{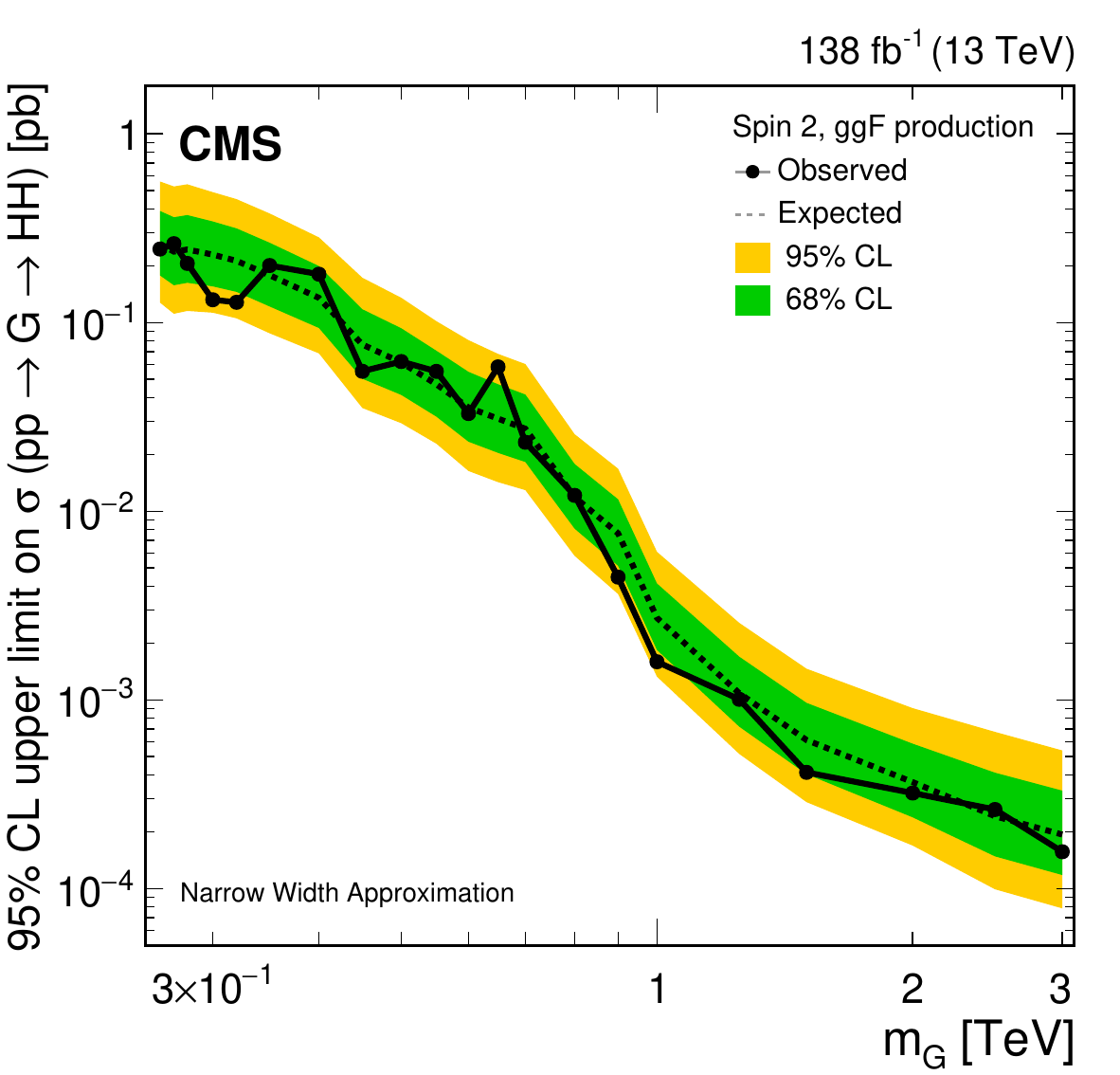}
  \caption{
    Search for $\PX\to\PH\PH$/$\PG\to\PH\PH$: Observed and expected 95\%~\CL upper 
    limits on the product of the cross section $\sigma$ for the production of a 
    (\cmsLeft) spin-0 resonance \PX and (\cmsRight) a spin-2 resonance \PG, via 
    gluon-gluon fusion, and the branching fraction \BR for the corresponding 
    \HH decay, as obtained from the combined likelihood analysis of all 
    contributing individual analyses presented in this report and shown in 
    Fig.~\ref{fig:Limits_on_HH}. In addition to the limit from the combined 
    likelihood analysis the 68 and 95\% central intervals for the 
    expected upper limits in the absence of a signal are shown as colored bands. 
  }
  \label{fig:Limits_on_HH_combo}
\end{figure}
Figure~\ref{fig:Limits_on_HH} shows the upper limits on $\sigma\BR$ as functions of the 
resonance mass for both spin hypotheses.
The exclusion in terms of $\sigma\BR$ 
extends down to 0.2\unit{fb} for both spin scenarios probed. The best sensitivity at low masses 
is obtained by the diphoton search, while at high masses the two searches with \PQb-tagged merged jets 
show the best sensitivity. 
The results of the statistical combination as described in Section~\ref{Sec:Combination} are 
shown as red lines. These combined results are presented again separately in Fig.~\ref{fig:Limits_on_HH_combo} 
along with the $\pm1$ and $\pm2$ \SD intervals on the expected limits. 
No deviation larger than 2 \SD from the expected limits is observed. 
Large improvements in sensitivity relative to the best individual channel are achieved 
in the range of $\mX\sim 0.5$--1\TeV, where many channels contribute with about the same weight 
to the combination. Below masses of 0.32\TeV and above 0.8\TeV, this combination gives the 
strongest observed limits to date on resonant \HH production. A recent combination 
of \HH searches performed by the ATLAS Collaboration can be found in Ref.~\cite{ATLAS:2023vdy}.

\subsection{The \texorpdfstring{$\PX\to\PY\PH$}{X->YH} decays}\label{Sec:Results_X_to_YH}

Three searches target the $\PX\to\PY\PH$ decay. 
Two are dedicated to lower masses with the \PH boson decaying to \GamGam or \tautau 
with two \PQb-tagged AK4 jets for the reconstruction of the \PY boson (as discussed in Section~\ref{Sec:Analysis_X_to_YH}). 
The search in the fully hadronic final state with two double-\PQb-tagged 
AK8 jets targets the high-mass regime. 
As the \PY boson decays to \bb in all cases considered, this allows for 
a direct comparison of the results from these three searches, 
without the assumption of a specific model. 
Furthermore, this makes a model-independent combination possible, where only the 
branching fractions of the \PH boson need to be taken into account. 

\begin{figure}[tbp]\centering
  \includegraphics[width=.8\textwidth]{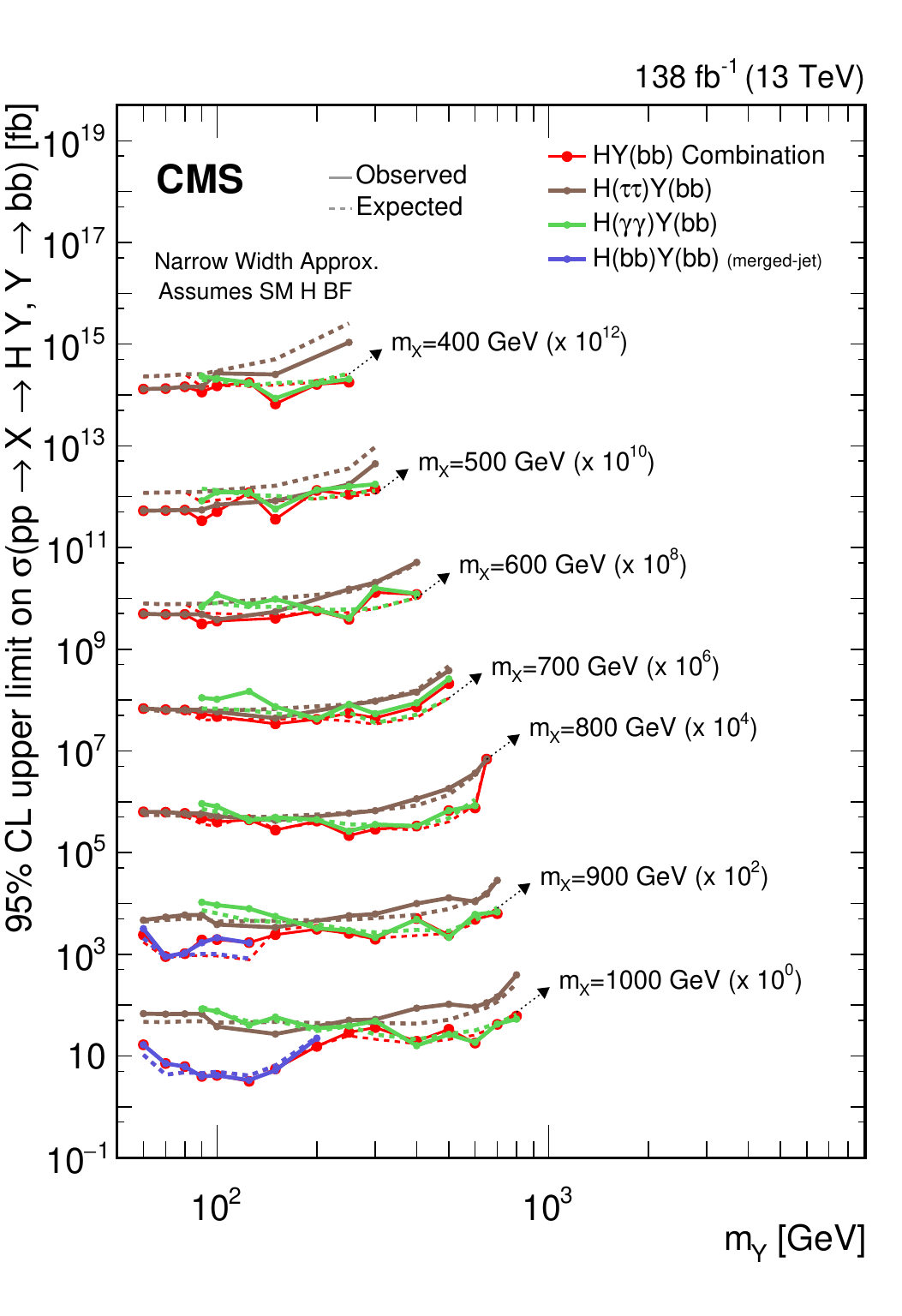}
  \caption{
    Search for $\PX\to\PY\PH$: Observed and expected upper limits, at 95\%~\CL, on the 
    product of the cross section $\sigma$ for the production of a resonance \PX 
    via gluon-gluon fusion and the branching fraction \BR for the $\PX\to\PY(\bb)
    \PH$ decay. For the branching fractions of the $\PH\to\tautau$, $\PH\to\gamma\gamma$
    and  $\PH\to\bb$ decays, the SM values are assumed.
    The results derived from the individual analyses presented in this 
    report and the result of their combined likelihood analysis are shown as 
    functions of \mY and \mX for $\mX\le1\TeV$. Observed limits are indicated by 
    markers connected with solid lines, expected limits by dashed lines. For presentation purposes, the limits 
    have been scaled in successive steps by two orders of magnitude, each. For 
    each set of graphs, a black arrow points to the \MX related legend.
  }
  \label{fig:XYH_combination}
\end{figure}
\begin{figure}[tbp]\centering
  \includegraphics[width=.8\textwidth]{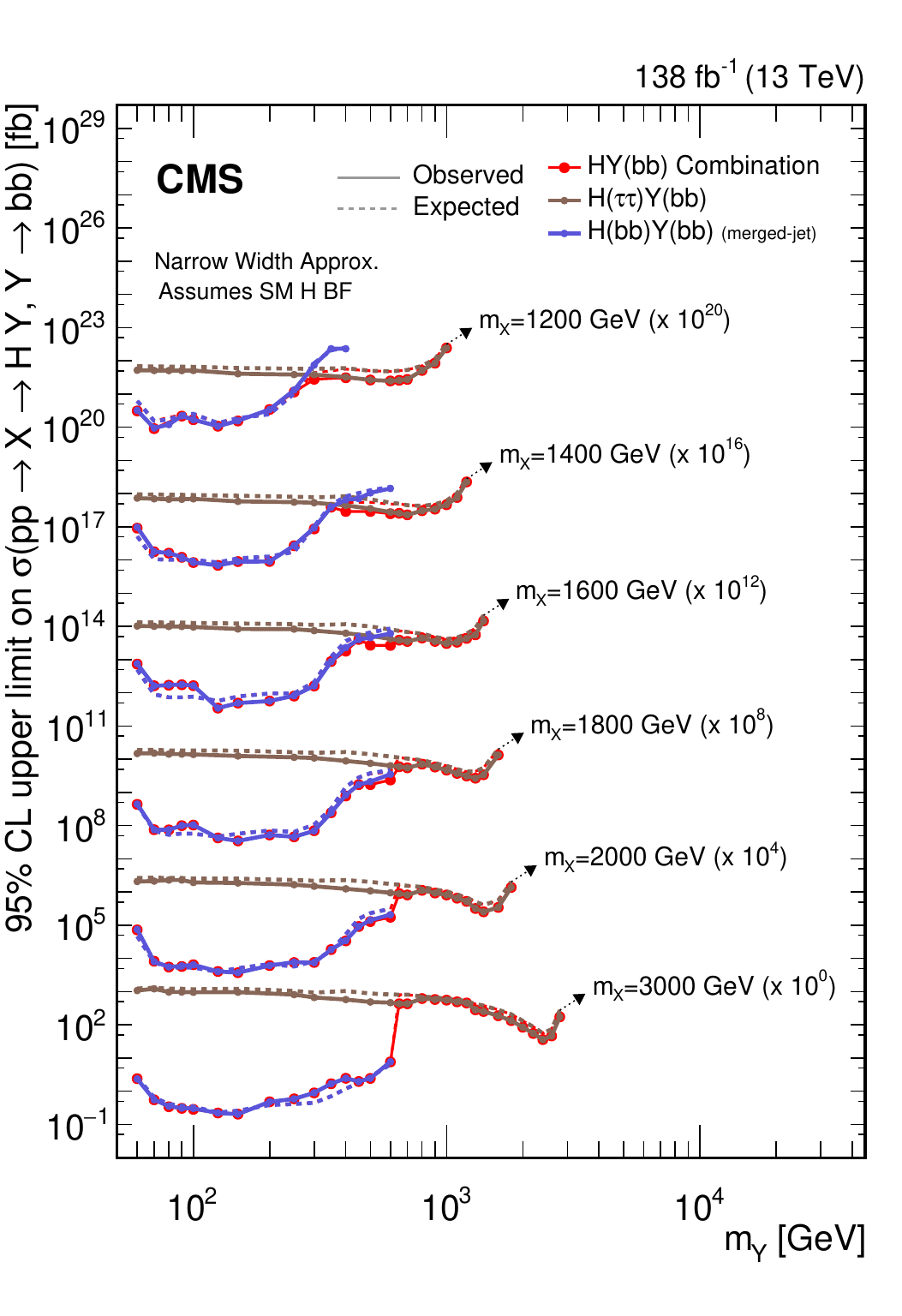}
  \caption{
    Search for $\PX\to\PY\PH$: Observed and expected upper limits, at 95\%~\CL, on the 
    product of the cross section $\sigma$ for the production of a resonance \PX 
    via gluon-gluon fusion and the branching fraction \BR for the $\PX\to\PY(\bb)
    \PH$ decay. For the branching fractions of the $\PH\to\tautau$
    and  $\PH\to\bb$ decays, the SM values are assumed.
    The results derived from the individual analyses presented in this 
    report and the result of their combined likelihood analysis are shown as 
    functions of \mY and \mX for $\mX\ge1.2\TeV$. Observed limits are indicated by 
    markers connected with solid lines, expected limits by dashed lines. For presentation purposes, the 
    limits have been scaled in successive steps by four orders of magnitude, each. 
    For each set of graphs, a black arrow points to the \MX related legend.
  }
  \label{fig:XYH_combination_2}
\end{figure}
Figures~\ref{fig:XYH_combination} and \ref{fig:XYH_combination_2} show
the upper limits on $\sigma \BR$ as functions of the \mY for $\mX \le 1\TeV$ and
for $\mX \ge 1.2\TeV$, respectively. The results have been achieved 
by adjusting each channel to the corresponding SM branching fraction 
of the \PH boson decay under consideration. No correction has been made for the
unknown branching fraction of $\PY\to\bb$, which is the same in all searches. 

At low \mX, the $\PY(\bb)\PH(\tautau)$ and $\PY(\bb) \PH(\GamGam)$ 
analyses provide the best sensitivity. For $\mX = 1\TeV$ and higher, 
the $\PY(\bb)\PH(\bb)$ in the merged jet topology dominates for small and medium values of \mY. 
At the largest values of \mY, however, approaching the kinematic limit, 
the sensitivity of the $\PY(\bb)\PH(\bb)$ analysis is reduced because the Lorentz boost of 
the \PY boson rest frame is too small for the fragmentation products of the two \PQb quarks 
to merge into a single jet.

The three analyses are statistically combined as described in Section~\ref{Sec:Combination}, 
and the resulting expected and observed limits are shown in Figs.~\ref{fig:XYH_combination} and 
\ref{fig:XYH_combination_2}. 
Covering the full mass grid, however, is beyond the scope of this Report. 
This combination is shown as an example for the given mass points, and a separate publication 
in the near future will include a full combination featuring a larger set of decay modes. 
The typical exclusion upper limits on $\sigma\BR$ are about 50, 5, and 0.3\unit{fb} for 
$\mX=0.5$, 1, and 3\TeV, respectively.
No excess larger than two \SD above the expected limit is observed at any of these mass points.
A two-dimensional representation of the experimental limits in the (\mX, \mY) parameter space is shown 
as part of the interpretation in Section~\ref{Subsubsec:Interpretations_NMSSM_TRSM}.

\section{Model-specific interpretation}\label{Sec:ResultsInterpretations}

We interpret the results of the individual searches and their combinations 
in specific models. The interpretations highlight the coverage of the 
analyses in the corresponding parameter space and show which regions 
are excluded by the current data.  
The first three subsections address how the measurements can constrain 
the parameter space of models with an extended Higgs sector, 
warped extra dimensions, and in a HVT framework. 
Section~\ref{Sec:Effects_finite_width_and_interference} 
is dedicated to studies going beyond the narrow-width approximation (NWA), 
where we investigate the effects of non-negligible resonance widths and 
interference.

\subsection{Extended Higgs sector models}\label{Sec:Interp_in_Extended_Higgs_sector}

\subsubsection{The MSSM}

The decays $\PX\to\HH$ and $\PA\to\ZH$ can have sizable branching fractions 
in models with two complex Higgs doublets. However, as discussed previously, the branching 
fractions get suppressed when approaching the alignment limit, where the \PH boson becomes SM-like. 
Searches for \HH and \ZH can nonetheless set important constraints in these models, 
in particular at low to intermediate values of \tanb, and for masses below or near 
the \ttbar threshold, $m_{\PX/\PA} \lesssim 350$\GeV.

\begin{figure}[tbp]
    \centering
    \includegraphics[width=0.8\textwidth]{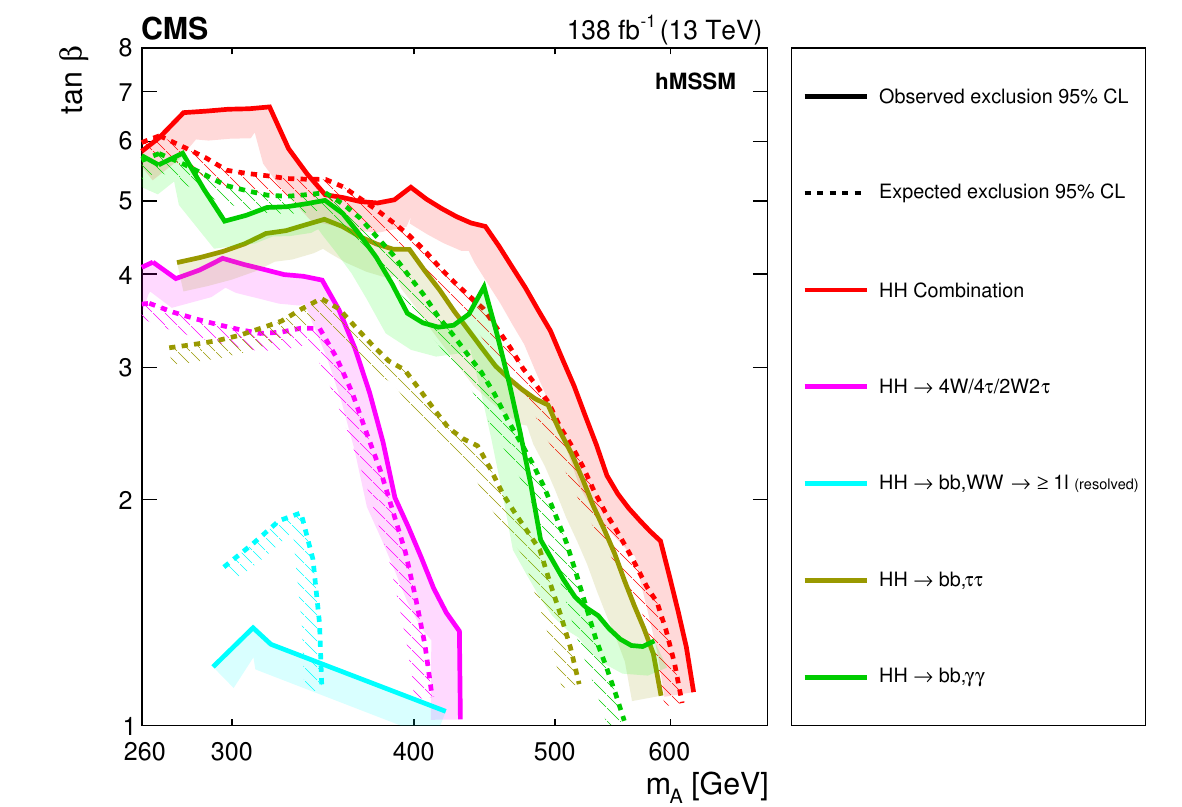}
    \includegraphics[width=0.8\textwidth]{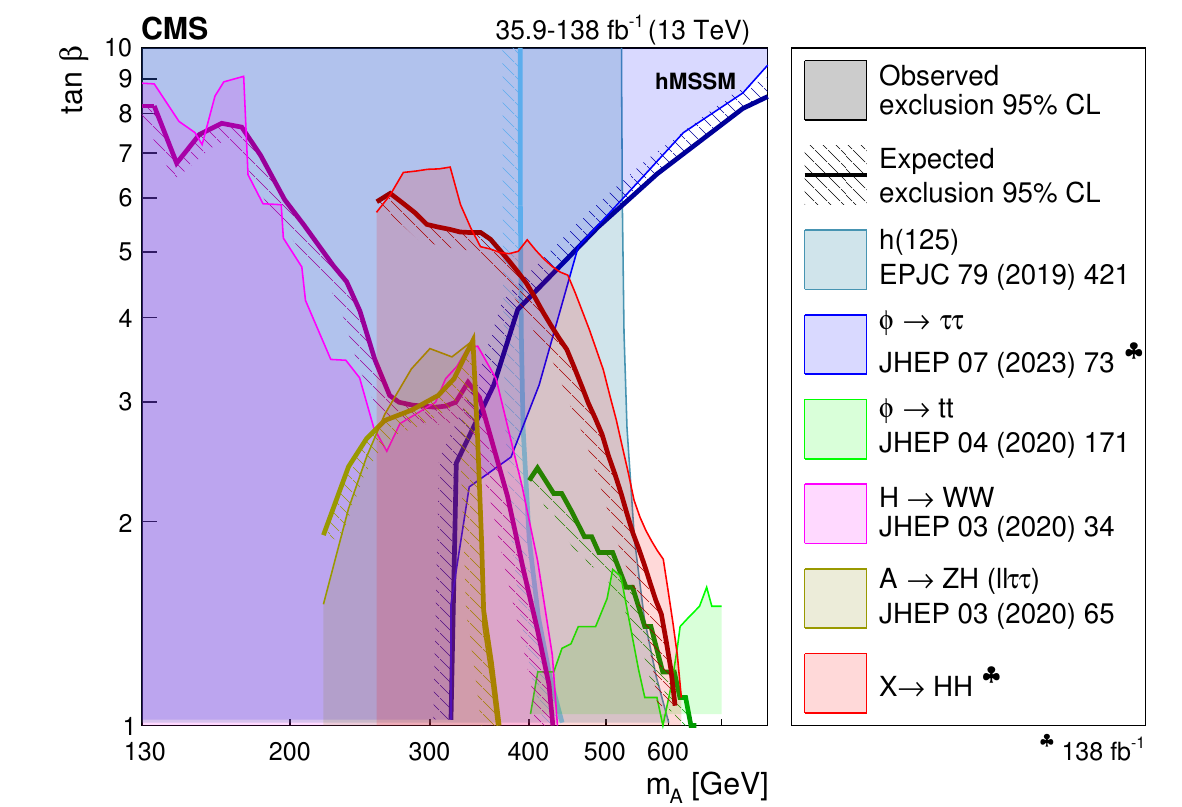}
    \caption{
      Interpretation of the results from the searches for the $\PX\to\HH$ decay, 
      in the hMSSM model. In the upper part of the figure, the observed and expected 
      exclusion contours at 95\%~\CL, in the (\mA, \tanb) plane, from the individual 
      \HH analyses presented in this report and their combined likelihood analysis 
      are shown. In the lower part of the figure, a comparison of the region 
      excluded by the combined likelihood analysis shown in the upper part of the 
      figure with selected results from other searches for the production of heavy 
      scalar bosons in the hMSSM, in \tautau~\cite{CMS:2022goy},
      \ttbar~\cite{CMS:2019pzc} and $\PW\PW$~\cite{CMS:2019bnu} decays is shown. 
      Also shown, are the results from one representative search for $\PA\to\ZH
      $~\cite{CMS:2019kca} and indirect constraints obtained from measurements of 
      the coupling strength of the observed \PH boson~\cite{CMS:2018uag}. Results 
      not marked by a club symbol are based on an integrated luminosity of $35.9
      \fbinv$.
    }
    \label{fig:hMSSM}
\end{figure}
Figure~\ref{fig:hMSSM} shows exclusion regions in the (\mA, \tanb) plane of the hMSSM~\cite{Djouadi:2013vqa,Djouadi:2013uqa,
Djouadi:2015jea,lhc_higgs_working_group_mssm_subgroup_2022_6793918}. 
For this and the following model interpretations, the version numbers of the corresponding tools are documented in
Ref.~\cite{lhc_higgs_working_group_mssm_subgroup_2022_6793918}.
The branching fractions are obtained with \textsc{HDECAY}~\cite{Djouadi:1997yw,Djouadi:2018xqq}.
The gluon fusion cross section is obtained with \textsc{SusHi}~\cite{Harlander:2012pb, Harlander:2016hcx}, 
which includes higher-order QCD corrections~\cite{Spira:1995rr,Harlander:2005rq,Harlander:2002wh, Anastasiou:2002yz, 
Ravindran:2003um,Harlander:2002vv, Anastasiou:2002wq} and EW effects from light quarks~\cite{Aglietti:2004nj, Bonciani:2010ms}. 
The $\PX\to\HH$ searches result in an exclusion for $\tanb \lesssim 6$ for \mA just above the 
\HH production threshold of 250\GeV, decreasing to $\tanb \lesssim 1$ for $\mA\approx600\GeV$. 
This is complementary to the exclusion regions from searches for fermionic decays, such as $\PA\to\tautau$, 
which exclude regions of large \tanb. 
The $\PA\to\ZH$ search in the $\PH\to\tautau$ channel provides sensitivity for $220 < \mA < 350\GeV$ and 
excludes regions below $\tanb = 3.6$ for $\mA \lesssim 330\GeV$. 
Compared to other direct searches, there is a unique sensitivity of the $\PX\to\HH$ searches for 
$\mA \gtrsim 450\GeV$ and $\tanb < 5$. 
At the same time, the constraints derived from the measurements of the \PH boson couplings are 
somewhat more stringent, albeit these place only indirect constraints on this model.

\begin{figure}[tbp]
  \centering
  \includegraphics[width=0.8\textwidth]{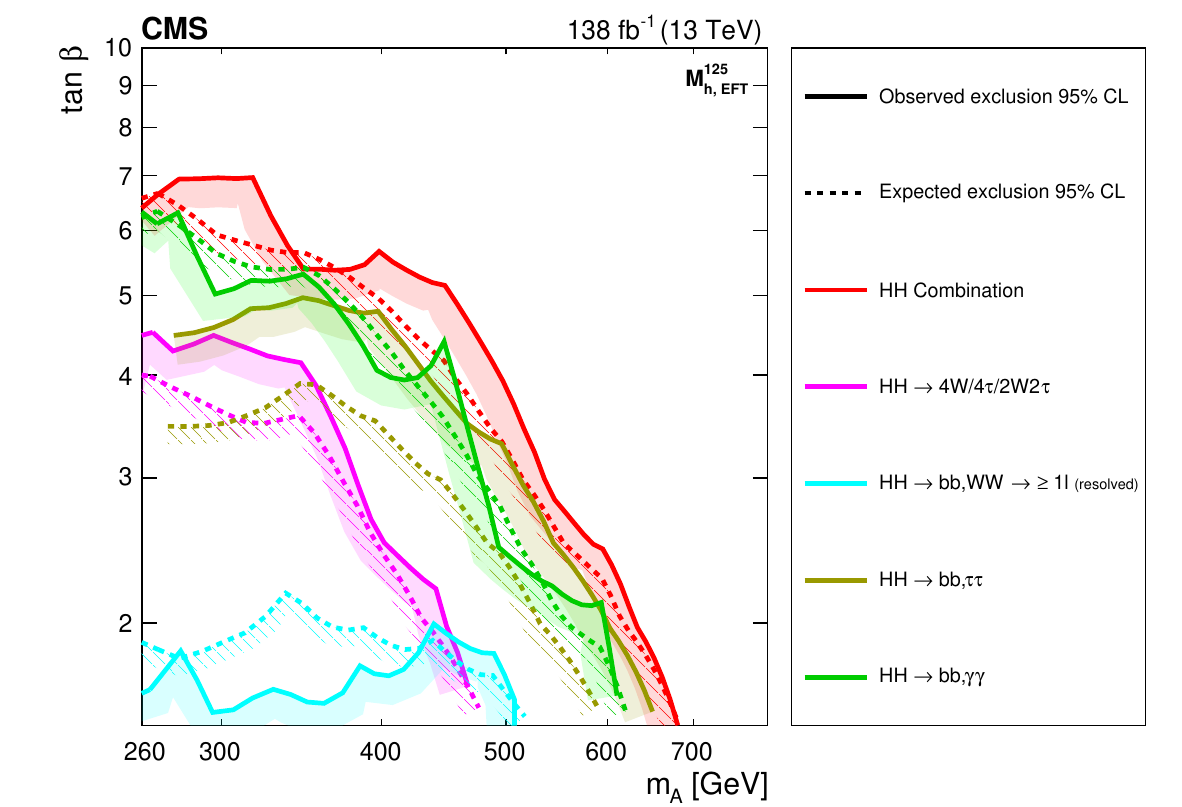}
  \includegraphics[width=0.8\textwidth]{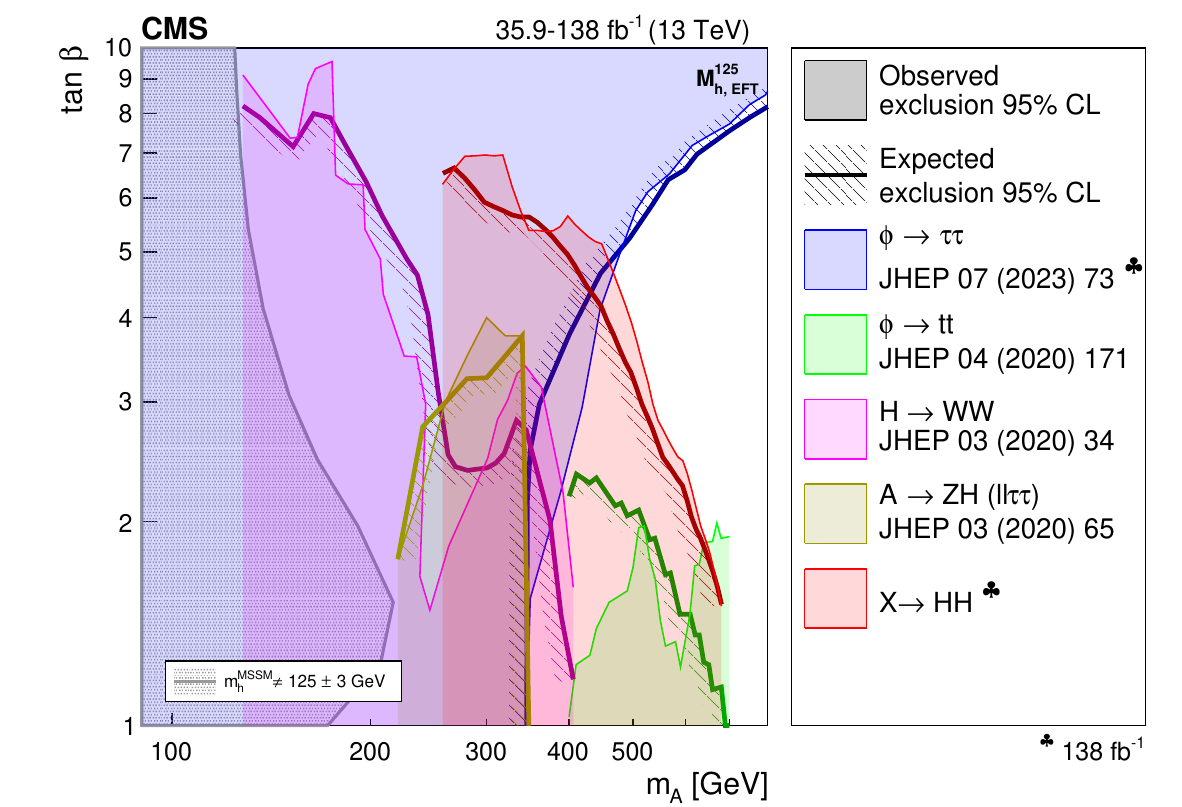}
  \caption{
    Interpretation of the results from the searches for the $\PX\to\HH$ decay, 
    in the \MhEFTScen benchmark scenario. In the upper part of the figure, the 
    observed and expected exclusion contours at 95\%~\CL are shown, in the (\mA, \tanb) 
    plane from the individual \HH analyses presented in this report and their 
    combined likelihood analysis. In the lower part of the figure, a 
    comparison of the region excluded by the combined likelihood analysis shown 
    in the upper part of the figure with selected results from other searches 
    for the production of heavy scalar bosons in the \MhEFTScen scenario, in 
    \tautau~\cite{CMS:2022goy}, \ttbar~\cite{CMS:2019pzc} and $\PW\PW$~\cite{
      CMS:2019bnu} decays is shown. Also shown, are the results from one 
    representative search for $\PA\to\ZH$~\cite{CMS:2019kca}. The parameter 
    region in which the mass of the lightest MSSM Higgs boson does not coincide 
    with 125\GeV within a 3\GeV margin is indicated by the dark hatched area. 
    Results not marked by a club symbol are based on an integrated luminosity 
    of $35.9\fbinv$.
  }
  \label{fig:MSSM_mh125}
\end{figure}
The frequently used \MhScen benchmark model is not very suitable for interpretations of results from $\PX\to\HH$ searches  
as these exclude regions at low \tanb where the SM-like scalar has a mass inconsistent with 125\GeV 
and thus with the observed \PH boson. Instead, we choose to interpret these results in the \MhEFTScen scenario~\cite{Bahl:2019ago,lhc_higgs_working_group_mssm_subgroup_2022_6793918}. 
Higgs boson masses and mixings are obtained with \textsc{FeynHiggs}~\cite{Heinemeyer:1998yj, Heinemeyer:1998np, 
Degrassi:2002fi, Frank:2006yh, Hahn:2013ria, Bahl:2016brp, Bahl:2017aev,Bahl:2018qog}. The branching fraction calculations 
make use of both \textsc{FeynHiggs}, \textsc{HDECAY}, and \textsc{PROPHECY4f}~\cite{Bredenstein:2006rh, Bredenstein:2006ha}. 
The cross section for gluon fusion production is obtained from the same tools and predictions as in the hMSSM scenario.
The resulting exclusion regions in the (\mA, \tanb) plane are shown in Fig.~\ref{fig:MSSM_mh125}. 
In this scenario, the parameter regions excluded by the \HH combination are found not to be in conflict with the measured 
\PH boson mass. For $\mA \gtrsim 400\GeV$ the results from the combination provide unique exclusions. 
Otherwise, the overall picture is similar to the hMSSM scenario.

\subsubsection{The 2HDM}

\begin{figure}[tbp]\centering
    \includegraphics[width=\cmsFigWidth]{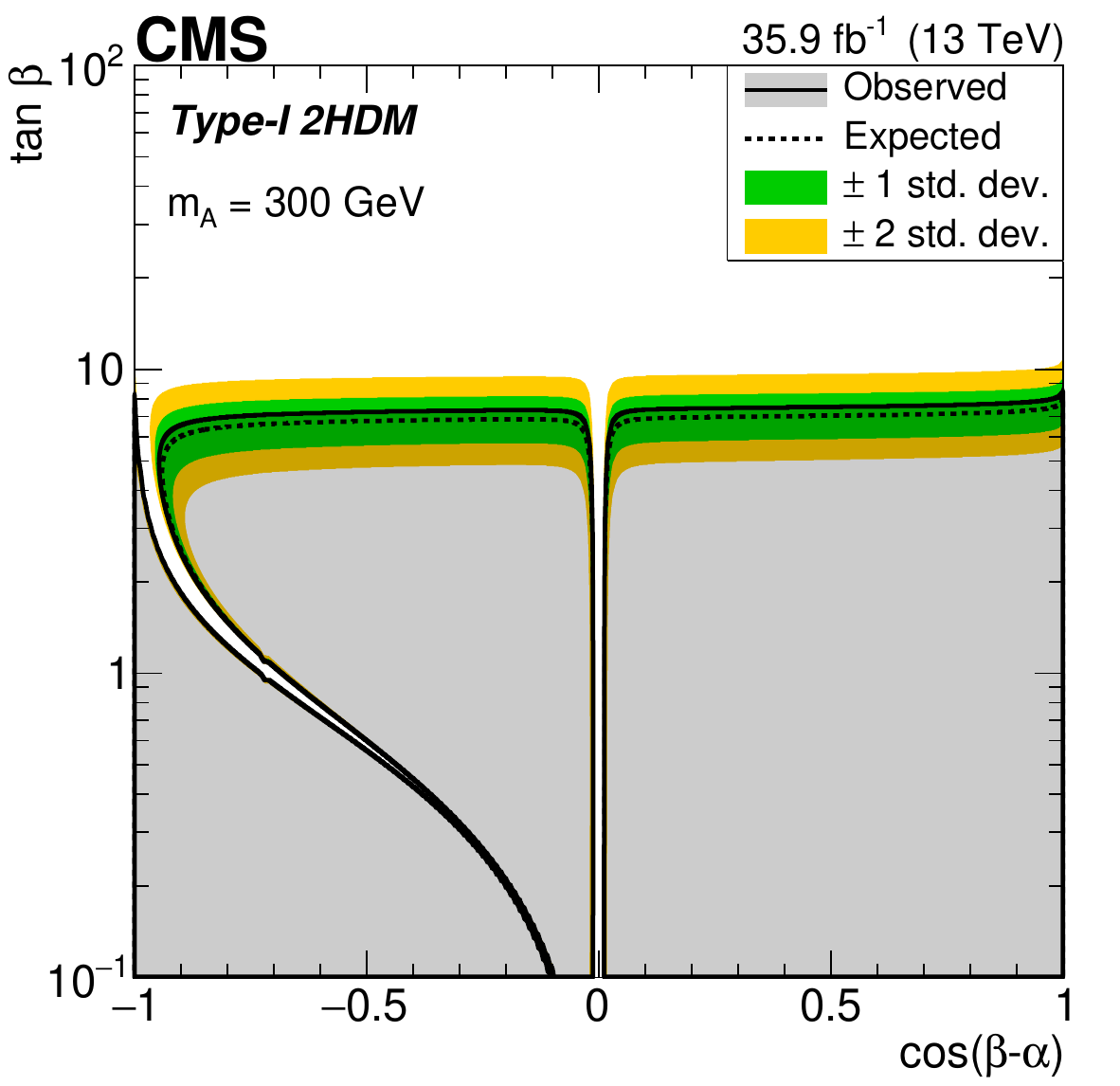}
    \includegraphics[width=\cmsFigWidth]{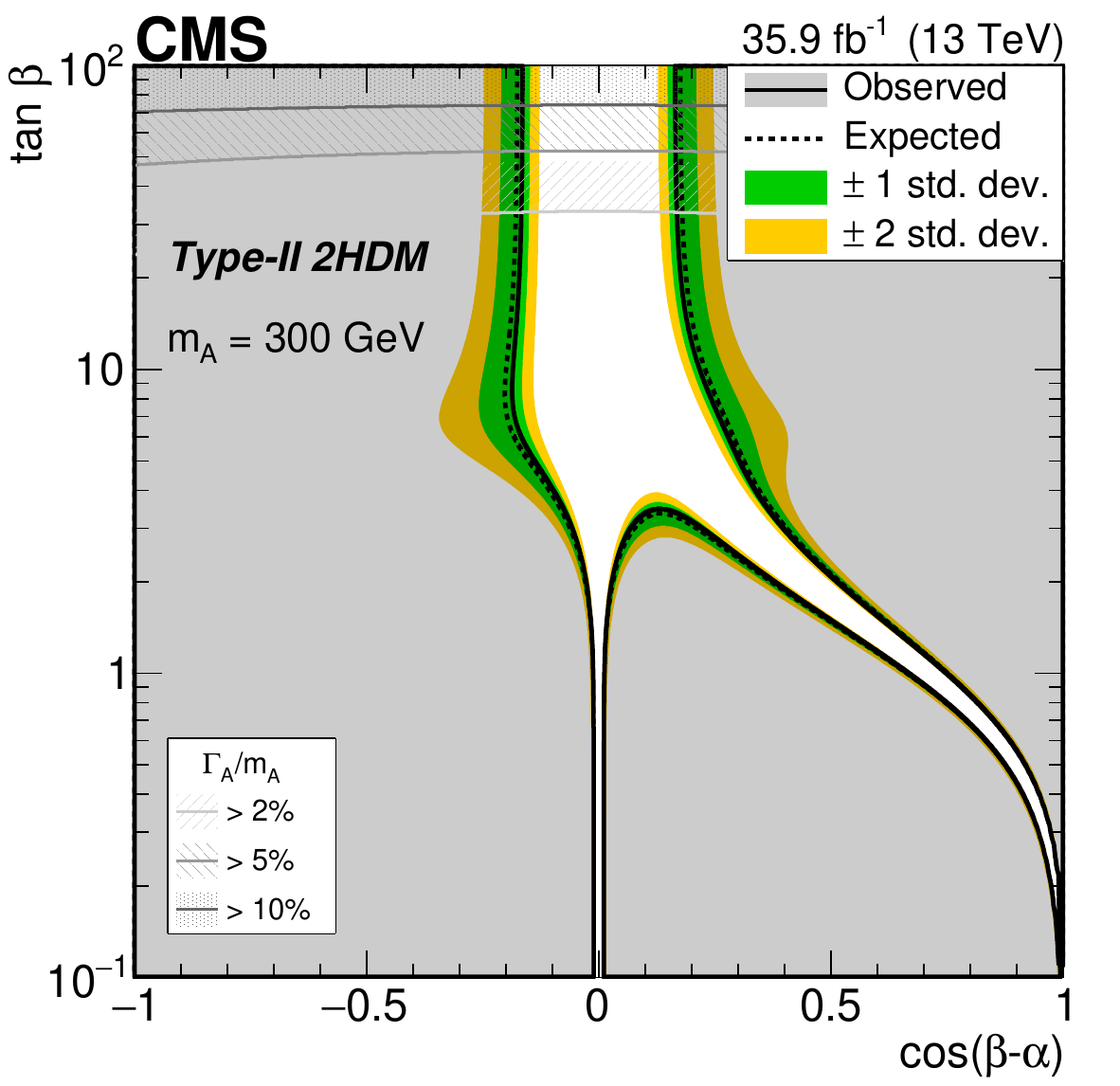}
    
    \includegraphics[width=\cmsFigWidth]{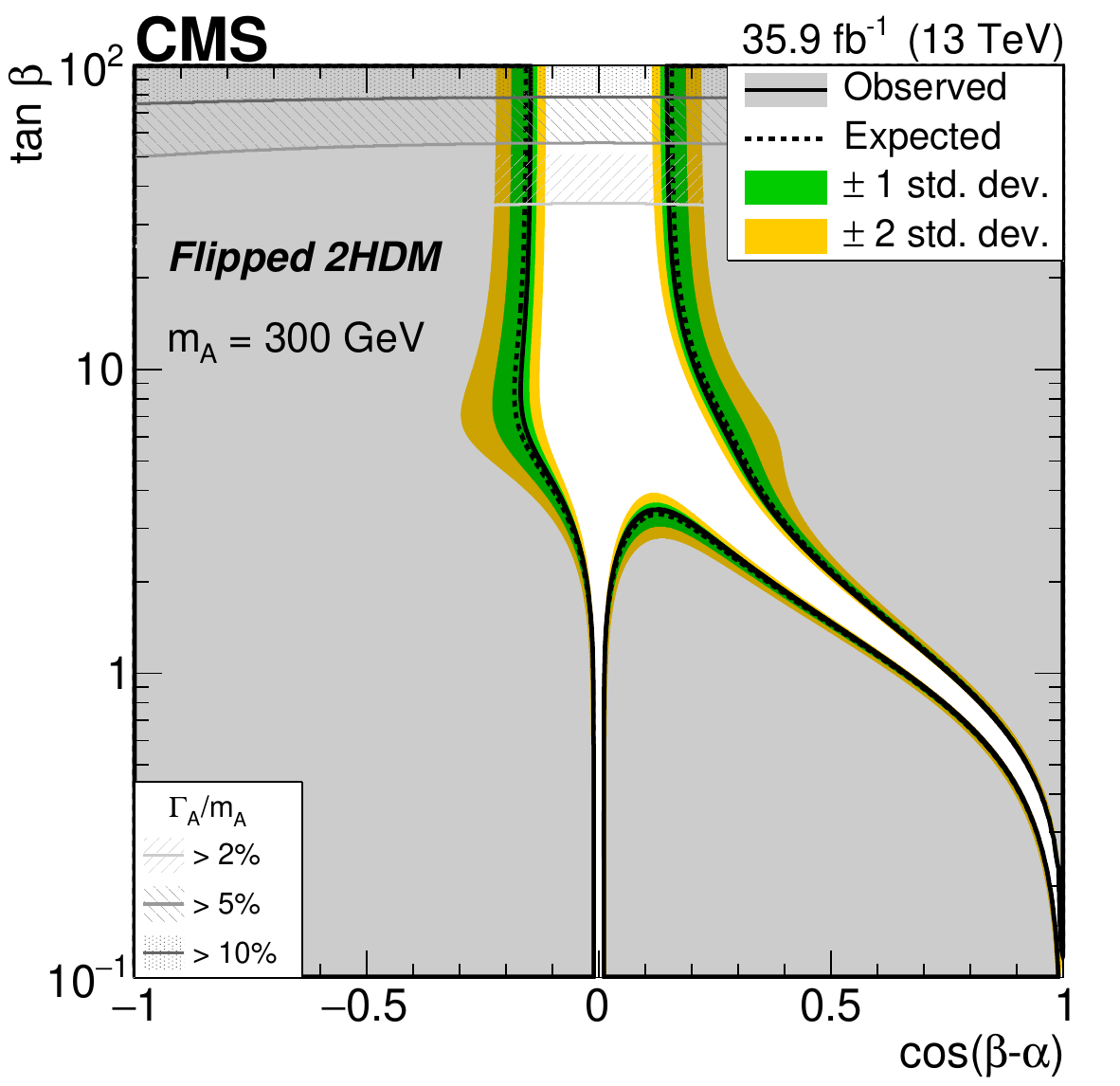}
    \includegraphics[width=\cmsFigWidth]{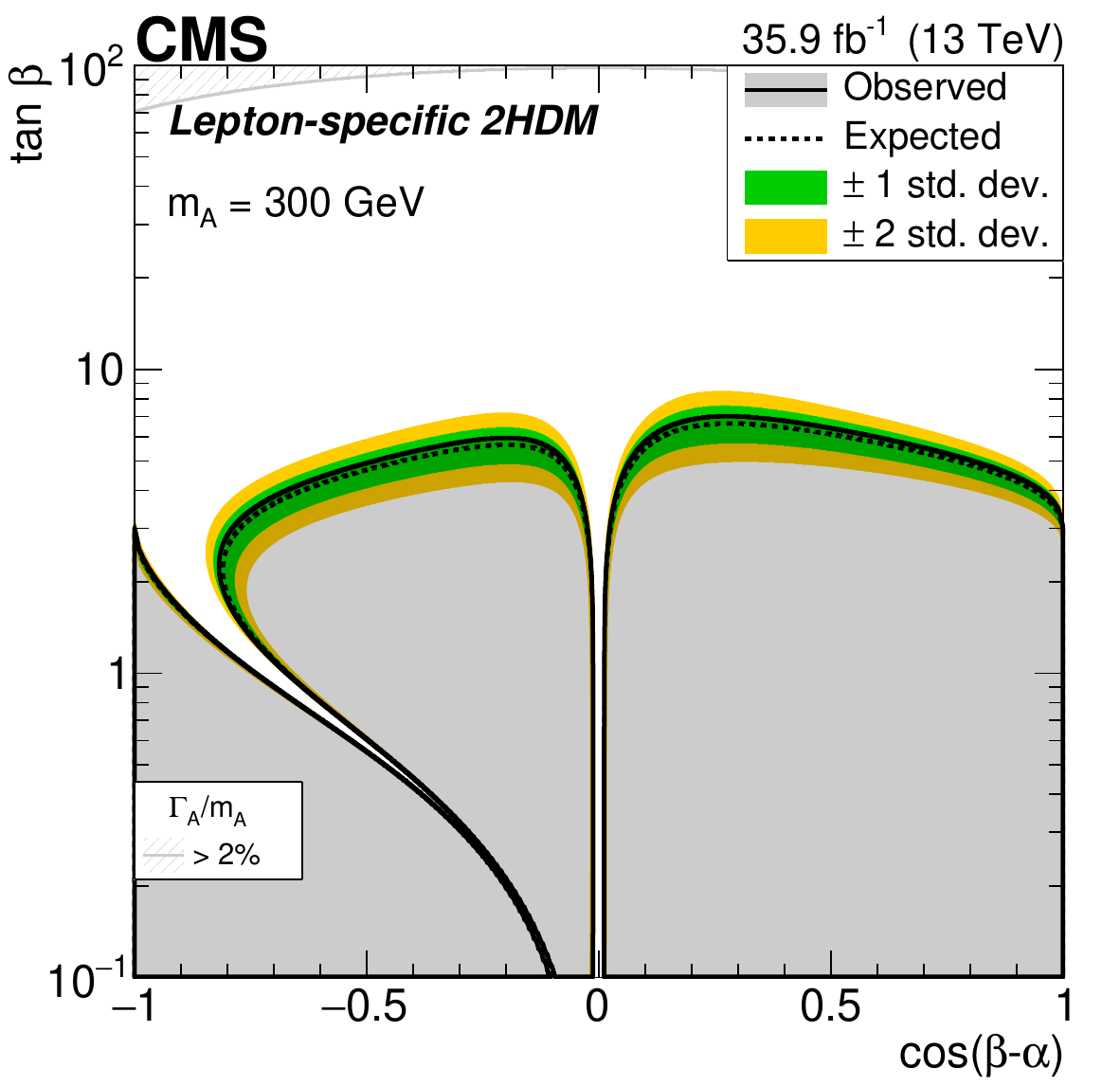}
    \caption{
      Interpretation of the results of the $\PA\to\PZ\PH(\bb)$ 
      analysis~\cite{CMS:2019qcx}, in the (upper \cmsLeft) Type~I, (upper 
      \cmsRight) Type~II, (lower \cmsLeft) flipped, and (lower \cmsRight) 
      lepton-specific 2HDM models. In each case observed and expected exclusion 
      contours at 95\%~\CL, in the plane defined by \cosba and \tanb, are shown. The excluded 
      regions are represented by the shaded gray areas. The 68 and 95\% central 
      intervals of the expected exclusion contours in the absence of a signal 
      are indicated by the green and yellow bands. Contours are derived from the 
      projection on the corresponding 2HDM parameter space for $\mA = 300\GeV$. 
      The regions of parameter space where the natural width of the \PA boson 
      $\Gamma_\PA$ is comparable to or larger than the experimental resolution 
      and thus the narrow-width approximation is not valid are represented by
      hatched gray areas. Figure from Ref.~\cite{CMS:2019qcx}.
    }
    \label{fig:2HDM_VHbb_cosba}
\end{figure}
Exclusion limits in the 2HDM are derived from the results of the search for $\Aboson\to\ZH(\Pb\Pb)$~\cite{CMS:2019qcx}. 
The 2HDM cross sections and branching fractions are computed with \textsc{2HDMC}~\cite{Eriksson:2009ws} 
and \textsc{SusHi}, respectively. The light \PH boson mass is set to 125\GeV and $\mX = \mHpm = \mA$ is used. 
The \PZ boson branching fractions are set to the measured values~\cite{PDG2022}.
Figure~\ref{fig:2HDM_VHbb_cosba} shows the constraints in the (\tanb, \cosba) plane 
for $\mA = 300\GeV$~\cite{CMS:2019qcx}. 
The search excludes nearly the whole region of low \tanb in all four 2HDM scenarios, except for a narrow region around $\cosba = 0$
for which the branching fraction goes to zero, another narrow region at negative \cosba 
for the Type~I and lepton-specific scenarios, and at positive \cosba for the other two scenarios. 
At high \tanb, the excluded region widens in the Type~II and flipped scenarios, 
whereas there is no sensitivity in the Type~I and lepton-specific scenarios because the production 
cross section for the \PA boson becomes too small.

\subsubsection{The NMSSM and TRSM models}
\label{Subsubsec:Interpretations_NMSSM_TRSM}

The searches for $\PX\to\PY\PH$ decays are interpreted in the NMSSM
and TRSM models.  In both models, there are various free parameters
besides the masses of the additional \PX and \PY bosons that affect
the cross sections and branching fractions.  To check whether the
searches are sensitive to a point in the (\mX, \mY) plane, a parameter
scan is performed to determine the maximally allowed cross section,
taking all previous constraints on the models into account.

In the NMSSM case, we obtain the maximally allowed cross section values
for $\sigma(\PX \to \PY\PH \to \bbbb)$ from the scans in
Ref.~\cite{Ellwanger:2022jtd}, which are based on version 5.6.2 of the
program \textsc{NMSSMTools}~\cite{Ellwanger:2022jtd,Ellwanger:2024etv}.
These numbers are divided by the corresponding branching fraction
$\BR(\Hbb)$ to obtain an approximation for the maximally allowed
values of $\sigma(\PX\to\PY(\bb)\PH)$.  Uncertainties arising from the
precision of the measured branching fractions of the \PH boson are
neglected.

\begin{figure}[tbp]
  \centering
  \includegraphics[width=\cmsFigWidth]{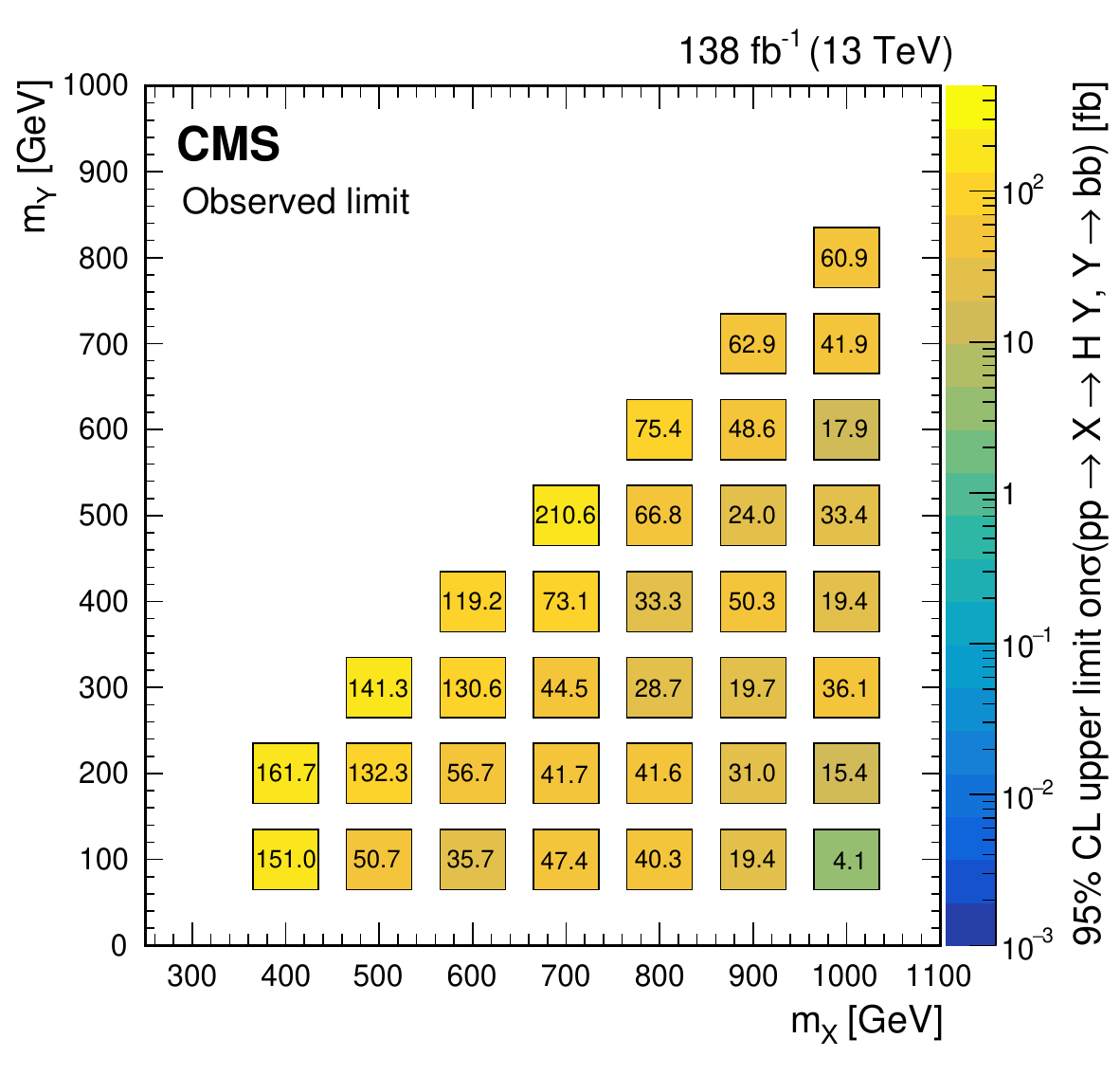}
  \includegraphics[width=\cmsFigWidth]{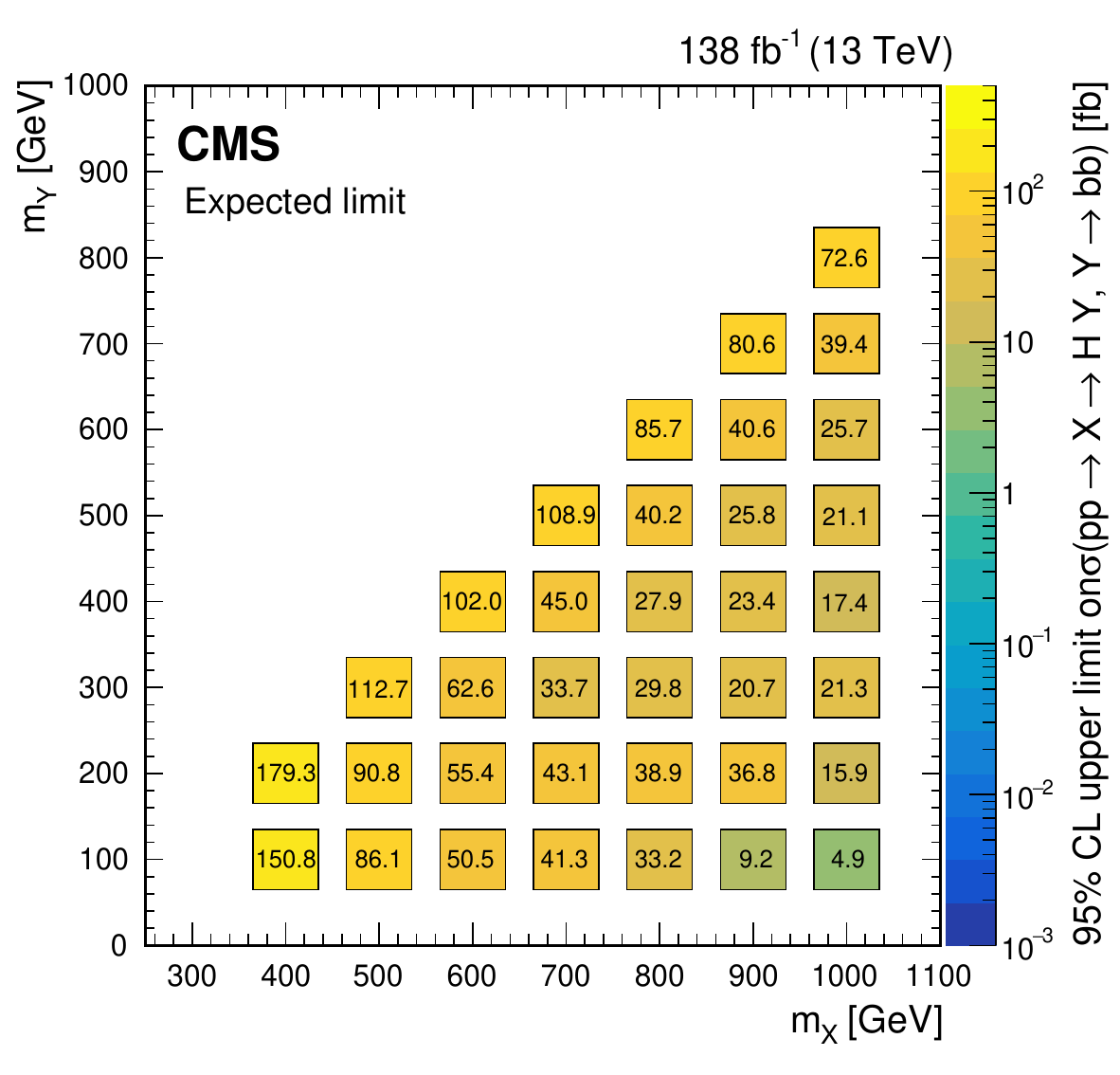}
  \includegraphics[width=\cmsFigWidth]{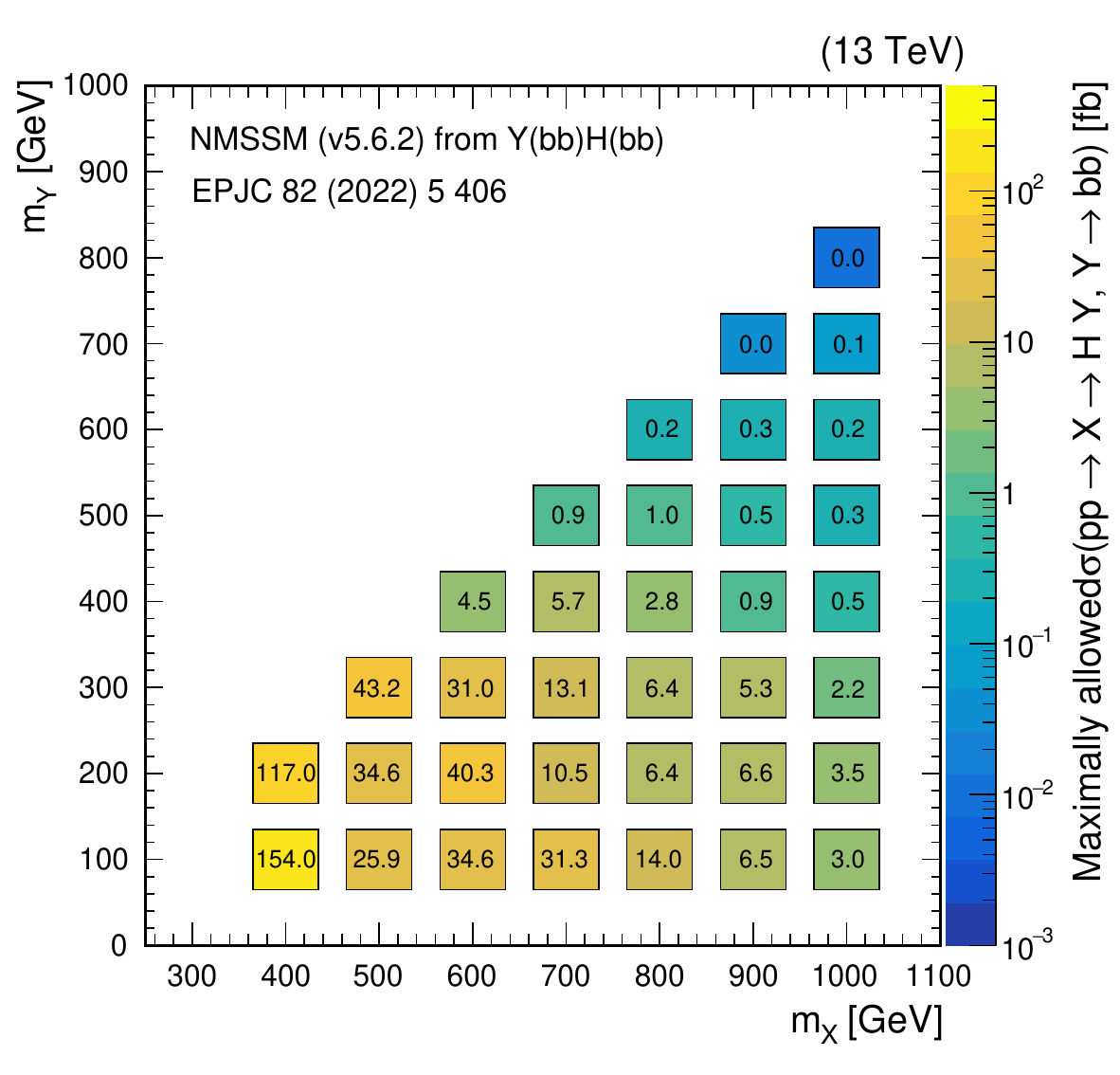}
  \caption{
    (Upper \cmsLeft) Observed and (upper \cmsRight) expected upper limits at 
    95\%~\CL, on the product of the cross section $\sigma$ for the production of 
    a resonance \PX via gluon-gluon fusion and the branching fraction \BR for 
    the $\PX\to\PY(\bb)\PH$ decay, as obtained from a combined likelihood analysis 
    of the individual analyses presented in this report and shown in 
    Fig.~\ref{fig:XYH_combination}. The results are presented in a plane 
    defined by \mX and \mY. The limits have 
    been evaluated in discrete steps corresponding to the centers of the boxes.
    The numbers in the boxes are given in~\unit{fb}.
    The corresponding maximally allowed values of $\sigma\BR$ in the NMSSM are 
    also shown for comparison (lower plot), as adapted from
    Ref.~\cite{Ellwanger:2022jtd}.
  }
  \label{fig:XYH_combination_mXmY}
\end{figure}

Figure~\ref{fig:XYH_combination_mXmY} shows the observed and expected
upper limits at 95\%~\CL on $\sigma\BR$ of the combined $\PX\to\PY\PH$
searches~(upper panel), together with the maximally allowed model
values~(lower panel).  While the experimental limits appear to touch
the model predictions in several places, there is not much additional
exclusion. This is expected because many relevant measurements,
including the CMS searches for $\PX \to \PY(\bb)\PH(\tautau)$ and
$\PX\to\PY(\bb)\PH(\bb)$ presented in this article, are already
accounted for in this version of \textsc{NMSSMTools}, which lowers the
maximally allowed NMSSM cross sections correspondingly. Therefore, no
new constraints are expected from these channels compared to those in
the original publications~\cite{CMS:2022kdi,CMS:2023boe,CMS:2022suh}.

Comparisons of the measured limits for $\PX\to\PY(\bb)\PH(\bb)$ in
merged final states with the maximally allowed TRSM values can be
found in Ref.~\cite{CMS:2022suh}. The measurement excludes part of the
allowed TRSM parameter space in a wedge-shaped region between $\mX
\approx 1000$--1300\GeV and around $\mY \approx 125\GeV$. An
interpretation of the $\PX \to \PY(\bb) \PH(\tautau)$ measurements
within the TRSM benchmark planes can be found in
Ref.~\cite{Robens:2023oyz}.

\subsubsection{The real-singlet extension}

The additional scalar boson \PX predicted in the real-singlet model has the same relative 
couplings to SM particles as the SM \PH boson. 
Most searches for $\PX\to\PH\PH$ assume that the width of the \PX boson is much smaller  
than the reconstructed mass resolution, such that the NWA holds.
We use the real-singlet model for a dedicated study of nonnegligible width and interference effects 
and present the results in Section~\ref{Sec:Effects_finite_width_and_interference}. 
The corresponding model interpretations for the $\PX\to\PH\PH$ combination in the real-singlet model are presented there.

\subsection{Warped extra dimensions} \label{Sec:Interp_in_Warped_Extra_Dimensions}

\begin{figure}[tbp]\centering
  \includegraphics[width=\cmsFigWidth]{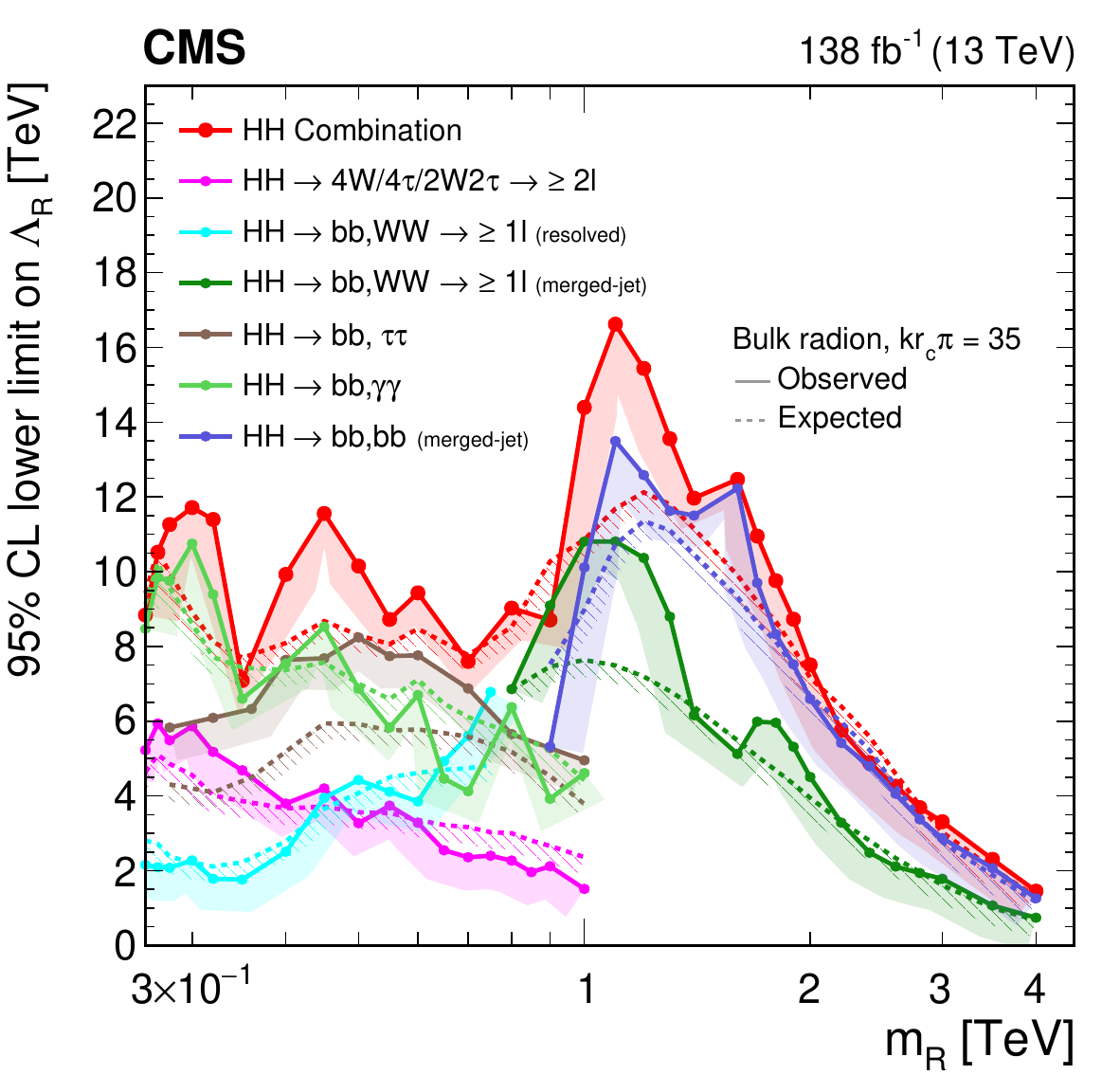}
  \includegraphics[width=\cmsFigWidth]{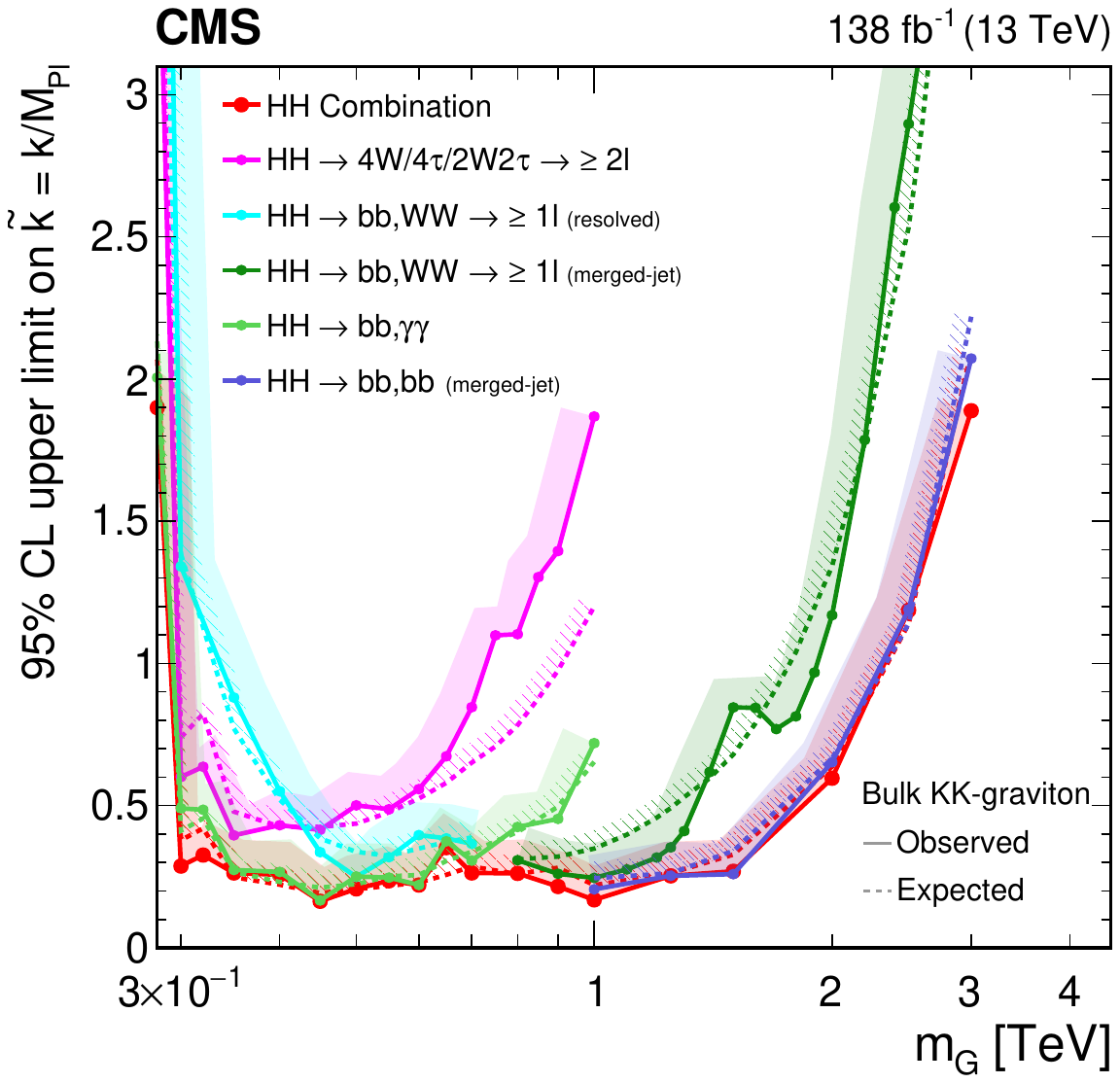}
  \caption{
    Observed and expected limits, at 95\%~\CL, on the parameters of 
    models with warped extra dimensions, as obtained from the $\PX\to\PH\PH$ 
    analyses presented in this report and their combined likelihood analysis. 
    Shown are lower limits (\cmsLeft) on the bulk radion ultraviolet 
    cutoff parameter \LambdaR, as a function of the radion mass 
    $m_{\PR}$, and upper limits (\cmsRight) on the parameter 
    $\tilde{k}$ of the spin-2 bulk graviton \PG, as a function of 
    $m_{\PG}$. Excluded areas are indicated by the direction of the hatching along 
    the exclusion contours. 
  }
  \label{fig:Int_WED}
\end{figure}
\begin{figure}[tbp]\centering
  \includegraphics[width=\cmsFigWidth]{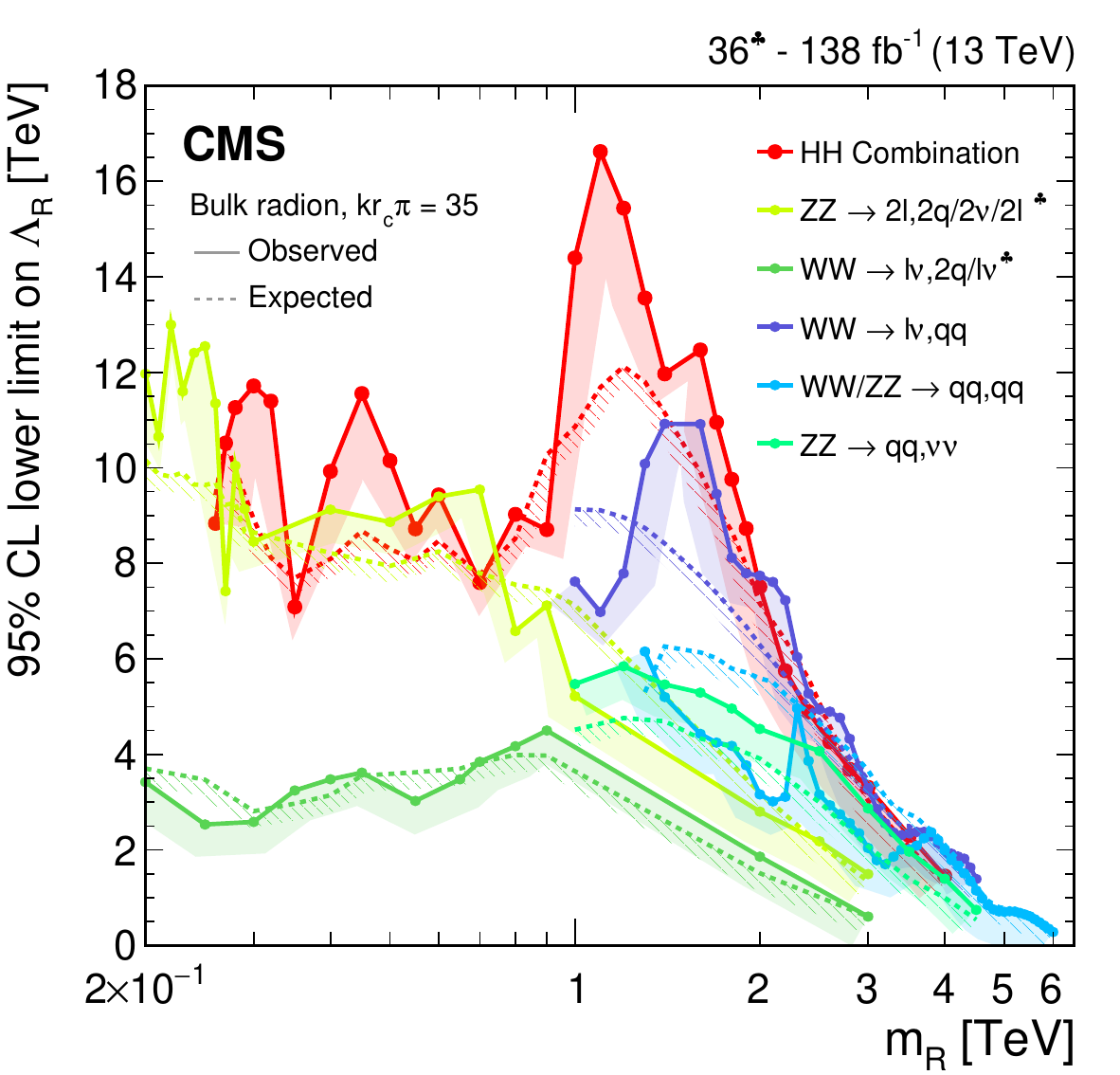}
  \includegraphics[width=\cmsFigWidth]{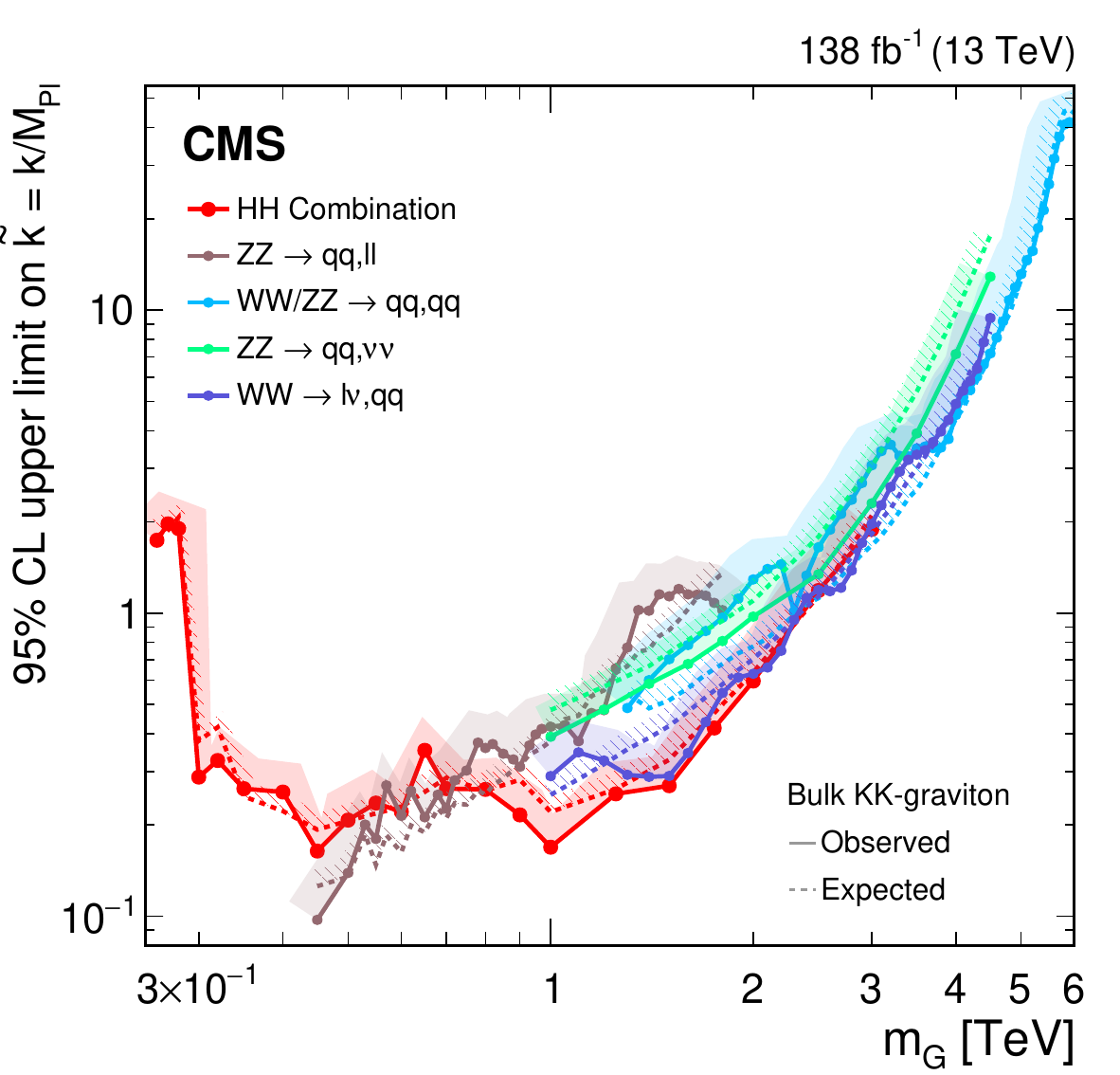}
  \caption{
    Observed and expected limits, at 95\%~\CL, on the parameters of 
    models with warped extra dimensions, as obtained from the combined likelihood 
    analysis of the individual $\PX\to\PH\PH$ analyses presented in this report 
    and shown in Fig.~\ref{fig:Int_WED}. The exclusion contours obtained 
    from the combined likelihood analysis are compared to similar exclusions 
    obtained from individual searches in the decays 
    $\PZ(\lep)\PZ(\qq/\PGn\PGn/\lep)$~\cite{CMS:2018amk}, 
    $\PW(\Pell\PGn)\PW(\Pell\PGn/\qq)$~\cite{CMS:2019bnu}, 
    $\PW(\Pell\PGn)\PW(\qq)$~\cite{CMS:2021klu}, 
    $\PV(\qq)\PV(\qq)$~\cite{CMS:2022pjv}, and 
    $\PZ(\PGn\PGn)\PZ(\qq)$~\cite{CMS:2021itu}, in case of the radion 
    interpretation, and from individual searches in the decays 
    $\PZ(\qq)\PZ(\lep)$~\cite{CMS:2021xor}, 
    $\PV(\qq)\PV(\qq)$~\cite{CMS:2022pjv}, 
    $\PZ(\PGn\PGn)\PZ(\qq)$~\cite{CMS:2021itu}, and 
    $\PW(\Pell\PGn)\PW(\qq)$~\cite{CMS:2021klu}, in the case of the graviton 
    interpretation. Excluded areas are indicated by the direction of the hatching 
    along the exclusion contours. 
  }
  \label{fig:Int_WED_all}
\end{figure}
The measured upper limits on resonant \HH production can also be interpreted 
in the context of WED models (as discussed in Section~\ref{Sec:Warped_Extra_Dimensions}). 
Figure~\ref{fig:Int_WED} (\cmsLeft) shows the lower limit on the bulk radion ultraviolet cutoff
parameter \LambdaR as a function of the radion mass $m_{\PR}$ for all
presented \HH analyses and their combination. 
The individual analyses with the best sensitivity are from the searches of 
$\PX\to\PH(\bb)\PH(\GamGam)$ for $\mX \lesssim 1\TeV$, and 
$\PX\to\PH(\bb)\PH(\bb)$ for $\mX \gtrsim 1\TeV$. 
In the regions $0.5 \lesssim \mX \lesssim 1\TeV$ and $1 \lesssim \mX \lesssim 1.5\TeV$, 
the $\PX\to\PH(\bb)\PH(\tautau)$ and $\PX\to\PH(\bb)\PH(\PW\PW)$ analyses contribute 
significantly to the combination. 
In the mass region below 1\TeV, the expected lower limit from the combination 
ranges from 8 to 10\TeV, with observed limits reaching up to 12\TeV. 
The strongest exclusion limits of about 12\TeV expected and 16\TeV observed are 
reached near $m_{\PR}=1.2\TeV$. The combination improves the sensitivity 
over the full mass range probed. 
Figure~\ref{fig:Int_WED} (\cmsRight) shows the corresponding
upper limits of the parameter $\tilde{k}$ of the spin-2 bulk graviton \PG.
The combination excludes values of $\tilde{k}$ larger than about 0.3 at 95\%~\CL 
for the large mass range $0.3 < m_{\PG} < 1.5\TeV$. 

We compare the limits obtained from the \HH combination with limits from searches 
for $\PX\to\PZ\PZ$~\cite{CMS:2021xor, CMS:2018amk, CMS:2021itu} and 
$\PX\to\PW\PW$~\cite{CMS:2019bnu, CMS:2021klu, CMS:2022pjv} in Fig.~\ref{fig:Int_WED_all}. 
The \HH combination is found to be very competitive, and it places 
stronger constraints on the WED models in some mass regions. 
For radions, shown on the \cmsLeft, the \HH combination 
shows about the same sensitivity as the $\PZ(\lep)\PZ(\qq/\PGn\PGn/\lep)$ final state~\cite{CMS:2018amk} 
for $m_{\PR} \lesssim 1\TeV$. The \HH combination has the best sensitivity in the 
region $1 < m_{\PR} < 2\TeV$, and for higher masses it has a comparable sensitivity 
as searches in final states from hadronic and semileptonic $\PW\PW$ decays~\cite{CMS:2021klu, CMS:2022pjv}.
For gravitons, the \HH combination places the best upper limits on 
$\tilde{k}$ for $250 < m_{\PG} < 450\GeV$ and $700 < m_{\PG} < 2000\GeV$.

\subsection{Heavy vector triplet models} \label{Sec:Interp_in_Heavy_Vector_Triplet}

The three searches for $\PX\to\PV\PH$ introduced in Section~\ref{Sec:Analysis_X_to_VH} probe for 
a new vector boson \PVpr (either \PWpr or \PZpr) in final states with 0, 1, and 2 leptons.
The resulting upper limits on $\sigma\BR$ presented in Section~\ref{Sec:Results_X_to_VH} 
are now interpreted in the HVT model. 
The theoretical cross sections are calculated at NLO in QCD with the models 
detailed in Ref.~\cite{Oliveira:2014kla,Pappadopulo:2014qza}.
The theory predictions with the corresponding $\BR(\PWpr \to \WH$) and $\BR(\PZpr \to \ZH$) 
in the models A, B, and C, where couplings of \PVpr to bosons are enhanced, 
have been shown in Figs.~\ref{fig:Limits_on_WH} and \ref{fig:Limits_on_ZH}. 
The upper limits on $\sigma\BR$ are translated into lower limits on the vector boson masses. 
The \PWpr and \PZpr masses are excluded up to 4.1 and 3.9\TeV, 
respectively, in model B interpretations.

\begin{figure}[tbp]\centering
  \includegraphics[width=0.4\linewidth]{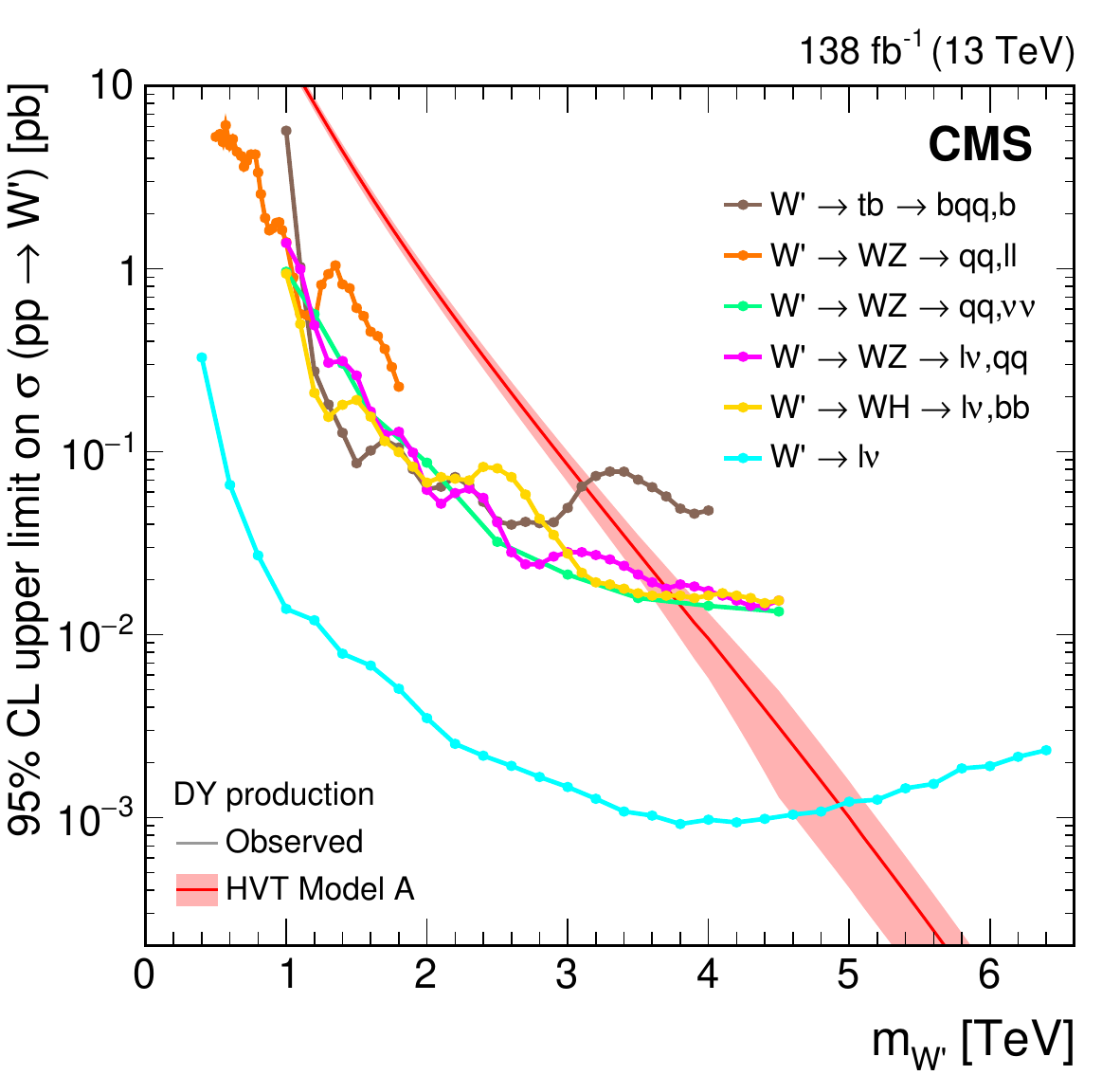}
  \includegraphics[width=0.4\linewidth]{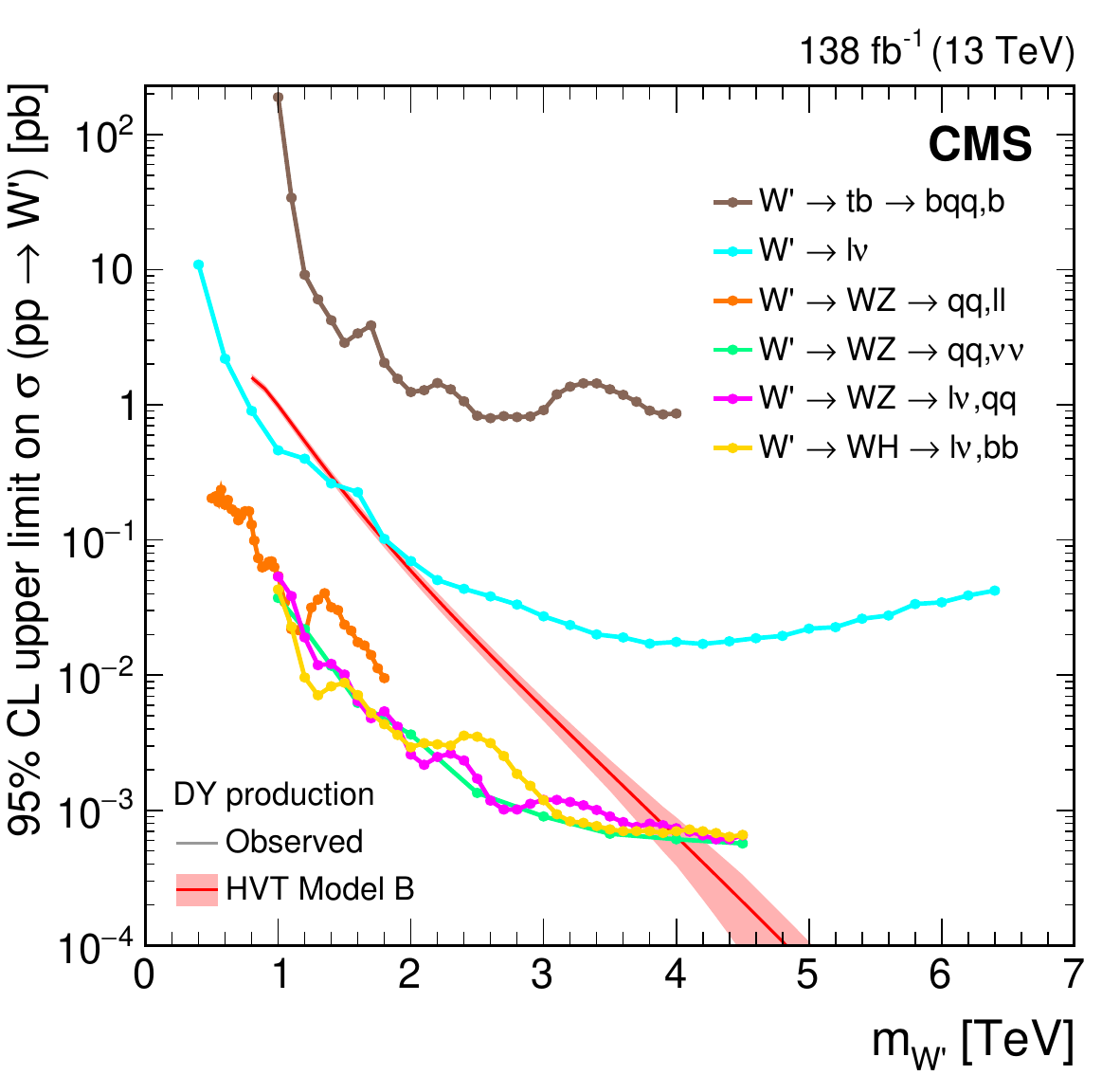} \\
  \includegraphics[width=0.4\linewidth]{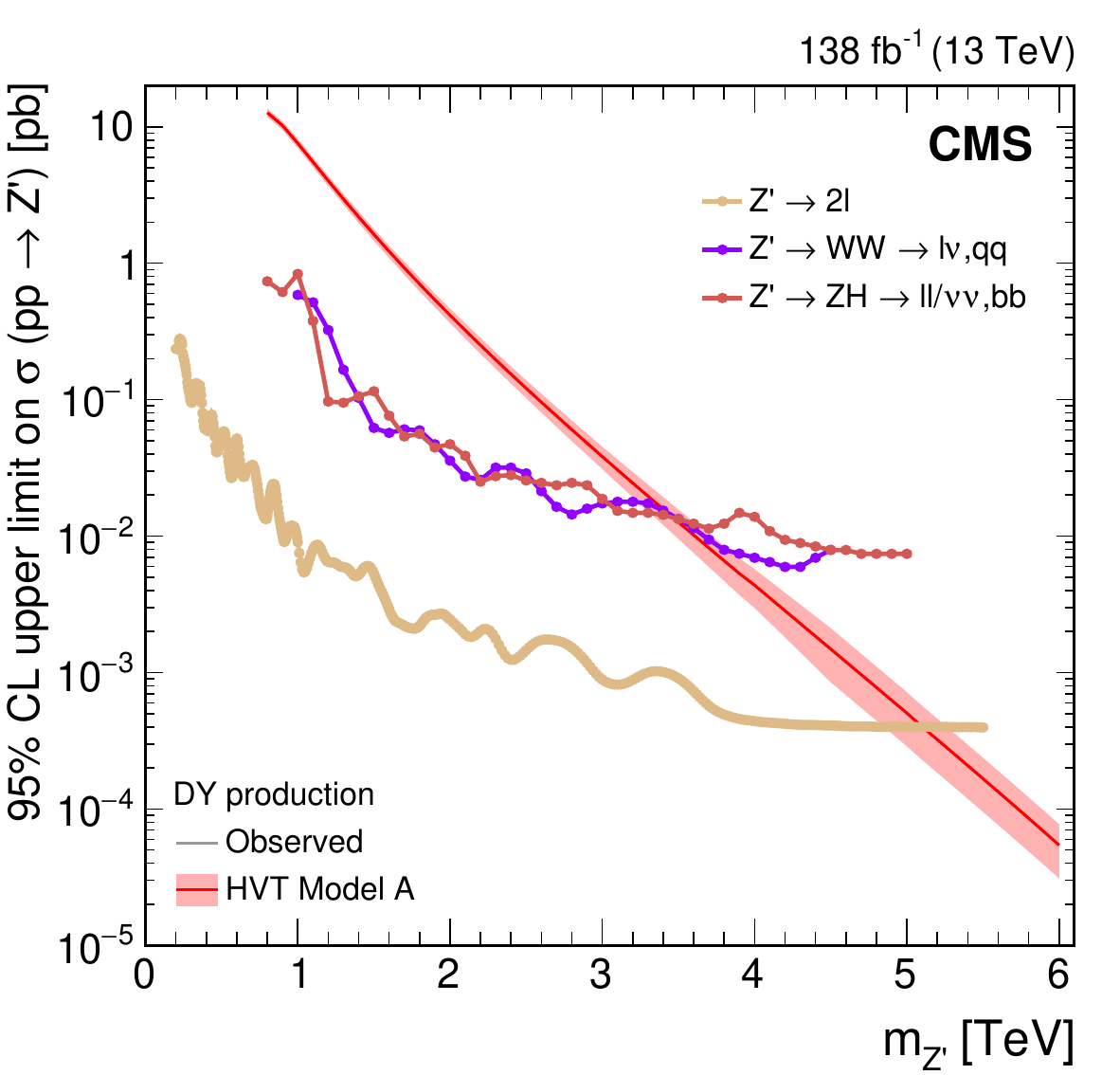} 
  \includegraphics[width=0.4\linewidth]{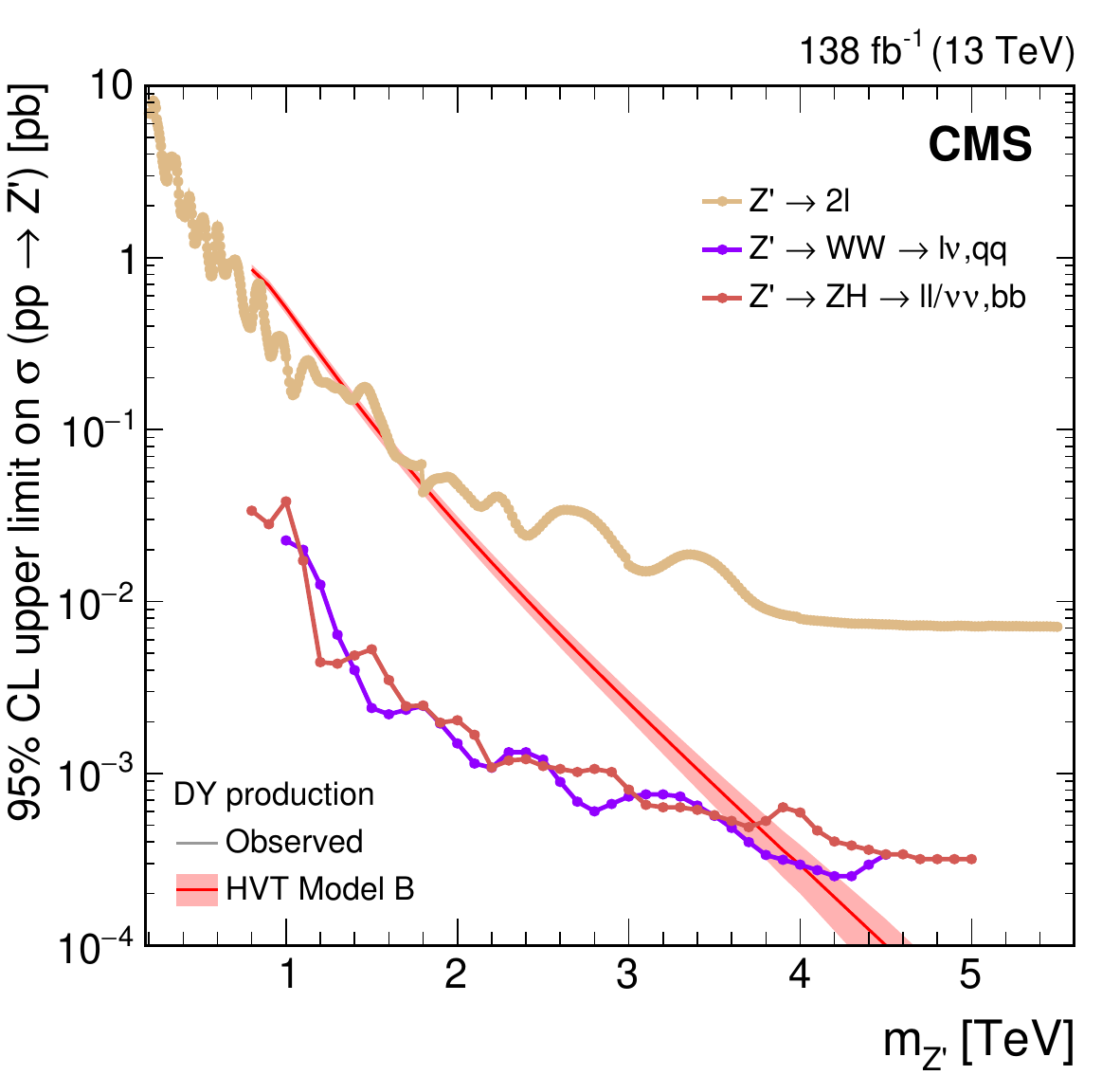} \\
  \includegraphics[width=0.4\linewidth]{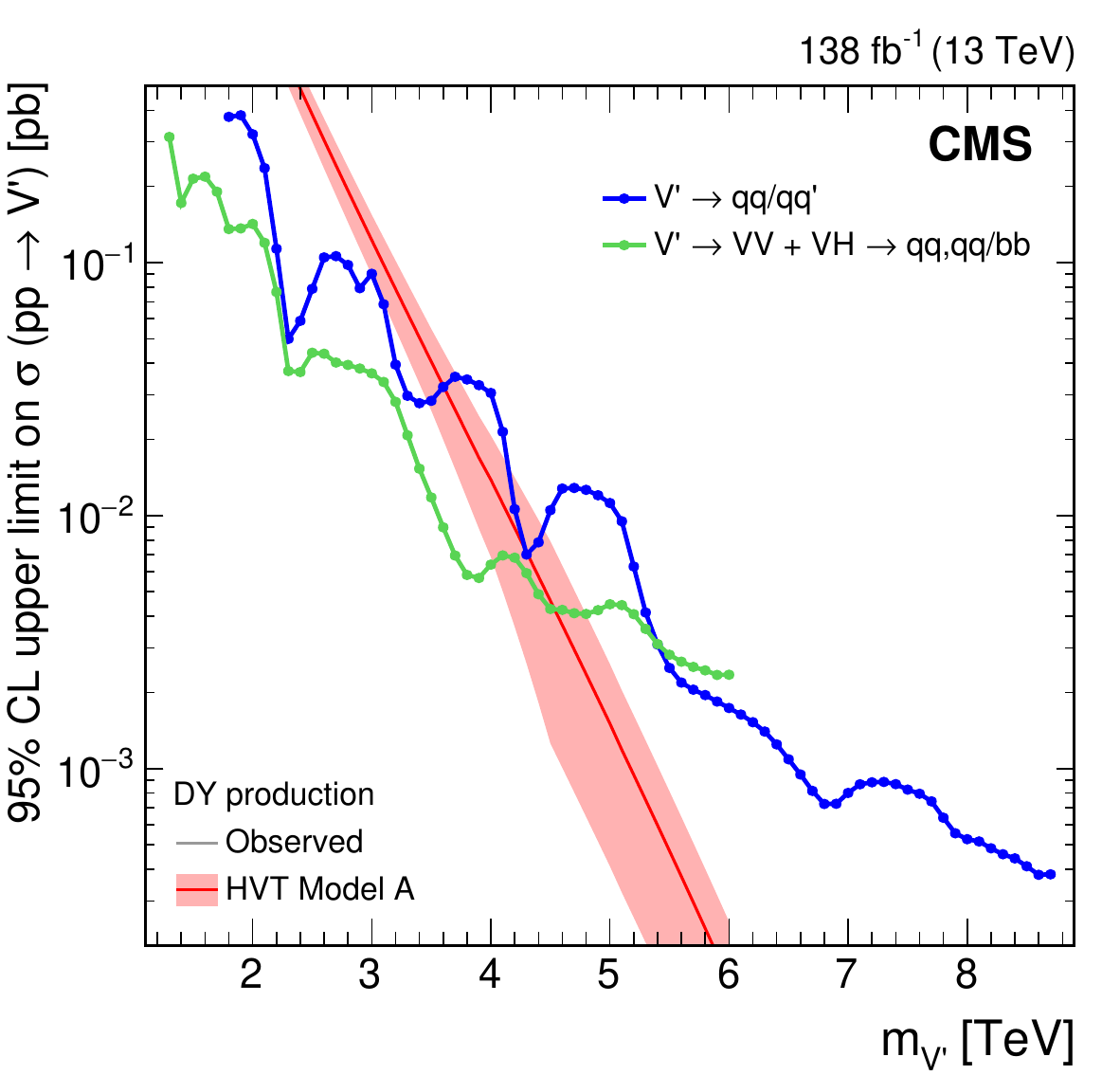}
  \includegraphics[width=0.4\linewidth]{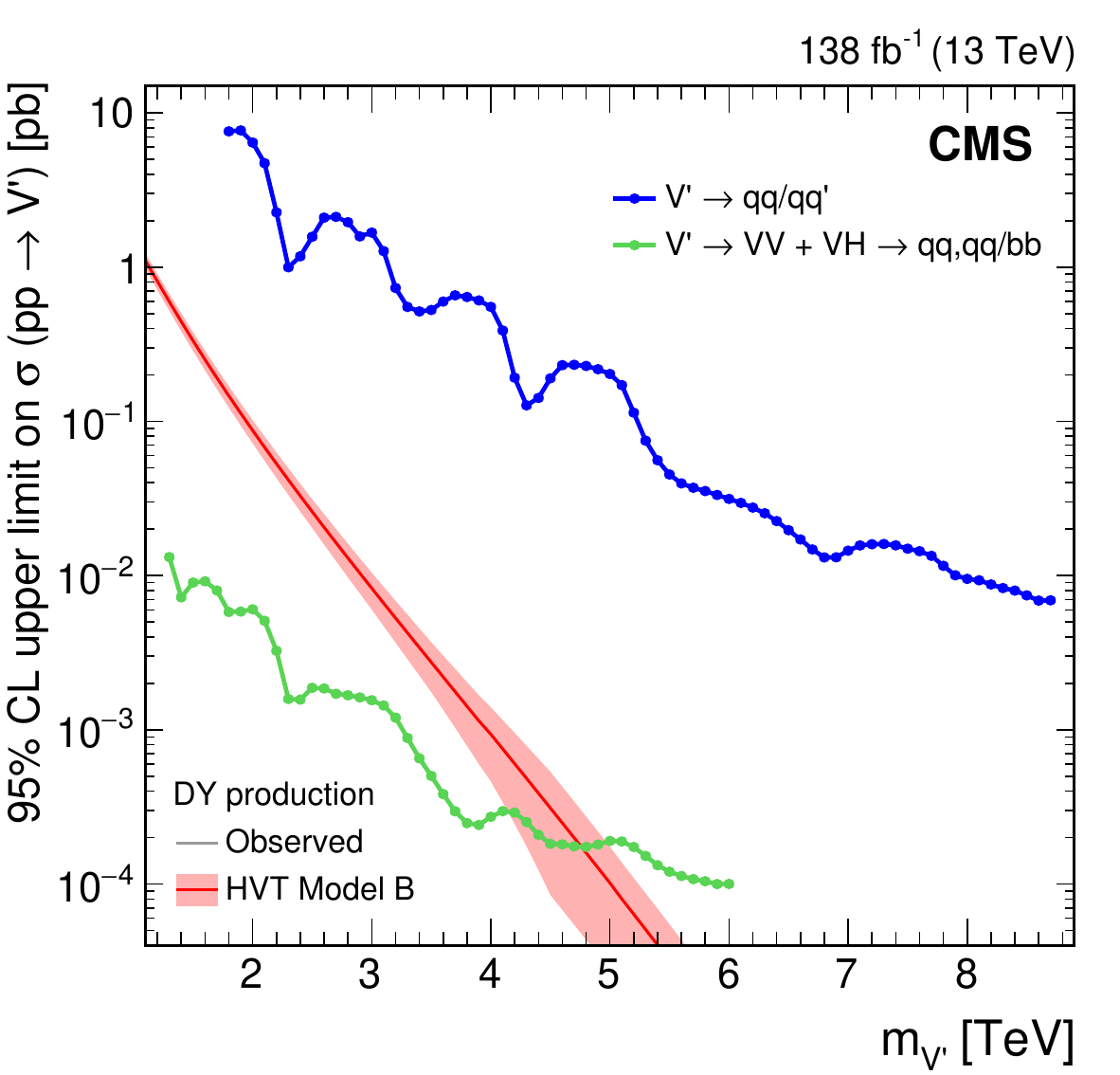}
  \caption{
    Observed upper limits, at 95\%~\CL, on the Drell--Yan production cross section 
    of (upper) \PWpr, (middle) \PZpr, and (lower) combined \PVpr spin-1 resonances 
    assuming branching fractions of the heavy vector triplet models (\cmsLeft) A 
    and (\cmsRight) B. The theory predictions from these models are also shown.
    Results from the $\PV\PH$~\cite{CMS:2021klu,CMS:2021fyk,CMS:2022pjv} and
    $\PV\PV$ channels~\cite{CMS:2021klu,CMS:2022pjv,CMS:2021xor,CMS:2021itu},
    as well as results from dijet~\cite{CMS:2019gwf}, $\PQt\PQb$~\cite{CMS:2021mux},
    \lep~\cite{CMS:2021ctt}, and $\Pell\PGn$~\cite{CMS:2022krd} final states 
    are shown for comparison.
  }
  \label{fig:Int_HVT_1D}
\end{figure}

Figure~\ref{fig:Int_HVT_1D} shows the upper limits on the DY production cross section of 
\PVpr in the \WH and \ZH channels, compared to those obtained from 
$\PV\PV$~\cite{CMS:2021klu,CMS:2022pjv,CMS:2021xor,CMS:2021itu} and fermion pair production 
channels~\cite{CMS:2021ctt, CMS:2021mux, CMS:2022krd, CMS:2019gwf} assuming branching fractions 
of the HVT models A and B. The corresponding theory predictions are overlaid. 
The all-jets channels are sensitive to both \PWpr and \PZpr production and are thus interpreted in 
combined \PVpr production. While in model A, searches for fermion pair production dominate the sensitivity, 
in model B, where couplings of \PVpr to bosons are large, the $\PV\PV$ and $\PV\PH$ searches are most sensitive.

\begin{figure}[tbp]\centering
  \includegraphics[width=\cmsFigWidth]{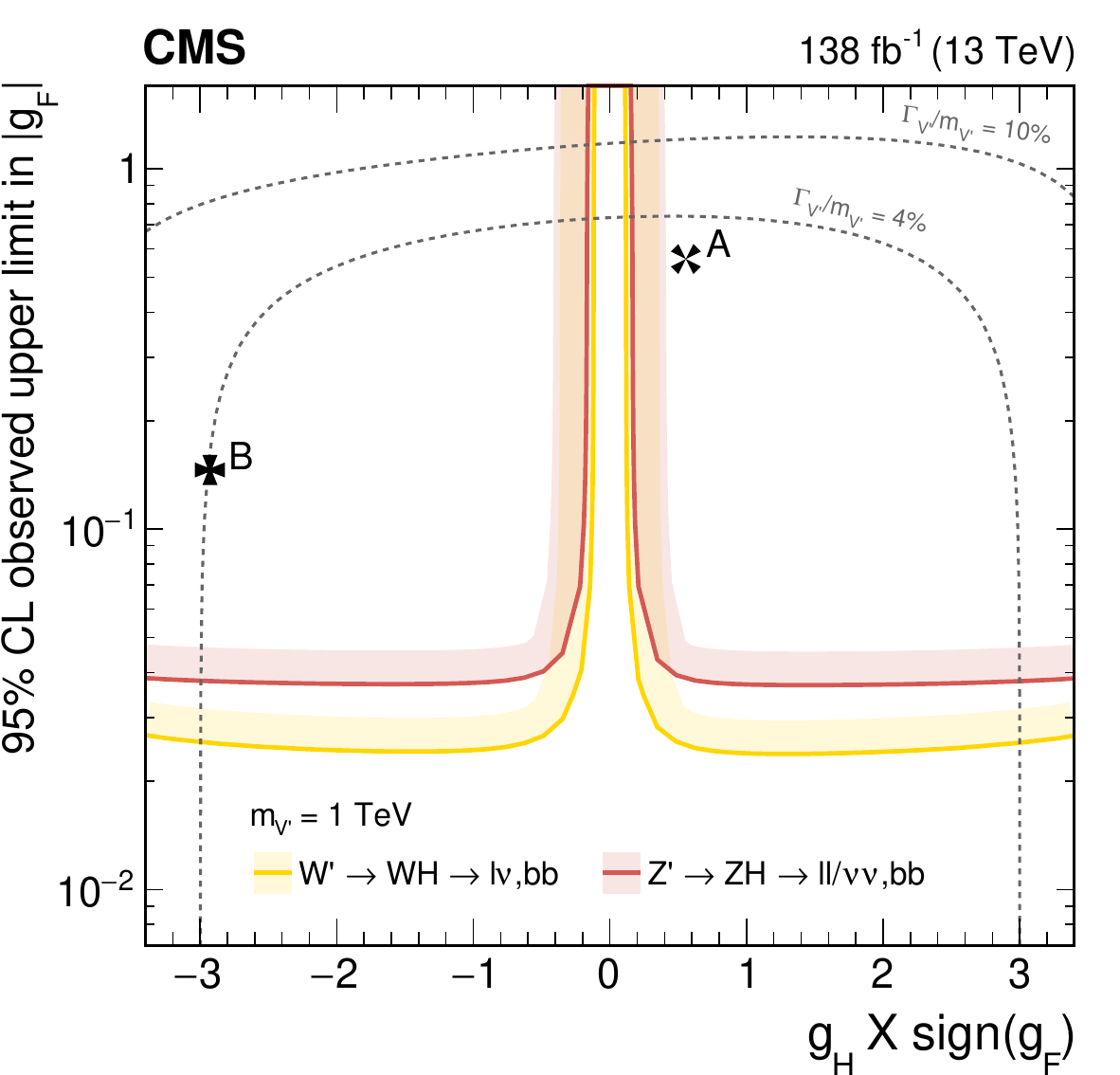}
  \includegraphics[width=\cmsFigWidth]{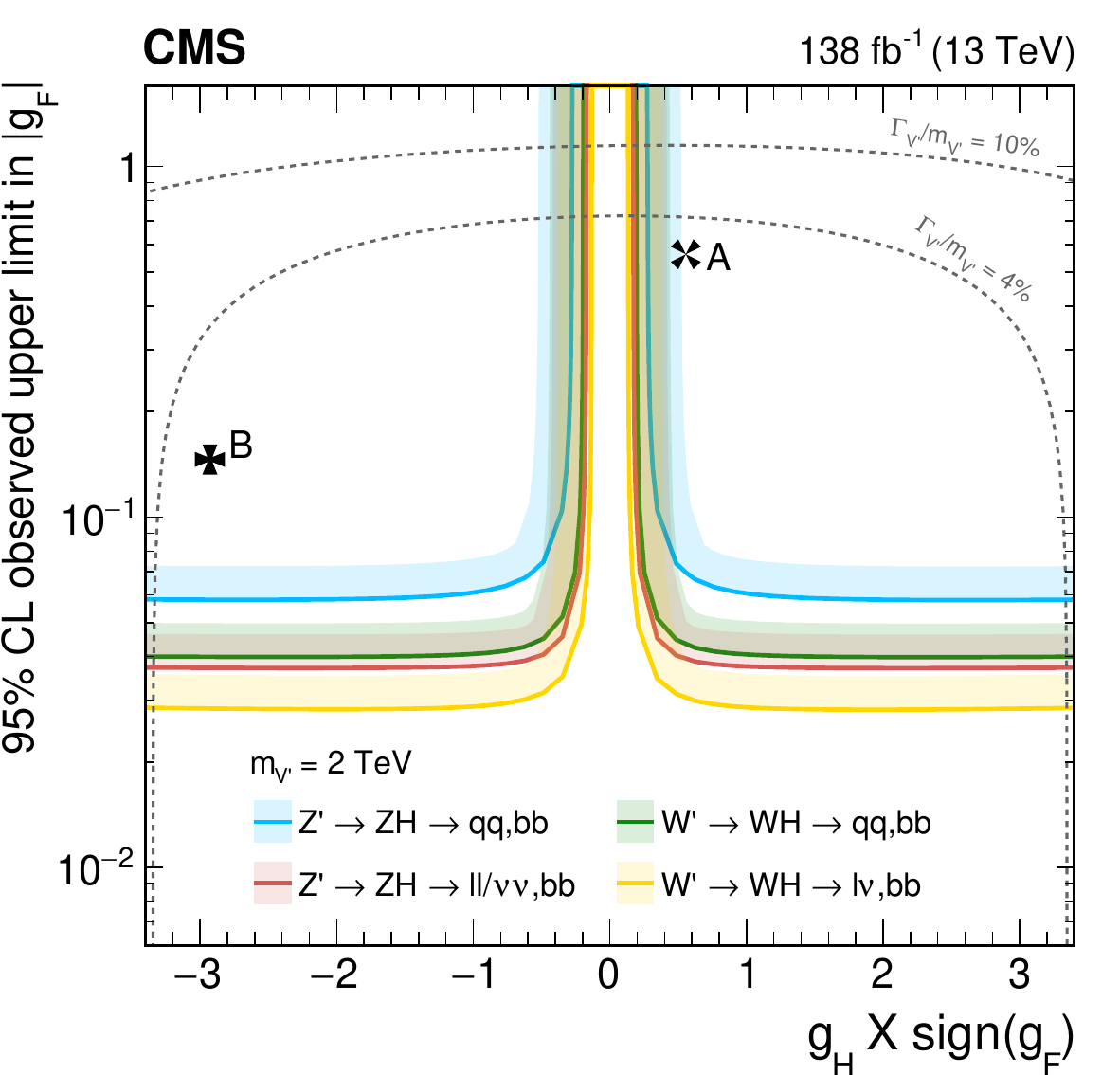}\\
  \includegraphics[width=\cmsFigWidth]{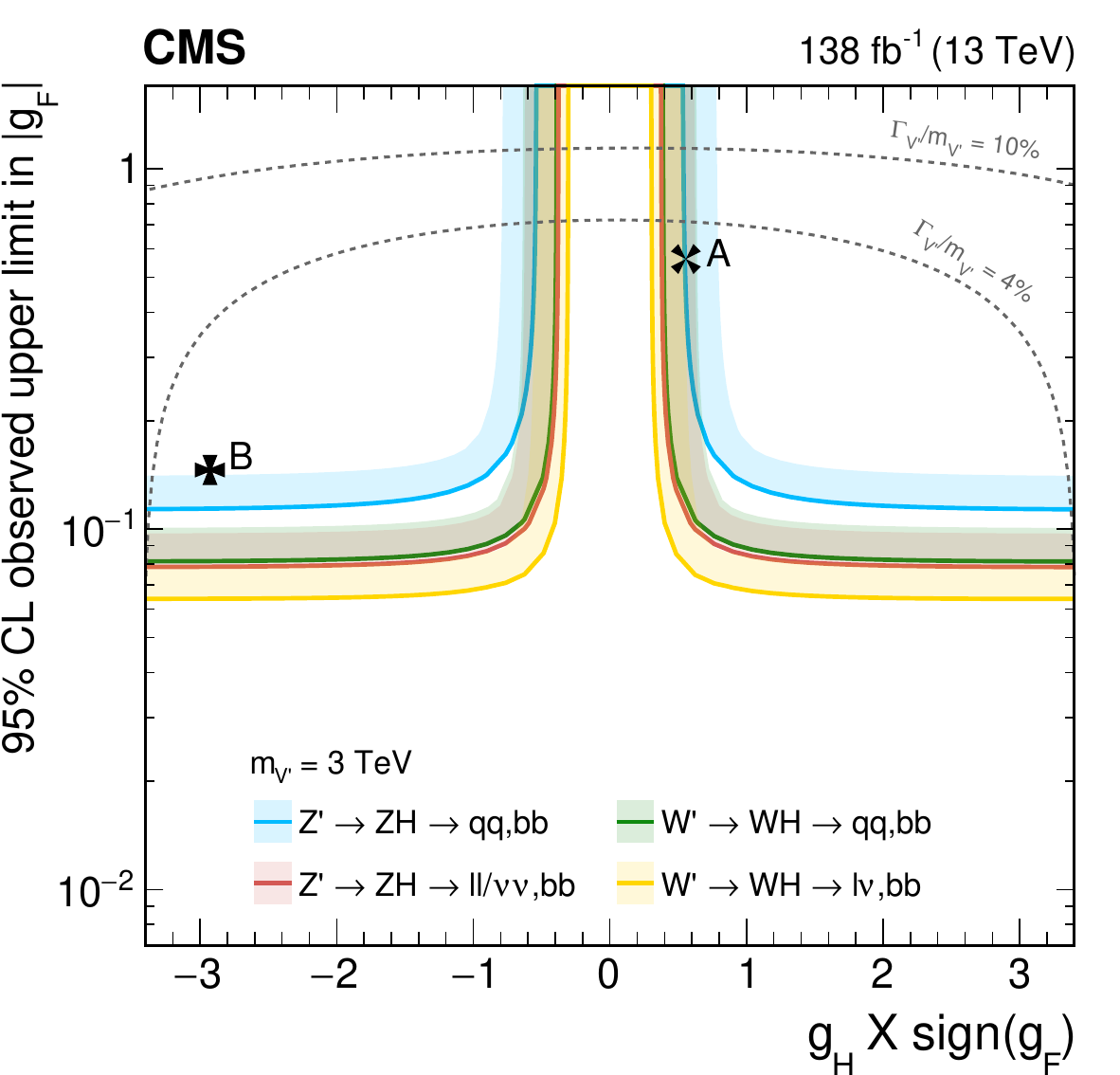}
  \includegraphics[width=\cmsFigWidth]{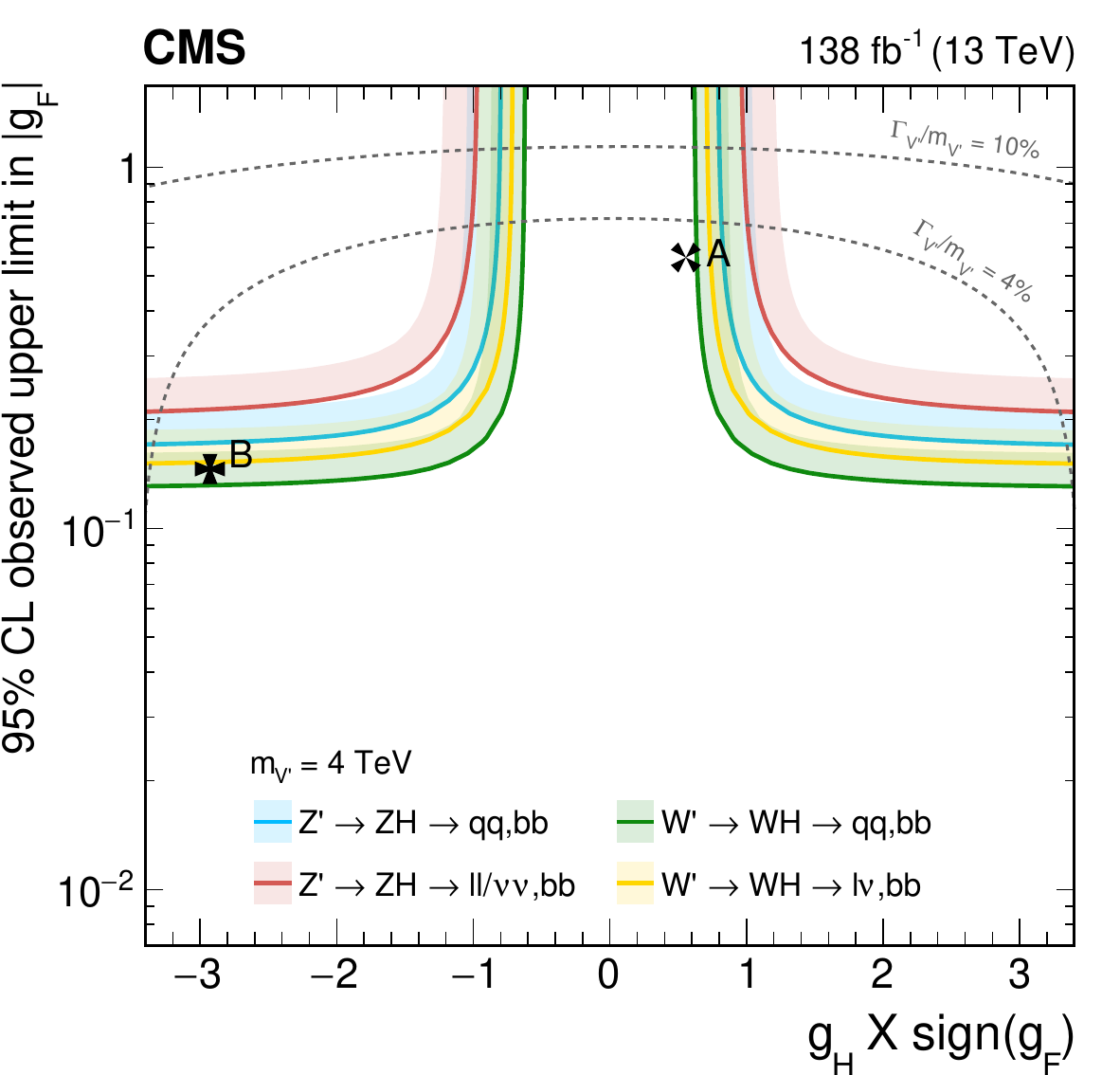}
  \caption{ 
    Observed upper limits, at 95\% \CL, on the \PVpr couplings \gF and \gH within 
    the HVT model for \PVpr masses of (upper \cmsLeft) 1, (upper 
    \cmsRight) 2, (lower \cmsLeft) 3, and (lower \cmsRight) 4\TeV, from DY 
    production, derived from \VH channels of Refs.~\cite{CMS:2021klu,CMS:2021fyk,
    CMS:2022pjv} discussed in this report. Excluded areas are indicated by the 
    direction of the shading along the exclusion contours. The dotted lines denote 
    coupling values above which the relative width of the resonance, $\Gamma_{\PVpr}
    /\MVpr$, exceeds 4 and 10\%.
    These dotted lines are to be compared with the experimental resolution
    to identify where the narrow width approximation no longer applies. The
    experimental resolution in final states with jets decreases as a
    function of resonance mass from ~7\% at 1\TeV to as low as ~4\% at
    4\TeV.
    The couplings corresponding to the 
    heavy vector triplet models A and B are indicated by cross markers.
  }
  \label{fig:Int_HVT1_VHonly}
\end{figure}

\begin{figure}[tbp]\centering
  \includegraphics[width=\cmsFigWidth]{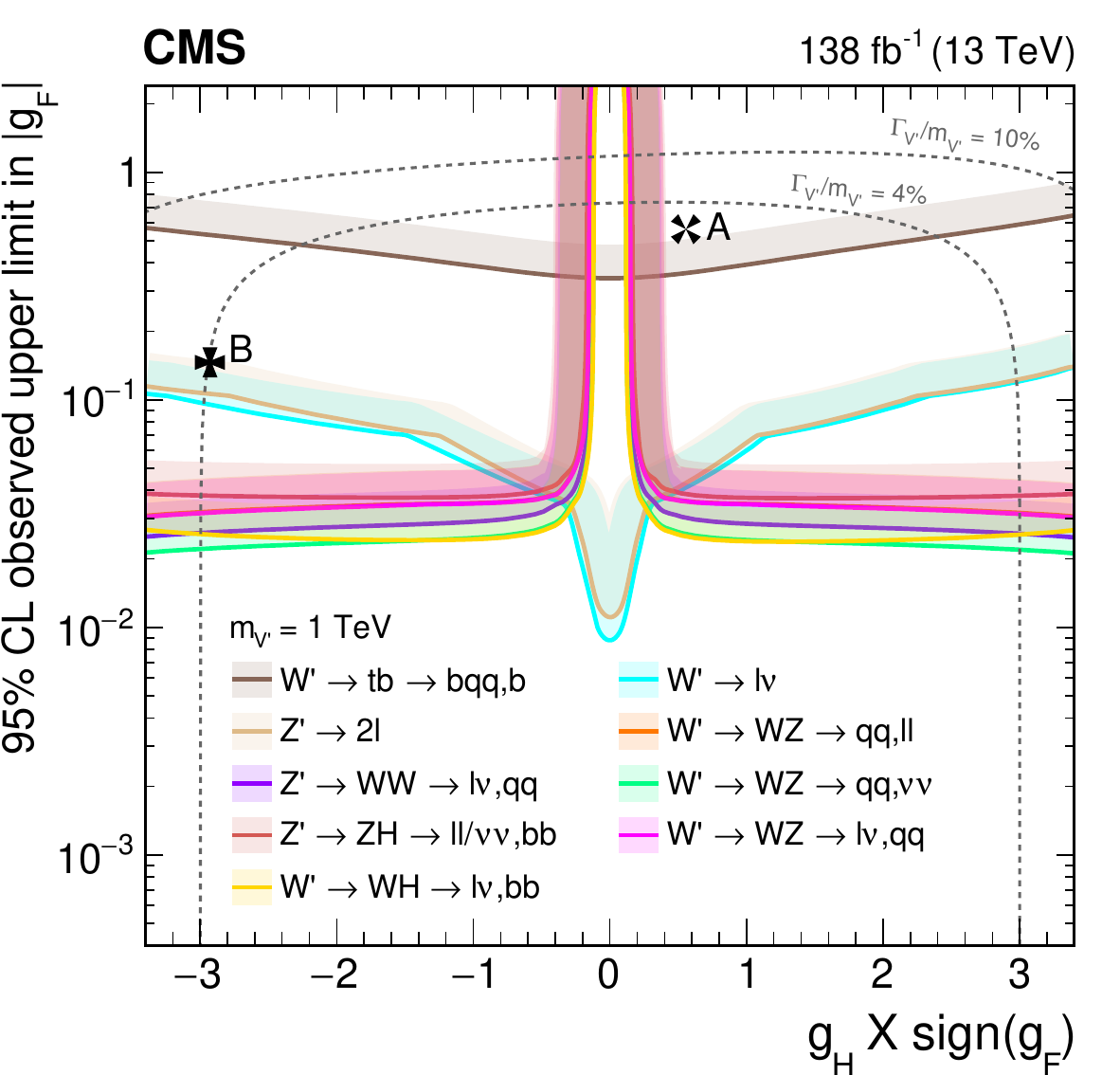}
  \includegraphics[width=\cmsFigWidth]{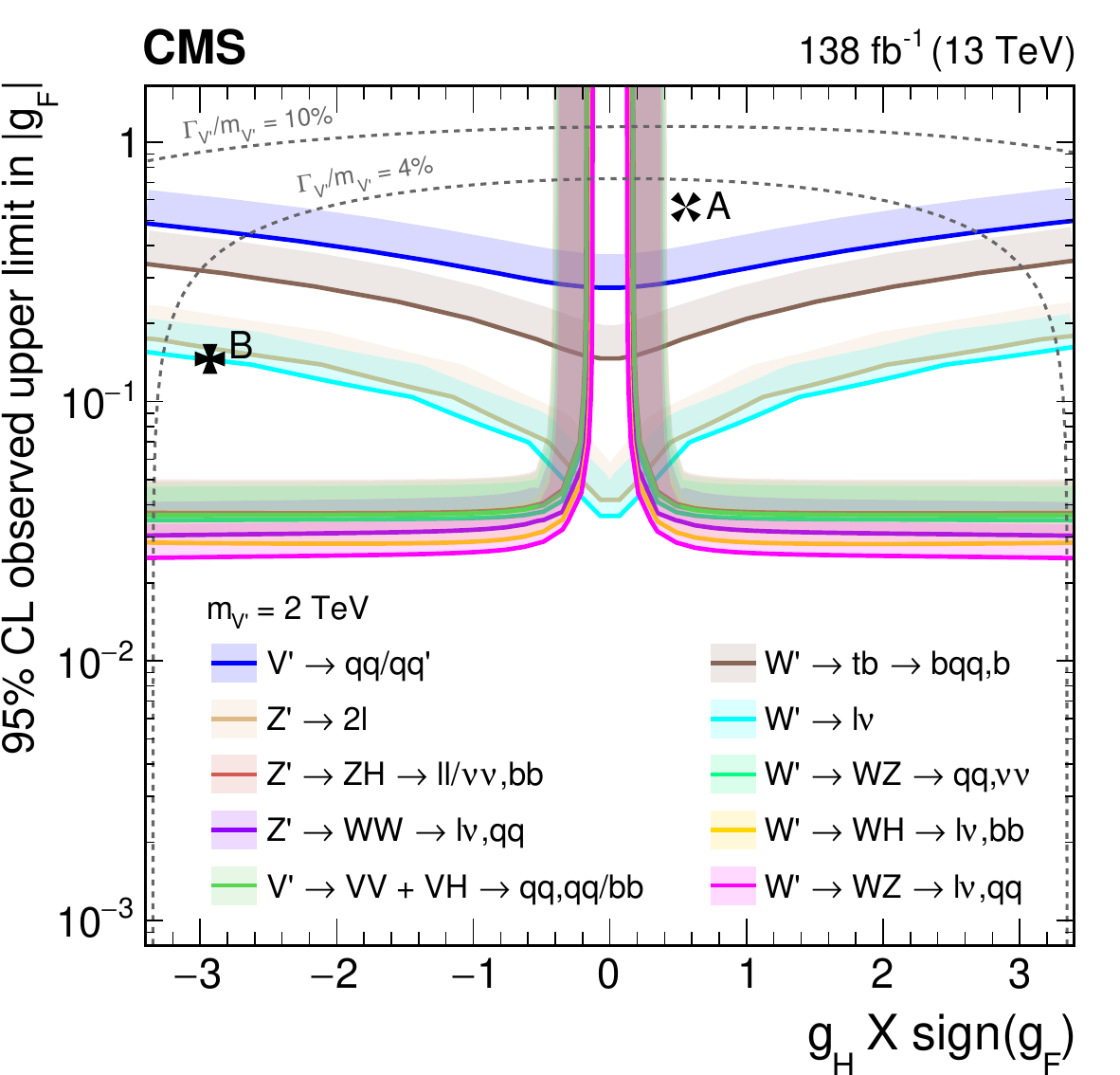}\\
  \includegraphics[width=\cmsFigWidth]{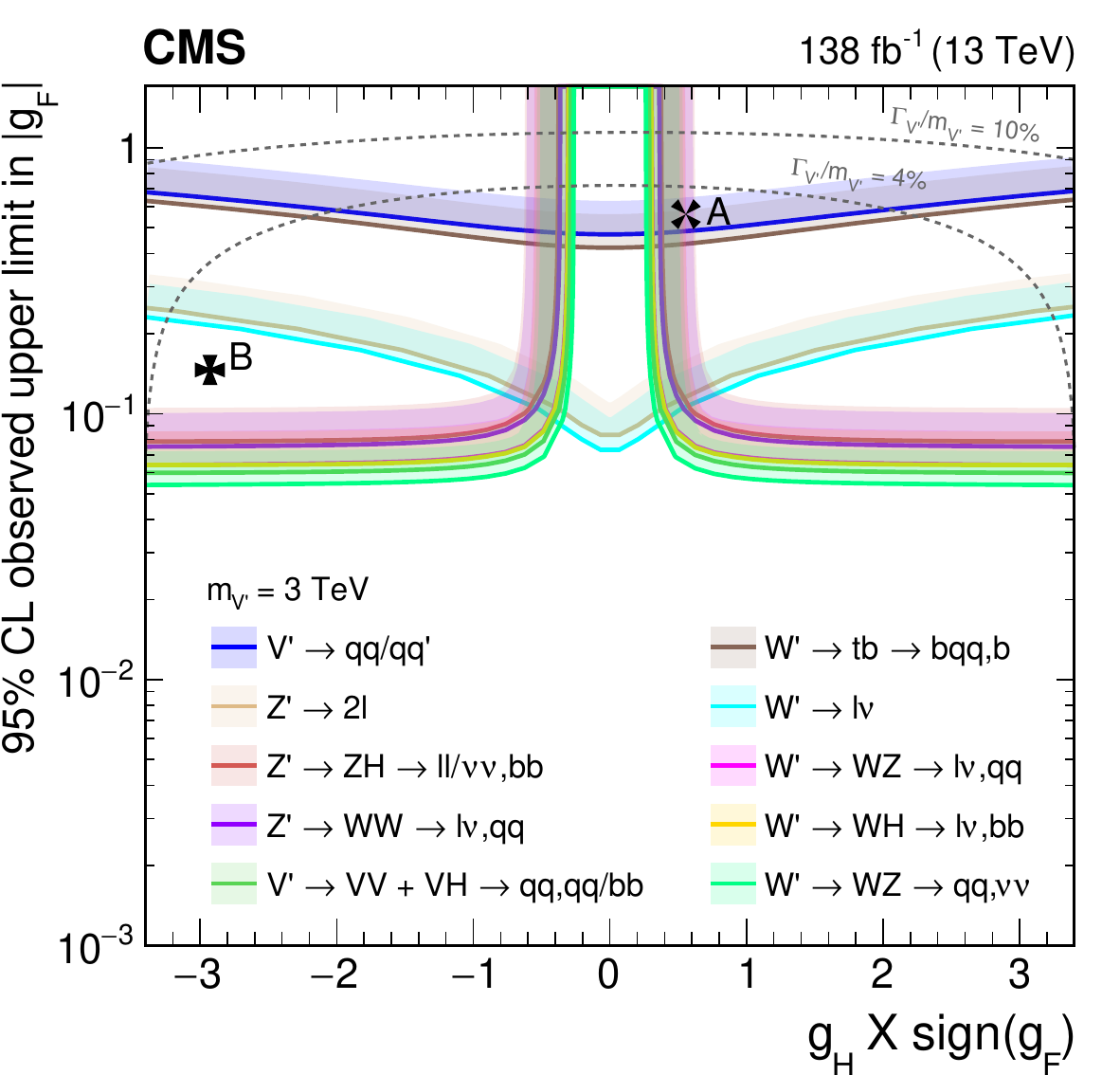}
  \includegraphics[width=\cmsFigWidth]{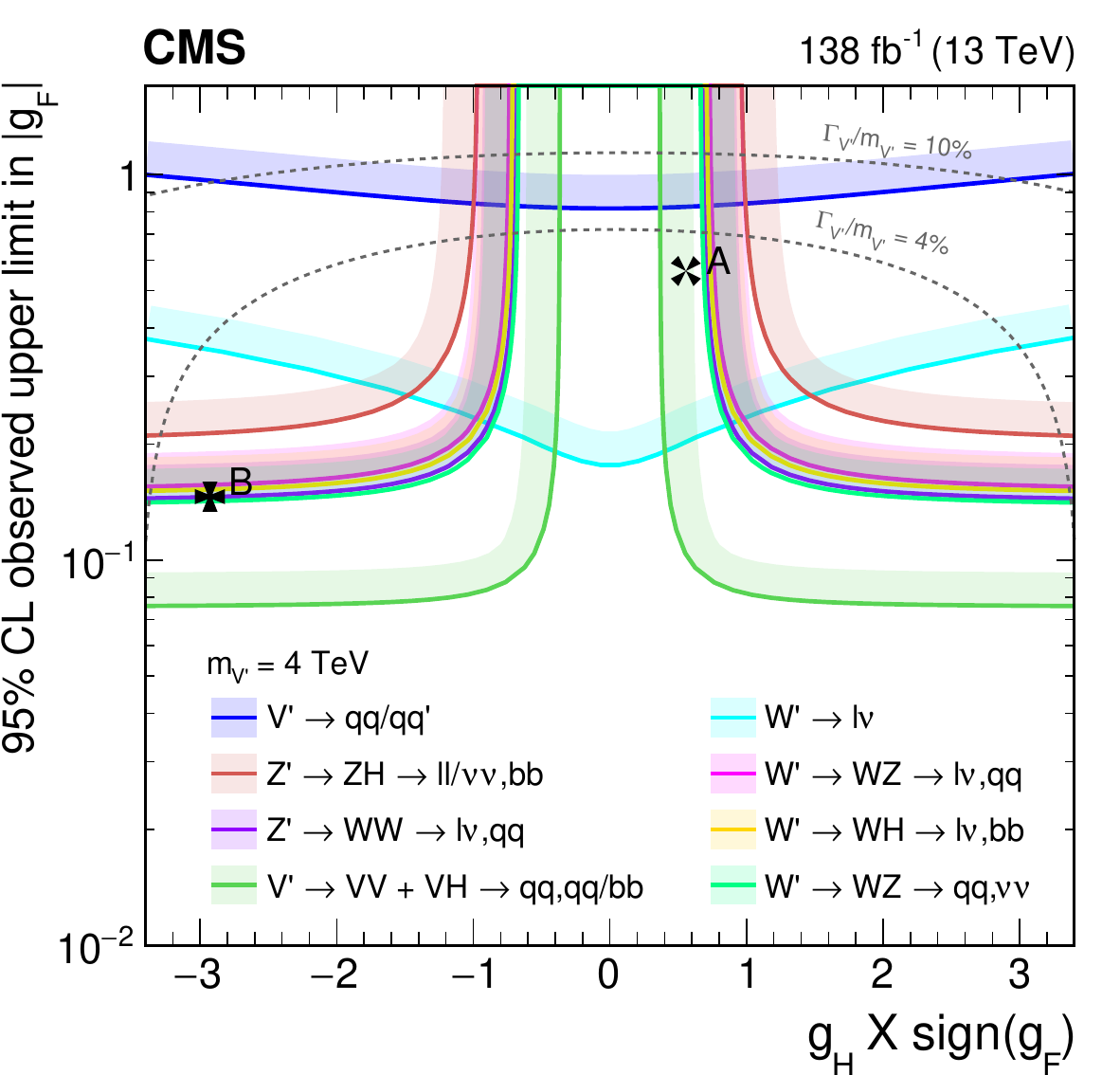}
  \caption{ 
    Observed upper limits, at 95\% \CL, on the \PVpr couplings \gF and \gH within 
    the HVT model for \PVpr masses of (upper \cmsLeft) 1, (upper 
    \cmsRight) 2, (lower \cmsLeft) 3, and (lower \cmsRight) 4\TeV, from DY 
    production, derived from \VH channels of Refs.~\cite{CMS:2021klu,CMS:2021fyk,
    CMS:2022pjv} discussed in this report and the $\PV\PV$ channels of 
    Refs.~\cite{CMS:2021klu,CMS:2022pjv,CMS:2021xor,CMS:2021itu}, as well as
    results from dijet~\cite{CMS:2019gwf}, $\PQt\PQb$~\cite{CMS:2021mux}, 
    \lep~\cite{CMS:2021ctt} and $\Pell\PGn$~\cite{CMS:2022krd} final states.
    Excluded areas are indicated by the direction of the shading along the 
    exclusion contours. The dotted lines denote coupling values above which 
    the relative width of the resonance, $\Gamma_{\PVpr}/\MVpr$, exceeds 4 
    and 10\%.
    These dotted lines are to be compared with the experimental resolution
    to identify where the narrow width approximation no longer applies. The
    experimental resolution in final states with jets decreases as a
    function of resonance mass from ~7\% at 1~\TeV to as low as ~4\% at
    4\TeV.
    The couplings corresponding to the 
    heavy vector triplet models A and B are indicated by cross markers.
  }
  \label{fig:Int_HVT1}
\end{figure}

\begin{figure}[tbp]\centering
  \includegraphics[width=\cmsFigWidth]{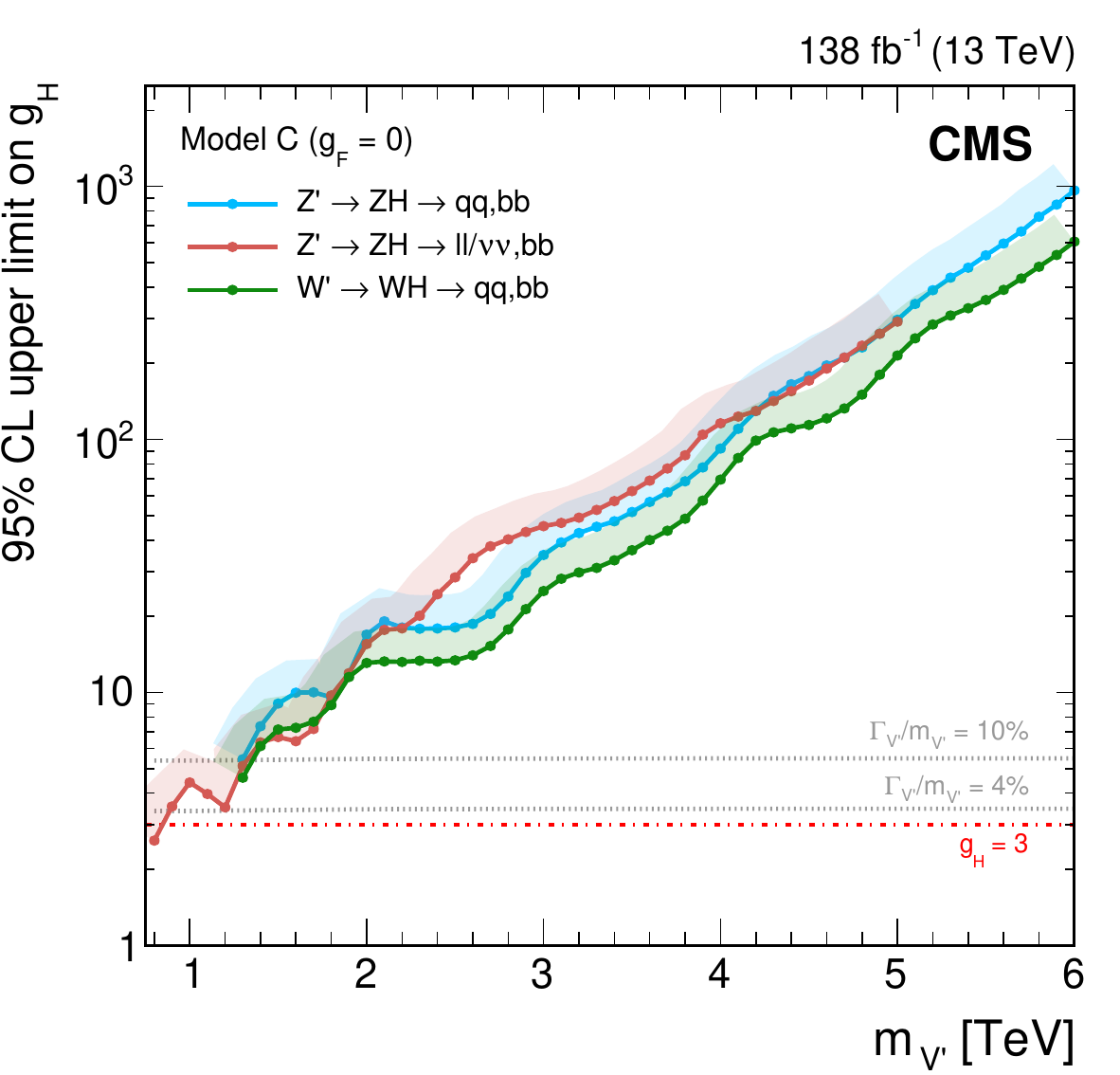}
  \includegraphics[width=\cmsFigWidth]{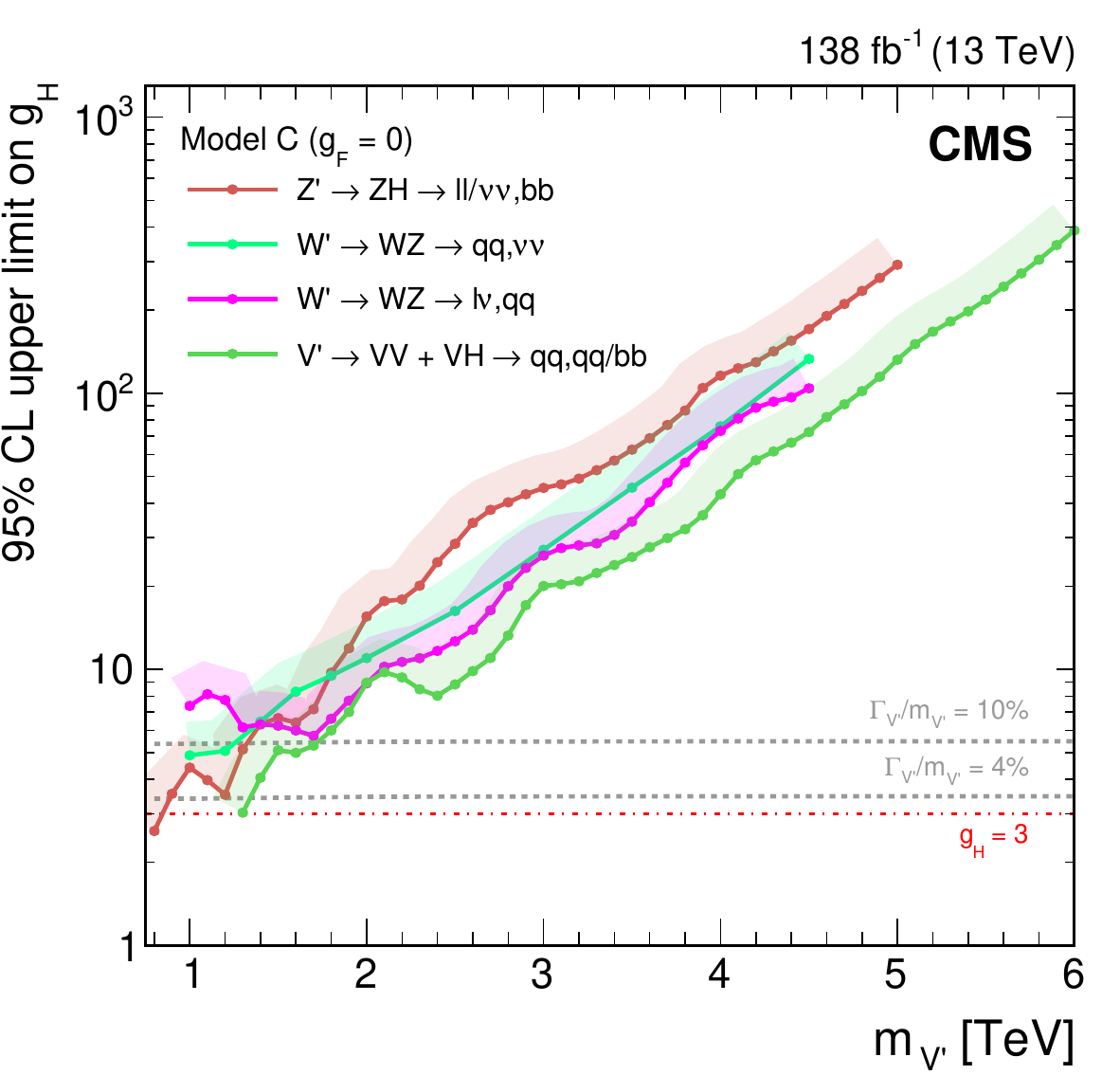}
  \caption{ 
    Observed upper limits, at 95\% \CL, on the coupling \gH within the heavy 
    vector triplet model, as a function of the \PVpr mass. The limits are shown 
    for the vector boson fusion production mode in the context of model C, in 
    which $\gF = 0$. The results are shown (\cmsLeft) for the \WH and \ZH 
    analyses of Refs.~\cite{CMS:2021klu,CMS:2021fyk,CMS:2022pjv}, individually, 
    and for a combination with the \WZ final states of Refs.~\cite{CMS:2021itu,
    CMS:2022pjv,CMS:2021klu} (\cmsRight), where the \WH and \ZH results from all-hadronic 
    final states have been combined with the corresponding $\PV\PV$ channels.
    The dotted lines denote coupling values above which the relative width of 
    the resonance, $\Gamma_{\PVpr}/\MVpr$, exceeds 4 and 10\%. 
    These dotted lines are to be compared with the experimental resolution
    to identify where the narrow width approximation no longer applies. The
    experimental resolution in final states with jets decreases as a
    function of resonance mass from ~7\% at 1~\TeV to as low as ~4\% at
    4\TeV.
  }
  \label{fig:Int_HVT2}
\end{figure}

For four resonance mass hypotheses, the cross section exclusion limits from DY production
are translated into two-dimensional upper limits on the coupling parameters for
fermions and bosons of the HVT model.
Figure~\ref{fig:Int_HVT1_VHonly} shows only the constraints from \VH production, 
while Fig.~\ref{fig:Int_HVT1} includes also $\PV\PV$ and fermion pair production channels for comparison.
The constraints from \VH searches are most stringent, apart from the region with small boson couplings, 
where the complementary searches with fermion final states provide stronger constraints.

In model C, where the \PVpr is produced exclusively via VBF, the data set is not sufficient to exclude couplings below
$\gH=3$ in any range of \MVpr. The corresponding results are shown in Fig.~\ref{fig:Int_HVT2}.

\subsection{Effects of finite width and interference in resonant \texorpdfstring{\HH}{HH} production}\label{Sec:Effects_finite_width_and_interference}

Most of the \HH and \YH analyses performed by the CMS Collaboration make use of the NWA, 
where the width of BSM particles is neglected and no interference with nonresonant 
Higgs boson pair production occurs.
However, in general, interference effects can strongly impact 
the \HH cross section~\cite{Heinemeyer:2024hxa,Basler:2019nas}.
The interference can be either constructive or destructive, enhancing or decreasing the \HH 
production rate~\cite{Carena:2018vpt,Grober:2017gut}, and have a nonnegligible effect 
in BSM exclusion limits.
We study the impact of the interference between nonresonant and resonant production in 
the inclusive $\pp\to\HH$ production, which can receive contributions from resonant $\PX\to\HH$ production. 
This work provides the first measure of interference effects, identifying phase space regions 
where the NWA is valid.
We use as a benchmark a simplified scenario based on the real-singlet model introduced in 
Section~\ref{Sec:AdditionalSinglets}, as it includes the smallest number of additional free 
parameters~\cite{OConnell:2006rsp}.
We note that interference effects are model dependent and may be different for other BSM scenarios.

For this specific study, we modify the singlet model by not imposing the $\mathbb{Z}_2$ symmetry. 
The $\mathbb{Z}_2$ symmetry precludes terms of odd powers of the additional singlet scalar field, which are known 
to be responsible for a stronger first-order EW phase transition~\cite{Papaefstathiou:2020iag,Papaefstathiou:2021glr}.
Exploiting EW symmetry breaking on the singlet model scalar potential, we are left with two mixing states.
After mass diagonalization, the identification of one of the states with the SM \PH boson 
reduces the number of uncorrelated parameters further from seven to five. 
The other state is associated with a new particle \PX. 
The couplings of the \PH boson and the \PX particle are given by
\begin{equation}
  \label{eq:singlet_couplings}
  g_{\PH kk} = g^{\text{SM}}_{\PH kk}\cos\alpha \quad \text{and} \quad g_{\PX kk} = -g^{\text{SM}}_{\PH kk}\sin\alpha, 
\end{equation}
where $\alpha$ is the mixing angle, and $k$ represents any SM particle.
For $\mX > 2 \mH$ the width of the \PX resonance can be calculated as 
\begin{equation}
  \label{eq:width}
  \widthGamma = \sin^{2}\alpha\,\Gamma^{\text{SM}}(m_{\PX}) + \frac{\lambda_{\HH\PX}^{2}\sqrt{1 - 4m_{\PH}^{2}/m_{\PX}^{2}}}{8\pi m_{\PX}},
\end{equation}
where \couplingLambda is the trilinear coupling between two \PH bosons and the new particle \PX, 
and $\Gamma^{\text{SM}}(m_{\PX})$ represents the width of a scalar boson of mass \mX with the same 
decay modes as the SM \PH boson.
The latter has been calculated by interpolating the values published in Ref.~\cite{deFlorian:2016spz}. 
In addition to $\alpha$, \mX and \couplingLambda, this singlet model also depends on the trilinear 
\PH coupling modifier $\kappaLambda \equiv \lambda_{\PH\PH\PH}/\lambda^{\text{SM}}_{\PH\PH\PH}$, and on an additional scalar coupling.

We use the \MGvATNLO generator version 2.9.7~\cite{Alwall:2014hca}, 
to simulate inclusive \HH events in the singlet model at LO. 
A custom universal \textsc{FeynRules}~\cite{FeynRules} output (UFO) model based on Ref.~\cite{Papaefstathiou:2020iag} 
adds a heavy scalar boson to the SM with couplings to SM particles as defined in Eq.~\eqref{eq:singlet_couplings}.
The samples are created according to the following parameter grid with $\kappaLambda = 1$:
\begin{itemize}
  \item \mX [\!\GeV{}]: 280, 300, 400, 500, 600, 700, 800, 900, 1000, 
  \item $\sin\alpha$: 0.00, 0.10, 0.20, 0.30, 0.40, 0.50, 0.60, 0.70, 0.80, 0.90, 0.95, 0.99, 
  \item \couplingLambda [\!\GeV{}]: $-$600, $-$500, $-$400, $-$300, $-$200, $-$100, $-$50, 0, 50, 100, 200, 300, 400, 500, 600, 
\end{itemize}
where \mX is chosen based on the signal samples used in the \HH combination presented in Section~\ref{Sec:Results_X_to_HH}.
The resonant, nonresonant, and total cross sections for each combination of grid points are generated separately.
We perform a parameter scan in the parameters \mX, $\sin\alpha$, and \couplingLambda of the interference ratio defined as 
\begin{equation}\label{eq:Rint}
  \Rint = \frac{\sigma^{\text{full}} - \left(\sigma^{\text{resonant-only}} + \sigma^{\text{nonresonant}}  \right)}{\sigma^{\text{resonant-only}} + \sigma^{\text{nonresonant}}}.
\end{equation}
We obtain the nonresonant cross section by setting the coupling $g_{\PX kk}$ defined in Eq.~\eqref{eq:singlet_couplings} to zero, 
and the resonant-only cross section by setting the coupling $g_{\PH kk}$ to zero. 
The variable \Rint provides information concerning the relative strength of the interference between the SM and 
BSM processes. The larger the deviation of \Rint from zero, the stronger the modification of the cross 
section due to the interference. We consider the gluon fusion production mode due to its dominant contribution 
to the cross section.
The UFO model and procedure are validated using the program \textsc{hpair}~\cite{Grober:2015cwa, Grober:2017gut}
where the results varying \kappaLambda in the nonresonant scenario are found to agree with the 
NLO predictions of Ref.~\cite{CMS:2022dwd}. 

\begin{figure}[htbp]
  \centering
  \includegraphics[width=1.5\cmsFigWidth]{Figure_041.pdf}
  \caption{
    Contours of the variable \Rint as defined in Eq.~(\ref{eq:Rint}) and discussed 
    in the text, in the ($\sin\alpha$, \couplingLambda) plane for the singlet 
    model with $\kappaLambda = 1$ and different resonance masses \mX between (upper
    \cmsLeft) 280 and (lower \cmsRight) 800\GeV. Contours are shown for \Rint values 
    of (dashed blue) $-0.2$, (solid blue) $-0.1$, (solid green) $+0.1$, and (dashed green) 
    $+0.2$. Regions that are excluded, at 95\%~\CL, from the combined likelihood 
    analysis of the \HH analyses presented in this report are indicated by red 
    filled areas. Dashed black lines indicate constant relative widths of 5, 10, 
    and 20\%.
  }
  \label{fig:extSec_singlet}
\end{figure}

Exact conclusions from this study naturally depend on the allowed size of \Rint and the relative width \widthRatio. 
In the following, we choose as benchmark points $\Rint=\pm10$ and $\pm20\%$, and $\widthRatio=5$, 10 and 20\%.
The corresponding contours and exclusion limits derived from the \HH combination in the singlet 
model are shown in Fig.~\ref{fig:extSec_singlet}. 

Contours of positive (green) and negative (blue) interference ratios are shown 
as solid
(for $\Rint=\pm10\%$) and dashed (for $\Rint=\pm20\%$) lines. 
They are found to swap positions at $\mX = 400\GeV$, 
likely because of the peak of the nonresonant \HH distribution.
The dotted lines denote coupling value combinations beyond which 
the relative width of the resonance, $\Gamma_{\PX}/\mX$, exceeds 5
and 10\%, respectively, implying the narrow width 
approximation not being accurate anymore.   
For a given \mX, the quadratic dependence of \widthGamma on both $\sin\alpha$ and \couplingLambda 
according to Eq.~\eqref{eq:width} leads to elliptical isolines of constant \widthRatio. 
The experimental bound from the \HH combination discussed in Section~\ref{Sec:Results_X_to_HH} 
is obtained from the 95\% \CL upper limit on $\sigma(\pp\to\PX)\BR(\PX\to\HH)$, 
with the \PX production cross section growing with increasing $\sin\alpha$, 
and $\BR(\PX\to\HH)$ growing with increasing \couplingLambda.
We note that large values of $\sin\alpha$, corresponding to regions where the \PH boson is less SM-like, 
also tend to be excluded by precision measurements of the \PH boson~\cite{Papaefstathiou:2020iag}.

\begin{figure}[!htb]
  \centering
  \includegraphics[width=\cmsFigWidth]{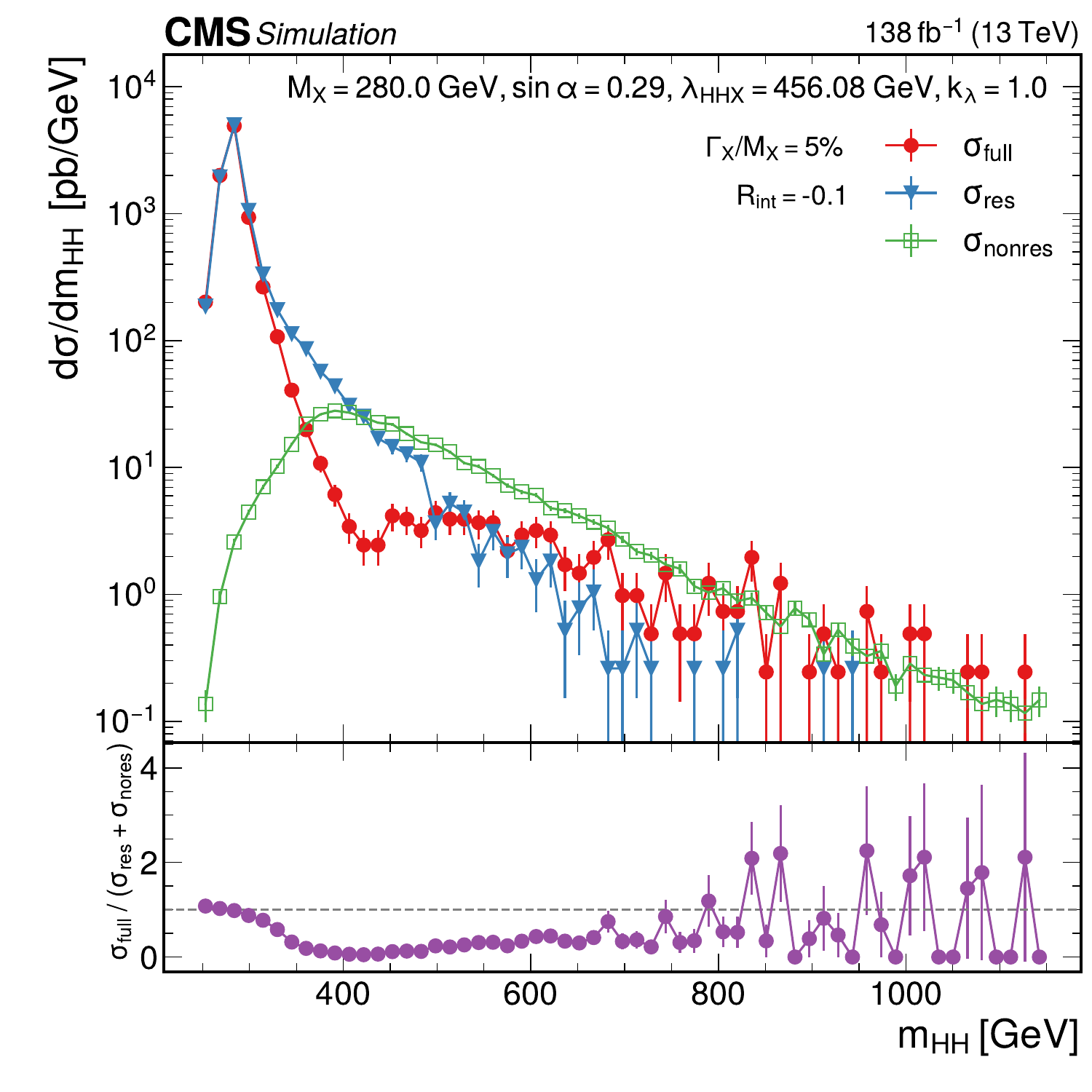}
  \includegraphics[width=\cmsFigWidth]{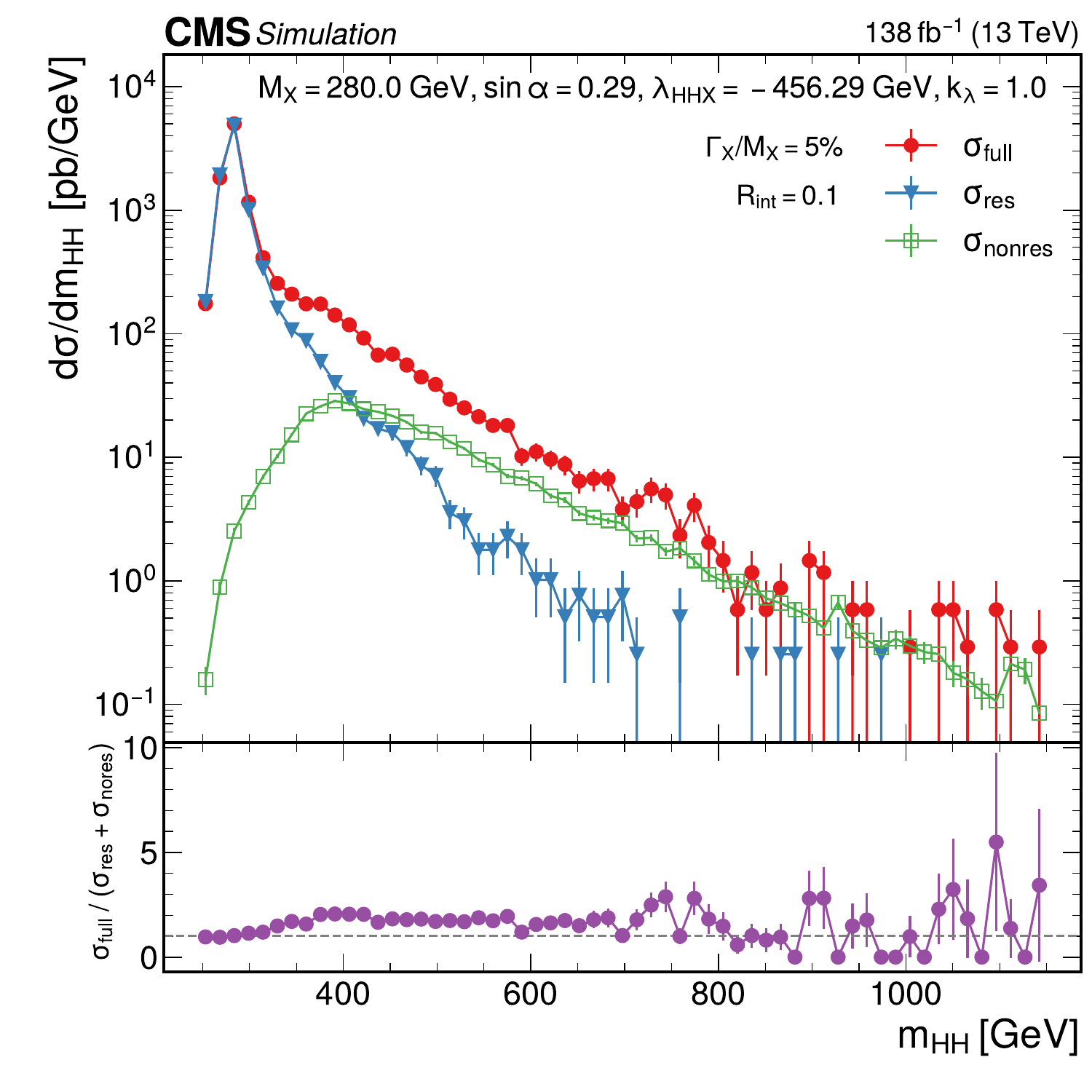}\\
  \includegraphics[width=\cmsFigWidth]{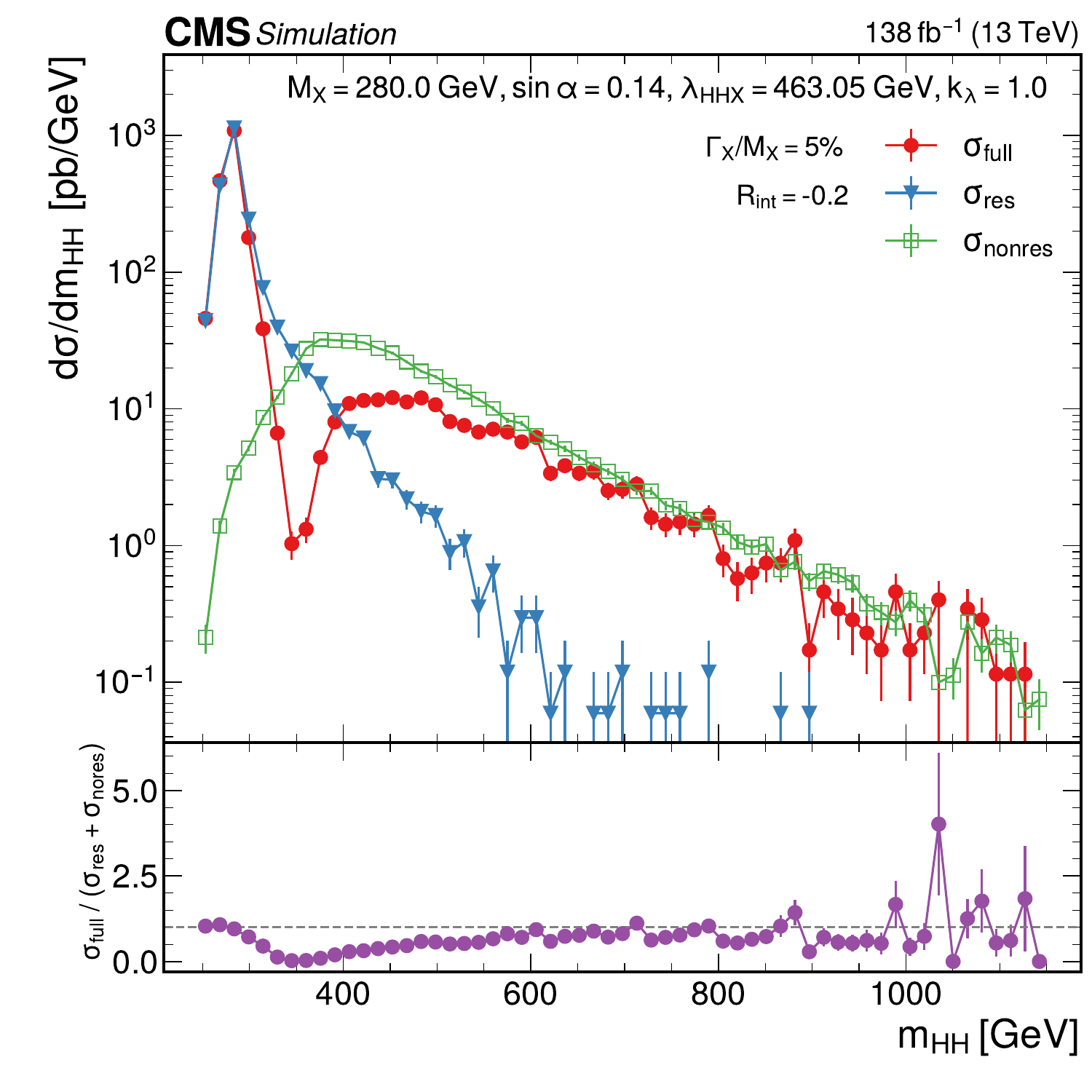}
  \includegraphics[width=\cmsFigWidth]{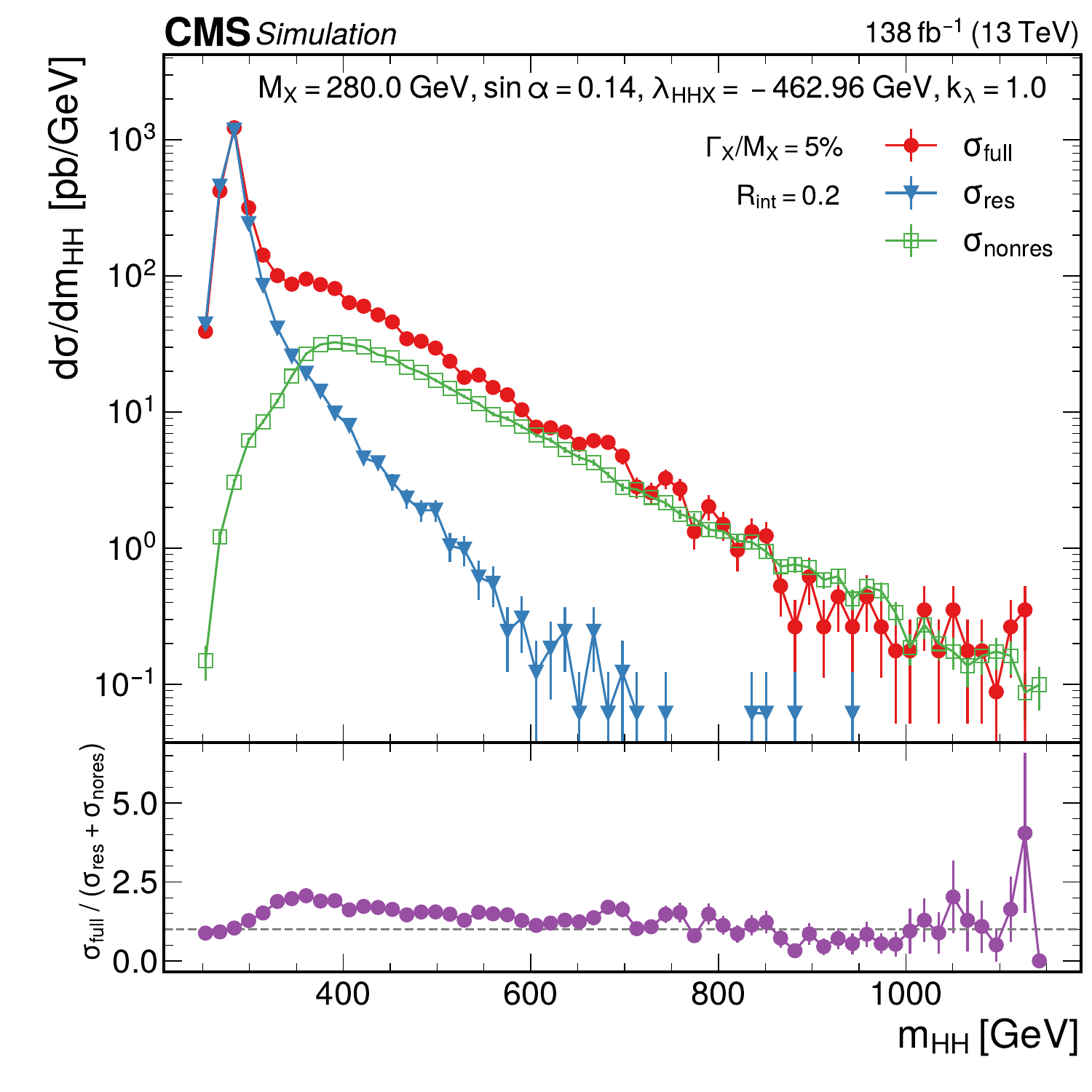}
  \caption{
    Expected differential cross sections for \HH production, as a function of 
    \mHH, for the real-singlet model with $\mX = 280\GeV$ and 
    $\widthRatio = 5\%$. The parameters $\sin\alpha$ and \couplingLambda have been 
    chosen such that (upper row) $\Rint=\pm 10\%$ and (lower row) $\Rint=\pm 20\%$, 
    for (\cmsLeft) negative and (\cmsRight) positive values of \Rint. The total 
    cross section for \HH production $\sigma^{\text{full}}$ (red line, labelled 
    as $\sigma_{\text{full}}$) is compared to the cross sections $\sigma^{\text{
    resonant-only}}$ (blue line, labelled as $\sigma_{\text{res}}$) and $\sigma^{
    \text{nonresonant}}$ (green line, labelled as $\sigma_{\text{nonres}}$) 
    considering only resonant and nonresonant production. In the lower panels the 
    ratio of $\sigma^{\text{full}}$ over $(\sigma^{\text{resonant-only}}+\sigma^{
    \text{nonresonant}})$ is shown.
  }
  \label{fig:lineshapes-280GeV}
\end{figure}

\begin{figure}[!htb]
  \centering
  \includegraphics[width=\cmsFigWidth]{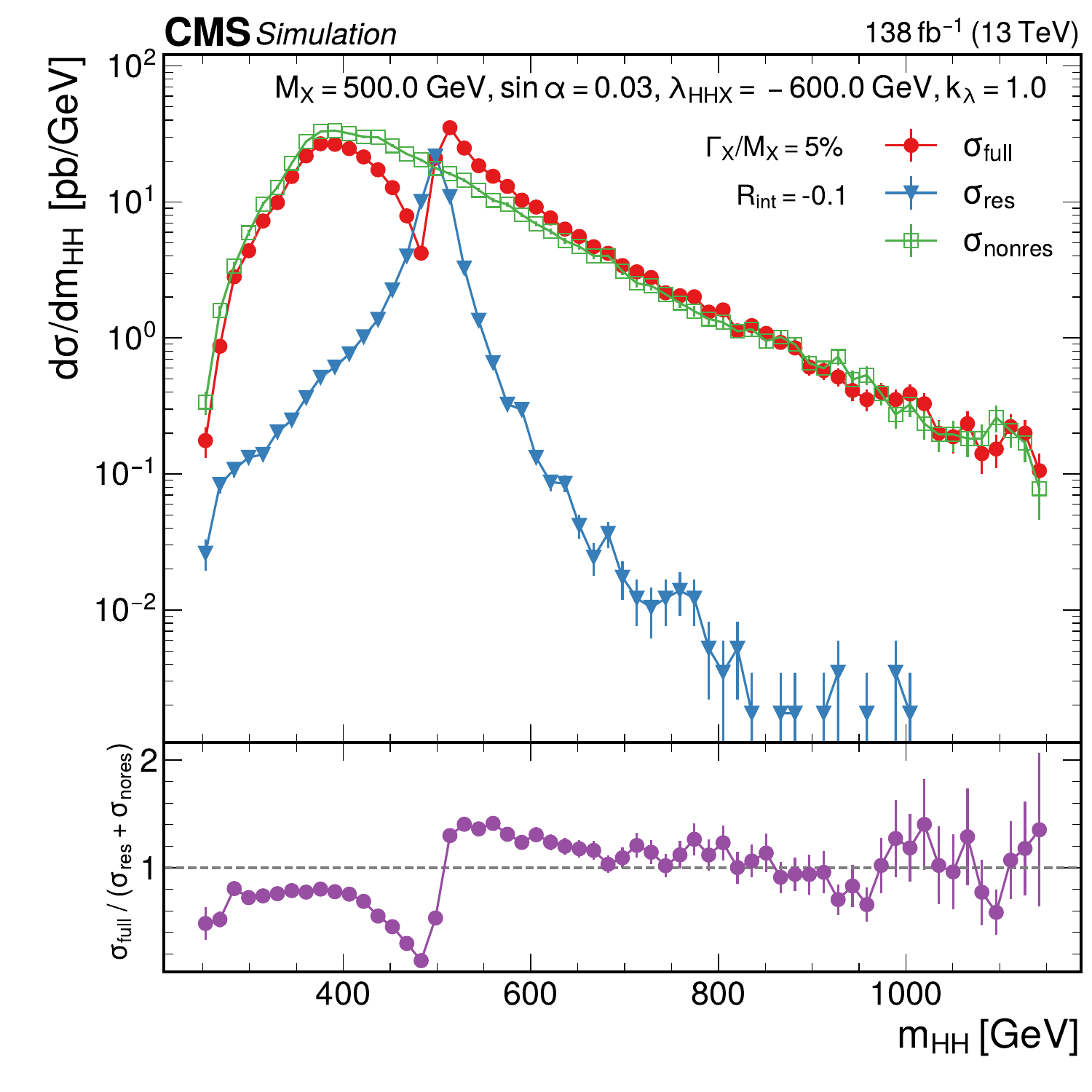}
  \includegraphics[width=\cmsFigWidth]{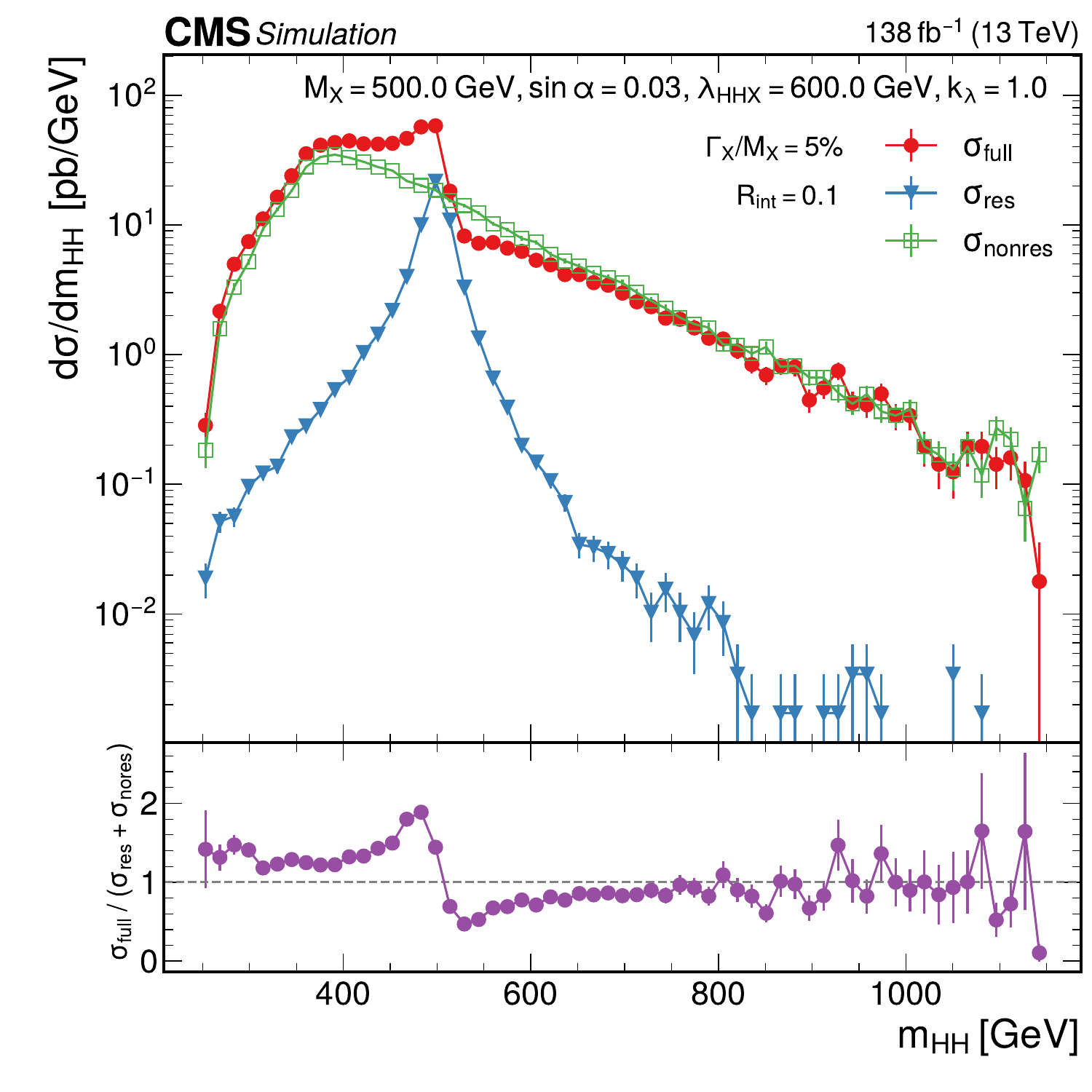}
  \includegraphics[width=\cmsFigWidth]{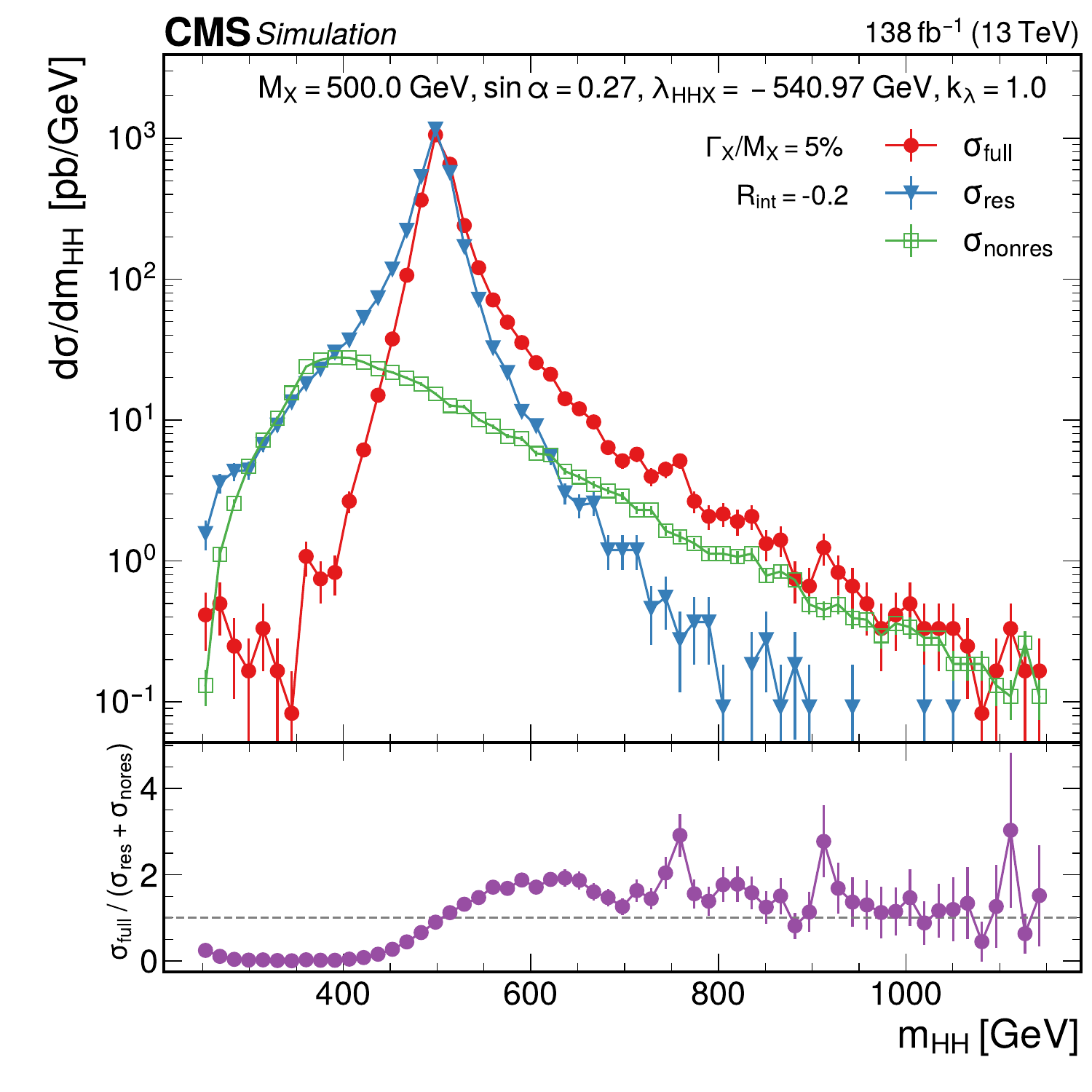}
  \includegraphics[width=\cmsFigWidth]{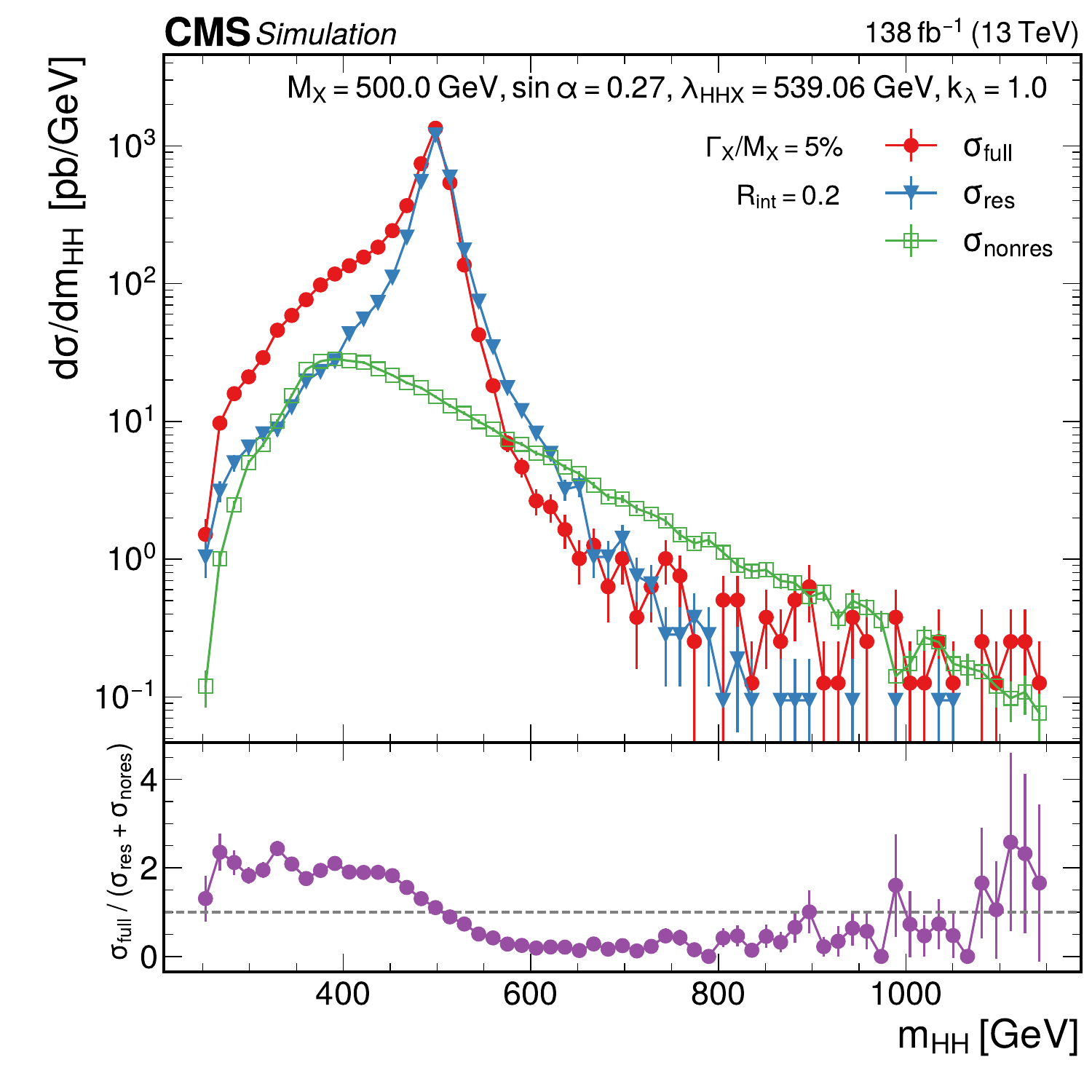}
  \caption{
    Expected differential cross sections for \HH production, as a function of 
    \mHH for the real-singlet model with $\mX = 500\GeV$ and 
    $\widthRatio = 5\%$.  The parameters $\sin\alpha$ and \couplingLambda have been 
    chosen such that (upper row) $\Rint=\pm 10\%$ and (lower row) $\Rint=\pm 20\%$, 
    for (\cmsLeft) negative and (\cmsRight) positive values of \Rint. The total 
    cross section for \HH production $\sigma^{\text{full}}$ (red line, labelled 
    as $\sigma_{\text{full}}$) is compared to the cross sections $\sigma^{\text{
    resonant-only}}$ (blue line, labelled as $\sigma_{\text{res}}$) and $\sigma^{
    \text{nonresonant}}$ (green line, labelled as $\sigma_{\text{nonres}}$) 
    considering only resonant and nonresonant production. In the lower panels the 
    ratio of $\sigma^{\text{full}}$ over $(\sigma^{\text{resonant-only}}+\sigma^{
    \text{nonresonant}})$ is shown.
  }
  \label{fig:lineshapes-500GeV}
\end{figure}

For most of the studied mass points, sizable interference ratios occur
only in parameter regions to which the current measurements are either
not yet sensitive, or at too large values of $\sin\alpha$. In
particular, for large resonance masses, where interference effects tend
to grow, they are far below the current sensitivity and might only
play a role when the full data set from HL-LHC becomes
available~\cite{Cepeda:2019klc}, as discussed in
Section~\ref{Sec:DiscoveryPotential_X_to_YH}. However, there are
regions at intermediate \mX where the interpretation of NWA-based
limits for \HH derived in the singlet model would solicit some care
already in the Run~2 combination (\eg, $\mX = 500\GeV$, $\sin\alpha =
0.2$ and $\couplingLambda = 400\GeV$). It is important to note that
such interpretations are generally model dependent.

The differential cross sections as a function of \mHH are shown for representative points from 
the ($\sin\alpha$, \couplingLambda) parameter space in 
Fig.~\ref{fig:lineshapes-280GeV} for $\mX=280\GeV$, and in Fig.~\ref{fig:lineshapes-500GeV} for $\mX=500\GeV$. 
The parameters are chosen such that $\widthRatio = 5\%$, which is well below the detector resolution for resonance masses below 1~\TeV,
and $\Rint=\pm10\%$ or $\pm$20\%, so that sizable interference effects are expected. 
The lineshapes show points in parameter space where the \Rint contours intersect with lines of 
constant $\widthRatio = 5\%$ in Fig.~\ref{fig:extSec_singlet}. 

The mass points of $\mX=280$ and 500\GeV have been chosen because these values are 
on the left- and right-hand side of the peak in the \mHH distribution for nonresonant SM \HH production. 
The total signal contribution of the resonance, including the interference effect, 
can be assessed as the difference between $\sigma^{\text{full}}$ (red graph) and $\sigma^{\text{nonresonant}}$ (green graph).
In the $\mX=280\GeV$ case, the resonance peak is at a mass where the non-resonant background is low in comparison; hence the central part of the peak is not much affected in its shape, and a classical bump hunt should still work. However, the total cross section is modified as specified by $\Rint$. For a precision measurement, which is not yet in our reach, a distortion of the signal shape, either a peak-dip or peak-tail pattern depending on the relative sign of the amplitudes, would have to be taken into account.
At $\mX=500\GeV$, in the top panels of Fig.~\ref{fig:lineshapes-500GeV}, the signal shape is found to be strongly modified by the interference effect. However, this occurs in a parameter region still relatively far away from the regions currently probed, as can be seen in Fig.~\ref{fig:extSec_singlet}. 
Although the expected interference effects clearly depend on the underlying model, 
they can be expected to be of mounting importance in the future as the LHC data set increases.

\section{Discovery potential at the HL-LHC}\label{Sec:DiscoveryPotential}

The HL-LHC~\cite{ZurbanoFernandez:2020cco} is planned to start in 2029 and aims to deliver a \pp collision 
data set corresponding to about 3000\fbinv of integrated luminosity in the baseline scenario, and up 
to 4000\fbinv in the ultimate scenario, at an unprecedented center-of-mass energy of 14\TeV. 
The CMS detector will be upgraded to cope with the large size of 140 (200) PU events on 
average for the baseline (ultimate) scenario. The upgraded detector will also meet the 
challenges from the adverse effects due to the radiation dose to which the detector components 
are exposed, which is one order of magnitude higher than at the current LHC. Furthermore, major 
improvements of the software for the online and offline event reconstruction are under development 
to fully exploit the potential of the upgraded detector. Searches for scalars \PX decaying to \HH or \YH 
are among the most relevant targets of research at the HL-LHC, and thus projection studies are very important 
to motivate the ongoing hardware and software upgrades. Meanwhile, such studies can provide an 
estimate of the sensitivity to the relevant BSM theories which can be achieved with the HL-LHC data.

This section describes the perspectives for the searches for \PX boson resonances decaying to \HH or \YH at the HL-LHC, 
in the most sensitive decay channels \bbgg, \bbtt, and \bbbb, in the baseline scenario of the HL-LHC. 
Using the combined likelihood method, individual channels are statistically combined to exploit their 
complementarity in sensitivity to different regions in parameter space of the tested BSM theories. 
The expected upper limits at 95\%~\CL on the cross sections of the BSM processes of interest are 
provided as functions of the masses of the BSM scalars. The expected exclusion in the parameters of 
the relevant BSM theories is estimated, as well as the expected discovery significance for benchmark BSM signals.

\subsection{Methodology for estimation of the discovery potential}\label{Sec:DiscoveryPotentialIngredients}

The projection studies are based on the resonant \HH and \YH searches in the most 
sensitive channels from the CMS Run~2 data set corresponding to an integrated luminosity of 138\fbinv, 
as summarized in Table~\ref{tab:inputchannels_proj}. Descriptions of the Run~2 \HH and \YH searches 
are given in the sections indicated in the table. 
\begin{table}[tb]
  \centering
	\topcaption{Searches for resonant \HH and \YH production considered for the projection study.}
	\begin{tabular}{ l  c  c }	
    Final state        & Reference          & Section \\ 
    \hline
    \bbtt              & \cite{CMS:2021yci} & \ref{Sec:HIG-20-014}\\
    \bbgg              & \cite{CMS:2023boe} & \ref{Sec:HIG-21-011}\\
    \bbbb (merged-jet) & \cite{CMS:2022suh} & \ref{Sec:B2G-21-003}\\
  \end{tabular}
  \label{tab:inputchannels_proj}
\end{table}

Using the same approach as studies in Ref.~\cite{Cepeda:2019klc},
searches using the Run~2 data set are projected to an integrated
luminosity of 3000\fbinv. Where appropriate, the signal cross sections
have been scaled to the center-of-mass energy of
14\TeV~\cite{deFlorian:2016spz}.
As the upgraded CMS detector will ensure a 
performance comparable to Run~2, the efficiency in the reconstruction and identification of photons, 
leptons, jets and \PQb jets, as well as the resolution in their energy and momentum measurements are 
assumed to be unchanged. The experimental sensitivity expected at the HL-LHC is derived using the 
following three systematic uncertainty scenarios. 
\begin{itemize}
  \item[S1:] All the systematic uncertainties are assumed to remain the same as in Run~2. This is 
  an over-conservative scenario because the CMS detector upgrade, the progress in the reconstruction 
  techniques, and the very large data set available for the experimental calibrations are expected to 
  bring a substantial reduction of several systematic uncertainties. Furthermore, progress in the theory 
  calculations is expected to reduce the uncertainties in the predictions. 
  \item[S2:] The theory uncertainties are halved, while the experimental uncertainties are set according 
  to the recommendations of Ref.~\cite{HLHELHCCommonSystematics}. 
  \item[\ \ \ Statistical only:] The results are derived considering only the statistical uncertainty in data. 
\end{itemize}

The projected results from the channels considered are statistically combined following the same 
procedure as adopted for the Run~2 combination which is described in Section~\ref{Sec:Combination}. 
In particular, the systematic uncertainties affecting multiple channels, such as the uncertainties 
in the luminosity and on the \PQb jet identification efficiency, are treated as correlated among 
all the input channels.

\subsection{Discovery potential for \texorpdfstring{$\PX\to\HH$}{X->HH}}\label{Sec:DiscoveryPotential_X_to_HH}

\begin{figure}[htbp]
  \centering
  \includegraphics[width=\cmsFigWidth]{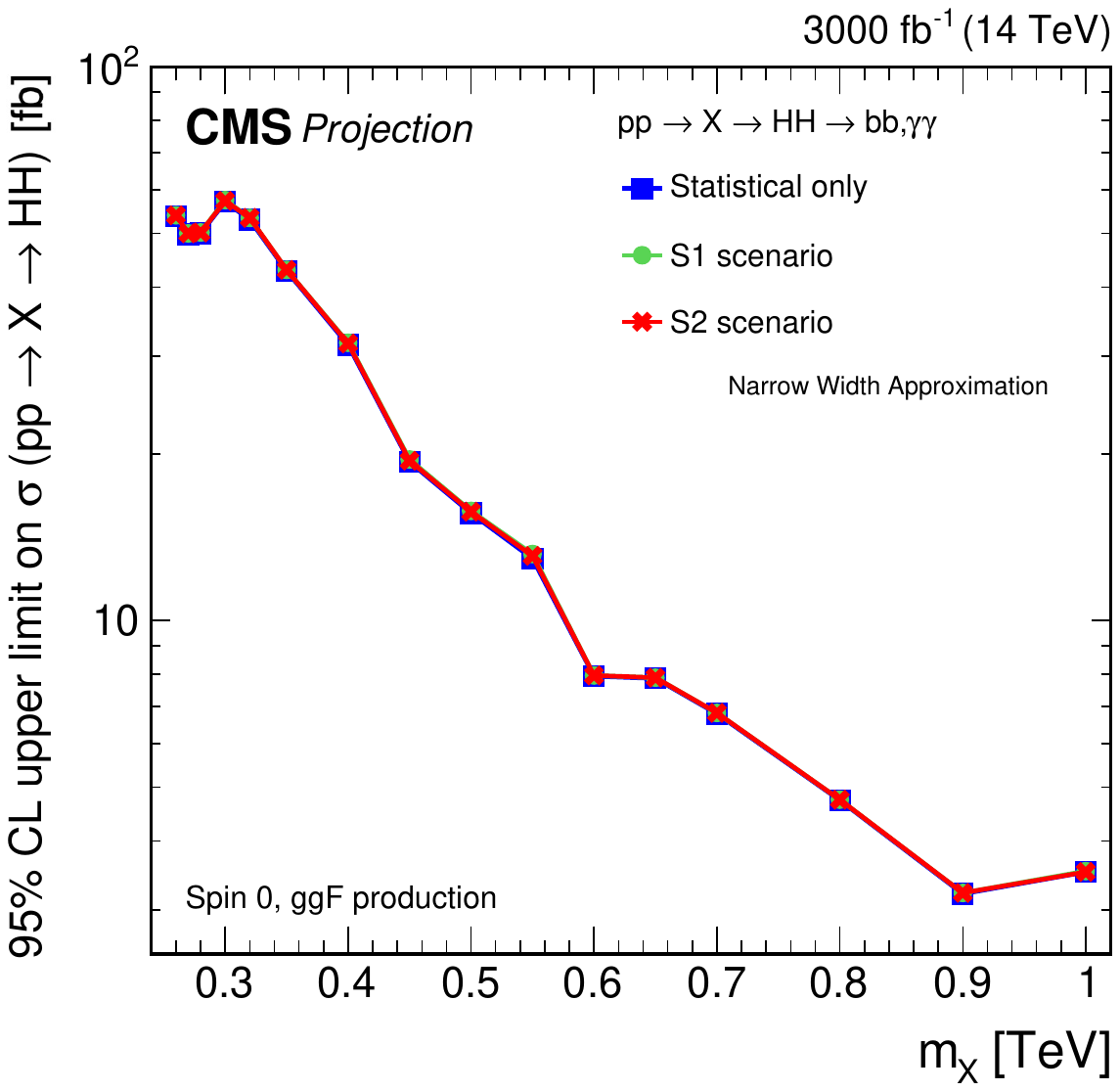}
  \includegraphics[width=\cmsFigWidth]{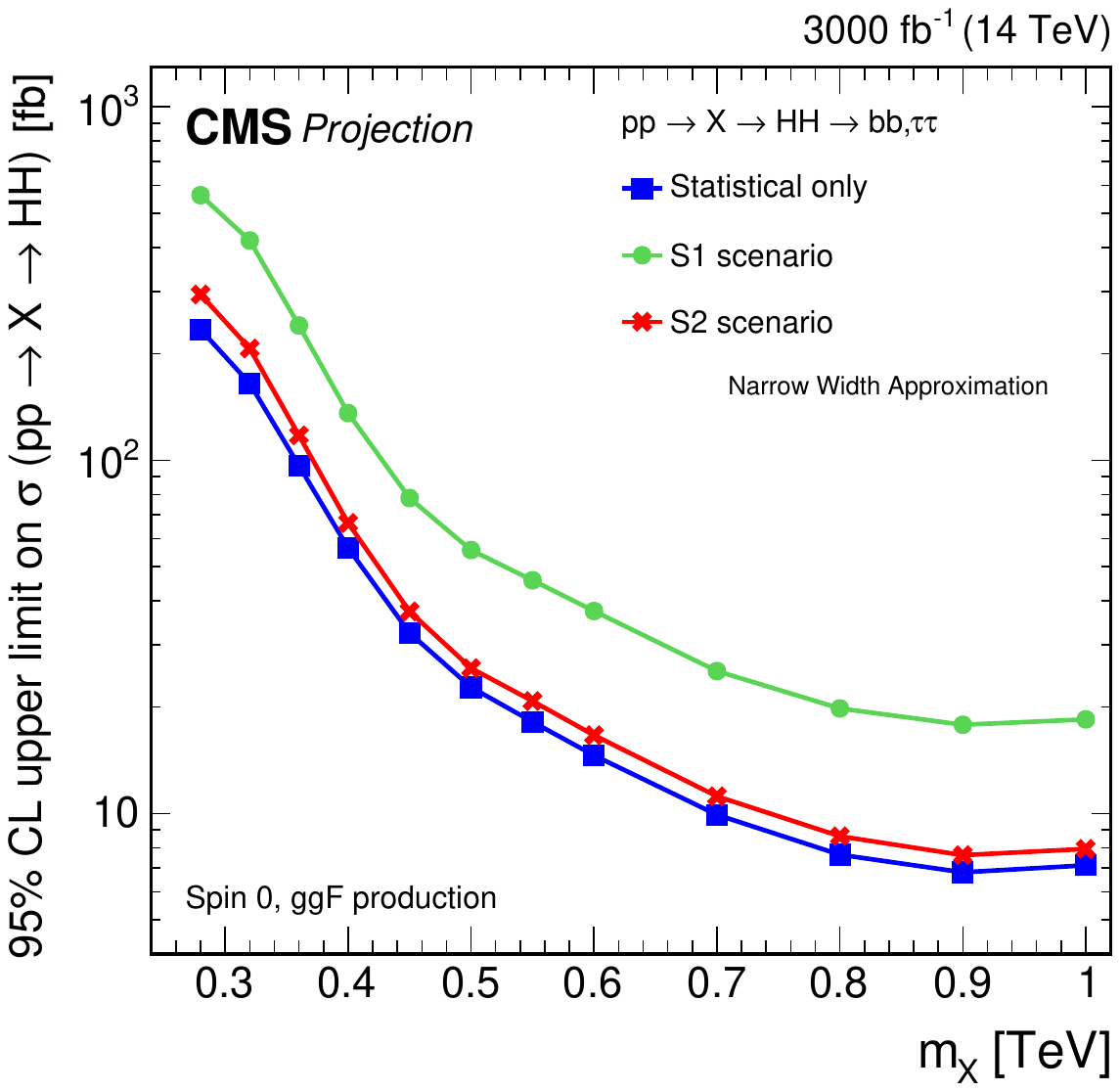}\\
  \includegraphics[width=\cmsFigWidth]{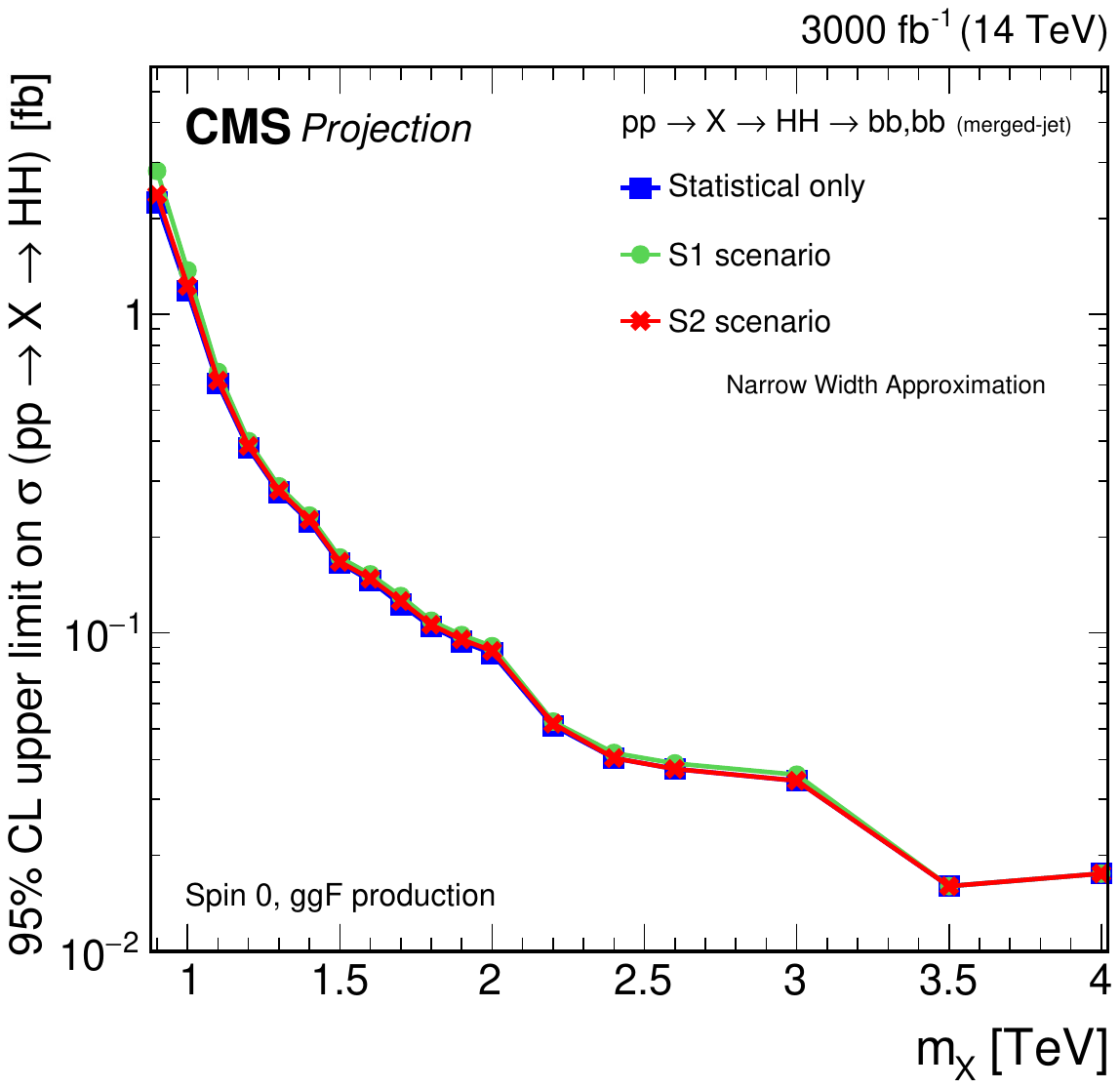}
  \caption{
    Expected upper limits at 95\%~\CL, on the product of the cross section for 
    the production of a spin-0 resonance \PX and the branching fraction $\BR(\PX
    \to\PH\PH)$, as functions of \MX from the (upper \cmsLeft) \bbtt~\cite{CMS:2021yci}, 
    (upper \cmsRight) \bbgg~\cite{CMS:2023boe},  and (lower) \bbbb with two merged 
    \bb jets~\cite{CMS:2022suh} analyses discussed in this report, projected to 
    an integrated luminosity of 3000\fbinv under the assumption of different 
    systematic uncertainty scenarios, as discussed in the text. All estimates 
    include the anticipated statistical uncertainties. 
  }
  \label{fig:XHH_projections_bychannel}
\end{figure}
The expected upper limits at 95\% \CL on the $\PX \to \HH$ cross section from the channels 
considered projected to 3000\fbinv are shown for the three different systematic uncertainty 
scenarios in Fig.~\ref{fig:XHH_projections_bychannel}. 
The projected upper limits from the \bbgg decay mode range between 60 and 3\unit{fb} for 
\MX within 300--1000\GeV. 
The overall impact of the systematic uncertainties on the \bbgg upper limits is below 1\% 
because of the small uncertainty on the background modeling thanks to the estimation procedure 
based on the fit to the data in sideband regions.  

The \bbtt channel provides upper limits on the cross section at 95\% \CL between 300 and 7\unit{fb} 
for \MX within 300--1000\GeV. The systematic uncertainty with the largest impact in the S1 scenario 
has its origin in the limited size of the MC simulation used for the background estimation. 
In the S2 scenario, the statistical uncertainties on the simulated events are assumed to be negligible 
and the main systematic uncertainties arise from the efficiencies of the \PQb jet and \PGt 
identification and misidentification. 

The \bbbb channel in the boosted regime, in the following simply referred to as \bbbb, covers \MX values between 900 and 4000\GeV, and is 
expected to provide upper limits between 0.04 and 0.02\unit{fb} at an integrated luminosity of 3000~\fbinv. 
The impact of the systematic uncertainties is very small, as the sensitivity of the analysis 
is mainly limited by the statistical uncertainty in the data.
\begin{figure}[tbp]
  \centering
  \includegraphics[width=\cmsFigWidth]{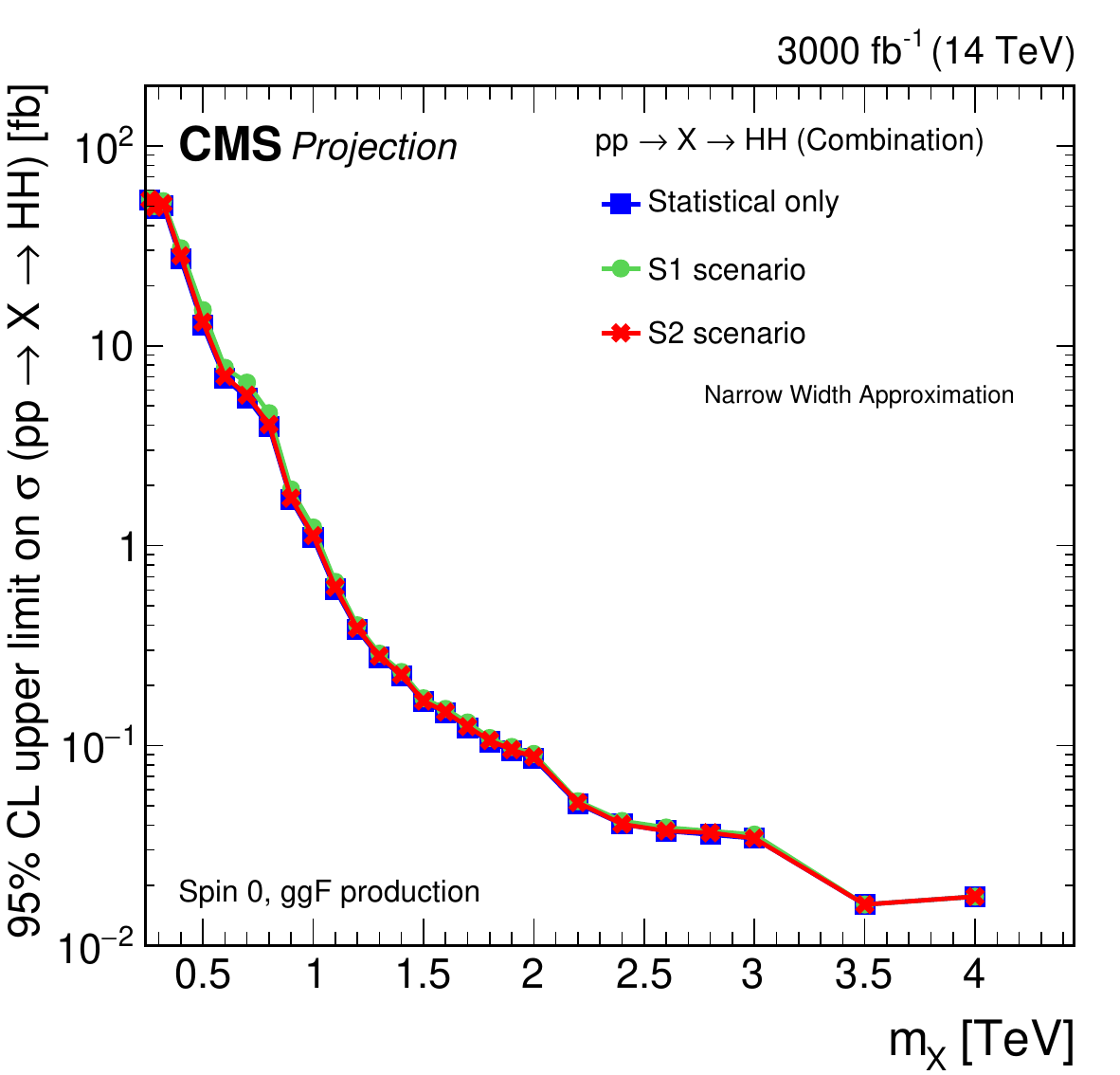}
  \includegraphics[width=\cmsFigWidth]{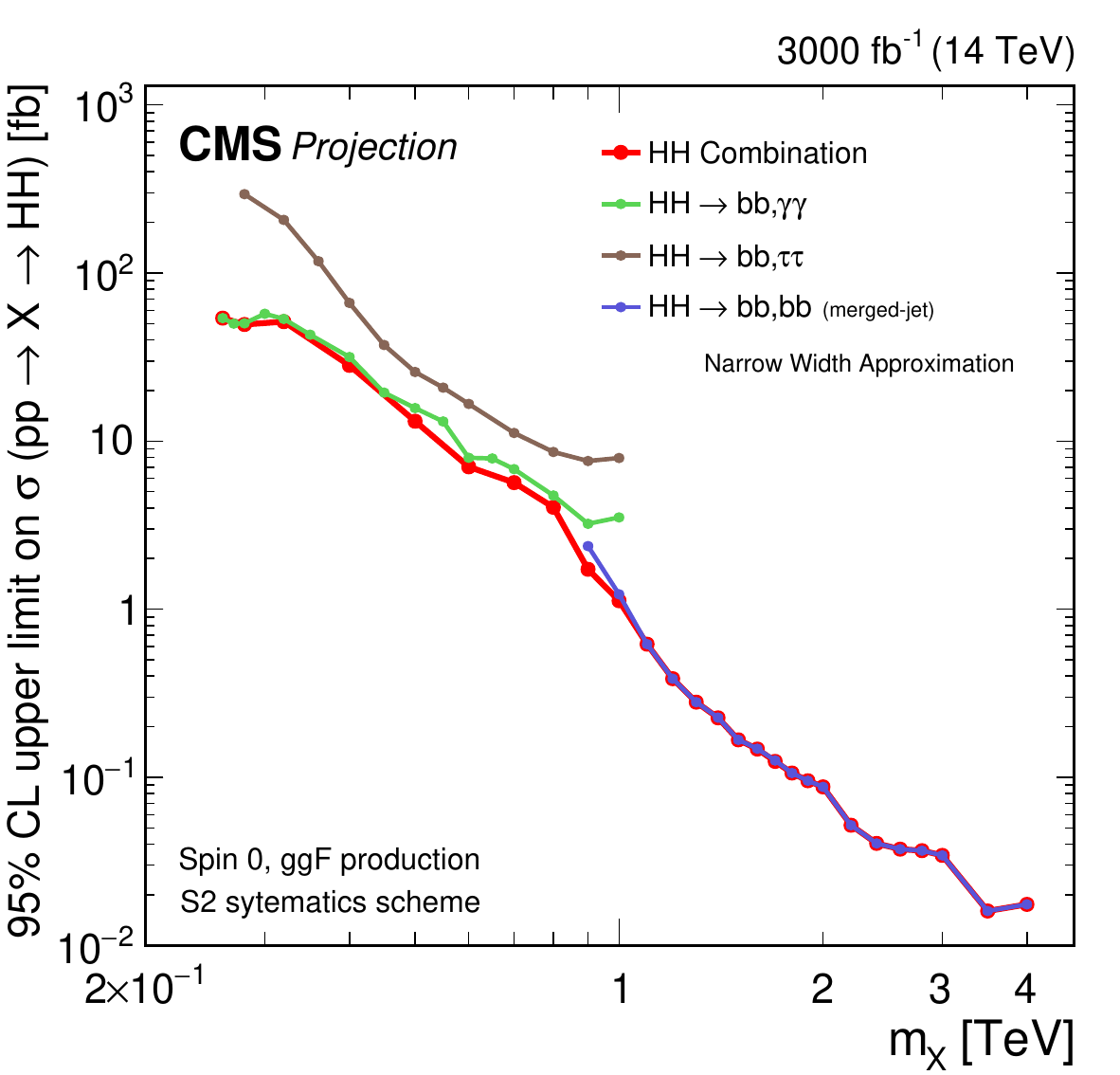}	
  \caption{
    Expected upper limits at 95\%~\CL, on the product of the cross section for 
    the production of a spin-0 resonance \PX and the branching fraction $\BR(\PX
    \to\PH\PH)$, as a function of \MX, for an integrated luminosity of 3000\fbinv 
    and the combination of the three analyses shown in 
    Fig.~\ref{fig:XHH_projections_bychannel}. Shown are the 
    effects of the different systematic uncertainty scenarios (\cmsLeft), and 
    the reach of the individual analyses for the S2 systematic scenario (\cmsRight). 
    All estimates include the anticipated statistical uncertainties.
  }
  \label{fig:XHHcomb_projections}
\end{figure}

The results of the combination of the resonant \HH searches considered are shown in Fig.~\ref{fig:XHHcomb_projections} 
in the different systematic uncertainty scenarios, and in comparison to the results from the individual channels. 
The \bbgg channel is found to dominate the sensitivity in the region $\MX < 500\GeV$; around 900\GeV the channel 
with the best sensitivity is \bbbb, followed by \bbgg and \bbtt. For $\MX > 1000\GeV$ the only channel 
considered is \bbbb which is expected to be the most sensitive in this kinematic region. 
Thanks to the small impact of the systematic uncertainties on the \bbgg and \bbbb channels, 
the differences between the three systematic uncertainty scenarios are rather small.

\subsubsection{Perspectives for the discovery of BSM benchmark signals}

\begin{figure}[!htbp]
  \centering
  \includegraphics[width=\cmsFigWidth]{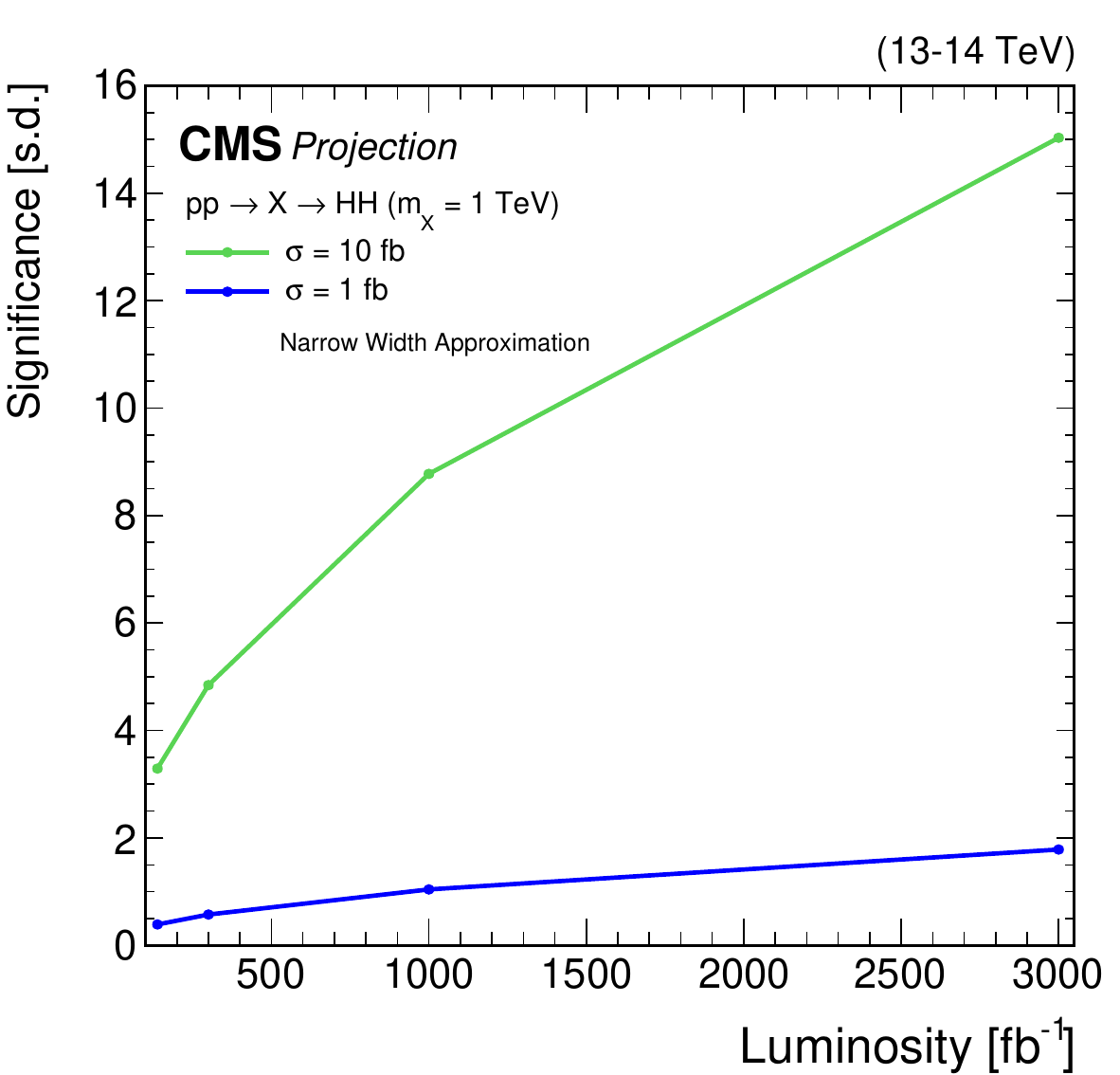}
  \caption{
    Expected discovery significance for a spin-0 resonance \PX with $\mX=1\TeV$ 
    and cross sections of 1 and 10\unit{fb}, obtained for the combined likelihood 
    analysis of the resonant \HH searches as discussed in Section~\ref{Sec:DiscoveryPotential} 
    and shown in Figs.~\ref{fig:XHH_projections_bychannel} 
    and~\ref{fig:XHHcomb_projections}, shown as function of the integrated 
    luminosity. 
  }
  \label{fig:XHHcomb_significance}
\end{figure}

The expected significance for the discovery of a benchmark BSM signal from a spin-0 resonance with a 
mass of 1\TeV is calculated for several signal cross sections and represented as a function of the integrated 
luminosity in Fig.~\ref{fig:XHHcomb_significance}. 
Based on the three channels considered in this projection, the significance of a signal of 
$\PX \to \HH$ with a cross section of 10\unit{fb} corresponds to about three standard deviations 
at Run~2, while an integrated luminosity of 300\fbinv would yield 4.8
standard deviations, indicating an attractive discovery potential
already for Run~3 and its combination with Run~2. The significance of
the same signal would reach about 15 standard deviations at 3000\fbinv. A signal with a cross section 
of 3\unit{fb} would be sufficient to reach an observation at the level of five standard deviations with 3000\fbinv.

\subsubsection{Perspectives for MSSM scenarios}

\begin{figure}[htbp]
  \centering
  \includegraphics[width=\cmsFigWidth]{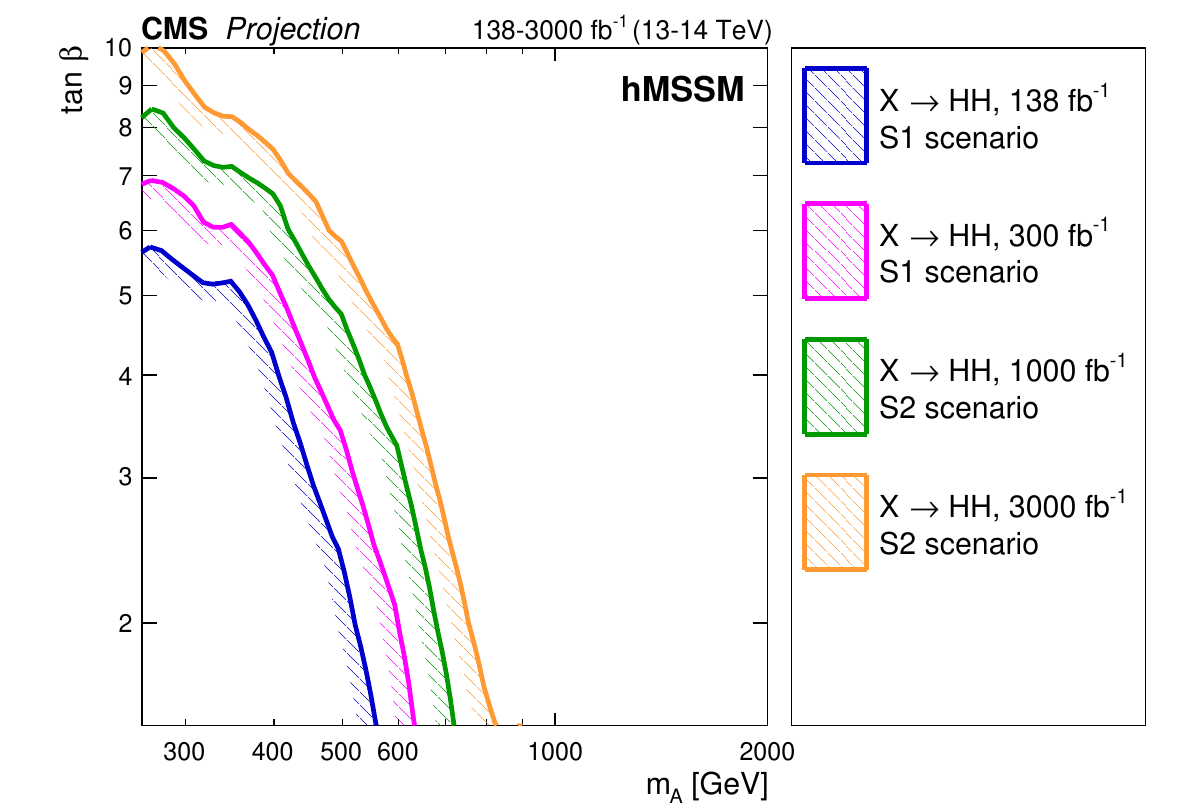}
  \includegraphics[width=\cmsFigWidth]{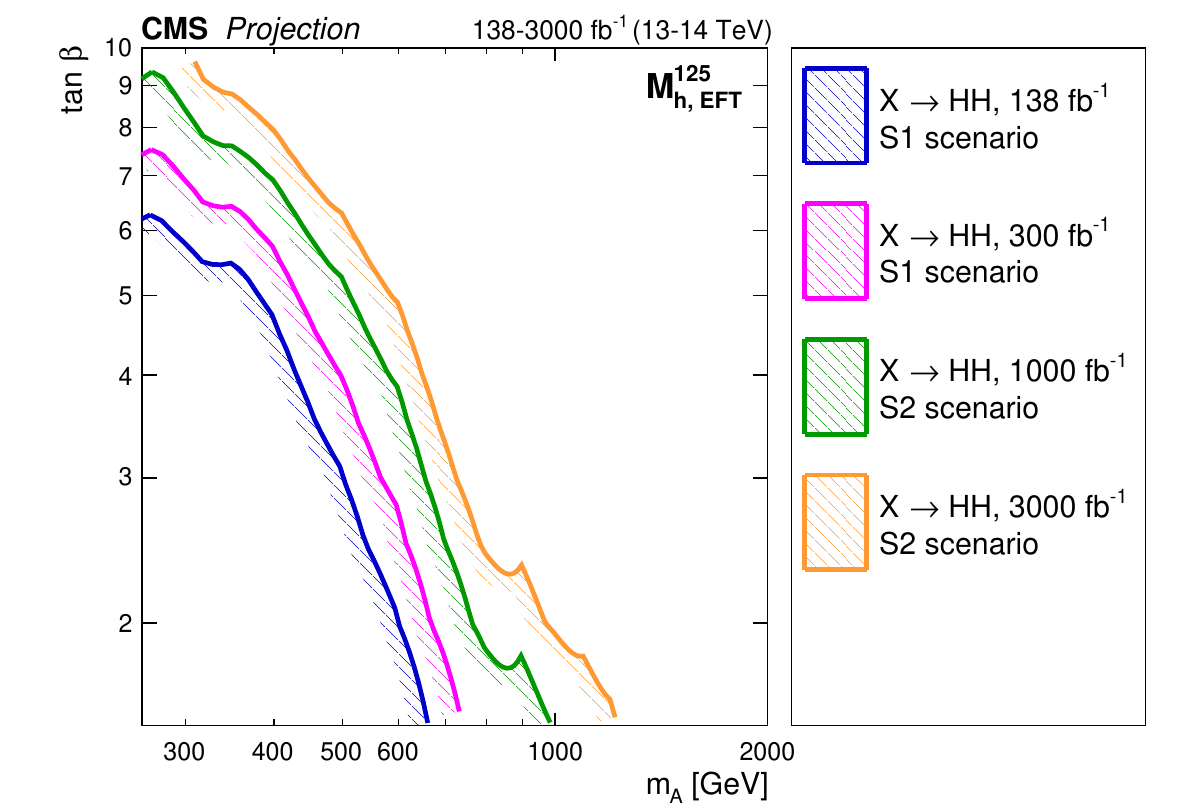}
  \caption{
    Expected exclusion contours at 95\%~\CL, in the (\tanb, \mA) plane of the 
    (\cmsLeft) hMSSM and (\cmsRight) \MhEFTScen scenarios obtained from the 
    combined likelihood analysis of the \HH searches discussed in 
    Section~\ref{Sec:Interp_in_Extended_Higgs_sector} and shown in 
    Figs.~\ref{fig:hMSSM} and~\ref{fig:MSSM_mh125}, for different integrated 
    luminosities and compared to the Run 2 result obtained at $\sqrt{s}=13\TeV$. 
    The projections assume $\sqrt{s}=14\TeV$.
  }
  \label{fig:MSSM_exclusion}
\end{figure}

The projected exclusion limits at 95\%~\CL of the hMSSM and \MhEFTScen benchmark scenarios from 
resonant \HH searches are shown in Fig.~\ref{fig:MSSM_exclusion}. 
The S1 systematic uncertainty scenario is used for the Run~2 result and conservatively also for the result with 300\fbinv, 
while the S2 systematic uncertainty scenario is used for the projected 1000 and 3000\fbinv results.  
Over the full accessible range in \tanb, the exclusion in \mA increases by about 250--300\GeV 
when moving from the Run~2 integrated luminosity to 3000\fbinv, for both the hMSSM and \MhEFTScen scenarios. 
This exclusion from the resonant \HH searches will complement the searches for \PX decaying to a 
pair of fermions or vector bosons. 

\subsubsection{Perspectives for the WED bulk scenario}

\begin{figure}[tbp]
  \centering
  \includegraphics[width=\cmsFigWidth]{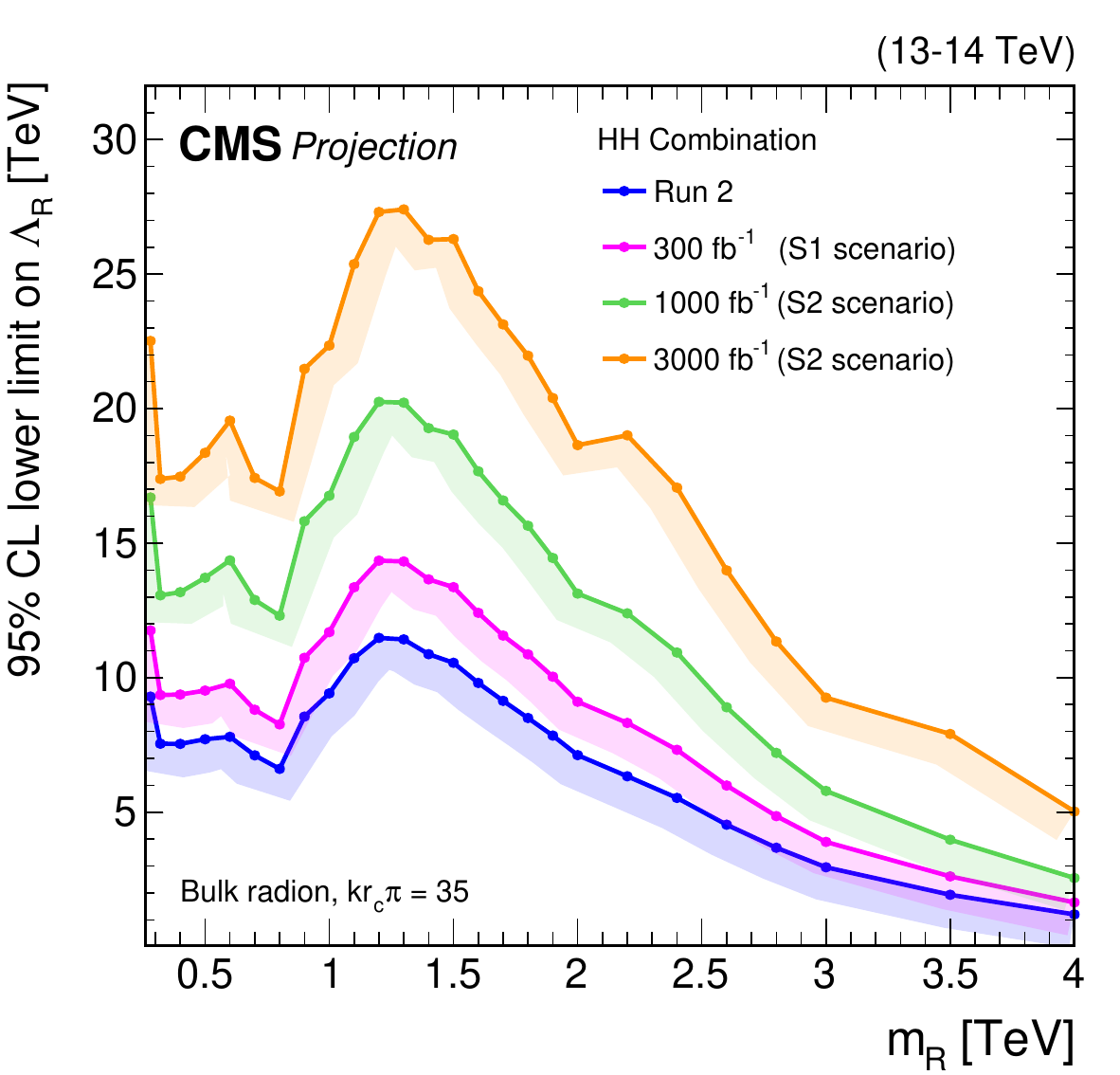}
  \caption{
    Expected lower limit at 95\%~\CL, on \LambdaR in the warped extra dimensions 
    bulk scenario for the production of a radion \PR, as a function of $m_{\PR}$. 
    The limits are derived from the combined likelihood analysis of the \HH 
    searches discussed in Section~\ref{Sec:Interp_in_Warped_Extra_Dimensions} 
    and shown in Fig.~\ref{fig:Int_WED}, for different values of the integrated 
    luminosity. Excluded areas are indicated by the direction of the hatching 
    along the exclusion contours.
  }
  \label{fig:Radion_exclusion}
\end{figure}

The expected lower limits at 95\%~\CL on the bulk radion parameter \LambdaR as a function of 
the radion mass $m_{\PR}$ are shown in Fig.~\ref{fig:Radion_exclusion}. The limits are obtained from the 
combination of resonant \HH searches in the WED bulk scenario. 
The S1 systematic uncertainty scenario is used for the Run~2 result and conservatively also for the result with 300\fbinv, 
while the S2 systematic uncertainty scenario is used for the projected 1000 and 3000\fbinv results. 
Over the full range in $m_{\PR}$, the limit on \LambdaR is expected to increase by a factor of at least two 
with the full HL-LHC data set. 

\subsubsection{Perspectives for the singlet scenarios}

In the singlet model of Section~\ref{Sec:Effects_finite_width_and_interference} with $\kappaLambda = 1$, 
limits are derived in the ($\sin\alpha$, \couplingLambda) plane from the combination of resonant \HH searches. 
Resonance masses between 280 and 800\GeV are probed using Run~2 data and projected to 
integrated luminosities corresponding to 300, 1000, and 3000\fbinv. 
Projected exclusion regions at 95\% \CL are shown in Fig.~\ref{fig:singlet_exclusion}. 
The HL-LHC dataset of 3000\fbinv has the potential to considerably expand the present exclusion regions in 
the ($\sin\alpha$, \couplingLambda) plane for all values of \mX. 
Compared to the present limits, the largest improvement is observed for large masses, $\mX=600\GeV$ and 
higher, where large regions of the ($\sin\alpha$, \couplingLambda) plane can be probed. 
\begin{figure}[tbp]
  \centering
  \includegraphics[width=1.5\cmsFigWidth]{Figure_049.pdf}
  \caption{
    Exclusion contours at 95\%~\CL, in the ($\sin\alpha$, \couplingLambda) plane 
    for $\kappaLambda = 1$ in the real-singlet model. These contours 
    are obtained from the combined likelihood analysis of the \HH searches 
    discussed in Section~\ref{Sec:Interp_in_Extended_Higgs_sector} for (upper 
    \cmsLeft to lower \cmsRight) $\mX = 280$, 400, 500, 600, 700, and 800\GeV. 
    The expected limits from the Run~2 dataset have been projected to integrated 
    luminosities of 300, 1000, and 3000\fbinv. Excluded areas are indicated by 
    the direction of the hatching along the exclusion contours.
  }
  \label{fig:singlet_exclusion}
\end{figure}

\subsection{Discovery potential for \texorpdfstring{$\PX\to$\YH}{X->YH}} 
\label{Sec:DiscoveryPotential_X_to_YH}

\begin{figure}[tbp]
  \centering
  \includegraphics[width=\cmsFigWidth]{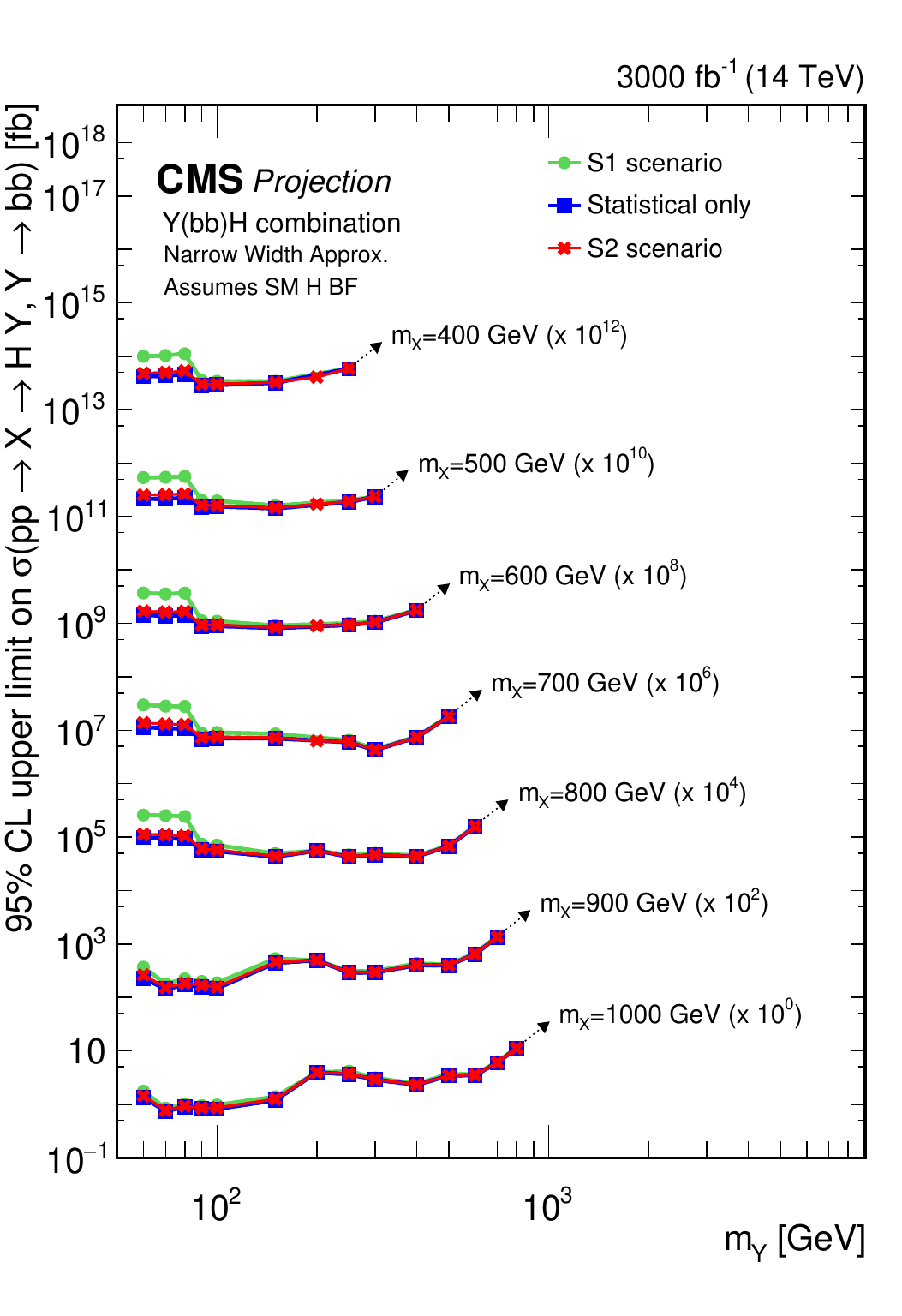}
  \includegraphics[width=\cmsFigWidth]{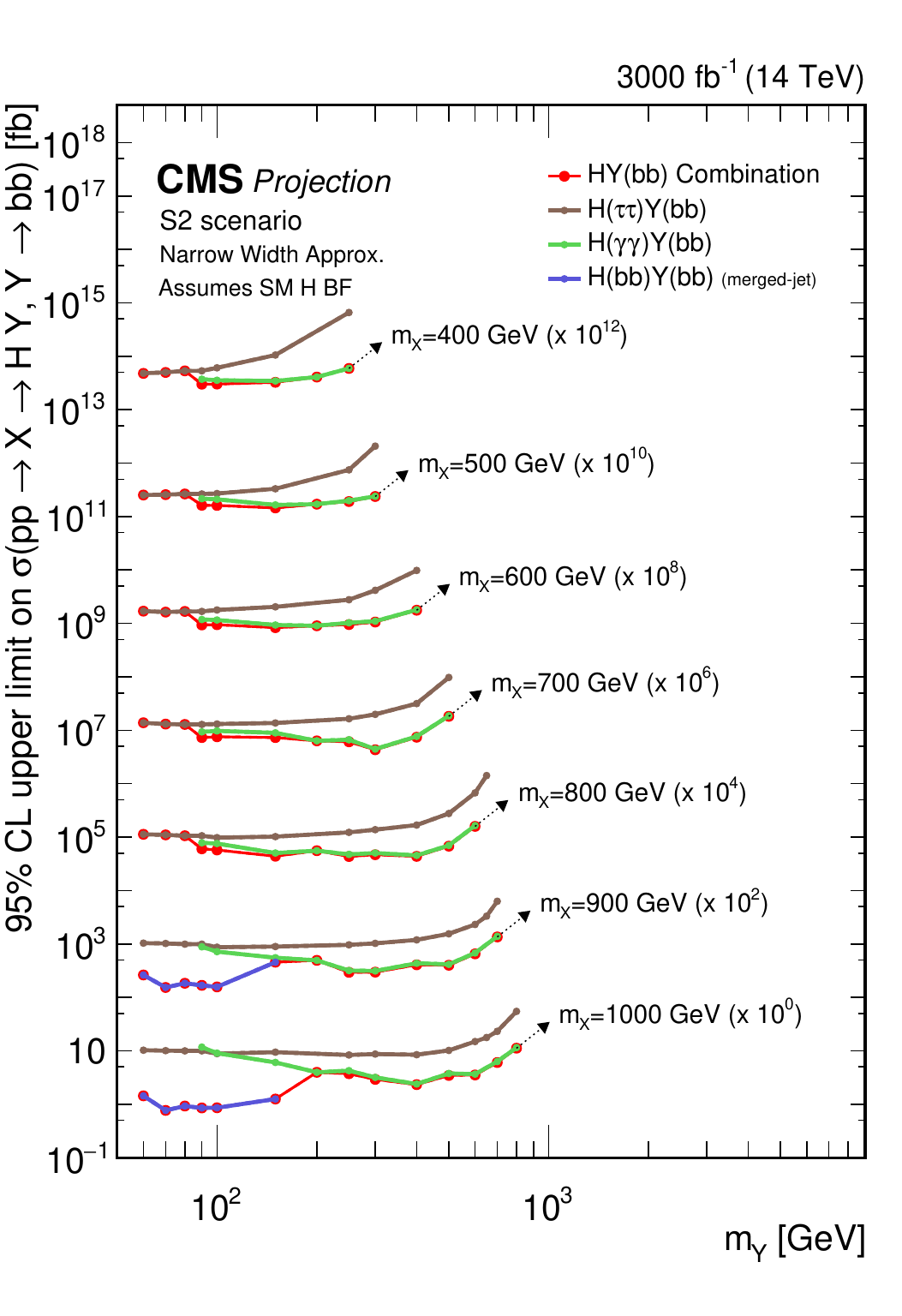}
  \caption{
    Expected upper limits at 95\%~\CL, on the product of the cross section 
    $\sigma$ for the production of a resonance \PX via gluon-gluon fusion and the 
    branching fraction \BR for the $\PX\to\PY(\bb)\PH$ decay, as functions of 
    \MY, for $\MX\leq 1\TeV$.
    For the branching fractions of the $\PH\to\tautau$, $\PH\to\gamma\gamma$
    and  $\PH\to\bb$ decays, the SM values are assumed.
   The limits are obtained from the combined likelihood 
    analysis of all analyses discussed in Section~\ref{Sec:Results_X_to_YH} and 
    shown in Fig.~\ref{fig:XYH_combination}, projected to an integrated 
    luminosity of 3000\fbinv. Shown are the projections for the 
    combined likelihood analysis for different systematic uncertainty scenarios (\cmsLeft), 
    and the projections for the combined likelihood 
    analysis and the individual contributing analyses assuming the S2 scenario (\cmsRight). 
    For presentation purposes, the limits have been scaled in successive 
    steps by two orders of magnitude. For each set of graphs, a black arrow 
    points to the \MX related legend.
  }
  \label{fig:XYHcomb_leq1000_projections}
\end{figure}

\begin{figure}[bp]
  \centering
  \includegraphics[width=\cmsFigWidth]{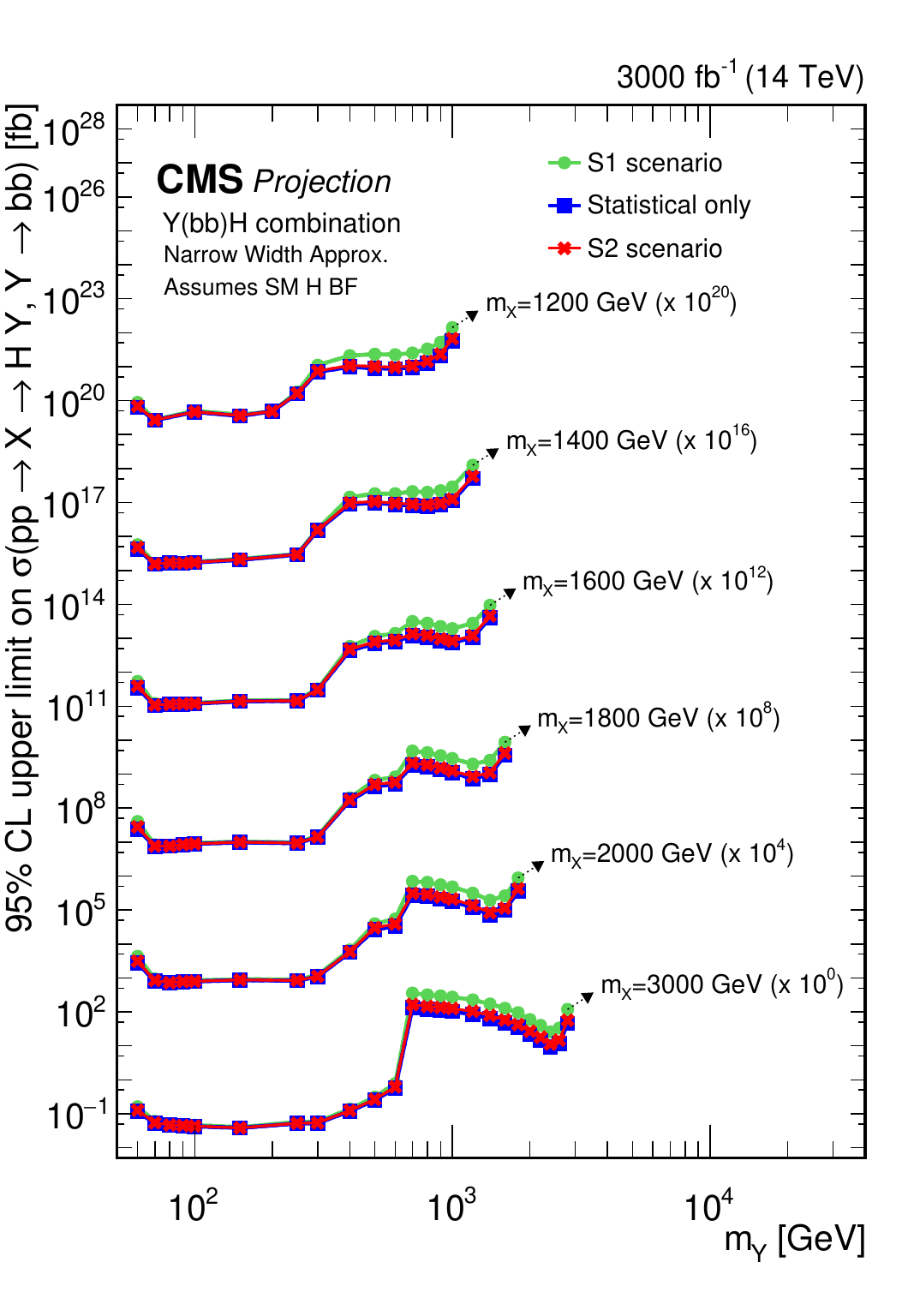}
  \includegraphics[width=\cmsFigWidth]{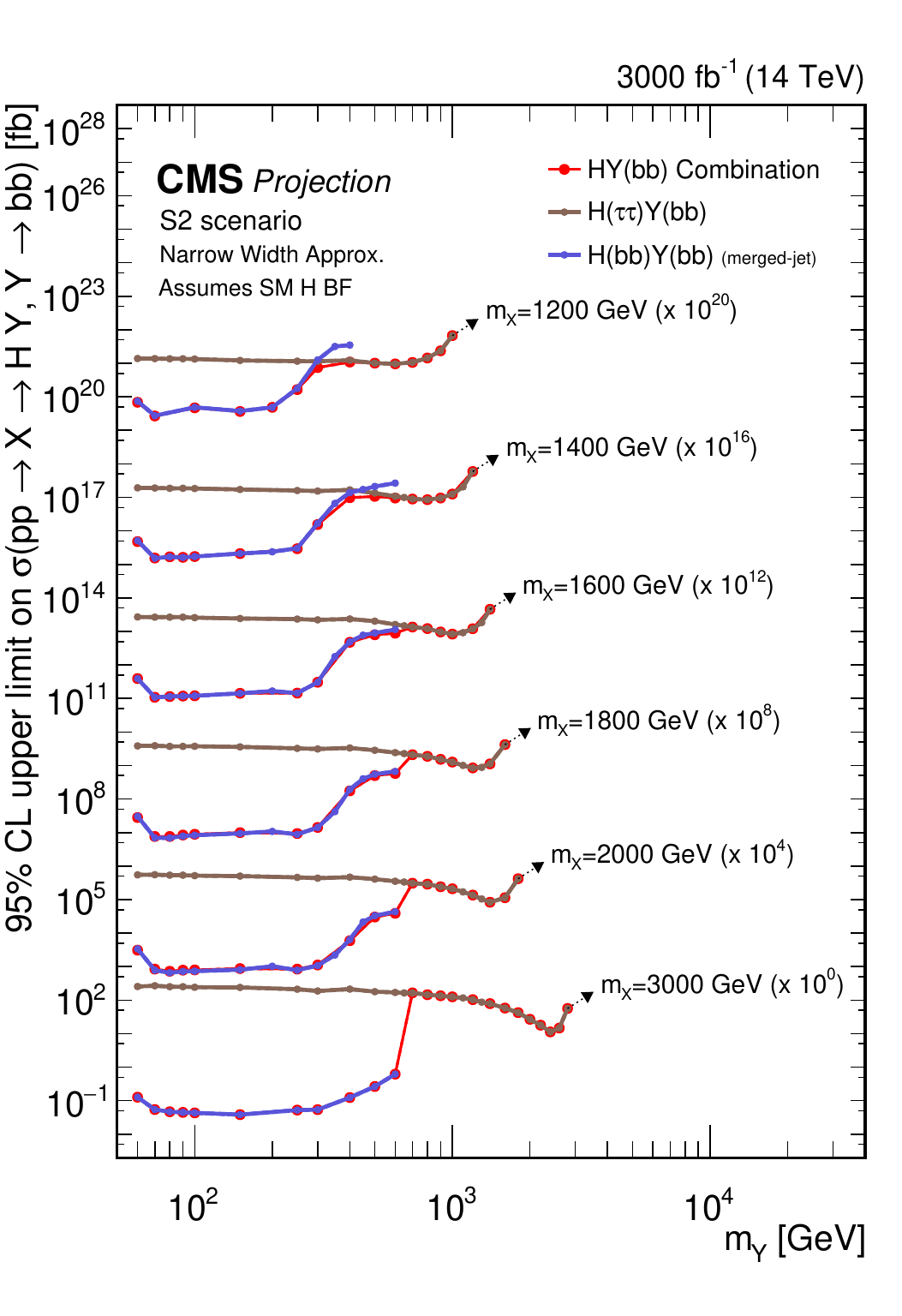}
  \caption{
    Expected upper limits at 95\%~\CL, on the product of the cross section 
    $\sigma$ for the production of a resonance \PX via gluon-gluon fusion and the 
    branching fraction \BR for the $\PX\to\PY(\bb)\PH$ decay, as functions of 
    \MY, for $\MX\geq 1.2\TeV$. 
    For the branching fractions of the $\PH\to\tautau$
    and  $\PH\to\bb$ decays, the SM values are assumed.
    The limits are obtained from the combined likelihood 
    analysis of all analyses discussed in Section~\ref{Sec:Results_X_to_YH} and 
    shown in Fig.~\ref{fig:XYH_combination_2}, projected to an integrated 
    luminosity of 3000\fbinv. Shown are the projections for the 
    combined likelihood analysis for different systematic uncertainty scenarios (\cmsLeft), 
    and the projections for the combined likelihood 
    analysis and the individual contributing analyses assuming the S2 
    scenario (\cmsRight). For presentation purposes, the limits have been scaled in successive 
    steps by four orders of magnitude. For each set of graphs, a black arrow 
    points to the \MX related legend.
  }
  \label{fig:XYHcomb_gt1000_projections}
\end{figure}

The upper limits on the cross section for $\PX \to \YH$ are also projected to an integrated luminosity of
3000\fbinv for the three systematic uncertainty scenarios. 
The projections are derived for the individual channels in the \bbgg, \bbtt, and \bbbb final states, 
and for a combination with the assumption of SM \PH boson branching fractions, 
where we use the same procedure as for the \HH projections. 
The differences between the upper limits in the S1, S2, and statistical-only
scenarios are analogous to the findings for the corresponding channels in the $\PX \to \HH$ projections.

The results of the $\PX \to \YH$ projections are presented in Figs.~\ref{fig:XYHcomb_leq1000_projections} 
and \ref{fig:XYHcomb_gt1000_projections} for \mX up to and above 1\TeV, respectively. 
The regions of the (\MX, \MY) parameter space with the largest ratios of $\MY/\MX$ correspond to a 
\PY particle with low transverse momentum, and can be probed with the \bbgg channel. 
In the regions with small ratios of $\MY/\MX$, the \PY particle receives a large Lorentz boost, 
such that the \bbbb boosted channel has the highest sensitivity and only this final state is considered. 
In the intermediate region, the \bbgg and \bbtt channels provide comparable sensitivity and about 
equal weight in the combination. 

\begin{figure}[tbp]
  \centering
  \includegraphics[width=\cmsFigWidth]{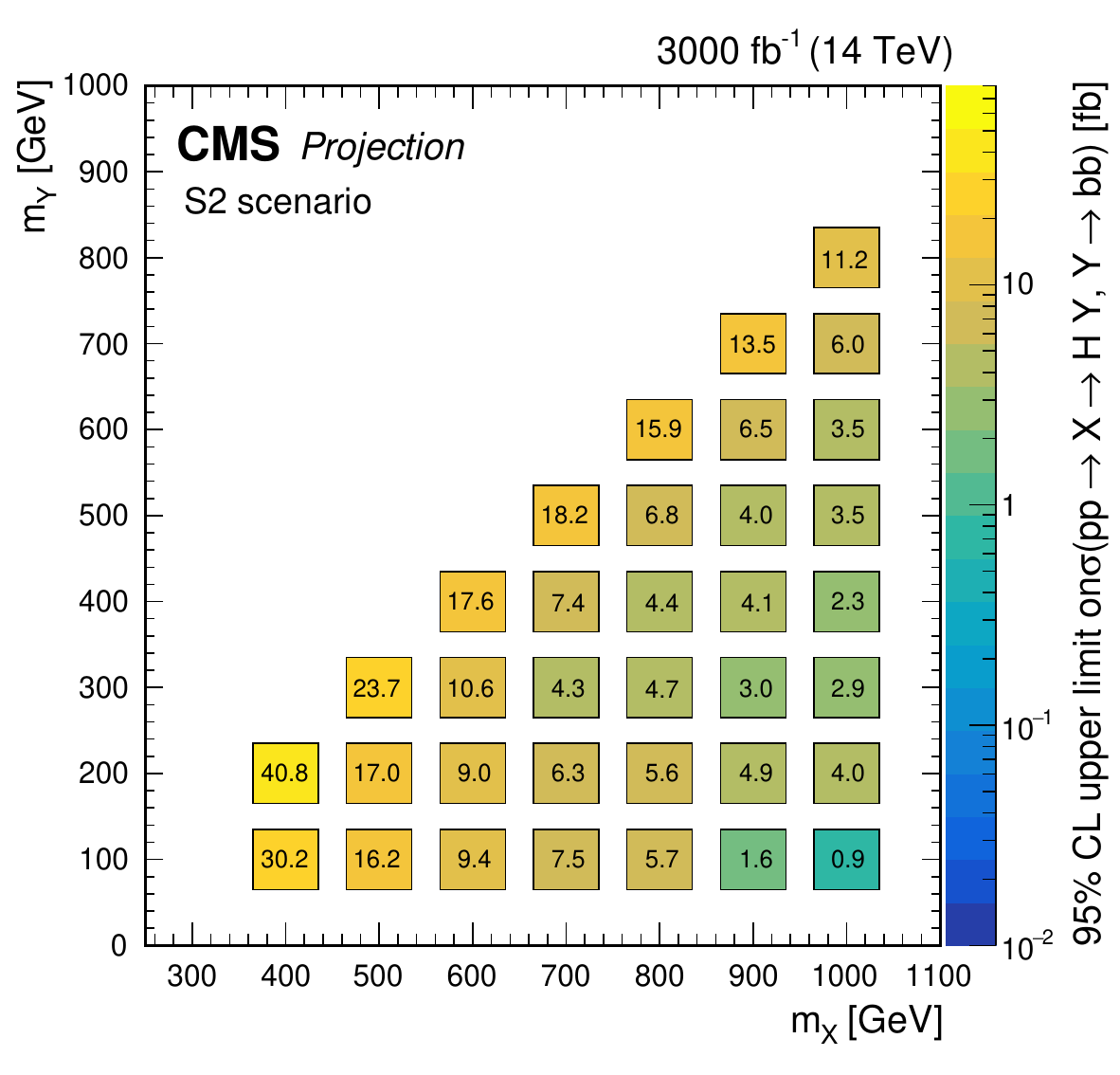}
  \caption{
    Expected upper limits at 95~\%~\CL on the product of the cross section 
    $\sigma$ for the production of a resonance \PX via gluon-gluon fusion and the 
    branching fraction \BR for the $\PX\to\PY(\bb)\PH$ decay, as obtained from 
    the combined likelihood analysis of the individual analyses presented in 
    Section~\ref{Sec:Results_X_to_YH} and Figure~\ref{fig:XYH_combination}. The 
    results are shown in the plane spanned by \mY and \mX for $\mX\le1\TeV$, and projected 
    to an integrated luminosity of 3000\fbinv, assuming the S2 systematic uncertainty scenario. 
    The numbers in the boxes are given in~\unit{fb}.
  }
  \label{fig:XYH_combination_mXmY_projection}
\end{figure}

Selected bins of the projections from the \YH combination are used for presenting expected upper limits as 
functions of \MX and \MY, and are shown in Fig.~\ref{fig:XYH_combination_mXmY_projection}. 
In comparison with Fig.~\ref{fig:XYH_combination_mXmY}, the improvement is clearly visible.

\subsubsection{Perspectives for the NMSSM and TRSM}

\begin{figure}[tbp]
  \centering
  \includegraphics[width=\cmsFigWidth]{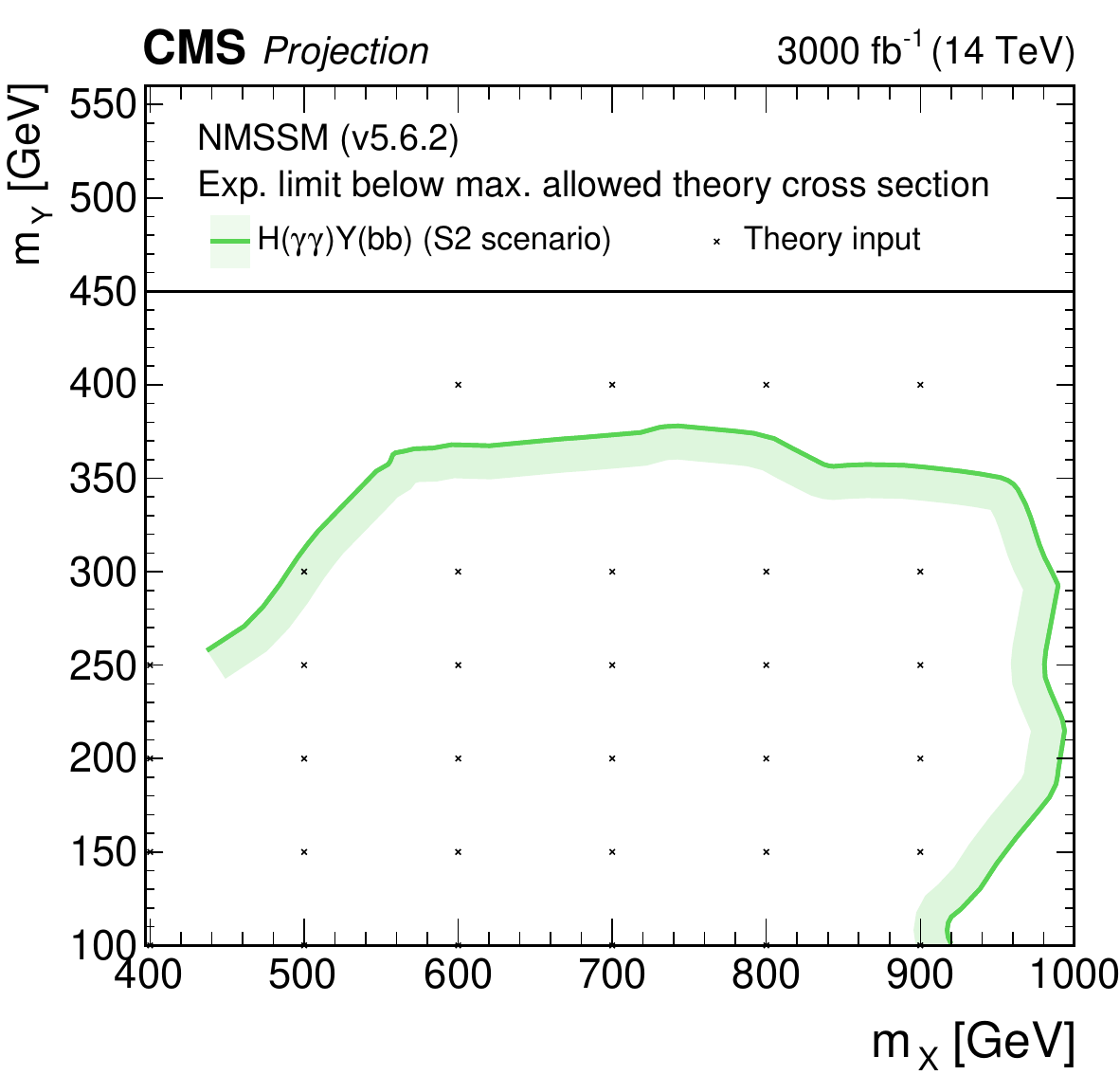}
  \includegraphics[width=\cmsFigWidth]{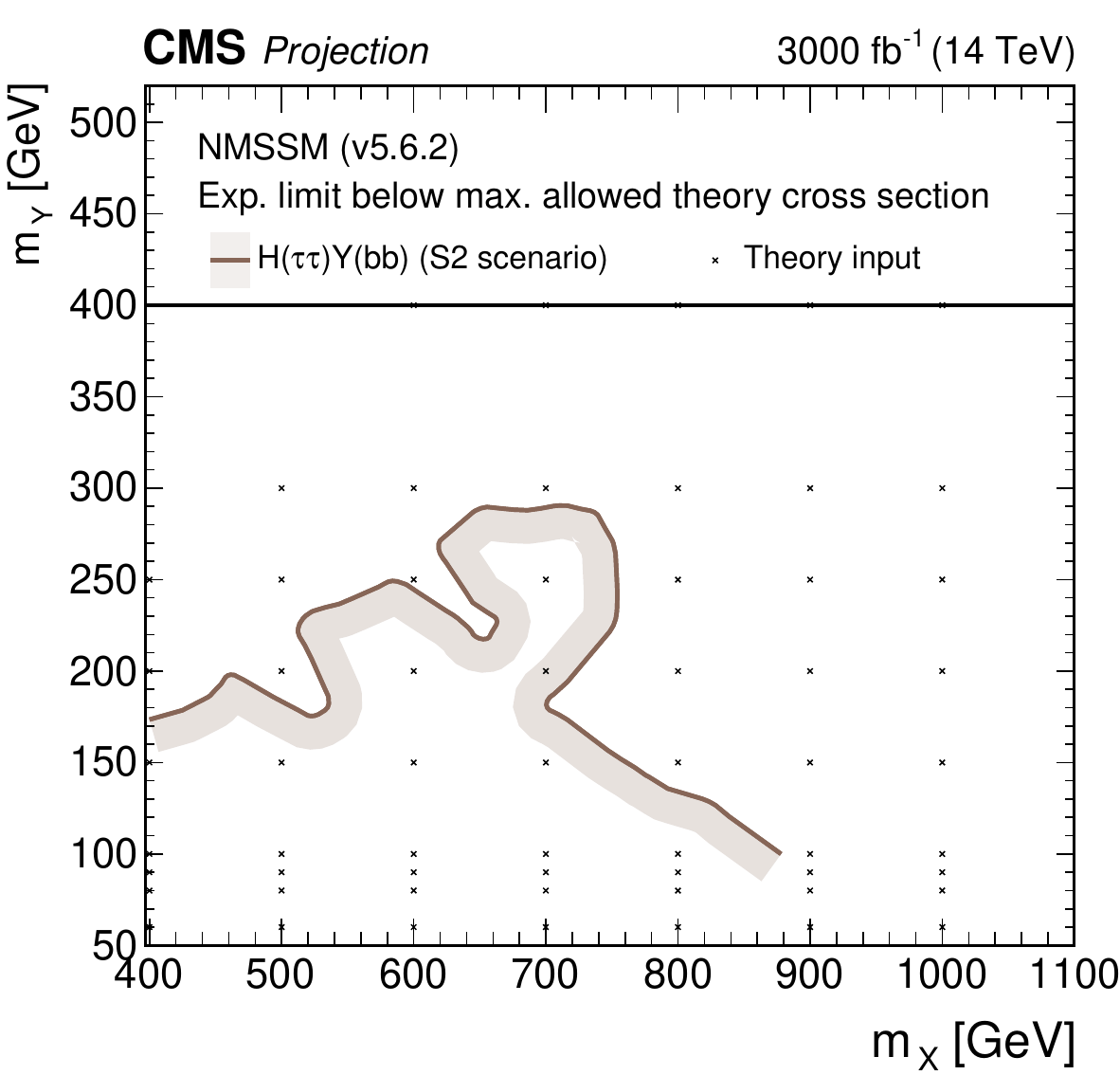}\\
  \includegraphics[width=\cmsFigWidth]{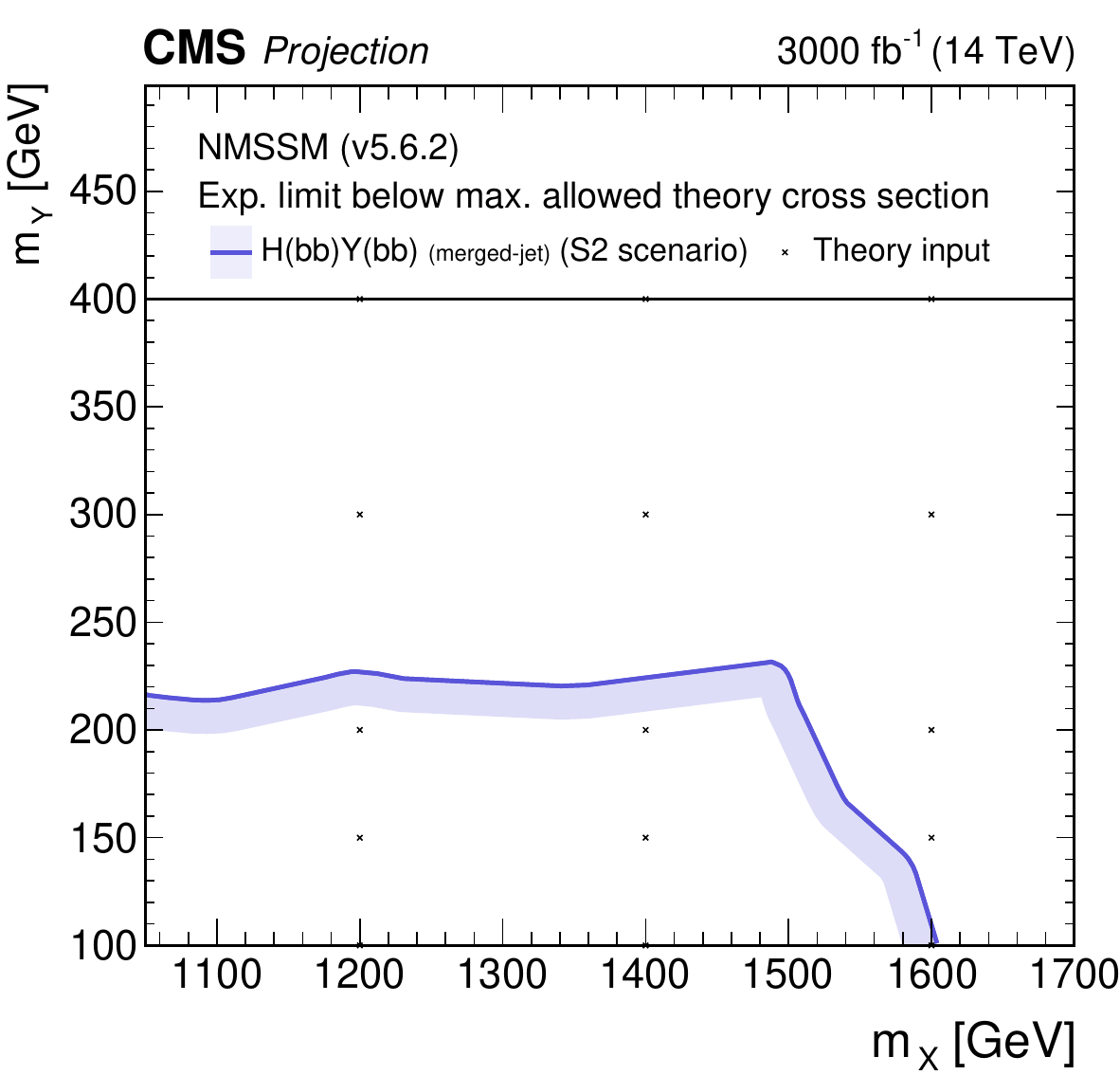}
  \includegraphics[width=\cmsFigWidth]{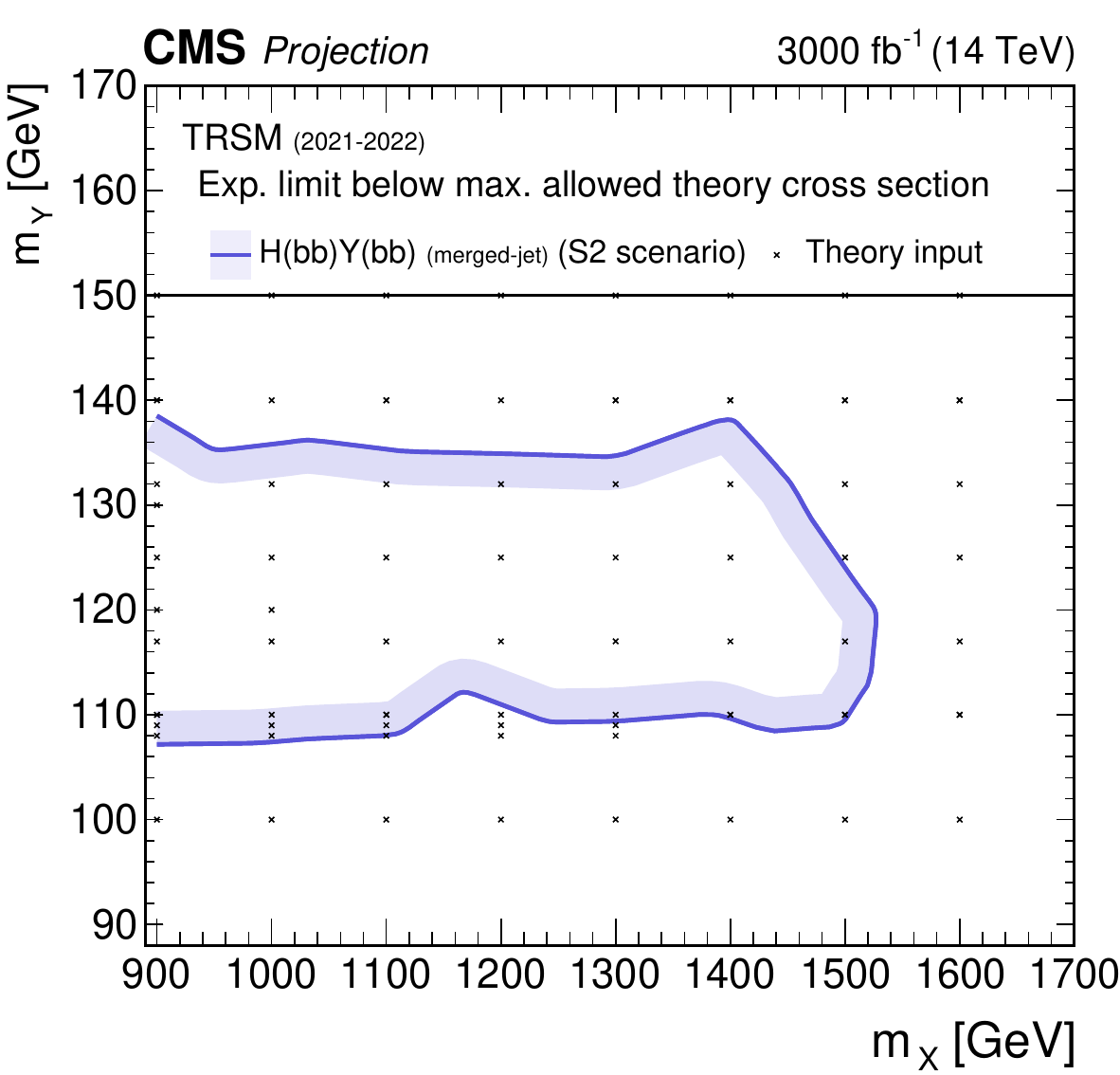}
  \caption{
    Interpretation of the upper limits at 95\%~\CL, on the product of the cross 
    section $\sigma$ for the production of a resonance \PX via gluon-gluon fusion 
    and the branching fraction \BR for the $\PX\to\PY(\bb)\PH$ decay, obtained 
    from the projections to an integrated luminosity of 
    3000\fbinv of the (upper \cmsLeft) $\PY(\bb)\PH(\gamma\gamma)$~\cite{CMS:2023boe}, (upper 
    \cmsRight) $\PY(\bb)\PH(\tautau)$~\cite{CMS:2021yci}, and (lower row) $\PY(\bb)\PH(\bb)$~\cite{CMS:2022suh} analyses, 
    assuming the S2 systematic uncertainty scenario. 
    The projected limits are mapped onto the (\MX, \MY) plane, and compared with the 
    maximally allowed cross sections of the NMSSM (\cmsLeft and upper \cmsRight),  
    and TRSM models (lower \cmsRight) discussed in 
    Section~\ref{Subsubsec:Interpretations_NMSSM_TRSM}. The points
    indicate the available theory predictions.
    The mass dependences of both the projected experimental limits 
    and the maximally allowed theory cross sections have been interpolated to obtain approximate
    exclusion contours.
    The NMSSM predictions based on \textsc{NMSSMTools} version 5.6.2
    are adapted from Ref.~\cite{Ellwanger:2022jtd}, whereas the TRSM 
    is described in Ref.~\cite{Robens:2019kga}.
    In both cases, the model predictions have been scaled to $\sqrt{s}=14\TeV$. 
  }
  \label{fig:YH_projections_NMSSM}
\end{figure}
We compare the maximally allowed cross sections predicted by the NMSSM model scans 
with the expected upper limits at 95\%~\CL on the $\PX \to \YH$ cross sections, projected to
an integrated luminosity of 3000\fbinv. The NMSSM model scans are obtained with \textsc{NMSSMTools} 
version 5.6.2~\cite{Ellwanger:2022jtd}, as described in Section~\ref{Subsubsec:Interpretations_NMSSM_TRSM}, and take 
many relevant experimental constraints from Run~2 into account. 
Figure~\ref{fig:YH_projections_NMSSM} shows the projected exclusion contours for the final states 
$\PY(\bb)\PH(\GamGam)$ (upper \cmsLeft), $\PY(\bb)\PH(\tautau)$ (upper \cmsRight), and $\PY(\bb)\PH(\bb)$ 
in the merged-jet topology (lower \cmsLeft). The maximized model
$\sigma \BR$ values may depend non-monotonically on \MY which can be
reflected in the contours. Substantial regions of the parameter space can be excluded 
in the probed mass region with $\MX=500$--1000\GeV and $\MY=100$--350\GeV, as well as up to $\MY=200\GeV$ 
for $\MX=1100$--1500\GeV. This indicates a huge increase in sensitivity for the HL-LHC compared to the results from Run~2. 
Similarly, we compare the predictions of the TRSM model~\cite{Robens:2019kga} with the projected results from the 
$\PY(\bb)\PH(\bb)$ channel in the merged-jet topology in Fig.~\ref{fig:YH_projections_NMSSM}~(lower \cmsRight), 
which results in a sizable exclusion region for $\MX=900$--1500\GeV and $\MY=110$--135\GeV.

\clearpage

\section{Summary} \label{Sec:Summary}

The analyses searching for the production of the Higgs (\PH) boson  
through decays of heavy resonances, performed by the CMS Collaboration 
using the Run~2 data set, are reviewed. This Report covers final states
with two bosons with at least one an \PH boson, namely
an \PH boson and a vector boson (\VH), a pair of \PH bosons (\HH), and an \PH boson
joined by a new boson \PY (\YH), where V represents a \PW or a \PZ boson.

The analyses cover a wide range of \PH boson decay modes,
in particular, decays into photons, \PQb quarks, \PGt leptons, and \PW
bosons. The \PY boson is exclusively searched for in \PQb quark final
states. Topologies involving both resolved and merged jet objects are
used to cover a wide range of the phase space. Multivariate methods are
employed in various ways to improve the performance.

The results are presented as summary plots which show the
sensitivity of all channels in direct comparison. For the \HH and \YH
final states, the results obtained by combining all decay channels 
are presented for the first time.

The results are interpreted in the context of various beyond-the-standard model 
scenarios for resonances decaying into \VH, \HH and \YH final states. 
These include various extended Higgs sector
models, warped extra-dimension models, and heavy vector triplet
models. The results from resonant \PH boson production searches are
compared with results from searches in other channels.

While all presented analyses assume the validity of the
narrow-width approximation, a dedicated study of the impact of finite
width and interference is performed for the first time in CMS 
for the real singlet extension of the standard model. This study shows the modification of
the \HH cross section and line shape in regions of the parameter space 
where the narrow-width approximation is not valid anymore. 

The expected sensitivity of the analyses in the \HH and \YH final states
is estimated for future data sets with integrated luminosities of 300,
1000, and 3000\fbinv, the last number corresponding to the baseline
scenario of the High-Luminosity LHC (HL-LHC) over its full lifetime. The expected upper
limits for resonant \HH production for the HL-LHC scenario
range from about 50\unit{fb} at a resonance mass of 300\GeV to nearly 
0.01\unit{fb} for masses of 3\TeV and above. The exclusions in terms of \tanb
in the hMSSM and \MhEFTScen scenarios are expanded by almost a
factor of two compared to the Run~2 data set.

This review shows how the specific strengths of many different
experimental signatures can be combined to chart very thoroughly the
territory where resonant Higgs boson production might reveal beyond the standard model physics, 
and gives a promising outlook towards the achievement
potential of future measurements in this sector.

\begin{acknowledgments}
\hyphenation{Bundes-ministerium Forschungs-gemeinschaft Forschungs-zentren Rachada-pisek} We congratulate our colleagues in the CERN accelerator departments for the excellent performance of the LHC and thank the technical and administrative staffs at CERN and at other CMS institutes for their contributions to the success of the CMS effort. In addition, we gratefully acknowledge the computing centers and personnel of the Worldwide LHC Computing Grid and other centers for delivering so effectively the computing infrastructure essential to our analyses. Finally, we acknowledge the enduring support for the construction and operation of the LHC, the CMS detector, and the supporting computing infrastructure provided by the following funding agencies: the Armenian Science Committee, project no. 22rl-037; the Austrian Federal Ministry of Education, Science and Research and the Austrian Science Fund; the Belgian Fonds de la Recherche Scientifique, and Fonds voor Wetenschappelijk Onderzoek; the Brazilian Funding Agencies (CNPq, CAPES, FAPERJ, FAPERGS, and FAPESP); the Bulgarian Ministry of Education and Science, and the Bulgarian National Science Fund; CERN; the Chinese Academy of Sciences, Ministry of Science and Technology, the National Natural Science Foundation of China, and Fundamental Research Funds for the Central Universities; the Ministerio de Ciencia Tecnolog\'ia e Innovaci\'on (MINCIENCIAS), Colombia; the Croatian Ministry of Science, Education and Sport, and the Croatian Science Foundation; the Research and Innovation Foundation, Cyprus; the Secretariat for Higher Education, Science, Technology and Innovation, Ecuador; the Estonian Research Council via PRG780, PRG803, RVTT3 and the Ministry of Education and Research TK202; the Academy of Finland, Finnish Ministry of Education and Culture, and Helsinki Institute of Physics; the Institut National de Physique Nucl\'eaire et de Physique des Particules~/~CNRS, and Commissariat \`a l'\'Energie Atomique et aux \'Energies Alternatives~/~CEA, France; the Shota Rustaveli National Science Foundation, Georgia; the Bundesministerium f\"ur Bildung und Forschung, the Deutsche Forschungsgemeinschaft (DFG), under Germany's Excellence Strategy -- EXC 2121 ``Quantum Universe" -- 390833306, and under project number 400140256 - GRK2497, and Helmholtz-Gemeinschaft Deutscher Forschungszentren, Germany; the General Secretariat for Research and Innovation and the Hellenic Foundation for Research and Innovation (HFRI), Project Number 2288, Greece; the National Research, Development and Innovation Office (NKFIH), Hungary; the Department of Atomic Energy and the Department of Science and Technology, India; the Institute for Studies in Theoretical Physics and Mathematics, Iran; the Science Foundation, Ireland; the Istituto Nazionale di Fisica Nucleare, Italy; the Ministry of Science, ICT and Future Planning, and National Research Foundation (NRF), Republic of Korea; the Ministry of Education and Science of the Republic of Latvia; the Research Council of Lithuania, agreement No.\ VS-19 (LMTLT); the Ministry of Education, and University of Malaya (Malaysia); the Ministry of Science of Montenegro; the Mexican Funding Agencies (BUAP, CINVESTAV, CONACYT, LNS, SEP, and UASLP-FAI); the Ministry of Business, Innovation and Employment, New Zealand; the Pakistan Atomic Energy Commission; the Ministry of Education and Science and the National Science Centre, Poland; the Funda\c{c}\~ao para a Ci\^encia e a Tecnologia, grants CERN/FIS-PAR/0025/2019 and CERN/FIS-INS/0032/2019, Portugal; the Ministry of Education, Science and Technological Development of Serbia; MCIN/AEI/10.13039/501100011033, ERDF ``a way of making Europe", Programa Estatal de Fomento de la Investigaci{\'o}n Cient{\'i}fica y T{\'e}cnica de Excelencia Mar\'{\i}a de Maeztu, grant MDM-2017-0765, projects PID2020-113705RB, PID2020-113304RB, PID2020-116262RB and PID2020-113341RB-I00, and Plan de Ciencia, Tecnolog{\'i}a e Innovaci{\'o}n de Asturias, Spain; the Ministry of Science, Technology and Research, Sri Lanka; the Swiss Funding Agencies (ETH Board, ETH Zurich, PSI, SNF, UniZH, Canton Zurich, and SER); the Ministry of Science and Technology, Taipei; the Ministry of Higher Education, Science, Research and Innovation, and the National Science and Technology Development Agency of Thailand; the Scientific and Technical Research Council of Turkey, and Turkish Energy, Nuclear and Mineral Research Agency; the National Academy of Sciences of Ukraine; the Science and Technology Facilities Council, UK; the US Department of Energy, and the US National Science Foundation.

Individuals have received support from the Marie-Curie programme and the European Research Council and Horizon 2020 Grant, contract Nos.\ 675440, 724704, 752730, 758316, 765710, 824093, 101115353,101002207, and COST Action CA16108 (European Union) the Leventis Foundation; the Alfred P.\ Sloan Foundation; the Alexander von Humboldt Foundation; the Belgian Federal Science Policy Office; the Fonds pour la Formation \`a la Recherche dans l'Industrie et dans l'Agriculture (FRIA-Belgium); the Agentschap voor Innovatie door Wetenschap en Technologie (IWT-Belgium); the F.R.S.-FNRS and FWO (Belgium) under the ``Excellence of Science -- EOS" -- be.h project n.\ 30820817; the Beijing Municipal Science \& Technology Commission, No. Z191100007219010 and USTC Research Funds of the Double First-Class Initiative No. YD2030002017 (China); the Ministry of Education, Youth and Sports (MEYS) of the Czech Republic; the Shota Rustaveli National Science Foundation, grant FR-22-985 (Georgia); the Hungarian Academy of Sciences, the New National Excellence Program - \'UNKP, the NKFIH research grants K 131991, K 133046, K 138136, K 143460, K 143477, K 146913, K 146914, K 147048, 2020-2.2.1-ED-2021-00181, and TKP2021-NKTA-64 (Hungary); the Council of Scientific and Industrial Research, India; ICSC -- National Research Centre for High Performance Computing, Big Data and Quantum Computing, funded by the EU NexGeneration program, Italy; the Latvian Council of Science; the Ministry of Education and Science, project no. 2022/WK/14, and the National Science Center, contracts Opus 2021/41/B/ST2/01369 and 2021/43/B/ST2/01552 (Poland); the Funda\c{c}\~ao para a Ci\^encia e a Tecnologia, grant FCT CEECIND/01334/2018; the National Priorities Research Program by Qatar National Research Fund; the Programa Estatal de Fomento de la Investigaci{\'o}n Cient{\'i}fica y T{\'e}cnica de Excelencia Mar\'{\i}a de Maeztu, grant MDM-2017-0765 and projects PID2020-113705RB, PID2020-113304RB, PID2020-116262RB and PID2020-113341RB-I00, and Programa Severo Ochoa del Principado de Asturias (Spain); the Chulalongkorn Academic into Its 2nd Century Project Advancement Project, and the National Science, Research and Innovation Fund via the Program Management Unit for Human Resources \& Institutional Development, Research and Innovation, grant B37G660013 (Thailand); the Kavli Foundation; the Nvidia Corporation; the SuperMicro Corporation; the Welch Foundation, contract C-1845; and the Weston Havens Foundation (USA).  
\end{acknowledgments}

\bibliography{auto_generated}

\clearpage

\appendix

\section*{Glossary}
\begin{tabular}{>{\bfseries}ll}
  A, a & \CP-odd Higgs bosons in extended Higgs sector models\\
  ATLAS & A Toroidal LHC Apparatus\\
  BDT & Boosted decision tree\\
  BSM & Beyond the standard model\\
  CL & Confidence level\\ 
  CMS & Compact Muon Solenoid \\
  \CP & Charge-parity (symmetry)\\
  CR & Control region\\
  DDT & Designing decorrelated taggers (procedure)\\
  DNN & Deep neural network\\
  DT  & Deep tau (identification algorithm)\\
  DY & Drell--Yan (process)\\
  ECAL & Electromagnetic calorimeter\\
  EFT & Effective field theory\\
  EW & Electroweak \\ 
  G   & Graviton\\
  ggF & Gluon-gluon fusion (production process)\\
  HCAL & Hadron calorimeter\\
  HH  & Higgs boson pair\\
  HL-LHC & High-Luminosity Large Hadron Collider \\
  hMSSM & Habeat MSSM (scenario)\\
  HVT & Heavy vector triplet (model)\\
  KK & Kaluza-Klein (graviton)\\
  LHC & Large Hadron Collider \\
  LO & Leading order\\
  MC & Monte Carlo (simulation)\\ 
  MSSM & Minimal supersymmetric standard model\\
  MVA & Multi-variate analysis\\
  NMSSM & Next-to-minimal supersymmetric standard model\\
  NLO & Next-to-leading order\\
  NN   & Neural network\\
  NWA & Narrow-width approximation\\
  N2HDM & Next-to-minimal 2HDM\\
  PF  & Particle flow (method of reconstructing particle candidates)\\
  pp & Proton-proton\\ 
  PU & Pileup\\
  PUPPI & Pileup-per-particle identification (algorithm)\\
  QCD & Quantum chromo-dynamics\\
  R   & Radion (graviscalar in the RS model), also distance in ($\Delta \eta, \Delta \phi$) space\\
  RS   & Randall--Sundrum (model)\\
  Run 2 & The second run of the LHC, during the years 2015--2018\\ 
  SD & Soft-drop (algorithm)\\
  SM & Standard model \\
  SR & Signal region\\ 
  SUSY & Supersymmetry\\
  $\mathbf{\sqrt{s}}$ & The center-of-mass energy\\ 
  TRSM & Two-real-singlet Model\\
\end{tabular}
\newpage
\begin{tabular}{>{\bfseries}ll}
  UFO & Universal FeynRules output\\
  V   & Vector boson (\PW or \PZ) \\
  VBF & Vector boson fusion  (production process)\\
  VH  & Vector plus Higgs boson (production process or decay channel)\\
  WED & Warped extra dimensions (model) \\
  2HDM & Two-Higgs-doublet Model\\
  2HDM+S & Two-Higgs-doublet-plus-additional-singlet model\\
\end{tabular}

\cleardoublepage \section{The CMS Collaboration \label{app:collab}}\begin{sloppypar}\hyphenpenalty=5000\widowpenalty=500\clubpenalty=5000
\cmsinstitute{Yerevan Physics Institute, Yerevan, Armenia}
{\tolerance=6000
A.~Hayrapetyan, A.~Tumasyan\cmsAuthorMark{1}\cmsorcid{0009-0000-0684-6742}
\par}
\cmsinstitute{Institut f\"{u}r Hochenergiephysik, Vienna, Austria}
{\tolerance=6000
W.~Adam\cmsorcid{0000-0001-9099-4341}, J.W.~Andrejkovic, T.~Bergauer\cmsorcid{0000-0002-5786-0293}, S.~Chatterjee\cmsorcid{0000-0003-2660-0349}, K.~Damanakis\cmsorcid{0000-0001-5389-2872}, M.~Dragicevic\cmsorcid{0000-0003-1967-6783}, P.S.~Hussain\cmsorcid{0000-0002-4825-5278}, M.~Jeitler\cmsAuthorMark{2}\cmsorcid{0000-0002-5141-9560}, N.~Krammer\cmsorcid{0000-0002-0548-0985}, A.~Li\cmsorcid{0000-0002-4547-116X}, D.~Liko\cmsorcid{0000-0002-3380-473X}, I.~Mikulec\cmsorcid{0000-0003-0385-2746}, J.~Schieck\cmsAuthorMark{2}\cmsorcid{0000-0002-1058-8093}, R.~Sch\"{o}fbeck\cmsorcid{0000-0002-2332-8784}, D.~Schwarz\cmsorcid{0000-0002-3821-7331}, M.~Sonawane\cmsorcid{0000-0003-0510-7010}, S.~Templ\cmsorcid{0000-0003-3137-5692}, W.~Waltenberger\cmsorcid{0000-0002-6215-7228}, C.-E.~Wulz\cmsAuthorMark{2}\cmsorcid{0000-0001-9226-5812}
% ...
\par}
\vskip\cmsinstskip
\dag:~Deceased\\
$^{1}$Also at Yerevan State University, Yerevan, Armenia\\
$^{2}$Also at TU Wien, Vienna, Austria\\
% ...
\end{sloppypar}
\end{document}